\documentclass[a4paper,onecolumn,11pt,accepted=2024-08-02]{quantumarticle}
\pdfoutput=1
\PassOptionsToPackage{usenames,dvipsnames}{xcolor}
\usepackage[utf8]{inputenc}
\usepackage[english]{babel}
\usepackage[T1]{fontenc}
\usepackage{amsmath}
\usepackage{hyperref}

\usepackage[numbers,sort&compress]{natbib}

\usepackage{tikz}
\usepackage{lipsum}

\usepackage{ragged2e}
\usepackage{array}
\usepackage{comment}

\usepackage{etoolbox}
\usepackage{bm,bbm}
\usepackage{physics}
\usepackage{amssymb}
\usepackage{amsmath}

\usepackage{multirow}
\usepackage{mathtools}
\usepackage{textgreek}
\usetikzlibrary{cd}
\usepackage[normalem]{ulem}

\usepackage{manfnt}
\usepackage{enumerate}
\usepackage[export]{adjustbox}
\usepackage{wasysym}
\usepackage{algorithm}
\usepackage{algpseudocode}

\definecolor{redgray}{rgb}{0.8, 0.5, 0.5}
\definecolor{bluegray}{rgb}{0.5, 0.5, 0.8}
\definecolor{greengray}{rgb}{0.5 0.63, 0.5}
\definecolor{amber}{rgb}{0.8, 0.6, 0.0}

\newcommand{\Er}{{{\rc{E_{r}}}}}
\newcommand{\Eg}{\gc{E_{g}} }
\newcommand{\Eb}{\bc{E_{b}} }
\newcommand{\Ey}{\yc{E_{y}} }

\newcommand{\rc}[1]{#1}
\newcommand{\bc}[1]{#1}
\newcommand{\gc}[1]{#1}
\newcommand{\yc}[1]{#1}
\newcommand{\rcg}[1]{{\color{gray}{#1}}}
\newcommand{\bcg}[1]{{\color{gray}{#1}}}
\newcommand{\gcg}[1]{{\color{gray}{#1}}}

\usepackage{soul}

\def\lp{\left (}
\def\rp{\right )}

\usepackage{dsfont}
\newcommand{\ttt}[1]{\texttt{#1}}
\newcommand{\id}{\mathds{1}}

\newcommand{\vol}{K}
\newcommand{\volr}{K_{\rc{r}}}
\newcommand{\volg}{K_{\gc{g}}}
\newcommand{\volb}{K_{\bc{b}}}
\newcommand{\voly}{K_{\yc{y}}}

\newcommand{\abelian}{G}

\newcommand{\be}{\begin{equation}}
\newcommand{\ee}{\end{equation}}

\def\gt{{\color{gray}\times}}
\def\tsig{\texttt{\textsigma}}

\def\rx{\rc{\texttt{rx}}}
\def\ry{\rc{\texttt{ry}}}
\def\rz{\rc{\texttt{rz}}}
\def\gx{\gc{\texttt{gx}}}
\def\gy{\gc{\texttt{gy}}}
\def\gz{\gc{\texttt{gz}}}
\def\bx{\bc{\texttt{bx}}}
\def\by{\bc{\texttt{by}}}
\def\bz{\bc{\texttt{bz}}}
\def\yz{\yc{\texttt{yz}}}

\def\fr{\rc{\texttt{f}_{\texttt{r}}}}
\def\fg{\gc{\texttt{f}_{\texttt{g}}}}
\def\fb{\bc{\texttt{f}_{\texttt{b}}}}
\def\frbar{\overline{\rc{\texttt{f}_{\texttt{r}}}}}
\def\fgbar{\overline{\gc{\texttt{f}_{\texttt{g}}}}}
\def\fbbar{\overline{\bc{\texttt{f}_{\texttt{b}}}}}

 \def\rgz{\rc{\ttt{r}}\gc{\ttt{g}}\ttt{z}}
 \def\gbz{\gc{\ttt{g}}\bc{\ttt{b}}\ttt{z}}
\def\rbz{\rc{\ttt{r}}\bc{\ttt{b}}\ttt{z}}

\def\rgx{\rc{\ttt{r}}\gc{\ttt{g}}\ttt{x}}
\def\rbx{\rc{\ttt{r}}\bc{\ttt{b}}\ttt{x}}
\def\ryx{\rc{\ttt{r}}\yc{\ttt{y}}\ttt{x}}
\def\gbx{\gc{\ttt{g}}\bc{\ttt{b}}\ttt{x}}
\def\gyx{\gc{\ttt{g}}\yc{\ttt{y}}\ttt{x}}
\def\byx{\bc{\ttt{b}}\yc{\ttt{y}}\ttt{x}}
\def\rgbyx{\rc{\ttt{r}}\gc{\ttt{g}}\bc{\ttt{b}}\yc{\ttt{y}}\ttt{x}}

\def\rxi{\rc{\texttt{rx}_\rc{\texttt{1}}}}
\def\ryi{\rc{\texttt{ry}_\rc{\texttt{1}}}}
\def\rzi{\rc{\texttt{rz}_\rc{\texttt{1}}}}
\def\gxi{\gc{\texttt{gx}_\gc{\texttt{1}}}}
\def\gyi{\gc{\texttt{gy}_\gc{\texttt{1}}}}
\def\gzi{\gc{\texttt{gz}_\gc{\texttt{1}}}}
\def\bxi{\bc{\texttt{bx}_\bc{\texttt{1}}}}
\def\byi{\bc{\texttt{by}_\bc{\texttt{1}}}}
\def\bzi{\bc{\texttt{bz}_\bc{\texttt{1}}}}
\def\rxii{\rc{\texttt{rx}_\rc{\texttt{2}}}}
\def\ryii{\rc{\texttt{ry}_\rc{\texttt{2}}}}
\def\rzii{\rc{\texttt{rz}_\rc{\texttt{2}}}}
\def\gxii{\gc{\texttt{gx}_\gc{\texttt{2}}}}
\def\gyii{\gc{\texttt{gy}_\gc{\texttt{2}}}}
\def\gzii{\gc{\texttt{gz}_\gc{\texttt{2}}}}
\def\bxii{\bc{\texttt{bx}_\bc{\texttt{2}}}}
\def\byii{\bc{\texttt{by}_\bc{\texttt{2}}}}
\def\bzii{\bc{\texttt{bz}_\bc{\texttt{2}}}}
\def\rxiii{\rc{\texttt{rx}_\rc{\texttt{3}}}}
\def\ryiii{\rc{\texttt{ry}_\rc{\texttt{3}}}}
\def\rziii{\rc{\texttt{rz}_\rc{\texttt{3}}}}
\def\gxiii{\gc{\texttt{gx}_\gc{\texttt{3}}}}
\def\gyiii{\gc{\texttt{gy}_\gc{\texttt{3}}}}
\def\gziii{\gc{\texttt{gz}_\gc{\texttt{3}}}}
\def\bxiii{\bc{\texttt{bx}_\bc{\texttt{3}}}}

\def\bziii{\bc{\texttt{bz}_\bc{\texttt{3}}}}
\def\rxiv{\rc{\texttt{rx}_\rc{\texttt{4}}}}

\def\rziv{\rc{\texttt{rz}_\rc{\texttt{4}}}}
\def\gxiv{\gc{\texttt{gx}_\gc{\texttt{4}}}}

\def\bxiv{\bc{\texttt{bx}_\bc{\texttt{4}}}}
\def\byiv{\bc{\texttt{by}_\bc{\texttt{4}}}}
\def\bziv{\bc{\texttt{bz}_\bc{\texttt{4}}}}

\def\yziii{\yc{\texttt{yz}_\yc{\texttt{3}}}}
\def\yzii{\yc{\texttt{yz}_\yc{\texttt{2}}}}
\def\yzi{\yc{\texttt{yz}_\yc{\texttt{1}}}}

\def\rxig {\rcg{\texttt{rx}_\rcg{\texttt{1}}}}

\def\gzig {\gcg{\texttt{gz}_\gcg{\texttt{1}}}}
\def\bxig {\bcg{\texttt{bx}_\bcg{\texttt{1}}}}

\def\bzig {\bcg{\texttt{bz}_\bcg{\texttt{1}}}}
\def\rxiig {\rcg{\texttt{rx}_\rcg{\texttt{2}}}}

\def\rziig {\rcg{\texttt{rz}_\rcg{\texttt{2}}}}
\def\gxiig {\gcg{\texttt{gx}_\gcg{\texttt{2}}}}

\def\gziig {\gcg{\texttt{gz}_\gcg{\texttt{2}}}}
\def\bxiig {\bcg{\texttt{bx}_\bcg{\texttt{2}}}}

\def\bziig {\bcg{\texttt{bz}_\bcg{\texttt{2}}}}

\def\gxiiig{\gcg{\texttt{gx}_\gcg{\texttt{3}}}}

\def\gziiig{\gcg{\texttt{gz}_\gcg{\texttt{3}}}}
\def\bxiiig{\bcg{\texttt{bx}_\bcg{\texttt{3}}}}

\def\bziiig{\bcg{\texttt{bz}_\bcg{\texttt{3}}}}
\def\rxivg {\rcg{\texttt{rx}_\rcg{\texttt{4}}}}

\def\gzivg {\gcg{\texttt{gz}_\gcg{\texttt{4}}}}
\def\bxivg {\bcg{\texttt{bx}_\bcg{\texttt{4}}}}

\def\bzivg {\bcg{\texttt{bz}_\bcg{\texttt{4}}}}

\def\im{{\texttt{1m}}}
\def\mi{{\texttt{m1}}}
\def\ei{{\texttt{e1}}}
\def\ie{{\texttt{1e}}}
\def\me{{\texttt{me}}}
\def\em{{\texttt{em}}}
\def\mm{{\texttt{mm}}}
\def\EE{{\texttt{ee}}}
\def\ff{{\texttt{ff}}}
\def\CC{\text{CC}}
\def\TC{\text{TC}}

\renewcommand{\Pr}{P_{\rc{r}} }
\newcommand{\Pg}{P_{\gc{g}} }
\newcommand{\Pb}{P_{\bc{b}} }
\newcommand{\Prg}{P_{\rc{r}\gc{g}} }
\newcommand{\Prb}{P_{\rc{r}\bc{b}} }
\newcommand{\Pry}{P_{\rc{r}\yc{y}} }
\newcommand{\Pgb}{P_{\gc{g}\bc{b}} }
\newcommand{\Pgy}{P_{\gc{g}\yc{y}} }
\newcommand{\Pby}{P_{\bc{b}\yc{y}} }

\newcommand{\Pbarrg}{P_{\overline{\rc{r}\gc{g}}} }
\newcommand{\Pbarrb}{P_{\overline{\rc{r}\bc{b}}} }
\newcommand{\Pbarry}{P_{\overline{\rc{r}\yc{y}}} }
\newcommand{\Pbargb}{P_{\overline{\gc{g}\bc{b}}} }
\newcommand{\Pbargy}{P_{\overline{\gc{g}\yc{y}}} }
\newcommand{\Pbarby}{P_{\overline{\bc{b}\yc{y}}} }

\begin{document}

\title{Quantum computation from dynamic automorphism codes}

\author{Margarita Davydova}
\thanks{These authors contributed equally.}
\affiliation{Department of Physics, Massachusetts Institute of Technology, Cambridge, MA 02139, USA}
\affiliation{Kavli Institute for Theoretical Physics, University of California, Santa Barbara, California 93106, USA}

\author{Nathanan Tantivasadakarn}
\thanks{These authors contributed equally.}
\affiliation{Walter Burke Institute for Theoretical Physics and Department of Physics, California Institute of Technology, Pasadena, CA 91125, USA}
\affiliation{Microsoft Quantum, Station Q, Santa Barbara, California, USA}

\author{Shankar Balasubramanian}
\thanks{These authors contributed equally.}
\affiliation{Center for Theoretical Physics, Massachusetts Institute of Technology, Cambridge, MA 02139, USA}

\author{David Aasen}
\affiliation{Microsoft Quantum, Station Q, Santa Barbara, California, USA}

\begin{abstract}
We propose a new model of quantum computation comprised of low-weight measurement sequences that simultaneously encode logical information, enable error correction, and apply logical gates. These measurement sequences constitute a new class of quantum error-correcting codes generalizing Floquet codes, which we call dynamic automorphism (DA) codes. We construct an explicit example, the DA color code, which is assembled from short measurement sequences that can realize all 72 automorphisms of the 2D color code. On a stack of $N$ triangular patches, the DA color code encodes $N$ logical qubits and can implement the full logical Clifford group by a sequence of two- and, more rarely, three-qubit Pauli measurements. We also make the first step towards universal quantum computation with DA codes by introducing a 3D DA color code and showing that a non-Clifford logical gate can be realized by adaptive two-qubit measurements.
\end{abstract}

\maketitle
\setcounter{tocdepth}{3}
\tableofcontents
\hrulefill

\section{{Introduction}}

Finding new ways of performing quantum computations fault-tolerantly is an important open question, both from practical and theoretical standpoints. To this end, a new family of quantum error-correcting codes has been recently proposed, known as Floquet codes~\cite{Hastings2021}. 
Floquet codes promote fault tolerance to a spacetime picture, which could be the most natural way to formulate fault tolerance~\cite{Gottesman2022opportunities,Delfosse2023}.
In fact, time naturally enters the picture when two-dimensional topological codes are considered in the presence of measurement errors. 
Normally, fault tolerance in such cases is achieved by repeating measurements of commuting stabilizers~\cite{Dennis_2002}. This defines a `trivial' process in spacetime because the codespace does not change under commuting measurements. Floquet codes, on the other hand, push the information forward in time by sequences of anticommuting (typically, weight-two Pauli) measurements,
which leads to a nontrivial evolution of the codespace from round to round, while also (ideally) being fault-tolerant.  Moreover, a nontrivial automorphism, such as the $\ttt e - \ttt m$ automorphism of the toric code~\cite{Hastings2021,Aasen2022}, can be periodically induced during the evolution of a Floquet code, hinting that something interesting can be done to logical information. In this paper, we unlock a new capability of Floquet codes by showing that the evolution of the logical information induced by time-dependent measurement sequences can, in fact, implement a desired quantum computation, all while error syndromes are being simultaneously extracted. Under this framework, quantum memory and quantum computation become two constituents of the same concept, similarly to measurement-based~\cite{Raussendorf2003,Raussendorf_2006,Briegel_2009,Robertsbrown} and fusion-based~\cite{Bartolucci2023Feb} quantum computation.

The first example of a Floquet code is the honeycomb code, introduced by Hastings and Haah~\cite{Hastings2021}. It is based on the two-body operators of the Kitaev honeycomb model~\cite{Kitaev_2005}. When the two-body operators are considered as gauge checks of a subsystem code, no logical qubits are encoded~\cite{Suchara_2011}. 
However, if these anticommuting sets of check operators\footnote{In a more general setting, it is not necessary for the measurements themselves to anticommute between different rounds, rather each measurement round is designed such that it does not reveal encoded logical information in the state. At the same time, measurements need to build non-trivial correlations enabling error correction. } are measured with a specific period-three schedule, the code dynamically generates logical qubits, and transforms between instantaneous codespaces equivalent to an effective toric code, all while preserving logical information. This code was also shown to have a threshold for both circuit-level noise as well as measurement errors~\cite{Gidney2021faulttolerant,Paetznick2023,Gidney2022benchmarkingplanar}.
An especially intriguing property of the honeycomb code is that it implements an $\ttt e - \ttt m$ automorphism of the underlying toric code topological order after every 3 rounds. This idea became the basis of the $\texttt{\ttt e - \ttt m}$ automorphism code~\cite{Aasen2022}, as well as other formal results~\cite{Sullivan23,Aasen2023} seeking to understand the origin of such automorphisms in Floquet codes.

Since the proposal of the honeycomb code, a large number of new examples and theoretical developments are being put forward at an ever-growing rate. In particular, a CSS (Calderbank-Shor-Steane) version of the honeycomb code was constructed in Refs.~\cite{Davydova22,brown_2022,Bombin23}, and it was noticed that the underlying subsystem code structure~\cite{Aasen2022,Davydova22} and periodicity of the measurement schedule~\cite{Davydova22,brown_2022} are not necessary to define a fault-tolerant Floquet code. Floquet codes whose instantaneous codespaces realize different topological codes were also proposed. 
 The color code ISG (instantaneous stabilizer group) obtained from one- and two-qubit measurements appears in two upcoming works, one of which uses a technique of ``Floquetifying'' that is based on the $ZX$-calculus approach~\cite{Townsend_23}, and the other is based on a subsystem code~\cite{Dua23}. 
A 3D Floquet toric code and a Floquet double semion code were constructed in Ref.~\cite{Bauer2023}, and fracton Floquet codes were constructed in Refs.~\cite{Davydova22,Zhang22}. Constructing boundaries of Floquet codes is challenging~\cite{Vuillot2021planar}. Though some progress has been made in this direction~\cite{Haah2022boundarieshoneycomb,brown_2022,Ellison23}, a general understanding is still lacking.
There have also been recent conceptual developments that helped deepen our understanding of Floquet codes. Several viewpoints have been proposed, including paths of Hamiltonians~\cite{Aasen2022}, other constructions from measurement-based and fusion-based quantum computation~\cite{Paesani22, Bombin23}, and a tensor-network path integral approach~\cite{Bombin23,Bauer2023}. Partial results towards a classification of Floquet codes based on unitary loops~\cite{Sullivan23} and measurement quantum cellular automata~\cite{Aasen2023} have also been developed.

Despite this progress, it is still not entirely clear what the best way of performing quantum computation with Floquet codes is. 
Methods which are standard in surface codes have been used and updated to the Floquet setting, such as lattice surgery~\cite{Haah2022boundarieshoneycomb} and braiding twist defects~\cite{Ellison23}. 
Another generic option is available when the instantaneous codespace of a Floquet code admits a transversal gate; then such a gate can be unitarily applied at these rounds.

One aspect that has not been utilized is the automorphisms of the topological code that can be implemented in one period of a Floquet code protocol\footnote{In general, an automorphism of a topological stabilizer code is a locality-preserving unitary map that preserves the codespace, meaning that it takes the stabilizer group back to itself and preserves locality. The automorphisms of the topological codes are closely related to the automorphisms of the corresponding topological order. The logical operators of the topological code will undergo analogous transformations as the anyon strings of the respective anyon theory.
}. For example, the $\ttt e - \ttt m$ automorphism occurs both in the honeycomb and the $\ttt e - \ttt m$ automorphism codes and consists of an exchange of $Z$ and $X$ ($\ttt e$ and $\ttt m$) logical strings of the same homology, which, in turn, corresponds to a logical gate. The appearance of the $\ttt e - \ttt m$ automorphism has been appreciated as a remarkable dynamical property of Floquet codes but has not been considered as a practically useful feature. Constructing a more general dynamic code where more kinds of automorphisms (i.e. more transformations of logical operators) can be implemented by measurement sequences could open a door to a new way to perform quantum computation intrinsic to Floquet codes.

\begin{figure} \centering
\begin{centering}
\includegraphics[scale=0.27]{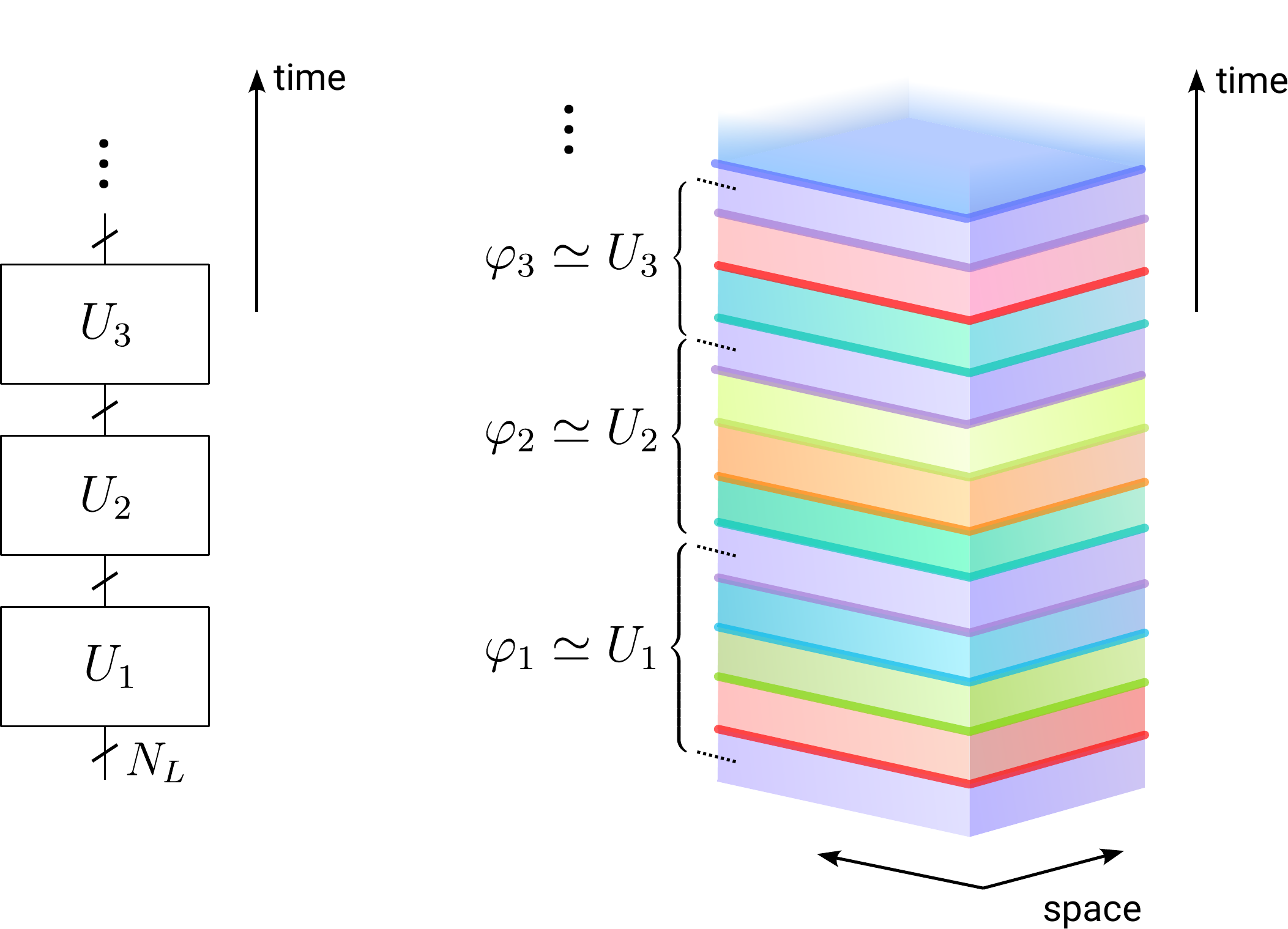}
\caption{A schematic of quantum computation with a dynamic automorphism code. On the left, we have a computation on $N_L$ logical qubits which is comprised of a sequence of unitaries $U_1$, $U_2$, $U_3$, etc. On the right, a dynamic automorphism code encoding $N_L$ logical qubits is executed by a sequence of automorphisms, $\varphi_1$, $\varphi_2$, $\varphi_3$, etc. The transition from one code to another is achieved by a round of few-qubit measurements shown by a darker layer, which pushes the logical information forward in time while measuring syndromes for error correction.
Every few rounds, the same codespace is realized, albeit with the underlying topological code having undergone a nontrivial automorphism. 
This automorphism has an action $U_i$ on the logical subspace. 
Thus, this sequence of automorphisms performs a quantum computation shown on the left.}
\label{fig:intro}
\end{centering}
\end{figure}

In this paper, we propose a way to execute quantum computation that is {\it native} to Floquet codes: in our model, a computation is realized by a sequence of automorphisms implemented by few-qubit measurement sequences. 
We develop a new class of quantum error-correcting codes capable of this, which we call {\it dynamic automorphism} (DA) codes\footnote{We use the name `dynamic' instead of `Floquet' to emphasize that periodicity is not necessary, and a dynamic automorphism code will in fact not be periodic for a generic computation.}. For a desired sequence of multi-qubit logical gates $\{U_1, U_2, U_3,... \} $ associated with a given computation, a dynamic automorphism code will be defined by a measurement sequence that applies a sequence of automorphisms $\{\varphi_1, \varphi_2, \varphi_3,... \} $ to the codespace, and each automorphism corresponds to a gate in the computation, i.e. $\varphi_i \simeq U_i$ (see Fig.~\ref{fig:intro}). For each such sequence, one can define an appropriate DA code. Importantly, these measurements can be simultaneously used to infer error syndromes for error correction.

The key results of this paper are: 
\begin{itemize}
 \item[--] Construction of an explicit example of a dynamic automorphism code, called the {\it dynamic automorphism color code} that can realize all 72 automorphisms of the 2D color code~\cite{Yoshida2015,Kesselring_2018} by two-qubit measurement sequences of length 5 or less (Sec.~\ref{sec:torus}).
 \item[--] A framework for understanding the nature of DA codes from the viewpoint of TQFTs (Sec.~\ref{sec:tqft_sec}).
 \item[--] A method to consistently put these codes on a lattice with boundaries and construction of a dynamic automorphism color code on a triangle with Pauli boundaries. From this construction, the full Clifford group on multiple logical qubits can be implemented by automorphisms of layers of such triangles (Sec.~\ref{sec:triangle}).
 \item[--] Proposal of the three-dimensional dynamic automorphism color code and a protocol for implementation of a non-Clifford gate with adaptive measurements, thus making a first step towards universal quantum computation using these codes (Sec.~\ref{sec:3DFCC}).
 \item[--] Error correction in most of the aforementioned constructions and a blueprint towards fault-tolerant quantum computation (Secs.~\ref{sec:EC} and ~\ref{sec:EC_triangle}).
\end{itemize}

\begin{figure} \centering
\includegraphics[width = 1\textwidth]{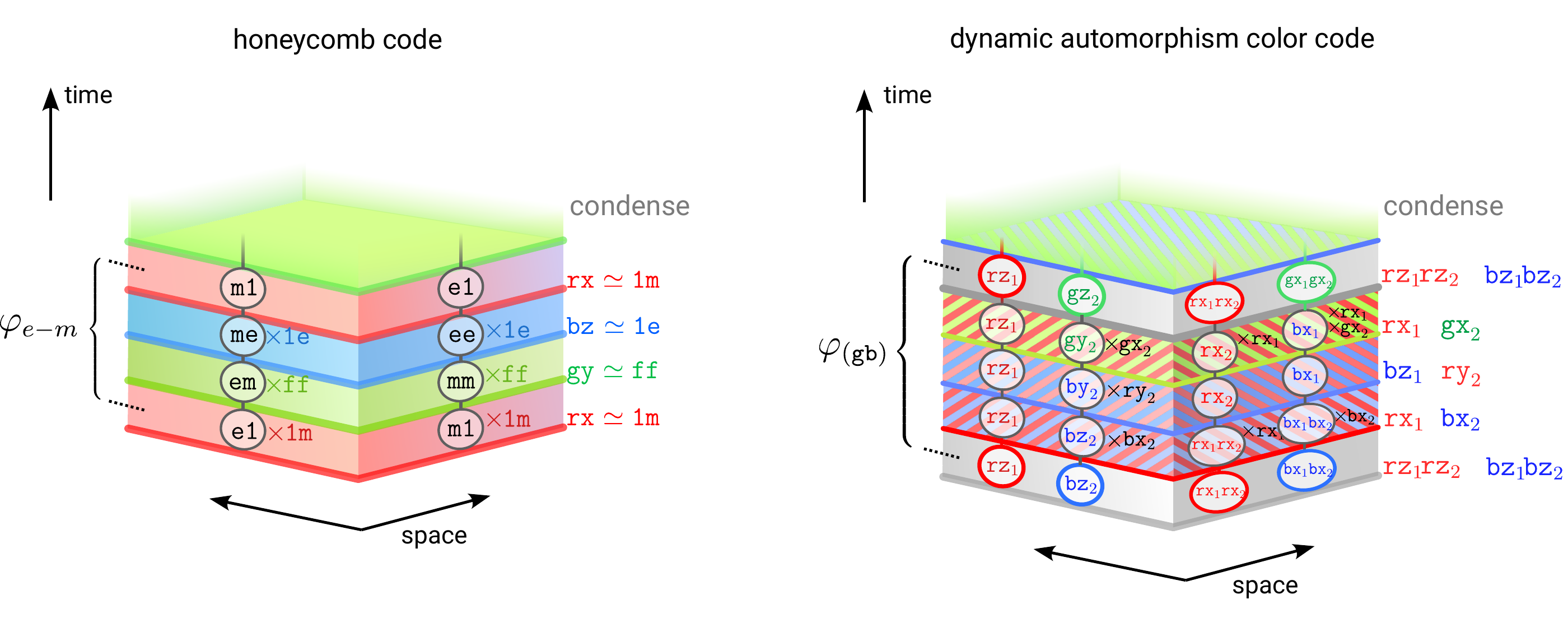}
\caption{(Left) The $\ttt e - \ttt m$ automorphism $\varphi_{\ttt e - \ttt m}$ implemented by a period of the honeycomb code shown from a sequence of condensations. For the honeycomb code, the parent theory is the color code which is equivalent to two copies of the toric code. The condensations are shown on the right from the stack of topological codes (the colored slabs) and the boson condensed at each round is written in color code notation and the double toric code notation (see Secs.~\ref{sec:II_intro} and \ref{sec:one_step_unfolding}). The colored slabs represent the condensed model living between each pair of subsequent condensations. Representatives of the two deconfined anyons at each round are shown in circles. These representatives sometimes have to be fused with a condensed anyon of the current round to survive to the next round (this is the same as saying that the logical information is conserved), which is shown by respective multiplication. Because of this update rule, the $\ttt {m1}$ and $\ttt{e1}$ anyons swap places upon a period of the protocol. (Right) An example of a measurement sequence for the DA color code, whose parent code is two copies of the color code (the subscript refers to the copy). We show one of the many sequences that are the building blocks of the full DA color code, which corresponds to a $\varphi_{(\ttt{gb})}$ automorphism. Two bosons are condensed at each round, listed on the right. We track a complete basis of four deconfined anyons at each time, shown in circles, and they are appropriately updated from round to round. At the first and last steps of the sequence, the effective theory is the color code, where we see that the blue and green anyons swap as a result of the evolution, corresponding to the action of the $\varphi_{(\ttt{gb})}$ automorphism.
}
\label{fig:intro2}
\end{figure}

Our approach is based on the formalism of anyon condensation from a parent code first introduced in the context of topological codes in Ref.~\cite{brown_2022}. The parent code in our case is two (three) copies of the 2D (3D) color code in the cases of two (three) spatial dimensions. As an illustration, an example of a sequence of anyon condensations permuting red and green anyon strings of the color code is shown in Fig.~\ref{fig:intro2} side by side with an analogous depiction of the $\ttt e - \ttt m$ automorphism of the honeycomb code. 

There are several natural questions related to putting our results into a broader context. For instance, what would be the advantage of applying gates using two-body measurements if the color code stabilizer group admits them as transversal unitaries?  In particular, for the codes proposed in this paper we provide a measurement-based implementation of logical operations which already admit an implementation as a transversal or finite-depth local unitary.  However, we believe that, at a minimum, we made conceptual progress by showing a new way to perform quantum computation that is native to Floquet/dynamic codes. Additionally, inspired by our construction, one could come up with an example of a dynamic code where implementing logical operations goes beyond implementing automorphisms of the instantaneous codespace. Finally, for an architecture where few-qubit measurements are native operations, such as photonic~\cite{Browne_2005,Bourassa_2021,Bartolucci2023Feb} or Majorana-based~\cite{Karzig2017} quantum computers, our method might be practically preferable over applying transversal gates. 

Another question one might ask is, what is the difference between the dynamic automorphism codes approach and the measurement-based quantum computation (MBQC) approach~\cite{Raussendorf2003,Raussendorf_2006,Briegel_2009,Robertsbrown}? These in fact can be shown to be equivalent~\cite{Bombin23}, though the mapping between them is nontrivial. However, the actual implementation and what the code and computation look like at each point in time can be drastically different, so one should view them as practically different ways of doing computation.

Aside from conceptual interest, the dynamic automorphism code approach appears quite attractive also because of the low spatial overhead compared to lattice surgery and because error detection that can be done in parallel with the computation itself. Our work invites exploration into searching for and optimizing fault-tolerant dynamic codes, which may yield new ways of performing universal quantum computation.

\subsubsection*{Summary of the results}

Let us sketch the general idea behind dynamic automorphism codes and summarize the content of the rest of the paper. For the DA codes studied in this paper, we leverage the concept of a parent code and anyon condensation proposed in~\cite{brown_2022}. In particular, for the two-dimensional DA color code, the parent code is two copies of the color code. Once the parent theory is chosen, one first must categorize all possible reversible condensations of the theory. 
Colloquially speaking, a pair of condensations from a parent theory is reversible if they define an one-to-one correspondence between a pair of respective (condensed)  child theories. Reversibility guarantees that it is possible to go between two child theories while preserving logical information.  This is a deep concept that we explore in Sec.~\ref{sec:tqft_sec} that goes even beyond dynamic automorphism codes. We can organize the structure of reversible condensations and child theories into an object we call the condensation graph. The vertices of the condensation graph correspond to condensations, and an edge between a pair of condensations is drawn whenever a pair is reversible. A closed path starts and ends at the same theory, and thus, can be labeled by an automorphism.

The condensation paths implementing automorphisms can be put in correspondence with measurement sequences for DA codes. Namely,  measurements of a given round in the sequence correspond to the operators performing respective condensation in the Hamiltonian picture. An example of such a condensation sequence for the DA color code is shown in Fig.~\ref{fig:intro2}. These sequences are good for building codes, as one can guarantee conservation of logical information between the measurement rounds (there is also an intimate relation to measurement quantum cellular automata (MQCA)~\cite{Aasen2023}). However, not all of these sequences will generate the codespace starting from an arbitrary product state (also referred to as ``dynamically generate logical qubits''), and furthermore, not all of the sequences that generate the codespace will be error-detecting. Thus, a further search among the candidate sequences is required to identify the error-correcting protocols, and it is an open question to understand what constraints this places on the possible automorphisms in a general DA code. However, as the example of the DA color code that we construct in this paper shows, this goal is certainly achievable. 

In three dimensions, the situation is more complicated. Reversible transitions, however, can also be defined. We find an example of a condensation graph that allows us to achieve a set of nontrivial automorphisms for the 3D DA color code. We also find a sequence of measurements that realizes a non-Clifford gate. Nevertheless, a richer set of measurement sequences and corresponding automorphisms might be still discovered in the future.

The rest of the paper is organized as follows. In Section~\ref{sec:torus}, we start by reviewing the background concepts, such as the 2D color code, its unfolding, and automorphisms. Then we introduce the main building block for dynamic automorphism color code. Namely, starting from a parent model, which is two copies of a color code, a reversible pair of condensations can be introduced which takes us between two child codes: an effective color code and two copies of the toric code.
Based on this, we construct measurement sequences implementing all generators of the automorphism group of the 2D color code. 

Section~\ref{sec:EC} discusses error correction for the 2D DA color code without boundaries. We utilize the detector formalism and construct a basis of detectors for the 2D DA color code, and show how to detect and correct independent Pauli errors. We introduce a trick we call padding, where non-error-correcting sequences can be turned into error-correcting ones by inserting additional sequences in between. This section provides what can be viewed as a foundation of error correction in general dynamic automorphism codes.

In Section~\ref{sec:tqft_sec}, we provide a general TQFT perspective on 2D DA color codes. We introduce the concept of reversible condensations between two child theories obtained from the same parent theory. Reversible condensations are the building blocks for constructing sequences that implement automorphisms of a child theory, and we show how to determine the automorphism implemented by a given sequence, as well as a way to design target sequences. Finally, we introduce a ``condensation graph'' which summarizes the space of possible condensation paths. 

In Section~\ref{sec:triangle}, we discuss how to incorporate boundaries in the 2D DA color code. We explain the principle behind constructing information-preserving boundaries. Next, we introduce measurement sequences for the DA color code generating the single-qubit Clifford group on a single triangle with Pauli boundaries using two-body measurements only. Finally, we consider two DA color code triangles and show a two- and, occasionally, three-qubit measurement sequence that realizes an entangling iSWAP gate between the layers. These protocols applied to $N$ triangles give a generating set of gates for the full Clifford group on $N$ logical qubits. We also address error correction in the presence of boundaries in this section.

Motivated by the goal of a model for universal quantum computation, in Section~\ref{sec:3DFCC} we construct a 3D DA color code that realizes the 3D color code instantaneous stabilizer group at certain steps of the measurement protocol, where the parent model is three copies of the 3D color code. We show the sequences implementing color permutation automorphisms. Finally, we show that with two-qubit non-Pauli measurements, we can implement a transversal $T$ gate. This takes us a step closer to universal quantum computation; to complete the task, one might consider a natural path of interfacing the 2D and 3D DA color codes akin to dimensional jump in color codes~\cite{Bombin2016}. Alternatively, one might consider applying a non-Clifford gate by treating the time dimension as a substitute to the third dimension, similarly to the just-in-time technique~\cite{Bombin2018,Brown2020,Scruby_2022}. 

Finally, in Section~\ref{sec:smallcodes} we discuss examples of small DA color codes.

\section{{Dynamic automorphism color code}}
\label{sec:torus}

The basic element of the dynamic automorphism color code is given by a sequence of two-qubit parity measurements\footnote{In sec.~\ref{sec:triangle}, we will also see an example where three-qubit measurements are needed to do an entangling gate between two logical qubits, albeit only at one boundary of the code during one measurement round only. }, some of which dynamically transition between a color code and a pair of toric code instantaneous stabilizer groups. 
The sequence of measurements can be chosen such that a nontrivial automorphism of the color code is induced, which permutes the color code logical strings and corresponds to a logical gate. 
The measurement sequences for the DA color code can be viewed as building blocks, each block realizing one of the automorphisms of the color code, and the blocks can be combined together to give sequences of automorphisms. We show that all automorphisms of the color code can be realized by such sequences.

The measurement sequences of the DA color code are similar in spirit to those of Floquet codes, where the $\ttt e - \ttt m$ automorphism has been shown to occur after every three rounds of measurements in the honeycomb code~\cite{Hastings2021} and in the $\ttt e - \ttt m$ automorphism code~\cite{Aasen2022}. Nevertheless, the previous Floquet codes such as in Ref.~\cite{Hastings2021,Aasen2022,Davydova22,brown_2022} execute too few automorphisms to be used for quantum computation.

More generally, the DA color code can be turned into a stack of codes each encoding a few logical qubits. Both inter-layer and intra-layer measurements have to be performed in this case (vertical connectivity between neighboring pairs of layers is sufficient, thus, the DA color code has relatively low connectivity). While achieving all automorphisms of a stack of color codes is a significant step forward, it, however does not yet give us the complete Clifford group of logical operations. In Sec.~\ref{sec:triangle}, we show how the full Clifford group can be attained in a planar geometry.

Laying the groundwork for our exploration, Ref.~\cite{brown_2022} introduced the concept of parent codes and connected Floquet codes to anyon condensation in these parent codes. 
Moreover, Ref.~\cite{Aasen2023} showed that the sequence of measurements also needs to satisfy a certain property of being `locally reversible' in order to preserve logical information. 
Hence, we combine the ideas and introduce the notion of ``reversible condensation'' which is discussed further in Sec.~\ref{sec:tqft_sec}. The reversibility of the condensation sequence is a necessary constraint that guarantees the conservation of logical information; however it is not sufficient to ensure the resulting code is error-correcting. Thus, we use reversible condensations to constrain the space of measurement sequences within which we further search for error-correcting codes.

The parent anyon theory for the DA color code is a color code bilayer\footnote{I.e. two independent color code copies/layers; not to be confused with the doubled color code~\cite{Bravyi2015doubled}.} and defines a respective parent topological code\footnote{Any model, from which each time step of the dynamic code can be obtained directly, constitutes an appropriate parent model. Our choice of two layers of the color code leads to nicest and the most physically transparent set of condensations, and leads to the lowest-weight (two-body) measurement realization when realized in microscopics.}. An appropriate condensation of two bosons in the parent anyon theory gives a child theory isomorphic to the color code. Following a path of such condensations, we can perform a cycle beginning and ending at the same code, wherein an automorphism has been applied.
From these condensation sequences, the measurement sequences for respective topological codes implementing the corresponding automorphisms can be designed straightforwardly, producing the building blocks for the dynamic automorphism color code.
In this section, we explore such sequences on a honeycomb lattice with periodic boundary conditions, i.e. on a torus (although it is relatively straightforward to generalize to other closed manifolds), where two-qubit measurements are sufficient to realize all 72 automorphisms of the color code. Open boundary conditions are addressed in Sec.~\ref{sec:triangle}.
The gates that are achieved by these sequences form an order-72 subgroup of the real Clifford group on the four logical qubits on a torus and do not present a particularly interesting set of logical gates from the perspective of quantum computation. However, later in Sec.~\ref{sec:triangle} we propose dynamic automorphism codes on a single triangle and a pair of triangles and show how to perform a generating set of automorphisms leading to the full Clifford group. 
In Sec.~\ref{sec:EC}, we show that our protocols, starting from an arbitrary product state generate the appropriate code, and will dynamically generate logical qubits in the same sense as the original honeycomb code~\cite{Hastings2021}. We also explore the error correction of these codes.

The rest of the section is structured as follows. In Subsec.~\ref{sec:II_intro}, we review the basic properties of the color code and its relation to two copies of toric code via a measurement induced unfolding map.
In Subsec.~\ref{sec:one_step_unfolding}, we introduce the foundation for the condensation sequences by considering a single condensation from the parent bilayer color code theory to a color code or an isomorphic bilayer toric code theory. We also show how condensing the same objects allows one to reversibly transition between child theories, and lastly, we show how to achieve the same result in topological codes by two-qubit measurements. 
Finally, we show the condensation and measurement sequences implementing automorphisms of the color code in Subsec.~\ref{sec:II_automorphisms}.

\subsection{Short review: automorphisms of the color code, unfolding, and condensations} \label{sec:II_intro}

The anyon theory of the color code (CC)~\cite{Bombin2012,Kubica_2015} can be summarized as follows. The nine (non-trivial) bosons of the color code can be denoted $\ttt{c}\tsig$, where $\ttt{c} = \ttt{\rc r},\ttt{\gc g},\ttt{\bc b} $ is the color label, and $\tsig = \ttt{x},\ttt{y},\ttt{z}$ is the flavor (Pauli) label. The nine bosons can be arranged conveniently into a ``Mermin-Peres magic square''~\cite{Mermin90,Peres91}:
\begin{equation}
{\large
\renewcommand{\rc}[1]{{\color{red}{#1}}}
\renewcommand{\bc}[1]{{\color{blue}{#1}}}
\renewcommand{\gc}[1]{{\color{ForestGreen}{#1}}}
\renewcommand{\yc}[1]{{\color{amber}{#1}}}
 \begin{tabular}{|c|c|c|}
 \hline
\rx & \ry & \rz \\ \hline
\gx & \gy & \gz \\ \hline
\bx & \by & \bz \\\hline
\end{tabular} \triangleq 
\begin{tabular}{|c|c|c|}
\hline
\im & \em & \ei \\ \hline
\mm & \ff & \EE \\ \hline
\mi & \me & \ie\\ \hline
\end{tabular}
\label{eq:magicsquare}
}
\end{equation}
where the notation on the left is the color code notation~\cite{Kesselring_2018}, and on the right the anyons of the two constituent toric codes are used to label the bosons of the color code. The following properties hold for the magic square:
\begin{enumerate}
 \item The product of three bosons in individual rows (same color label) or columns (same flavor label) is the trivial anyon;
 \item The mutual statistics (full braid) of two bosons that share the same row or column is trivial. Otherwise, their mutual statistics is $-1$.
\end{enumerate}
There are also six additional fermions in the color code which we will call $\texttt{f}_\texttt{c}$ and $\overline{\texttt{f}_\texttt{c}}$ for $\ttt{c} = \ttt{\rc r},\ttt{\gc g},\ttt{\bc b}$\footnote{The fermions can be expressed as products of pairs of bosons that braid nontrivially, and the six fermions are
\begin{align*}
 \fr &= \tt{mf} \triangleq \gx\otimes \bz = \gy\otimes \rz = \rx\otimes \by & \frbar&= \tt{fe} \triangleq \bx\otimes \gz = \by\otimes \rz = \rx\otimes \gy \\
 \fg &= \tt{f1} \triangleq \bx\otimes \rz = \by\otimes \gz = \gx\otimes \ry & \fgbar&= \tt{1f} \triangleq \rx \otimes\bz = \ry \otimes\gz = \gx \otimes\by \\
 \fb &= \tt{ef} \triangleq \rx\otimes \gz = \ry\otimes \bz = \bx\otimes \gy &\fbbar &= \tt{fm} \triangleq \gx\otimes \rz = \gy\otimes \bz = \bx\otimes \ry\\
\end{align*}}. This naming convention comes from the fact that the color code is also equivalent to the double of the three-fermion theory (3F), namely, $\CC \simeq \text{3F} \boxtimes \overline{\text{3F}}$ (see Sec.~\ref{sec:condensation_complex} where the relation between the fermions and bosons is explored in more detail). It is also equivalent to two copies of the toric code~\cite{Bombin2012,Kubica_2015}, which we address later on.

The topological order describing the color code admits nontrivial automorphisms. Given an anyon theory $\mathcal{M}$, an automorphism of the anyon theory is a map $\varphi: \mathcal{M} \to \mathcal{M}$ that permutes anyons of the theory while preserving both fusion and braiding rules of the theory.
As an example, if the underlying topological order is $D(\mathbb{Z}_2)$, e.g. $\TC = \{ \ttt{1}, \ttt{e}, \ttt{m}, \ttt{f} \},$ the only nontrivial automorphism is the $\ttt e - \ttt m$ permuting one: $\varphi_{\ttt e - \ttt m}(\texttt{e}) = \texttt{m}$, $\varphi_{\ttt e - \ttt m}(\texttt{m}) = \texttt{e}$, $\varphi_{\ttt e - \ttt m}(\texttt{1}) = \texttt{1}$, and $\varphi_{\ttt e - \ttt m}(\texttt{f}) = \texttt{f}$. 
The color code has a richer structure $\text{CC} \simeq \TC \boxtimes \TC$ and has a larger number of nontrivial automorphisms. These automorphisms are associated with the symmetries of the magic square. In particular, the fusion rules between three bosons with the same color and the mutual statistics between a pair of bosons of the same color is left invariant if the color label is permuted. This indicates permuting rows of the magic square corresponds to a ``color'' automorphism. Similarly, permutations of the columns of the magic square preserve fusion and braiding and correspond to a ``flavor'' automorphism. Finally, the mutual statistics and fusion rules are left invariant under transposition of the magic square (i.e. mirror symmetry with respect to the diagonal of the magic square). Together, the automorphisms form a group $(S_3\times S_3)\rtimes \mathbb{Z}_2$: the two $S_3$ subgroups come from permuting rows and columns, and the $\mathbb{Z}_2$ subgroup corresponds to transposing the magic square.

We will also extensively use the concept of anyon condensation~\cite{Kong2014anyon,Burnell18} and its application to the color code~\cite{brown_2022} throughout the paper. Condensation of a (deconfined) anyon can be thought of as a way to transition out of a topologically ordered phase to a topologically ordered phase with fewer anyons. The reason is two-fold: (1) the condensed anyon will confine anyons which braid non-trivially with it, and (2) the condensed anyon mediates a non-trivial equivalence between the anyons of the parent theory. To simplify matters, we review the concept of anyon condensation for Abelian anyon theories. Given an Abelian ``parent'' anyon theory $\mathcal{M}$, consider a subgroup (under fusion operation) of bosons $\mathcal A$ where any pair of anyons $\ttt a_1, \ttt a_2 \in \mathcal A$ has trivial mutual braiding statistics (we will often abbreviate this as ``braid trivially''). When we condense the anyons in this subgroup, anyons that do not braid trivially with some anyon in $\mathcal A$ become \textit{confined}. The resulting ``child'' anyon theory $\mathcal{C}_{\mathcal A}$ therefore has anyons which are labeled by equivalence classes $[\ttt{c}] = \{\ttt{c} \otimes \ttt{a}: \ttt{a} \in \mathcal{A}\}$, the representative of the  equivalence class is $\ttt{c} \in \mathcal{M}$, and of course $\ttt{c}$ must braid trivially with the condensate to be deconfined. 
In particular, the anyons in $\mathcal A$ correspond to the vacuum sector of the child theory. In general, there are multiple choices for condensible anyons $\mathcal A$ and multiple possible condensed theories. One may also choose $\mathcal A$ such that no anyon braids trivially with $\mathcal A$ -- this is known as a Lagrangian subgroup, and condensing a Lagrangian subgroup of an anyon theory results in a trivial theory.

Transitions between two different child theories $\mathcal{C}_1$ and $\mathcal{C}_2$ found by condensing sets of bosons $\mathcal{A}_1$ and $\mathcal{A}_2$ can be implemented by starting within the condensed theory $\mathcal{C}_1$, lifting to the parent theory $\mathcal{M}$, and further condensing a different set of anyons $\mathcal{A}_2$ to transition into $\mathcal{C}_{2}$. When such a transition between two different child theories $\mathcal{C}_1$ and $\mathcal{C}_2$ exists, it is called a ``reversible condensation'', which is a concept explored further in Sec.~\ref{sec:tqft_sec} if the two child theories are isomorphic. Briefly, a reversible condensation corresponds to two compatible condensations of a parent theory such that the two child theories are isomorphic, and the isomorphism can be computed explicitly using the condensation data.

When it comes to concrete lattice realizations, a given topological code can be obtained by taking the fixed-point Hamiltonian whose low-energy theory is respective anyon theory and defining the individual terms in the Hamiltonian to be the stabilizers of the code. 
The color code stabilizer group can be defined on any planar graph that is simultaneously trivalent and tri-colorable~\cite{Bombin2006,Bombin2009Nov}.  We place qubits at its vertices (we pick colors $r,g,b$ in accordance with the colors of the bosons) and define the stabilizer group:
\be
\mathcal{S}(\text{CC}) = \left \langle P_{r} (X), P_{g} (X), P_{b} (X), P_{r} (Z), P_{g} (Z), P_{b} (Z)\right \rangle
\ee
where each $P_{c = r,g,b} (\sigma)$ stands for the group of plaquette stabilizers of a given color, and $\langle ..\rangle$ stands for a group generated by the objects inside of it. Each plaquette stabilizer is given by a product of Pauli operators (their type is shown as an argument) acting on the vertices at the boundary of the plaquette of a given color. The labeling of the logical operators $\mathcal{L}(\ttt{c}\tsig)$ of the topological code are inherited from the Hamiltonian model and are strings of Pauli operators of a given color (i.e. along the edges of a given color). In the Hamiltonian picture, these strings transport respective anyons $\ttt{c}\tsig$ along homologically nontrivial cycles.

The equivalence between the color code and the two copies of the toric code, i.e. $\text{CC} \simeq \TC \boxtimes \TC$~\cite{Bombin2012,Kubica_2015} is shown in the double toric code notation for the color code bosons on the right-hand side of Eq.~\eqref{eq:magicsquare}. It identifies the bosons in the color code with bosons within two copies of a toric code. In fact, there exists an {\it unfolding} procedure, wherein an unfolding unitary can be applied to the stabilizer group of the color code that turns it into two decoupled copies of the toric code~\cite{Kubica_2015}. There are 72 ways to unfold the color code into a pair of toric codes, corresponding to applying the symmetries of the magic square to one side of Eq.~\eqref{eq:magicsquare}. Moreover, unfolding can also  beperformed in the presence of boundaries (later in Sec.~\ref{sec:boundary_micro}, we will use a measurement-induced unfolding of a triangular color code into rectangular toric code with a domain wall along the diagonal).

Finally, how might we realize anyon condensation on a lattice? Assuming that the anyon theory of interest has a microscopic Hamiltonian realization, the simplest way to condense a certain boson of the theory is to add terms to the physical Hamiltonian that correspond to hopping this boson. For example, in the toric code, condensing an $\texttt{e}$ charge corresponds to adding a strong transverse field to the Hamiltonian that couples to the hopping operator $X$ for this charge, thus driving it to a trivial phase. In the color code, condensing a boson of a certain flavor $\sigma$ and color $c$ corresponds to adding the two-body operator $E_c(\sigma)$ to the Hamiltonian. 
In the case of topological codes, we can start with a parent code and the operators that we measure are the operators that are used to condense respective anyons in the Hamiltonian picture. This defines a transition from the parent stabilizer code into a new ``child'' code where the measured operator is part of the stabilizer group and the other elements are suitable products of the original stabilizers which commute with the hopping operator.

\begin{figure}[!t] \centering
\vspace{0pt}
\centering
\vspace{0pt}
\includegraphics[width= 1\columnwidth]{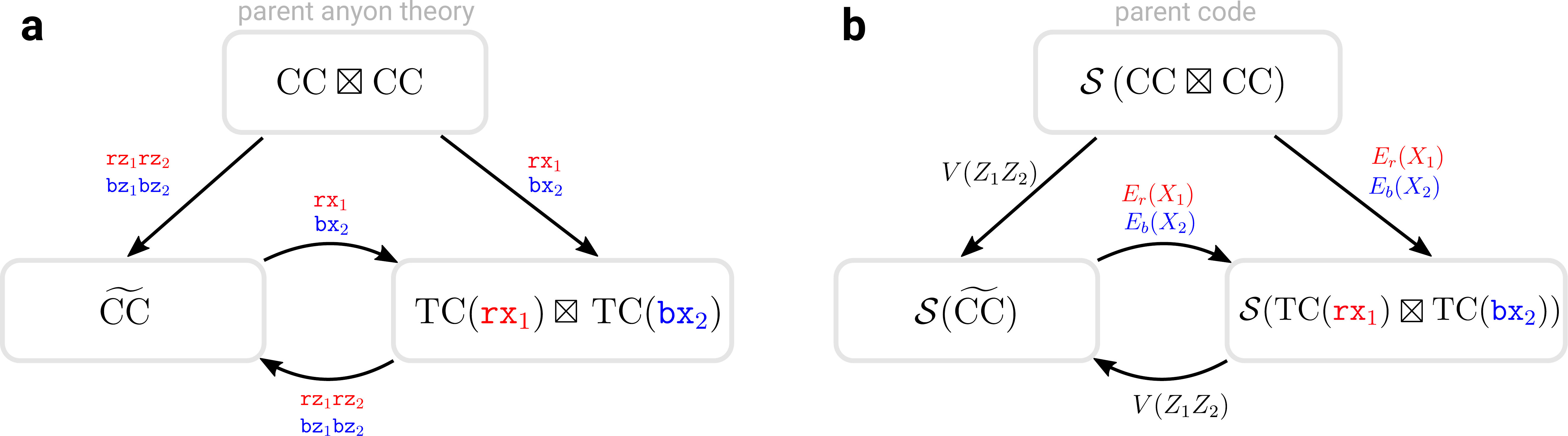 }
\caption{ (a) Two ways of going from the parent theory to child theories by condensations. Condensing $\rzi \rzii$ and $\bzi \bzii$ (and hence their product $\gzi \gzii$) leads to the effective color code child model $\widetilde{\text{CC}}$. Condensing $\rxi$ and $\bxii$ produces a child model that is two toric codes, $\TC(\rxi) \boxtimes \TC(\bxii)$. The two child theories are connected by condensing the same set of anyons (which are confined in the child theory), as shown by curvy arrows. It is necessary that the pair of condensations between two child theories are developed more formally in Sec.~\ref{sec:TQFTreview}.
(b)~Analogous transitions can be performed in topological codes by measurements, starting from a bilayer color code as a parent code. The same set of measurements can be used to go between the child codes as they are reversible stabilizer groups~\cite{Aasen2023}.}
\label{fig:3a}
\end{figure}
\subsection{The parent code, reversible condensations, and measurements} \label{sec:one_step_unfolding}

The construction of measurement sequences for the dynamic automorphism color code naturally follows from the picture of anyon condensation from a parent code, a perspective first introduced in Ref.~\cite{brown_2022}. There, the parent code was chosen to be a color code, and condensation to a single toric code was obtained using two-body measurements. A periodic sequence of measurements was then performed to implement the CSS honeycomb code (called the ``Floquet color code'' in Ref.~\cite{brown_2022}).
In the corresponding anyon theory, this sequence translates into condensing one of the bosons from the CC$\simeq$TC$\boxtimes$TC parent anyon theory (the color code~\cite{Bombin2006}) to obtain the toric code (TC) chil anyon theory~\cite{Kitaev2003}. 

In our construction, we choose the parent anyon theory to be two copies of the color code, $\CC \boxtimes \CC$. From this parent code, one can condense two independent (and braiding trivially) bosons and obtain the CC$\simeq$TC$\boxtimes$TC child topological order. There are two distinct types of condensations that we consider. The first one (which we call ``interlayer condensation'') consists of condensing two objects, each being a fusion of two identical bosons, one from each parent color code. The second (which we call ``intralayer condensation'') consists of condensing a single boson from each CC from the parent theory separately. 
As we will see later, in the language of the codes, this yields effective color codes or two effectively decoupled toric codes\footnote{Of course, there exist more options, such as condensing the Lagrangian subgroup of one of the copies of the color code, which will leave us with the other color code copy. We will also ignore the possibility of condensing bosons that are products of fermions in each layer. As we will shortly see, despite these restrictions, we can design condensation sequences that can implement all 72 automorphisms of the color code.}. Thus, we refer to the resulting child theories as the ``effective color code'' $\widetilde{\text{CC}}$ and $\text{TC}(\ttt{a}_\ttt{1})\boxtimes\text{TC}(\ttt{a}_\ttt{2})$ (where $\ttt{a}_{\ttt{1,2}}$ are the two independent condensed bosons of the parent theory). The two choices of condensation from the parent theory are summarized in Fig.~\ref{fig:3a}(a).

Starting from the parent $\CC \boxtimes \CC$ theory, we introduce the first type of condensation which we call ``interlayer'', namely we condense $\rzi\rzii \triangleq \rzi\otimes \rzii$ and $\bzi\bzii\triangleq \bzi\otimes \bzii $, corresponding to $(\ei)_\texttt{1} (\ei)_\texttt{2}$ and $(\ie)_\texttt{1}(\ie)_\texttt{2}$ in the toric code notation. We denote the layer/copy within the parent code where the respective boson belongs by the subscript $1,2$, and we call the resulting child theory the effective color code $\widetilde{\CC}$. Note that the tilde indicates that the anyons of the condensed color code are products of anyons of the parent color codes. The remaining deconfined bosons are
\begin{equation} \label{eq:effCC_bosons}
{%
\renewcommand{\rc}[1]{{\color{red}{#1}}}
\renewcommand{\bc}[1]{{\color{blue}{#1}}}
\renewcommand{\gc}[1]{{\color{ForestGreen}{#1}}}
\renewcommand{\yc}[1]{{\color{amber}{#1}}}
\begin{tabular}{|c|c|c|}
\hline
$\widetilde{\rx}$ & $\widetilde{\ry}$ & $\widetilde{\rz\,}$ \\ \hline
$\widetilde{\gx}$ & $\widetilde{\gy}$ & $\widetilde{\gz \,}$ \\ \hline
$\widetilde{\bx}$ & $\widetilde{\by}$ & $\widetilde{\bz \,}$\\
\hline
\end{tabular} \sim 
 \begin{tabular}{|c|c|c|}
 \hline
$\rxi\rxii$ & $\ryi\rxii$ & $\rzi$ \\ \hline
$\gxi\gxii$ & $\gyi\gxii $& $\gzi$ \\ \hline
$\bxi\bxii$ &$ \byi\bxii$ &$ \bzi$\\ \hline
\end{tabular}
}
\end{equation}
where on the left, we label the effective bosons of the $\widetilde{ \text{CC}}$ theory. One can confirm that the above anyons obey the fusion and braiding rules of the color code topological order. 
We emphasize that every anyon of the parent theory that remains deconfined after the condensation forms an equivalence class that can be obtained by fusion with the anyons that were condensed. For example, the equivalence class of {$\widetilde{\rz}$} consists of four anyons: $\{\rzi, \rzii, \gzi\bzii, \bzi\gzii \}$, and the equivalence is set by multiplying by $\rzi\rzii$ and $\bzi\bzii$. A particular representative for each of these anyons in the parent theory notation is shown in the table above on the right.

The second type of condensation corresponds to condensing two independent bosons, which, according to the labeling we have chosen, corresponds to separately condensing each color code of the parent theory into a toric code. This is akin to performing two honeycomb or CSS honeycomb-type condensations in parallel~\cite{brown_2022}; we refer to these as ``intralayer'' condensations. For example, condensing $\rxi$ in the first layer, and $\bxii$ in the second layer maps $\CC \boxtimes \CC \rightarrow \TC(\rxi) \boxtimes \TC(\bxii)$, where the representatives of the remaining toric code in each layer can be chosen to be
{\renewcommand{\rc}[1]{{\color{red}{#1}}}
\renewcommand{\bc}[1]{{\color{blue}{#1}}}
\renewcommand{\gc}[1]{{\color{ForestGreen}{#1}}}
\renewcommand{\yc}[1]{{\color{amber}{#1}}}
\begin{align}
\TC(\rxi) &= \{ \ttt{1}, \ttt{e}_\ttt{1}, \ttt{m}_\ttt{1}, \ttt{f}_\ttt{1} \}, & \ttt{e}_\ttt{1} &= \rz_\ttt{\rc1} \triangleq (\ei)_\ttt{1}, & \ttt{m}_\ttt{1} &= \bx_\ttt{\bc1} \triangleq (\mi)_\ttt{1}, \\
\TC(\bxii)& = \{ \ttt{1}, \ttt{e}_\ttt{2}, \ttt{m}_\ttt{2}, \ttt{f}_\ttt{2} \},& \ttt{e}_\ttt{2} &= \bz_\ttt{\bc2} \triangleq (\ie)_\ttt{2}, & \ttt{m}_\ttt{2} &= \rx_\ttt{\rc2} \triangleq (\im)_\ttt{2}.
\end{align}}Again, each equivalence class allows four representatives. We remark that although the toric codes appear decoupled, the representatives might contain anyons from both layers when viewed as an anyon in the parent theories. For example, the equivalence class of $\ttt{e}_\ttt{1}$ contains $\{{\rzi},$ $ {\ryi}$, ${\rzi\bxii}, {\ryi\bxii}\}$.

For the construction of the DA color code, we will assume that we always use the same specific condensation $\rzi\rzii$, $\bzi\bzii$ to obtain the effective color code $\widetilde{\CC}$ while utilizing the large number of ways to condense to $\TC(\ttt{a}_1) \boxtimes \TC(\ttt{a}_2)$ in order to be able to engineer various automorphisms. The condensations we'll be using for our sequences have to be reversible, in the sense that every consequent pair is a reversible pair of condensations. A pair of condensations is reversible if they define an isomorphism between the pair of corresponding child theories, and this concept is explored in greater depth in Sec.~\ref{sec:tqft_sec}.
The condensations between these two child theories $\widetilde{\CC} \leftrightarrow 
 \TC(\ttt{a}_1) \boxtimes \TC(\ttt{a}_2)$ can be performed both ways (reversibly) if the color code bosons $\ttt{a}_1$, $\ttt{a}_2$ are of different colors and either $(X,X)$, $(Y,Y)$, $(X,Y)$ or $(Y,X)$ flavors. We will also consider transitions between the child theories $\TC(\ttt{a}_1) \boxtimes \TC(\ttt{a}_2) \leftrightarrow \TC(\ttt{a}'_1) \boxtimes \TC(\ttt{a}'_2)$. In this case, the transitions are independent within each layer and follow the rules of the Floquet color code~\cite{brown_2022}. Namely, the bosons $\ttt{a}_1$ and $\ttt{a}'_1$, and similarly $\ttt{a}_2$ and $\ttt{a}'_2$, have to braid nontrivially. 

\begin{figure} \centering
\vspace{0pt}
\centering
\vspace{0pt}
\includegraphics[width= 0.26\columnwidth]{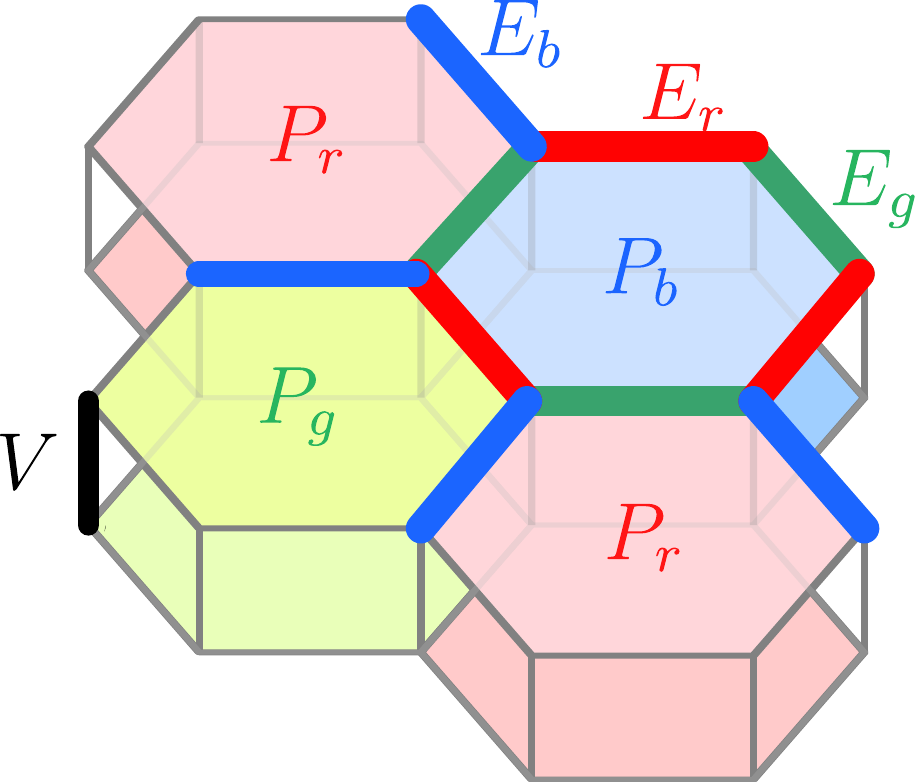}
\caption{A fragment of two layers of the honeycomb lattice displaying our notation convention. The vertex/interlayer terms are labeled $V$. The plaquettes and edges of the lattice are three-colorable and some of the red, green, and blue-colored edges $E_{r,g,b}$ are shown in layer 1. The plaquettes $P_{r,g,b}$ are labeled in the top layer only. }
\label{fig:3a0}
\end{figure}

Thus far, our discussion of the color code has not involved any microscopic model. For the purposes of designing a practical quantum code, we can realize two copies of the color code on a three-colorable graph with two qubits per vertex (two layers). We will restrict our discussion to the honeycomb lattice for simplicity. We will also assume that at the starting point of each measurement, the stabilizer group has already been prepared, and later we will show that the DA color code can also generate the stabilizer group (dynamically generate logical qubits~\cite{Hastings2021}). 

The anyon theories that we discuss above all have a fixed-point lattice Hamiltonian realization on a honeycomb lattice; condensations can be performed in the Hamiltonian picture by adding terms that hop respective anyons and taking the limit where these terms are large. The local terms of these lattice models generate the stabilizer groups of respective topological codes that bear the same name. The terms that are added in the Hamiltonian picture to perform condensations from one model to another correspond to measurements of these operators in the topological code, which achieve transitions between different codes.

\begin{figure}[t] \centering
\vspace{0pt}
\centering
\vspace{0pt}
\includegraphics[width= 0.8\columnwidth]{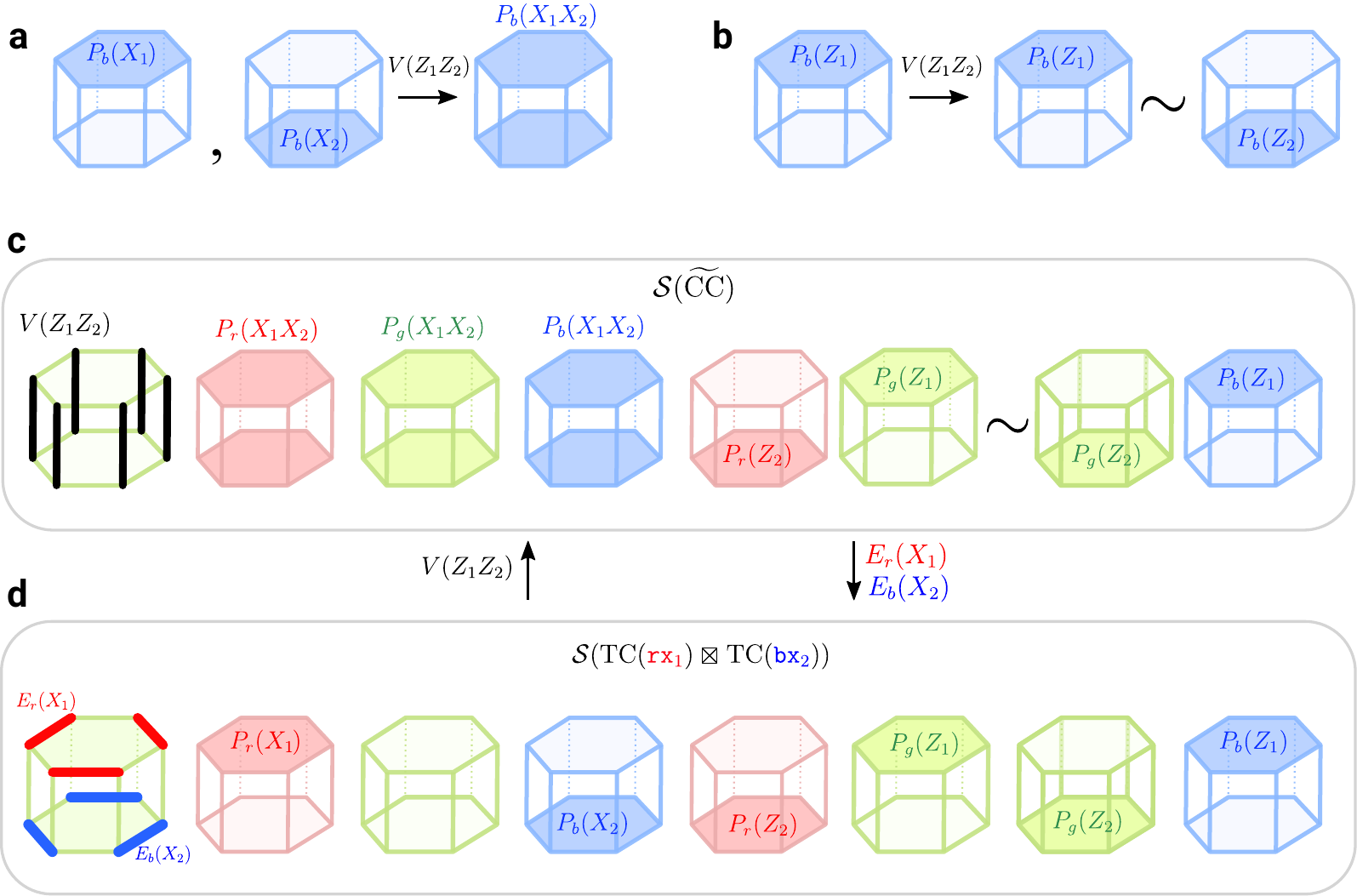}
\caption{(a) The $X$ plaquette stabilizers of two copies of the color code have to be multiplied to become a single stabilizer within the effective $\widetilde{\CC}$ when $V(Z_1 Z_2)$ measurements are performed (only blue plaquette is shown). (b) The $Z$ plaquette stabilizers of two copies of the color code become identified under $V(Z_1 Z_2)$ measurements. (c,d) The transition between the stabilizer generators of the $\mathcal S(\widetilde{\CC})$ and those of the $\mathcal S(\TC(\rxi) \boxtimes \TC(\bxii))$ under respective measurements. The $Z$ - flavored plaquettes in panels (b) and (c, green plaquette) are shown twice but in fact, correspond to equivalent stabilizers up to $V (Z_1 Z_2)$ checks. There are no standalone plaquette stabilizers of $X$ flavor in panel (d) because such stabilizers are contained in edge checks. }
\label{fig:3a1}
\end{figure}

We now discuss the stabilizer groups for the parent and child theories. The stabilizer group of the bilayer color code is a tensor product of stabilizer groups of the color code in each layer $\ell = 1,2$:
\be
\begin{split} \label{eq:S_CCxCC}
\mathcal{S}(\CC \boxtimes \CC) = \langle P_c(X_\ell), P_c(Z_\ell) \rangle_{c \in \{r,g,b\}, \ell \in \{ 1,2 \}}
\end{split}
\ee
First, let us consider the earlier introduced condensation $\CC \boxtimes \CC \rightarrow \widetilde{\CC}$, and show how to find an appropriate measurement that implements a transition between stabilizer groups of respective topological codes, i.e. $\mathcal{S}(\CC \boxtimes \CC) \rightarrow \mathcal S (\widetilde{\CC})$, where the second stabilizer group will be explained shortly. From considering the Hamiltonian picture, one finds that the needed measurement is $Z_1 Z_2$ at each vertex, i.e., a two-qubit measurement on a vertical/interlayer link in the bilayer honeycomb lattice. We denote this measurement by $V( Z_1 Z_2)$, where `$V$' stands for `vertex' or `vertical' (the schematic of the lattice together with our notation is shown in Fig.~\ref{fig:3a0}). Starting with the stabilizer group of the parent code and measuring $V (Z_1 Z_2)$, one arrives at the stabilizer group
\be \label{eq:Ceff}
\begin{split}
\mathcal{S}(\widetilde{\CC}) = \langle V(Z_1Z_2), P_{c}(X_1 X_2), P_{c}(Z_1) \rangle_{c \in \{r,g,b\} },
\end{split}
\ee
where $P_{c}(X_1 X_2)$ stands for a product of $X$-plaquettes of color $c$ in each layer (i.e. a weight-12 plaquette term). 
This stabilizer group is obtained from the parent one shown in Eq.~\eqref{eq:S_CCxCC} by adding the measured operators to it (we avoid the $\pm$ signs that would reflect the measurement outcomes but keep in mind that they have been recorded and are implicitly present throughout), and keeping all the elements of the parent stabilizer group that commute with these measurements. For example, $P_c(X_1)$ and $P_c(X_2)$ plaquettes of the parent stabilizer group had to be multiplied in order to commute with the measured operators, see Fig.~\ref{fig:3a1}(a). The plaquette operators $P_c(Z_1)$ become equivalent to $P_c(Z_2)$ upon multiplying by $V(Z_1Z_2)$ around the plaquette, see Fig.~\ref{fig:3a1}(b). The stabilizer group $\mathcal S(\widetilde {\CC})$ is exactly the color code stabilizer group if we consider each pair of qubits at each vertex as a single effective qubit fused by $V(Z_1Z_2)$ measurements. 
In other terms, each qubit pair forms a [[2,1,1]] stabilizer code defined by a $V(Z_1Z_2)$ stabilizer. The effective logical operators of this code are $X_{\text{eff}} = X_1 X_2 \sim Y_1 Y_2$ and $Z_{\text{eff}} = Z_1 \sim Z_2$. In essence, Eq.~\ref{eq:Ceff} describes the concatenation of [[2,1,1]] codes with the color code.

We remark that there there are 9 different interlayer measurements $V(\sigma_1\sigma_2)$ where $\sigma_j=X,Y,Z$ that we can perform, each achieving different effective color codes (it is also possible to perform a vertex Pauli measurement in a single layer, which will trivialize one of the color code copies and keep another; we do not consider this option here). In the condensation picture, measurement of $V(\sigma_1\sigma_2)$ corresponds to condensing $\ttt{r}\tsig_\ttt{\rc 1}\ttt{r}\tsig_\ttt{\rc 2}$ and $\ttt{b}\tsig_\ttt{\bc 1}\ttt{b}\tsig_\ttt{\bc 2}$. This can be seen by noting that the product of $V(Z_1Z_2)$ on two vertices bordering an edge colored $c$ is equal to $E_c(Z_1)E_c(Z_2)$, which is precisely the hopping operator for $\ttt{c}\tsig_\ttt{\rc 1}\ttt{c}\tsig_\ttt{\rc 2}$. As we emphasized before, in this paper we will only focus on the choice $V(Z_1Z_2)$, which already allows us to design sequences implementing all 72 automorphisms of the color code. Thus, whenever we refer to $\mathcal S(\widetilde{\CC})$, we mean the code obtained by measuring $V(Z_1Z_2)$ at every vertex.

The other type of measurement transition that we consider corresponds to condensing an independent boson in each layer, i.e. $\CC \boxtimes \CC \rightarrow \TC(\rxi) \boxtimes \TC(\bxii)$. To go between the stabilizer groups of respective topological codes, i.e. $\mathcal S(\CC) \boxtimes \mathcal S(\CC) \rightarrow \mathcal S(\TC(\rxi) \boxtimes \TC(\bxii))$, one has to measure two-body checks $\Er( X_1)$ and $\Eb( X_2 )$ on the red edges of the first layer, and blue edges of the second layer, see Fig.~\ref{fig:3a0}. The resulting stabilizer group is
\be \label{eq:TCeff}
\begin{split}
\mathcal{S}(\TC(\rxi) \boxtimes \TC(\bxii)) = &\langle \Er( X_1), \Pr(X_1), \Pg(Z_1),\Pb(Z_1) \rangle
\\
\times &\langle \Eb(X_2), \Pb(X_2), \Pr(Z_2),\Pg(Z_2) \rangle,
\end{split}
\ee
which is similarly obtained by adding the measured operators and keeping only stabilizers of the parent code that commute with these. We can view the measurements $\Er(X_1)$ and $\Eb(X_1)$ as defining [[2,1,1]] codes on the red and blue edges of the respective layers. 
Using the logical qubits of the [[2,1,1]] codes as effective qubits, the stabilizer group \eqref{eq:TCeff} defines two toric codes on triangular superlattices with vertices centered at red and blue plaquettes of the honeycomb lattice respectively~\cite{Hastings2021}. The transition $\CC \boxtimes \CC \rightarrow \TC(\ttt{a}_1) \boxtimes \TC(\ttt{a}_2)$ for other colors and flavors of condensed bosons can be worked out analogously. 

Similarly to reversible transitions between the child anyon theories in the condensation picture, we can define locally reversible transitions by measurements. Reversible transitions conserve logical information and preserve the rank of the stabilizer group~\cite{Aasen2023}. 
Conservation of logical information means that at each measurement round, there exists a complete set of representatives of logical operators (complete set of logical strings) that also commutes with the next round and thus survives to that round. In all the examples considered in this paper, reversible condensations are translated into reversible measurements. For example, in order for one to go in the direction $\mathcal S (\widetilde{\CC}) \rightarrow \mathcal{S}(\TC(\ttt{c}_\ttt1 \sigma_\ttt{1}) \times \TC(\ttt{c}_\ttt2 \sigma_\ttt{2}))$, it is necessary that the measurements $E_{c_1}(\sigma_1)$ and $E_{c_2}(\sigma_2)$ anticommute with $V (Z_1 Z_2)$\footnote{In terms of anyons, this is equivalent to the condition that $\ttt{c}_\ttt1\tsig_\ttt{1}$ and $\ttt{c}_\ttt2\tsig_\ttt{2}$ must braid non-trivially with $\rzi\rzii$ and $\bzi\bzii$.}, which means that $\sigma_1$ and $\sigma_2$ must be $X$ or $Y$ and $c_1 \neq c_2$. This tells us that there exist $24 = 6 \times 4$ ways to turn the given effective color code into two copies of the toric code. An example of the stabilizer updates for a specific choice of transition between $\widetilde{\CC}$ and $\TC(\rxi) \times \TC(\bxii)$ is shown in Fig.~\ref{fig:3a1}(c,d).

The effect of the measurements that perform $\mathcal S (\widetilde{\CC}) \rightarrow \mathcal{S}(\TC(\ttt{c}_\ttt1 \sigma_\ttt{1}) \times \TC(\ttt{c}_\ttt2 \sigma_\ttt{2}))$ is equivalent to the action of an unfolding unitary. We will denote such an unfolding unitary by $\hat U_{\ttt{c}_\ttt1\tsig_\ttt{1},\ttt{c}_\ttt2\tsig_\ttt{2}}$. 
Note that this operator is different from the conventional unfolding unitary between a color code and two copies of the toric code~\cite{Kubica_2015} because of the double layer structure needed for our measurement implementation and the fact that the unitary is induced by a locally reversible measurement. 
We can construct this unitary explicitly using the prescription of Ref.~\cite{Aasen2023}. 
Consider, for concreteness, an example $\widetilde{\CC} \rightarrow \TC(\rxi) \times \TC(\bxii)$. For each green (i.e. a complementary color) plaquette, let us enumerate the vertices clockwise as numbers 1 through 6. It is possible to find a basis for the measurements that have support on the $i^\text{th}$ green plaquette such that they form conjugate pairs, namely $A_i = \{ A_{i,j}\}$ and $B_i = \{B_{i,j}\} $ for $j=1,\ldots,5$ are such that $A_{i,j} B_{i',j'} = (-1)^{\delta_{ii'} \delta_{jj'}}B_{i',j'}A_{i,j}$\footnote{Concretely, for the $i^\text{th}$ green plaquette and the $k^\text{th}$ qubit around that plaquette, denote the Pauli operators $\sigma^{(i,k)}_\ell$ where $\sigma=X,Y,Z$ and $\ell=1,2$ is the layer index. Then, we may choose
\begin{align*}
 A_{i,j} &= \pm \prod_{k=1}^j Z_1^{(i,k)}Z_2^{(i,k)}, & B_{i,j} &= \pm X_{(3+(-1)^j)/2}^{(i,j)} X_{(3+(-1)^j)/2}^{(i,j+1)},
\end{align*}
where $\pm$ signs accommodate the random measurement outcomes for each operator such the resulting $A$($B$) type operator becomes a stabilizer.}.
Then the unfolding unitary can be written as 
\be
\hat U_{\rxi,\bxii} = \bigotimes _i \bigotimes_{j =1}^5 \frac{A_{i,j} + B_{i,j}}{\sqrt{2}}.
\ee
Conversely, performing the measurement $V(Z_1Z_2)$ ``folds'' the two copies of toric code back into the color code $\widetilde{\CC}$, and is equivalent to the action of the unitary $\hat U^{-1}_{\rxi,\bxii}$.

To conclude, we have shown how to perform reversible transitions between the stabilizer groups of the effective color code and two decoupled copies of the toric code. In the next subsection, we will show that this tool, together with honeycomb-type sequences in the decoupled toric code layers, is sufficient to generate all 72 automorphisms of the color code.

\subsection{Measurement protocols and automorphisms} \label{sec:II_automorphisms}

Assume that we start with an effective color code anyon theory $\widetilde {\text{CC}}$ (obtained as an appropriate condensation from the parent bilayer color code) and perform a reversible condensation into a pair of toric codes. We can then evolve each of these toric codes by reversible condensations that are similar in spirit to that of two individual honeycomb or CSS honeycomb codes (for which it is sufficient to condense a boson in each new round that braids nontrivially to that of the previous round), and finally arrive at a different pair of toric codes that can be folded by measurement into the effective color code $\widetilde {\text{CC}}$. The total sequence forms a cycle that can be summarized as:
\be
\begin{matrix}
t=0 & & t=1 & &... & & t=T-1 & & t=T\\ 
\widetilde{\CC} & \rightarrow & \TC(\ttt{c}_\ttt1\tsig_\ttt{1}) \boxtimes \TC(\ttt{c}_\ttt2\tsig_\ttt{2}) & \rightarrow &... & \rightarrow & \TC(\ttt{c}_\ttt1'\tsig_\ttt{1}') \boxtimes \TC(\ttt{c}_\ttt2'\tsig_\ttt{2}') & \rightarrow & \widetilde {\CC}
\end{matrix}
\ee
As a consequence of such a closed loop, the final theory can be related to the initial one by an automorphism. Thus, by appropriate choices of condensation paths, one might be able to controllably implement the automorphisms of the color code. We show that such sequences exist, and by choosing only those sequences that translate into weight-2 measurements, it is possible to achieve all 72 automorphisms of the color code with pairwise measurements.

\begin{figure}[t] \centering
\vspace{0pt}
\centering
\vspace{0pt}
\includegraphics[width= 0.5\columnwidth]{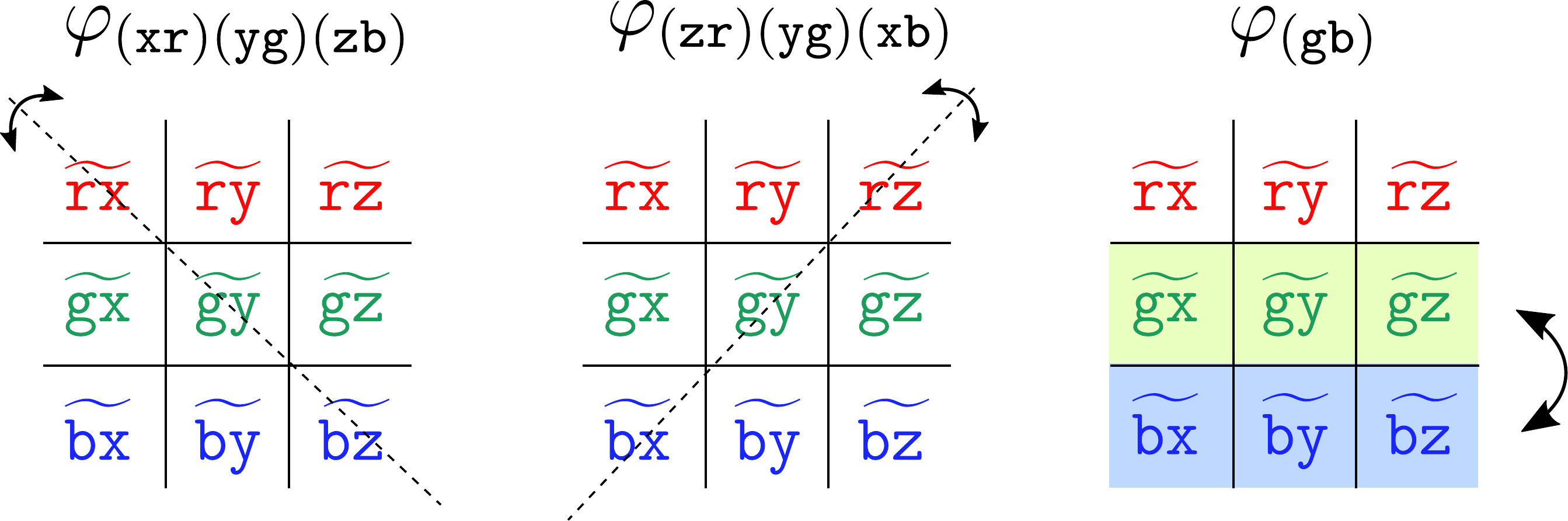}
 \caption{ A choice of three generating symmetries of the magic square, two mirror symmetries $\varphi_{(\texttt{xr})(\texttt{yg})(\texttt{zb})}$ and $\varphi_{(\texttt{zr})(\texttt{yg})(\texttt{xb})}$, and the row permutation $\varphi_{(\texttt{gb})}$ that generate all automorphisms of the color code. }
\label{fig:3b}
\end{figure}

\begin{table}[!b]
{
\renewcommand{\rc}[1]{{\color{red}{#1}}}
\renewcommand{\bc}[1]{{\color{blue}{#1}}}
\renewcommand{\gc}[1]{{\color{ForestGreen}{#1}}}
\renewcommand{\yc}[1]{{\color{amber}{#1}}}
 \centering
 \resizebox{\textwidth}{!}{\begin{tabular}{|c|c|c|c|c|c|c| c| }
 \hline
 \multirow{2}{*}{Aut/Gate} & \multirow{2}{*}{Cond./meas.} & \multicolumn{6}{c|}{Sequence}  \\ \cline{3-8} 
 & & $t=$0 & 1 & 2 & 3 & 4 & 5 \\
 \hline
 \multirow{2}{*}{$\varphi_{(\texttt{xr})(\texttt{yg})(\texttt{zb})}$} & \multirow{2}{*}{cond.} & $\rzi \rzii$ & $\rxi$ & $\gyi$ & $\bzi$ & $\rxi$ & $\rzi \rzii$ \\
 & & $\bzi \bzii$ & $\bxii$ & $-$ & $\gzii$ & $\bxii$ & $\bzi \bzii$ \\
 \hline
 \multirow{2}{*}{$ (H_1 \otimes H_3) \text{SWAP}_{13}$} & \multirow{2}{*}{ meas.} &\multirow{2}{*}{ $V(Z_1Z_2)$} & $\Er(X_1)$ & $\Eg(Y_1)$ & $\Eb(Z_1)$ & $\Er(X_1)$ & \multirow{2}{*}{ $V(Z_1Z_2)$} \\
 & & & $\Eb(X_2)$ & $-$ & $\Eg(Z_2)$ & $\Eb(X_2)$ & \\
 \hline \hline
 \multirow{2}{*}{$\varphi_{(\texttt{zr})(\texttt{yg})(\texttt{xb})}$} & \multirow{2}{*}{cond.} & $\rzi \rzii$ & $\rxi$ & $-$ & $\bzi$ & $\rxi$ & $\rzi \rzii$ \\
 & & $\bzi \bzii$ & $\bxii$ & $\ryii$ & $\gzii$ & $\bxii$ & $\bzi \bzii$ \\
 \hline
 \multirow{2}{*}{$ (H_2 \otimes H_4) \text{SWAP}_{24}$} & \multirow{2}{*}{meas.} & \multirow{2}{*}{ $V(Z_1Z_2)$} & $\Er(X_1)$ & $-$ & $\Eb(Z_1)$ & $\Er(X_1)$ 
 & \multirow{2}{*}{ $V(Z_1Z_2)$} \\
 & & & $\Eb(X_2)$ & $\Er(Y_2)$ & $\Eg(Z_2)$ & $\Eb(X_2)$& \\
 \hline \hline
 \multirow{2}{*}{$\varphi_{(\texttt{gb})}$} & \multirow{2}{*}{cond.} & $\rzi \rzii$ & $\rxi$ & $\bzi$ & $\rxi$ & $\rzi \rzii$ & $-$ \\
 & & $\bzi \bzii$ & $\bxii$ & $\ryii$ & $\gxii$ & $\bzi \bzii$ & $-$ \\
 \hline
 \multirow{2}{*}{$\text{CNOT}_{12}\text{CNOT}_{34}$} & \multirow{2}{*}{meas.} & \multirow{2}{*}{ $V(Z_1Z_2)$} & $\Er(X_1)$ & $\Eb(Z_1) $ & $\Er(X_1)$ & \multirow{2}{*}{ $V(Z_1Z_2)$} & $-$ \\
 & & & $\Eb(X_2)$ & $\Er(Y_2)$ & $\Eg(X_2)$ & & $-$ \\
 \hline
 \end{tabular}}
 }
 \caption{Condensation sequences for the generators of the 72 automorphisms the color code and respective measurement sequences derived from them. The first column also displays the gate acting on four logical qubits of the effective code on the torus. The `$-$' stands for nothing being performed in a given round.}
 \label{tab:generators}
\end{table}

As discussed in Subsec.~\ref{sec:II_intro}, we can translate condensation sequences to measurement sequences on the topological codes. Because we limit ourselves to two-qubit measurements, there are only 24 choices for the initial unfolding and similarly 24 choices for the final folding measurement rounds, which were summarized in the previous subsection.
The states of the initial (at round $t=1$) and final (at round $t=T-1$) toric code pairs are related by an isomorphism that can be written as a unitary $\hat U_{\text{seq}}$. The total action of the measurement sequence that occurred between two color code steps is equivalent to an overall application of a unitary:
\be
\hat U =\hat U _{\ttt{c}_\ttt1'\tsig_\ttt{1}',\ttt{c}_\ttt2'\tsig_\ttt{2}'} ^{\dagger} \hat U_{\text{seq}} \hat U_{\ttt{c}_\ttt1\tsig_\ttt{1},\ttt{c}_\ttt2\tsig_\ttt{2}}.
\ee

Thus, as a consequence of performing such a measurement sequence, a unitary $\hat U$ corresponding to some automorphism $\varphi \in \texttt{Aut}(\widetilde {\CC})$ will be applied to the color code.

\begin{table}[!b]
\vspace{0pt}
\centering
\vspace{0pt}
\includegraphics[width= 0.9\columnwidth]{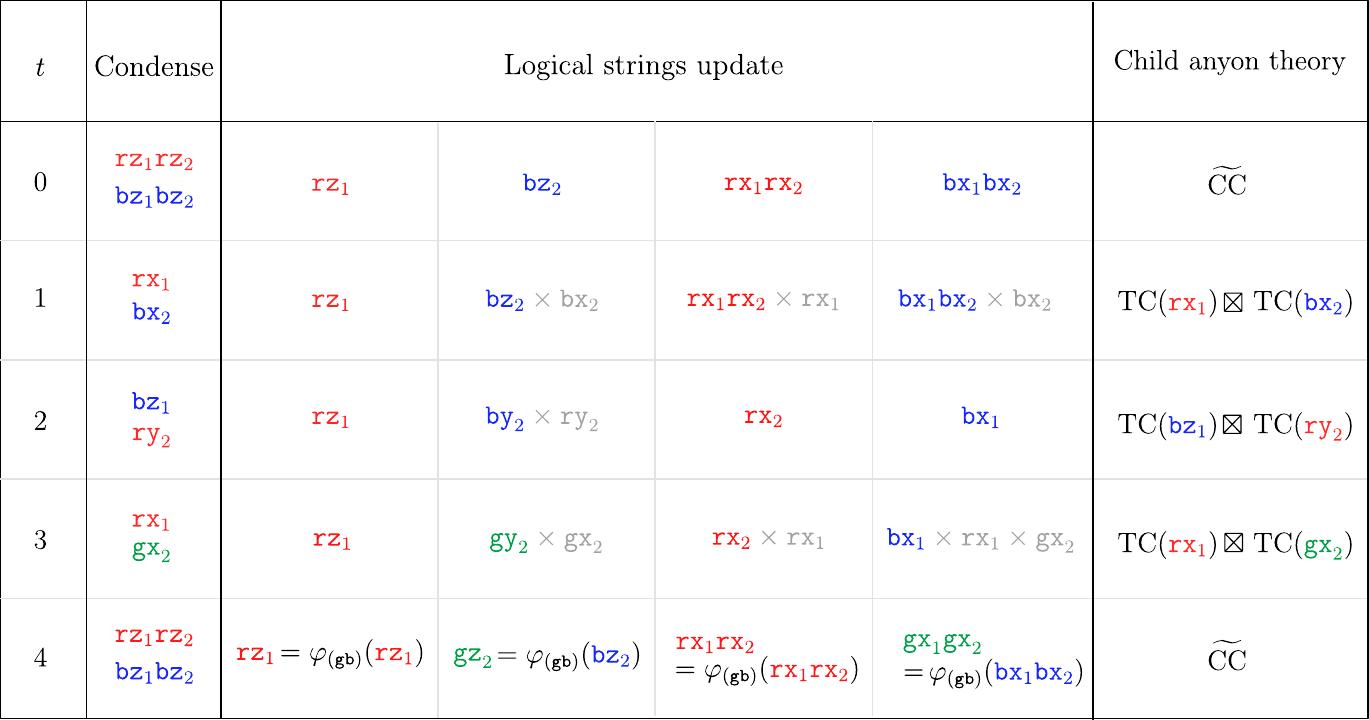}
\caption{Summary of the condensation protocol implementing $\varphi_{(\texttt{gb})}$ automorphism from Table~\ref{tab:generators}. The update of each type of logical string is shown in the respective column, each cell containing one representative from one class of the logical string. Whenever this representative needs to be fused with the anyon(s) condensed in the current round in order to remain a logical string in the next round (i.e. to braid trivially with the next condensation), such a fusion is shown in gray. Finally, upon completing the sequence, we identify the automorphism of the color code that was applied to the logical strings. The last column shows the child anyon theory at each round.
 }
\label{fig:3b2}
\end{table}

Let us first demonstrate how to produce the three automorphism generators, from which any other automorphism can be obtained as their product. The group of automorphisms of the color code can be conveniently depicted via its action on the magic square of the color code. In particular, all symmetries that preserve the braiding and fusion data can be generated by row permutations (dubbed `color symmetries'), column permutations (`Pauli flavor symmetries'), and diagonal reflections (`color-Pauli flavor symmetry'). We choose the three generators of the symmetries to be the two diagonal reflections $\varphi_{(\texttt{xr})(\texttt{yg})(\texttt{zb})}$ and $\varphi_{(\texttt{zr})(\texttt{yg})(\texttt{xb})}$, and row permutation $\varphi_{(\texttt{gb})}$, where the subscript denotes the permutation cycles of the rows and columns. One way to see that the above automorphisms generate all other automorphisms is to transcribe the automorphisms of $\widetilde{\CC}$ shown in Fig.~\ref{fig:3b} in terms of two copies of the toric code according to Eq.~\eqref{eq:magicsquare}. Then we immediately see that the automorphism can be thought of as the following transformations on the anyons
\begin{equation}
\begin{split}
 \varphi_{(\texttt{xr})(\texttt{yg})(\texttt{zb})}&: \ \ \ \ \ \ \ \ei\leftrightarrow \mi, \ \ie \leftrightarrow \ie, \im \leftrightarrow \im \\
 \varphi_{(\texttt{zr})(\texttt{yg})(\texttt{xb})}&: \ \ \ \ \ \ \ \ie \leftrightarrow \im,\ \ei \leftrightarrow \ei, \ \mi \leftrightarrow \mi \\
 \varphi_{(\texttt{gb})}&: \ \ \ \ \ \ \ \mi \leftrightarrow \mm, \ \ie \leftrightarrow \EE \ 
 \end{split}
\end{equation}
which generates a group of order 72. We keep track of the action of each automorphism in its subscript, where the colors and Pauli flavors that are changed are kept track of in a cycle notation.

There are many ways to realize the same automorphism as a sequence of condensations. We choose the sequences for the automorphism generators that, when translated into measurement sequences, have the nicest properties for the error correction (which is addressed in Sec.~\ref{sec:EC}), and are therefore the most practically useful. These are presented in Table~\ref{tab:generators}. The sequences for $\varphi_{(\texttt{xr})(\texttt{yg})(\texttt{zb})}$ and $\varphi_{(\texttt{zr})(\texttt{yg})(\texttt{xb})}$ take 5 rounds and the $ \varphi_{(\texttt{gb})}$ sequence takes 4 rounds.

\begin{figure}[!b] \centering
\vspace{0pt}
\centering
\vspace{0pt}
\includegraphics[width= 1\columnwidth]{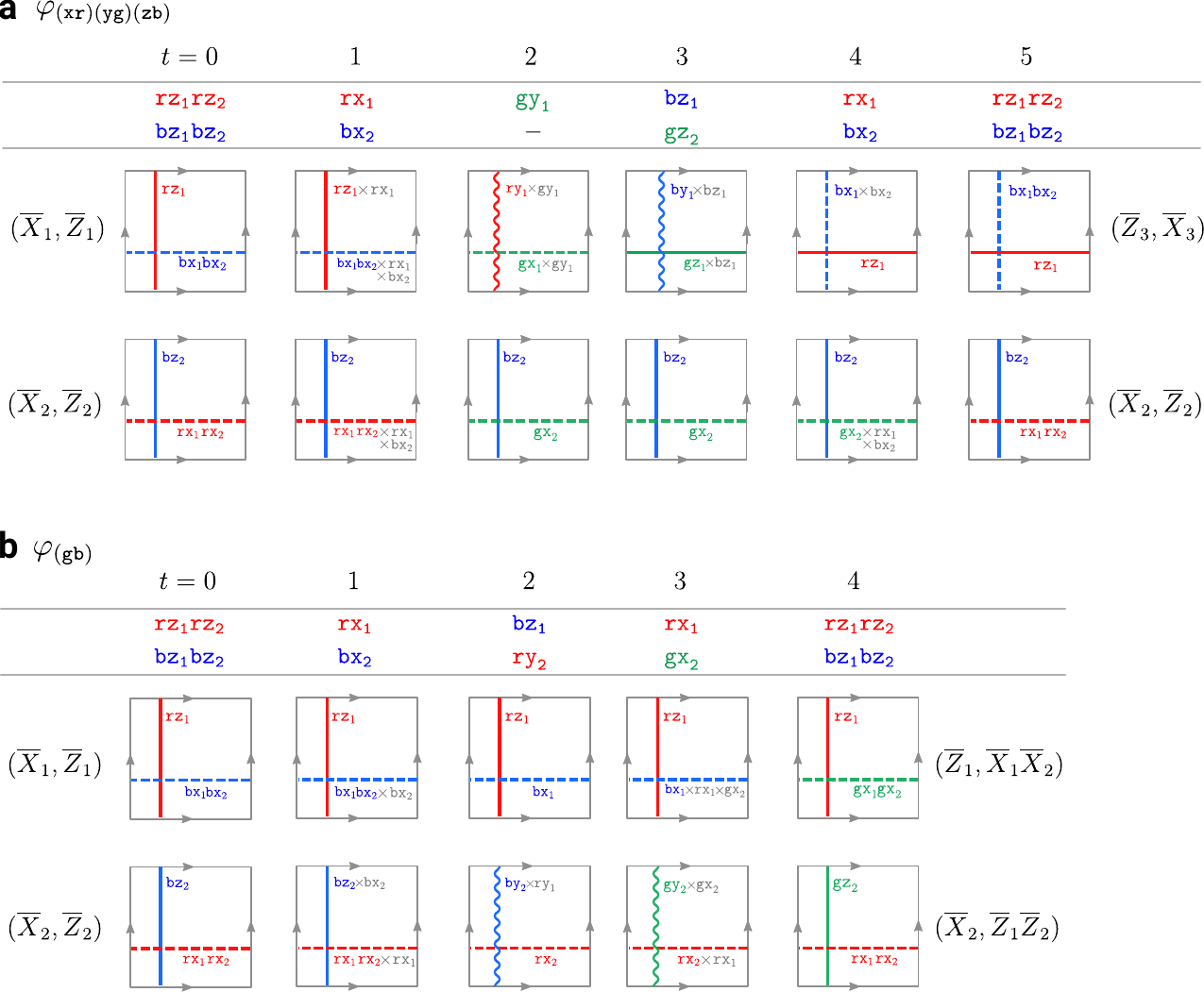}
\caption{The update of the logical strings throughout rounds of condensations. An instance of each string is passed from round to round which is shown in panels (a) and (b) for $\varphi_{(\texttt{xr})(\texttt{yg})(\texttt{zb})}$ and $\varphi_{(\texttt{gb})}$ automorphisms, respectively. When the logical operator has to be multiplied by the current round's checks, the multiplication is shown in gray. By comparing the logical strings after a period of measurements we are able to identify the logical gate that the measurement sequence implements. The remaining $\varphi_{(\texttt{xr})(\texttt{yg})(\texttt{zb})}$ automorphism can be understood by analogy with $\varphi_{(\texttt{xr})(\texttt{yg})(\texttt{zb})}$.}
\label{fig:3b22}
\end{figure}

Let us start by examining how one of the above condensation sequences works: in particular, the sequence $\varphi_{(\texttt{gb})}$. This is summarized in Table~\ref{fig:3b2}. For each deconfined anyon in the child theory (which generates logical strings), we show a representative from its equivalence class in a given column at each step. We track the anyons by labeling them in the parent theory language. As a reminder, the equivalence class is defined by fusion with the bosons condensed at that round. Because the protocols that we choose have the property that each pair of consecutive condensations is reversible by design, it is always possible to find a representative at a given round that will braid trivially with the condensation of the next round. This is the requirement for conserving a full set of logical strings throughout the measurement sequence. If a representative shown in the current round has to be multiplied by a condensed boson in order to braid trivially with the condensation of the next round, it is shown in gray in Table~\ref{fig:3b2}. For example, let us follow the $\bxi \bxii$ anyon of round $t=0$. It braids trivially with the condensations of round $t=1$, and thus, it is taken to the next round as is. However, it does not braid trivially with one of the condensations in round 2, which is fixed by fusion with the trivial logical $\bxii$ at round $t=1$ producing a different representative $\bxi \bxii \times \bxii \simeq \bxi$. This anyon braids trivially with the condensation of the next round, $t=3$, and thus is taken to that round unchanged. Lastly, in order to be taken to the last round of the sequence, $t=4$ that completes the cycle, we need to fuse it with both condensations of round 3, obtaining a $\bxi \times \rxi \times \gxii \simeq \gxi \gxii$ anyon at round 4. Thus, the overall change of this anyon consists of being multiplied by some of the condensed objects over the course of the condensation sequence. One can explicitly check that the total outcome corresponds to applying the $\varphi_{(\ttt{gb})}$ automorphism to all the anyons in the theory. The update rules are explained in much more depth in Sec.~\ref{sec:tqft_sec}. One can also verify that the shown representatives in the table indeed form a basis of anyons for the child theory at each round, which is shown in the last column of the table.

The transformation of the logical strings according to these update rules is also shown pictorially in Fig.~\ref{fig:3b22} for two of the generating automorphisms, $\varphi_{(\texttt{xr})(\texttt{yg})(\texttt{zb})}$ and $ \varphi_{(\texttt{gb})}$. The third automorphism is not shown because it is analogous to the first one up to an appropriate change in colors and layers.

\begin{table}[!t]
\vspace{0pt}
\centering
\vspace{0pt}
\includegraphics[width= 0.5\columnwidth]{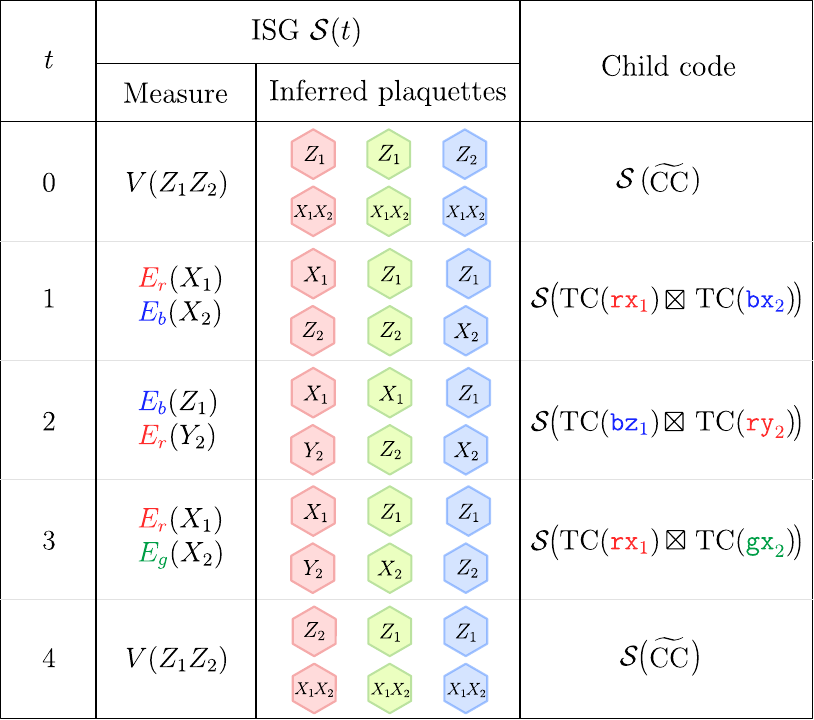}
\caption{(a) 
Evolution of the effective code and its logical strings throughout the measurement sequence for the $\varphi_{(\texttt{gb})}$ automorphism. The table shows the measurements used to obtain respective child code and the plaquette part of the ISG at each measurement round. The plaquette stabilizers are red, green, and blue colored hexagons, and the flavor for each plaquette is indicated inside of it.   
}
\label{fig:3b3}
\end{table}

These condensation sequences can be used to design measurement sequences that induce a related evolution in topological codes, which are shown in Table~\ref{tab:generators}. Let us analyze an example corresponding to the $\varphi_{(\ttt{gb})}$ automorphism, and understand the measurement sequence in more detail, which is shown in Table~\ref{fig:3b3}. The table shows the measurement sequence and the evolution of the ISG throughout it indicating the instantaneous stabilizer group at each step. 
We assume that we have initialized the system with an effective color code ISG (see Eq.~\ref{eq:Ceff}) at $t=0$. 
The ISG is updated from round to round according to Fig.~\ref{fig:3a1}. Namely, a plaquette in the ISG at round $r$ either commutes with all the measurements of round $r+1$ (in which case it is taken to the next round) or anticommutes with some of the measurements at round $r+1$. 
In the latter case, the plaquette needs updating. A new plaquette is formed by multiplying the plaquette at round $r$ by the check operators of round $r$ so that it commutes with the measurements of the next round. 
The local update rule is guaranteed by the local reversibility of the measurement paths. 
The effective ISGs at intermediate rounds are equivalent to that shown in Eq.~\eqref{eq:TCeff}. The label $Z_1/Z_2$ on some of the plaquettes means that both $Z_1$ or $Z_2$ plaquettes are in the ISG but they are equivalent up to $V( Z_1Z_2)$ checks. During the next round, both $Z_1$ and $Z_2$ plaquettes become independent elements of the ISG.

\begin{figure}[t] \centering
\vspace{0pt}
\centering
\vspace{0pt}
\includegraphics[width= 0.65\columnwidth]{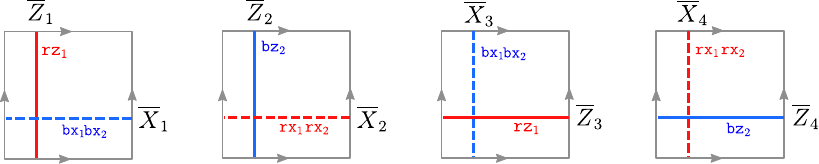}
\caption{Our convention for the correspondence between the anyon strings and the logical Pauli operators for 4 logical qubits that they generate on a torus. }
\label{fig:3b21}
\end{figure}

The evolution of the logical strings\footnote{As a reminder, a logical string $\mathcal{L}(\ttt{c} \tsig)$ is a path on edges of color $\ttt{c}$ where $\tsig$ is applied to each vertex on the path. } of the codes throughout the measurement sequence completely mirrors that of the logical strings during the condensation sequence shown in Table~\ref{fig:3b2}, except that logical strings become a Pauli string of corresponding flavor and color, and the fusion with condensed anyons is replaced with multiplication by measurements of the current round. 
As a consequence, the logical strings undergo a transformation corresponding to the automorphism implemented by the measurement sequence. Each such action can be translated into a logical gate once we adopt a basis for the logical qubits, and is different depending on the manifold and the boundary conditions. We show our convention for the torus geometry in Fig.~\ref{fig:3b21}. This way we obtain that the generating automorphisms correspond to the following gates:
\begin{align}
 \varphi_{(\texttt{xr})(\texttt{yg})(\texttt{zb})} &\to (H_1 \otimes H_3)\text{SWAP}_{13} \nonumber \\
 \varphi_{(\texttt{zr})(\texttt{yg})(\texttt{xb})} &\to (H_2 \otimes H_4)\text{SWAP}_{24} \nonumber \\
 \varphi_{(\texttt{gb})} &\to \text{CNOT}_{12}\text{CNOT}_{34}
\end{align}
which are precisely the gates indicated in Table~\ref{tab:generators} for the toric geometry. 

Finally, it is possible to achieve not only the three generators shown above but also all 72 automorphisms by short sequences of length 5 or less rounds. Throughout the paper, for various purposes (such as finding the best sequences for a system with boundaries and error correction), we will still need to introduce other sequences that have different lengths that achieve the same automorphisms. We provide exhaustive tables presenting examples showing the implementation of automorphisms of the color code by short measurement sequences in Appendix~\ref{app:72}.

\section{{Error correction for dynamic automorphism codes}}\label{sec:EC}

In this section, we address error correction in the DA color code on a torus. This requires us to further develop the method of spacetime detectors for these codes, and construct a basis of detectors for sequences implementing the generators of the automorphism group. 
We also conjecture that the code has an extensive distance and a threshold for an appropriately generalized minimum-weight perfect-matching decoder (see discussion in the last subsection). We discuss the error correction of planar dynamic automorphism codes in Sec.~\ref{sec:EC_triangle}.

The rest of this subsection is structured as follows. 
In Subsec.~\ref{sec:EC_padding} we explain how a given set of generating automorphisms can be padded such that the total set of realized automorphisms is unchanged, but the measurement sequence becomes manifestly error-correcting. We do so by introducing additional ``padding'' sequences that insert an identity automorphism between each automorphism that is implemented by the code. In Subsec.~\ref{sec:EC_dyn_gen} we explain in which sense our codes dynamically generate logical qubits. In Subsec.~\ref{sec:EC_error_basis}, we introduce a simplified error basis which does not affect the generality of our claims but simplifies the analysis, and we assume that this basis is used thereafter. Subsec.~\ref{sec:EC_honeycomb_detectors} discusses simple examples of spacetime detectors which are used for error correction in the honeycomb code. In Subsec.~\ref{sec:EC_detectors}, we introduce the basis of detectors for the DA color code and explain how they work in detail. We also show how any single-qubit Pauli error is detected and corrected. Finally, in Subsec.~\ref{sec:EC_decoder}, we discuss the question of fault-tolerance of the DA color code.

\subsection{Padding sequences} \label{sec:EC_padding}

The DA color code is capable of implementing an arbitrary sequence of automorphisms of the color code. This is naturally accomplished by combining measurement sequences corresponding to each automorphism in the sequence, and we assume that the automorphisms are broken down to products of generators (summarized in Table~\ref{tab:generators}). 
On a torus, these automorphisms do not furnish the entire Clifford group but a certain subgroup isomorphic to $(S_3 \times S_3) \rtimes \mathbb{Z}_2$. However, considering the DA color code on a torus is useful in order to set the stage for the next section where we construct such codes first on a triangle (which encodes a single logical qubit) and then on a pair of triangles, which allows us to realize automorphisms corresponding to a full Clifford group on two logical qubits. Before doing so, we would like to ask whether a DA color code implementing a given sequence of automorphisms on a torus can be designed to be error-correcting. As we will see, constructing error-correcting sequences will require some new techniques that we will introduce in this section, namely inserting ``padding'' sequences and detectors for the DA color code.

\begin{figure}[b] \centering
\vspace{0pt}
\centering
\vspace{0pt}
\includegraphics[width= 0.9\columnwidth]{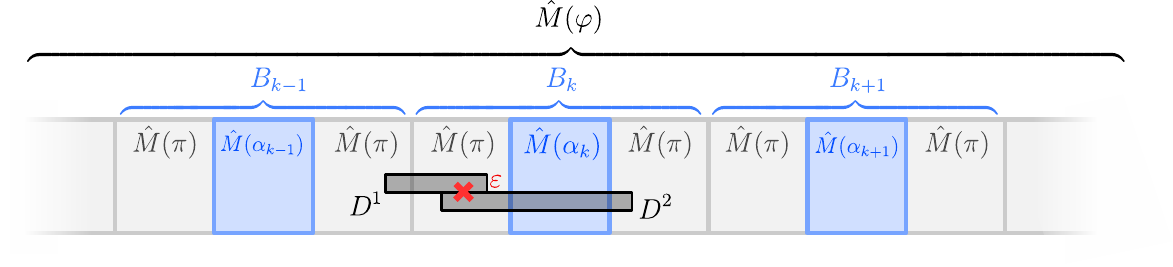}
\caption{A schematic of the padding procedure for the measurement sequence $\hat{M}(\varphi)$, where padding sequences $\hat{M}(\pi)$ (gray boxes) are applied between consecutive automorphisms $\hat{M}(\alpha_k)$ (blue boxes) that together realizing the desired quantum circuit. The red cross shows an arbitrary error that occurred at some location in spacetime. The temporal support of two detectors $D^{1,2}$ (to be defined more rigorously later on) that light up in response to this error is schematically shown as dark rectangles. 
The block operators $B_{k}$ discussed in the text are labeled. As we find, all the detectors in our basis are contained either within one block $B_k$ or within two neighboring paddings $\hat{M}(\pi)\circ \hat{M}(\pi)$.}
\label{fig:3c}
\end{figure}

In our discussion, we will draw a parallel between error correction in our codes and the known error correction protocols for the honeycomb and the CSS honeycomb codes. The known decoding scheme for both of these codes is somewhat similar to that of a toric code, which is seen from their decoding graph upon appropriately rotating it in spacetime. As a constituent step in error correction, one uses four measurement rounds in the honeycomb code and two in CSS honeycomb code to form a detector. 
 {\it Spacetime detectors}, colloquially speaking, are collections of measurements whose product has a fixed value in the absence of any errors, see Ref.~\cite{Delfosse2023} and references therein. These are generally spacetime objects, involving measurements at different locations and different rounds, which allow us to locate spacetime locations of errors; in what follows we refer to them as simply ``detectors''.

In the honeycomb and CSS honeycomb codes, the detectors have a simple form. The measurements forming the detectors can be combined into inferred plaquettes, a product of which at different times should give a constant value. The detectors of different colors (marked by the colors of the plaquette in their spatial support) and different Pauli flavors are identical in shape, albeit shifted in space and time. 
Their decoding graphs, derived from a spacetime lattice of detectors, are similar to the decoding graph of a regular toric code. The decoding graphs for even and odd timesteps are decoupled, and each of them is a rotated cubic lattice. For each single-qubit Pauli error a pair of detectors `light up' (i.e. the value of the detector flips its sign), which allows one to detect the spacetime location of the edge where the error occurred as well as its flavor. This is sufficient information to detect and correct a single-qubit error (this is discussed in more detail later on). For the depolarizing noise error model, the minimum-weight matching decoder utilizes information from lit-up detectors to apply a corrective operation mapping a set of spacetime error chains to a set of homologically trivial (and thus inconsequential) loops. Using the properties of the toric code decoder, both the honeycomb code and the CSS honeycomb code possess an error threshold against depolarizing noise~\cite{Gidney2021faulttolerant,Paetznick2023,Gidney2022benchmarkingplanar,brown_2022}.

In contrast to the honeycomb codes, the measurement sequences appearing in the DA color code are more complicated, and moreover, are generically not periodic as they can represent generic compilations of logical Clifford operations. 
As a result, we will need to introduce a few additional techniques. 
The first is a trick of introducing {\it padding}, which corresponds to inserting an additional measurement sequence between every two automorphisms implemented within the DA color code. 
This additional sequence implements the trivial automorphism, and thus, does not affect the operation of the DA color code, but allows for a simpler structure of detectors. The sequence with a padding is shown schematically in Fig.~\ref{fig:3c}.

\begin{table}[!t]
{\renewcommand{\rc}[1]{{\color{red}{#1}}}
\renewcommand{\bc}[1]{{\color{blue}{#1}}}
\renewcommand{\gc}[1]{{\color{ForestGreen}{#1}}}
\renewcommand{\yc}[1]{{\color{amber}{#1}}}
 \centering
 \begin{tabular}{|c|c|c|c|c|c|c|}
 \hline
 \multirow{2}{*}{Aut} & \multicolumn{6}{c|}{Sequence}  \\ \cline{2-7} 
 & $t=$0 & 1 & 2 & 3 & 4 & 5\\
 \hline
 \multirow{2}{*}{$\varphi_{(\texttt{rb})(\texttt{xz})}$} & \multirow{2}{*}{ $V(Z_1Z_2)$} & $\Er(X_1)$ & $\Eg(Y_1)$ & $\Eb(Z_1)$ & $\Er(X_1) $ & \multirow{2}{*}{ $V(Z_1Z_2)$} \\
 & & $\Eb(X_1)$ & $\Er(Y_2) $ & $\Eg(Z_2)$ & $\Eb (X_2)$ & \\
 \hline
 \end{tabular}
 \caption{The measurement sequence applying the $\pi = \varphi_{(\texttt{rb})(\texttt{xz})}$ automorphism. We insert this sequence twice, which realizes ``padding'' in all DA color code protocols. As a result, the detectors are easier to analyse. }
 \label{tab:padding}
}
\end{table}

There are many sequences that one might choose for the padding, and the one implementing an identity gate/automorphism $\mathbb I \otimes \mathbb I$ is the most natural one. We can choose a special sequence that implements the automorphism $\pi= \varphi_{(\texttt{rb})(\texttt{xz})}$ shown in Table~\ref{tab:padding}, and insert it twice in a row, which amounts to a trivial automorphism $\pi \circ \pi = \mathbb I$. 
The rounds between toric code ISG steps resemble a honeycomb code protocol on each toric code layer, which makes this measurement sequence especially simple. In what follows, we refer to each sequence realizing $\pi$ as a {\it padding sequence}.

Given an automorphism $\varphi \in \texttt{Aut}(\widetilde{\CC})$, we wish to implement, we can always decompose it into a sequence of the generators of the automorphism group listed in Table~\ref{tab:generators}. We denote this decomposition $\varphi = \prod_{k} \alpha_{k}$, where each $\alpha_k$ corresponds to one of the generators of $\texttt{Aut}(\widetilde{\CC})$. We denote $\hat M(\varphi)$ to be a measurement sequence implementing $\varphi$. 
Upon inserting padding sequences the automorphism becomes $\varphi = \prod_{k} \alpha_{k} = \left [ \prod_{k} \left (\pi \right )^2 \alpha_{k} \right] \left (\pi\right )^2 $, which can be used to generate the measurement sequence
\begin{equation} \label{eq:padded}
\begin{split}
 \hat{M}(\varphi) &= \left [ \prod_{k = 1} \left (\hat{M}\left (\pi\right ) \right )^2 \circ \hat{M}(\alpha_k) \right] \circ \left (\hat{M}\left (\pi\right ) \right )^2 \\
 &= \hat{M}\left (\pi\right ) \circ \left [\hat{M}\left (\pi\right ) \circ \hat{M}\left (\alpha_1\right ) \circ\hat{M}\left (\pi\right ) \right] \circ \left [\hat{M}\left (\pi\right ) \circ \hat{M}\left (\alpha_2\right ) \circ \hat{M}\left (\pi\right ) \right] \circ... \\
 &... \circ\left [ \hat{M}\left (\pi\right ) \circ \hat{M}\left (\alpha_m\right ) \circ
 \hat{M}\left (\pi\right ) \right] 
 \circ \hat{M}\left (\pi\right )
 \end{split}
\end{equation}
Each sequence of the form $B_k = \hat{M}\left (\pi\right ) \circ \hat{M}\left (\alpha_k\right ) \circ
\hat{M}\left (\pi\right )$ will be called a {\it ``padded sequence''}. We use the symbol `$\circ$' to denote that the measurement sequences are applied sequentially in time.

Due to padding the measurement sequences, the following useful simplifications occur regarding the structure of detectors:
\begin{itemize}
\item Each detector fits either within a single padded block $B_k$ (for some $k$) or within two consecutive paddings $\hat{M}\left (\pi\right ) \circ \hat{M}\left (\pi\right )$. 
\item For any single-qubit Pauli error $\varepsilon$ there exists a set of detectors $\{{D} \}_\varepsilon$ that allows us to detect this error in space and time and correct it\footnote{We omit a subtlety here that we discuss in detail in Subsec.~\ref{sec:EC_detectors}}. Moreover, the set $\{{D} \}_\varepsilon$ can be chosen such that {\it all} detectors in $\{{D} \}_\varepsilon$ have support {\it only} within a particular padded block $B_k$.
\end{itemize}

Intuitively, one can think of the padding procedure as surgically inserting honeycomb code measurement sequences into a larger code; given that the detectors of the honeycomb code are well-behaved, the overarching code inherits some of their structure. 
That being said, we would like to note that it is possible that the sequences producing automorphisms are error-correcting in a stronger sense; that is, padding might not be necessary, and error correction might be possible with a particular choice of short automorphism sequences. It would be interesting to see if there is a way to ensure error correction of the DA color code without introducing padding by appropriate choice of sequences for the automorphism generators.

\subsection{Dynamically generated logical qubits} \label{sec:EC_dyn_gen}

Similarly to the previously known Floquet codes, the DA color code can dynamically generate logical qubits, i.e. generate a state that is in the codespace of a topological stabilizer group even if the initial state is a product state or a maximally mixed state. This dynamical generation reflects the code's ability to measure all the elements of the full topological stabilizer group during a period of measurements. This is clear from the fact that a codespace cannot be generated by a measurement sequence without measuring each of its stabilizers.

The sequences for the automorphism generators shown in Table~\ref{tab:generators} do not generate the full topological stabilizer group. Each of them generates almost the entire stabilizer group of the effective color code \eqref{eq:Ceff} after a single period has been run, and no additional plaquettes will be added if the sequence is repeated. More specifically, the sequence $\varphi_{(\texttt{xr})(\texttt{yg})(\texttt{zb})}$ measures all the needed ISG elements apart from $P_b(X_1X_2)$ plaquettes. The sequence $\varphi_{(\texttt{zr})(\texttt{yg})(\texttt{xb})}$ measures everything apart from $P_r(X_1X_2)$ plaquettes, and $\varphi_{(\texttt{gb})}$ measures everything but $P_r(Z_1)$. Interestingly, any pair of sequences following one another will end up measuring the entire ISG of the effective color code. We remark that this is solely a consequence of our choice of automorphism measurement sequences, and there might exist longer sequences implementing automorphisms that dynamically generate logical qubits without padding. 

At the same time, the padding sequence shown in Table~\ref{tab:padding} generates the entire topological stabilizer group already after 5 rounds thanks to the honeycomb protocols in each layer being capable of generating the ISG of each of the toric codes after 4 rounds. Therefore, it is clear that upon padding, the DA color code will measure the stabilizer plaquettes more frequently, because even if some plaquette wasn't measured within a given generating automorphism, it's guaranteed to be measured within subsequent padding.

\subsection{Simplified error basis and measurement errors} \label{sec:EC_error_basis}
Throughout the rest of the paper, when treating error chains we assume that errors are first decomposed into an error basis that we introduce below. This allows us to significantly simplify the analysis of the error correction in the DA color code. 

For each interval between two subsequent rounds, we can define a basis for Pauli errors determined by the flavor of the current and the flavor of the next round in each layer $\ell$. Specifically, suppose that in layer $\ell$ we perform measurements of flavor $\sigma_n^\ell$ at time $t=n$ and flavor $\sigma_{n+1}^\ell$ at time $t=n+1$, and consider an error that occurs right after time $t=n$. If the flavor of the error is equal to $\sigma_n^\ell$, then it does not commute past the measurement of time $t=n+1$ since $\sigma_{n+1}^\ell \ne \sigma_n^\ell$. We will denote such an error $\varepsilon^\ell_n$ (we suppress the index indicating the spatial location of the error for brevity). In contrast, if the flavor of the error is equal to $\sigma_{n+1}^\ell$, then it does commute past the measurement of time $t=n+1$ and we can equivalently treat it as if it had happened after $t=n+1$, and we can denote such an error $\varepsilon^\ell_{n+1}$. Lastly, if the flavor of the error is neither that of $t=n$ nor $t=n+1$, then it can be decomposed as a product $\pm i \varepsilon_{n+1}^\ell \varepsilon_n^\ell $. That is, we can treat such an error as equivalent (up to a phase factor) to an error $\varepsilon_n^\ell$ occurring right after $t=n$ and an error $\varepsilon_{n+1}^\ell$ occurring right after $t=n+1$. This covers all three possible flavors for a single-qubit Pauli error. 

As a toy example, consider three qubits (1,2,3) in a single layer and two rounds of measurements: $X_1X_2$ at round 1 and $Z_2 Z_3$ at round 2. Consider an error of flavor $f$ after round 1 on qubit 2 that we call $\varepsilon_{f,2}$. Let us decompose this error into a simplified basis. First, note that the action of the two rounds of measurements on the wavefunction $\ket{L_0}$ before round 1 can be summarized as
\be
\Pi_{ZZ} \varepsilon \Pi_{XX} \ket{L_0} = \left (\frac{\mathbb I + (-1)^{m_2} Z_2Z_3}{2}\right ) \varepsilon_{f,2} \left (\frac{\mathbb I + (-1)^{m_1} X_1X_2}{2}\right ) \ket{L_0},
\ee
where $(-1)^{m_{1(2)}}$ are the measurement outcomes at the first and second rounds, and $\Pi_{XX,ZZ}$ are the projection operators. Then, for three possible flavors of Pauli errors, we have: 
\be
\begin{split}
&f = X \ \, \Rightarrow \ \ \Pi_{ZZ} \varepsilon_{f,2} \Pi_{XX} =\Pi_{ZZ} X_2 \Pi_{XX},
\\
&f = Z \ \ \Rightarrow \ \ \Pi_{ZZ} \varepsilon_{f,2} \Pi_{XX} =Z_2 \Pi_{ZZ} \Pi_{XX},\\
&f = Y \ \ \, \Rightarrow \ \ \Pi_{ZZ} \varepsilon_{f,2} \Pi_{XX} =-i Z_2 \Pi_{ZZ} X_2 \Pi_{XX} \text{ and } \varepsilon_{Y,2} \triangleq -i \varepsilon_{(Z,2)}\varepsilon_{(X,2)}.
\end{split}
\ee

Let us now address measurement outcome errors. Because all the measurements in our protocols on a torus are two-qubit, to be able to properly correct a single qubit error $\varepsilon_n^\ell$ we only need to find the spacetime location for the edge where the error has occurred, similarly to the honeycomb codes. The flavor of the single-qubit recovery operator is known from the timestamp $n$, i.e. it is the flavor of the check at round $n$ in that layer. If we accidentally correct the wrong qubit on the edge, this effect can be absorbed by the two-qubit edge measurement (which can at most multiply the total wavefunction by $-1$, and is thus inconsequential). For a composite error $\varepsilon_n^{\ell}\varepsilon_{n+1}^{\ell}$, as we later show, we are able to locate the respective edges at round $n$ and at round $n+1$, which, by the same argument, is sufficient to perform the correction. 

We must also deal with measurement errors. For the DA color code, similarly to Floquet codes, measurement errors are equivalent to correlated errors which also have to be decomposed according to the rules above.
More specifically, they are equivalent to a scenario with perfect measurements but where each faulty edge is replaced with a two single-qubit errors of a different flavor supported on this edge, one occurring right before and the other right after the measurement. Note that this is also true during the $V(Z_1Z_2)$ measurement rounds.

To illustrate this, consider the previous toy example and a pair of errors $\varepsilon$ and $\varepsilon'$ that occurred on the same qubit (say, qubit 2) after rounds 1 and 2, respectively. To mimic the measurement outcome error of the $Z_2Z_3$ measurement, their flavor has to be the same and the error has to anticommute with $Z_2Z_3$. For concreteness, consider $\varepsilon = \sigma_1$ and $\varepsilon' = \sigma_1$. Then $\varepsilon' \Pi_{ZZ} \varepsilon = \varepsilon' \left (\frac{\mathbb I + (-1)^{m_2} Z_2Z_3}{2}\right )\varepsilon = \varepsilon' \varepsilon \left (\frac{\mathbb I - (-1)^{m_2} Z_2Z_3}{2}\right ) = \left (\frac{\mathbb I - (-1)^{m_2} Z_2Z_3}{2}\right ) $. The resulting expression is as if no error has occurred but the measurement outcome of the $Z_2Z_3$ measurement has been flipped.

\subsection{Detectors: honeycomb codes example} \label{sec:EC_honeycomb_detectors}

Having clarified the error model, we are ready to introduce detectors~\cite{Paetznick2023}. A detector $D^i$ corresponds to the $i$-th row of an $\mathbb F_2$-valued matrix $D^i_{k}$, where $k$ is an index labelling spacetime locations of measurements in the code. More precisely, if $D^i_k = 1$, then we say that the measurement at spacetime location $k$ contributes to detector $D^i$; else, if $D^i_k = 0$, then the measurement at spacetime location $k$ does not contribute to $D^i$.  We define $m_k$ to be a vector of measurement outcomes which are also $\mathbb F_2$-valued (i.e. $m_k = 0$ if the measurement outcome at spacetime location $k$ was $+1$ and $m_k = 1$ if the outcome was $-1$).  We define the {\it value} of a detector via
\be
\text{value}(D^i) = \sum_{k} D^{i}_{k}m_k,
\ee
where the summation runs over all spacetime locations for the measurements.  Not any matrix $D^i_k$ defines a ``valid'' detector -- a detector is valid iff its value is independent of the measurement outcomes in the absence of noise.

When viewing the DA color code in spacetime, the qubits at each timestep live on the vertices of the spacetime lattice, which is a time-like stack of bilayer honeycomb lattices. According to our previous discussion, an error $\varepsilon_n^\ell$ will live on a time-like edge connecting the location of the error at time $n$ and $n+1$. We define the {\it support} of a detector to be the set of time-like edges such that an error occurring on any of these edges can affect the value of the detector. 
The precise description of the spacetime locations and respective flavors that the detector can detect depends on the choice of basis of errors. To simplify this description, we assume that all error chains are initially decomposed according to the error model introduced in the last subsection. Lastly, we emphasize that the support of the detector is a set of measurements in contradistinction to the support of a measurement, which is a set of qubits. 

We say that a plaquette operator is \textit{inferred} at a given round whenever its eigenvalue can be deduced by multiplying measurements of the current --- and possibly prior --- rounds.  We assign detectors to be of different colors, based on the colors of the constituent plaquettes that they detect. Denote the $\mathbb F_2$-valued measurement outcome of a plaquette by $\mathcal P_{c_j} = \sum_{ k \in \partial c_j} m_k$, where $c_j$ is a plaquette of color $c$ at spacetime location $j$. For DA color codes, it is possible to find a different matrix $\widetilde D^i_j$ such that the value of each detector becomes $\text{value}(D^i) = \sum_j \widetilde D^i_j \mathcal P_{c_j}$. Now, instead of individual measurements, $\widetilde D^i_{k}$ specifies which inferred plaquettes contribute to the value of $D^i$-th detector. We will colloquially say that a detector `lights up' whenever its value changes due to an error.

\begin{figure}[!b] \centering
\centering
\includegraphics[width= 0.9\columnwidth]{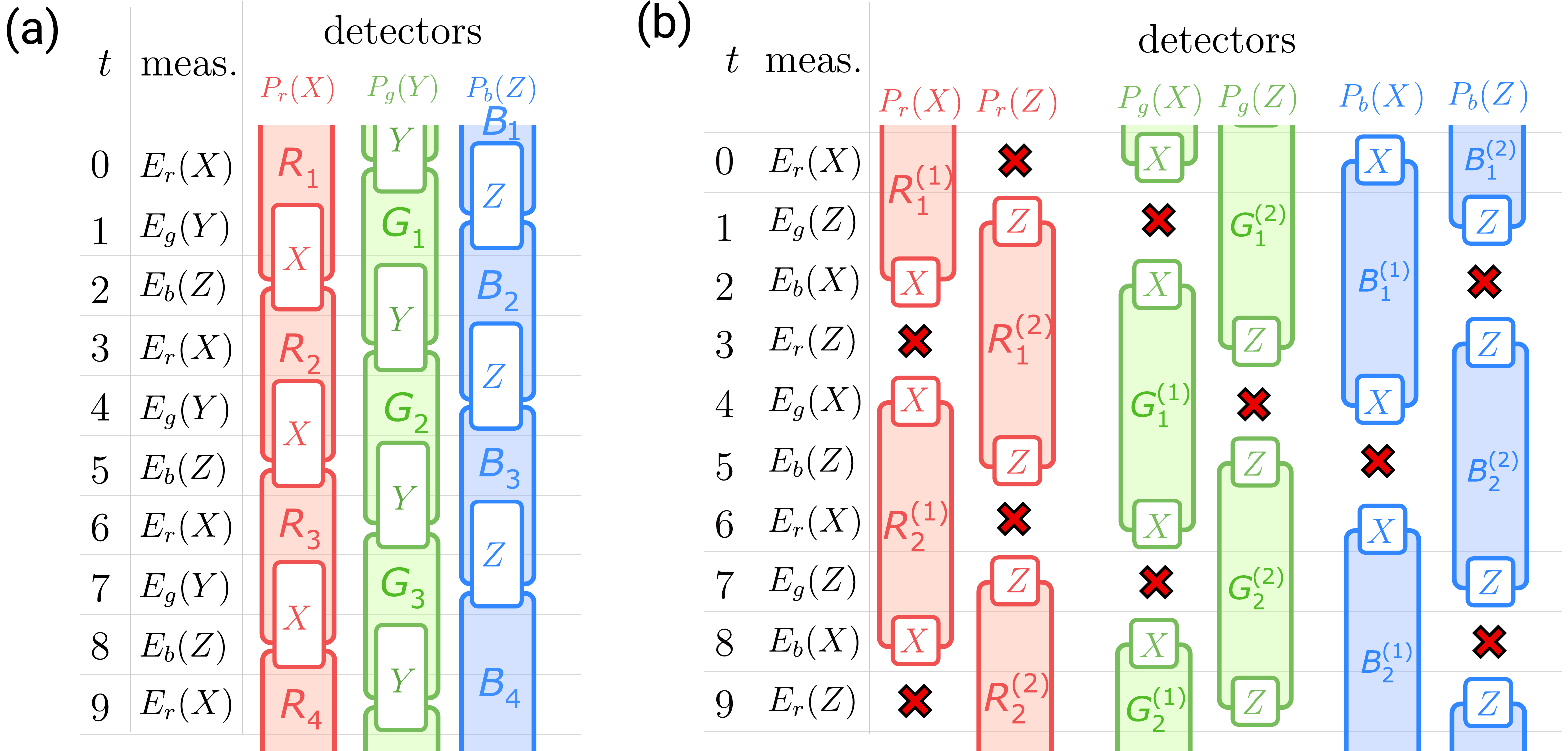}
\caption{Detectors for (a) honeycomb code and (b) CSS honeycomb code. Each colored shape marks the spacetime volume within a detector, types of which are labeled by a single plaquette of the respective color. The white boxes show the times when the plaquettes used to determine the value of the detector are inferred and are placed on the measurement rounds which are used to infer its value. The red crosses mark the rounds where the measurements randomize the values of plaquettes entering a given type of detector.}
\label{fig:3d1}
\end{figure}

Before proceeding to discuss the detectors in the DA color code, we first explain the method for constructing (and displaying) detectors in the honeycomb and CSS honeycomb codes. For clarity of presentation, we use the Kekule-Kitaev version of the honeycomb model, which is equivalent to the honeycomb code up to a depth-1 unitary circuit~\cite{Aasen2022}. 
The detectors for the honeycomb codes are shown schematically in Fig.~\ref{fig:3d1}. Each basis detector is labeled by a plaquette of a given color. For example, a detector labeled by a certain red plaquette is formed from values of that red plaquette inferred from measurements at several different times. 
This is similarly the case for detectors of green and blue plaquettes. In addition, each detector is labeled by a spatial location of the plaquette, which we suppress for brevity.

In Fig.~\ref{fig:3d1}(a), we denote the spacetime volume inside red, green, and blue detectors by red-, green-, and blue-shaded regions. It is designed in a way that the timelike edges of the flavor that anticommutes with the flavor of the detector (shown in the white boxes) in this spacetime volume belong to the support of the detector. 
The white boxes mark events in time when the corresponding plaquette (of the flavor given by the label of the box) has been measured. For example, the measured the values of the $R_i$ detectors in Fig.~\ref{fig:3d1}(a) can be found as
\be
\text{value} (R_i) = \mathcal P_r(Y)\big|_{3i-5} +\mathcal P_r(Z)\big|_{3i-4}+ \mathcal P_r(Y)\big|_{3i-2} +\mathcal P_r(Z)\big|_{3i-1}.
\ee
In the above equation, $\mathcal P_c(\sigma)\big|_{t=t_0}$ is the inferred value (which as a reminder, takes the values 0 or 1) of the plaquette of color $c$ and Pauli flavor $\sigma$ at round $t_0$, and the index of the spatial location is suppressed; all operations are performed in $\mathbb{F}_2$. Recalling that we are using the simplified error basis, we can further group the measurements and express the detector as follows:
\be \label{eq:Ri2}
\text{value} (R_i) = \mathcal P_r(X)\big|_{3i-4}+ \mathcal P_r(X)\big|_{3i-1},
\ee
and similarly for the green and blue plaquette detectors. In the above equation, we used the fact that $\mathcal P_r(Y)\big|_{3i-5} + \mathcal P_r(Z)\big|_{3i-4}$ allows us to infer the value of $\mathcal P_r(X)$ at $t = 3i-4$. It is more convenient to keep track of combined plaquettes in this fashion because (i) this grouping makes it clear why the given plaquette is not erased at the intermediate measurement rounds (for example $\Er(X)$ measurement rounds for $R_i$ plaquettes) which is necessary in order for their value to be inferred twice, and (ii) upon using the simplified error basis the detector will only light up for errors of a different Pauli flavor than the plaquette indicated on this detector. 

To summarize, each detector's value in Fig.~\ref{fig:3d1}(a) is equal to a product of two values of the same plaquette which had been inferred at two different times. In the absence of any errors, this product should give a predetermined constant. 
Similarly, Fig.~\ref{fig:3d1}(b) shows the basis of detectors for the CSS honeycomb code. One immediate difference is that there are twice as many kinds of detectors and the temporal support of these detectors is longer than that of those for the honeycomb code. Furthermore, there are `gaps' between the detectors, caused by measurements at certain rounds that randomize values of plaquettes of a certain type. Such events are labeled by a red cross. Another difference with the honeycomb code is that in the CSS honeycomb code, it only takes 1 round of measurements to infer the value of each type of plaquette.

\begin{figure}[!b] \centering
\centering
\includegraphics[width= 0.9\columnwidth]{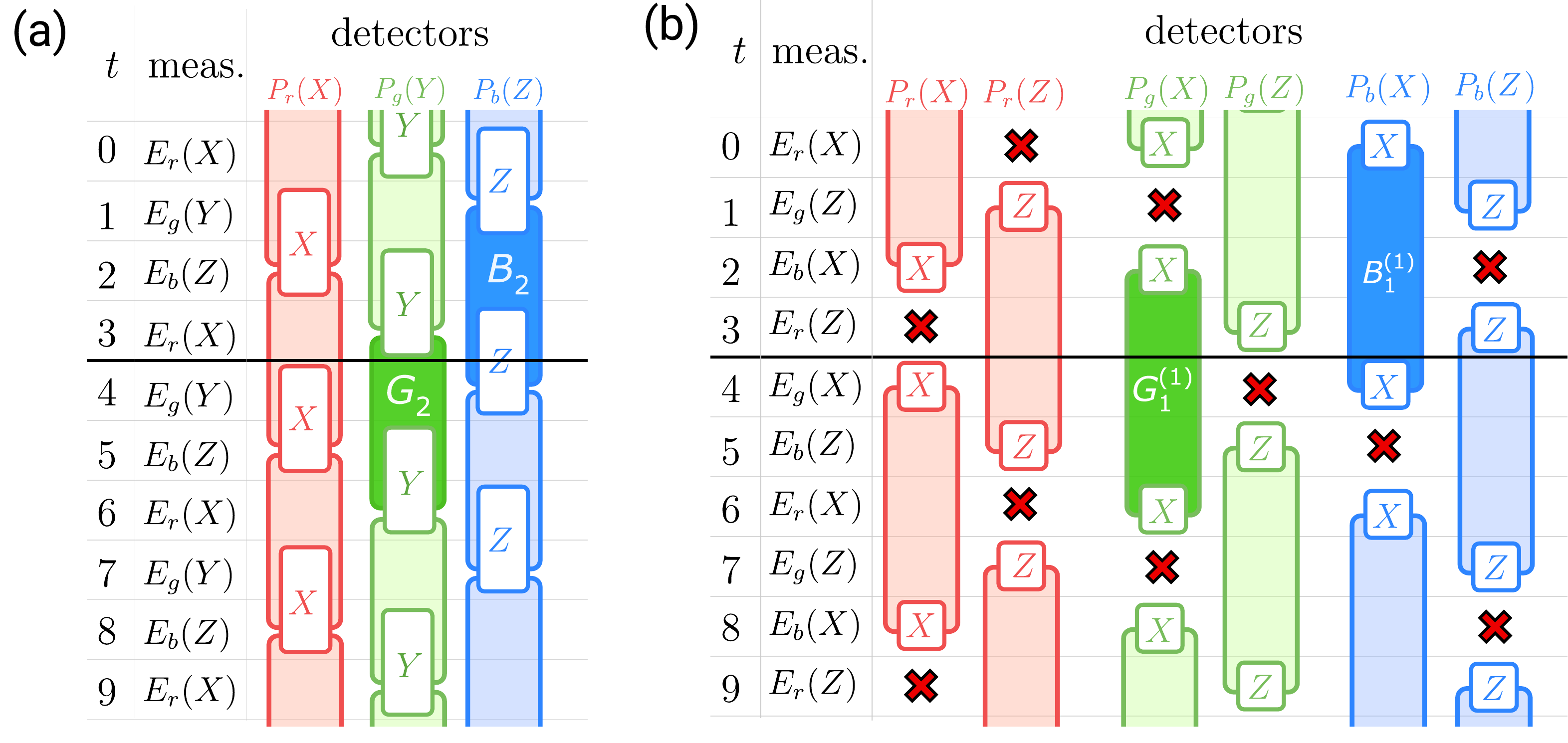}
\caption{Detectors that change the sign to $-1$ 
 (`light up') in response to an error $\varepsilon_3$ that occurred after round 3 (shown by a thick horizontal line) belong to the classes that are shaded darker. According to our error notation, the error flavor is $X$ in (a) and $Z$ in (b). The detectors that are lit up are (a) the $G_2$ and $B_2$ for the honeycomb code and (b) $G_1^{(1)}$ and $B_1^{(1)}$ for the CSS honeycomb code. The intersection of the supports of the pair of detectors that are lit up reveals information about the spacetime location of the error that is sufficient for its correction. }
\label{fig:3d2}
\end{figure}

Having discussed detectors for both the honeycomb and CSS honeycomb codes, we now turn to error correction. Assume that a single-qubit error has occurred, and recall that we assume the particular error basis introduced earlier. 
In the honeycomb code, each detector only lights up in response to errors after either odd or even rounds, and thus the decoding graph splits into even and odd sublattices. In the CSS honeycomb code, $Z$($X$) flavored detectors also detect errors after even(odd) rounds only and thus, belong to different decoding graphs. Assume an error occurred after round $t=3$. According to the error model, it is an $X$($Z$)-flavored error in honeycomb and CSS honeycomb protocols, respectively. For this error, exactly two detectors will be violated. For the honeycomb code, these detectors are labeled $G_2$ and $B_2$ in Fig.~\ref{fig:3d2}(a). The intersection of the spacetime support of these two detectors indicates the time stamp when the error has occurred and its spatial location up to an edge (red edge in this case). Similarly, for a $Z$ error after round 3 in the CSS honeycomb code, the detectors $G_1^{(1)}$ and $B_1^{(1)}$ light up, as shown in Fig.~\ref{fig:3d2}(b). Similarly, their spacetime intersection reveals the spatial edge and the time stamp of the error.

Next, one constructs a decoding graph in spacetime where the vertices correspond to detectors and the edges are drawn between the pairs of detectors that light up together in response to the errors. 
 In (2+1)D, 
for each logical string $\mathcal L^{(i)}$ we can define the logical membrane: $\mathcal L_0^{(i)} \cup \mathcal L_1^{(i)} \cup ... \cup \mathcal L_T^{(i)}$, where $T$ is the runtime of the protocol and the logical string $\mathcal L_t^{(i)}$ is defined at time $t$. The difference between $\mathcal L_{t}^{(i)}$ and $\mathcal L_{t+1}^{(i)}$ is an element of the ISG at round $t$, and this element is determined uniquely by the circuit. It defines the spacetime evolution of a single logical operator. 
If the error rate is smaller than the threshold $p_{c}$, the probability that an error loop after correction anticommutes with one or more logical membranes (implying that a logical error has occurred), asymptotically goes to zero in the thermodynamic limit. 

The decoding graph of the honeycomb codes formed by connecting pairs of detectors is a cubic lattice with the $(x,y,t) 
 = (0,0,1)$ axis rotated in the $(x,y,t) = (1,1,2)$ direction. The errors in the honeycomb codes are corrected up to error chains that look like spacetime loops of Pauli operators on the decoding graph \cite{Hastings2021}. The problem of decoding the honeycomb codes is identical to that for the toric code~\cite{Hastings2021,Bauer2023}, where a minimum-weight perfect matching algorithm allows one to find a correction that will turn error chains into closed loops. These loops are undetectable but inconsequential, i.e. they commute with the stabilizer group and all the logical membranes (moreover, they can be decomposed in such a way that they act like check operators on the wavefunction). 

\subsection{Detectors in the dynamic automorphism color code} \label{sec:EC_detectors}

Having reviewed the honeycomb codes from the perspective of detectors, we now consider the error correction procedure for the DA color code.

First, we find a basis of detectors for the padded sequences of automorphism generators $\hat M(\pi) \circ \hat M(\alpha) \circ \hat M(\pi)$, as well as for the padding sequence $\hat M(\pi) \circ \hat M(\pi)$. The measurement sequences entering these combinations are shown in tables~\ref{tab:generators} and \ref{tab:padding}. Without padding, the detector's support can `leak' into the following measurement sequence of the next automorphism. 
This makes finding a complete set of detectors needed for error correction a difficult task, as we may have to investigate all possible compositions of automorphism sequences.
To avoid this, we use padding, which guarantees that each detector terminates within each padded sequence or is contained entirely between two paddings, which drastically simplifies the analysis of error correction.

The construction of a basis of detectors can be algorithmically achieved as follows. We first decompose all error chains according to the simplified basis introduced earlier. 
All detectors are formed by combinations of measurements at a few different rounds, and for the DA color code, these measurements can be combined into a plaquette of a given color at each round. Thus, each detector is labeled by a plaquette of a single color. We also need to determine the flavors of the errors that the detector can sense, at each round and location. However, our choice of error basis has a nice property that each detector can be labeled by exactly one Pauli flavor (in each layer) and it detects precisely flavors of errors that anticommute with it. Thus, each detector can be labeled by a $P_{c}(\sigma_1 \sigma'_2)$. When constructing a basis of detectors below, we find many detectors of different colors with measurements supported on one layer only, as well as detectors between the two layers, which we label as $P_c(X_1X_2)$, $P_c(Y_1Y_2)$, $P_c(X_1Y_2)$, $P_c(Y_1X_2)$ and $P_c(Z_1Z_2)$. These bilayer plaquettes are chosen so that they commute with $V(Z_1Z_2)$ checks. 
While the above notation contains sufficient information do define all detectors for the sequences considered in this paper, in general one should follow the more general approach of Ref.~\cite{Delfosse2023}, which we revert back to when considering boundaries much later in the paper.

We may construct detectors corresponding to each of the plaquette types using the following recipe:
 \begin{enumerate}
 \item Find all spacetime coordinates of instances when a plaquette of the form $P_{c}(\sigma_1)$, $P_{c}(\sigma_2)$, $P_c(X_1X_2)$, $P_c(Y_1Y_2)$, $P_c(X_1Y_2)$, $P_c(Y_1X_2)$, or $P_c(Z_1Z_2)$ can be inferred from measurements. It is possible to use several rounds of measurement to infer a value of a single plaquette, and the measurements in different layers can occur at different times. 
 \item For each of the plaquettes, mark the times when these plaquettes are randomized by anticommuting measurements of a given round. 
 \item Find combinations of plaquette measurements whose product would have a constant value in the absence of any errors, similarly to example in Eq.~\eqref{eq:Ri2}. These combinations form detectors. Step (2) is used to determine the regions where the support of the detector cannot take place.
 \end{enumerate}

\begin{figure}[!b] \centering
\centering
\includegraphics[width= 0.9\columnwidth]{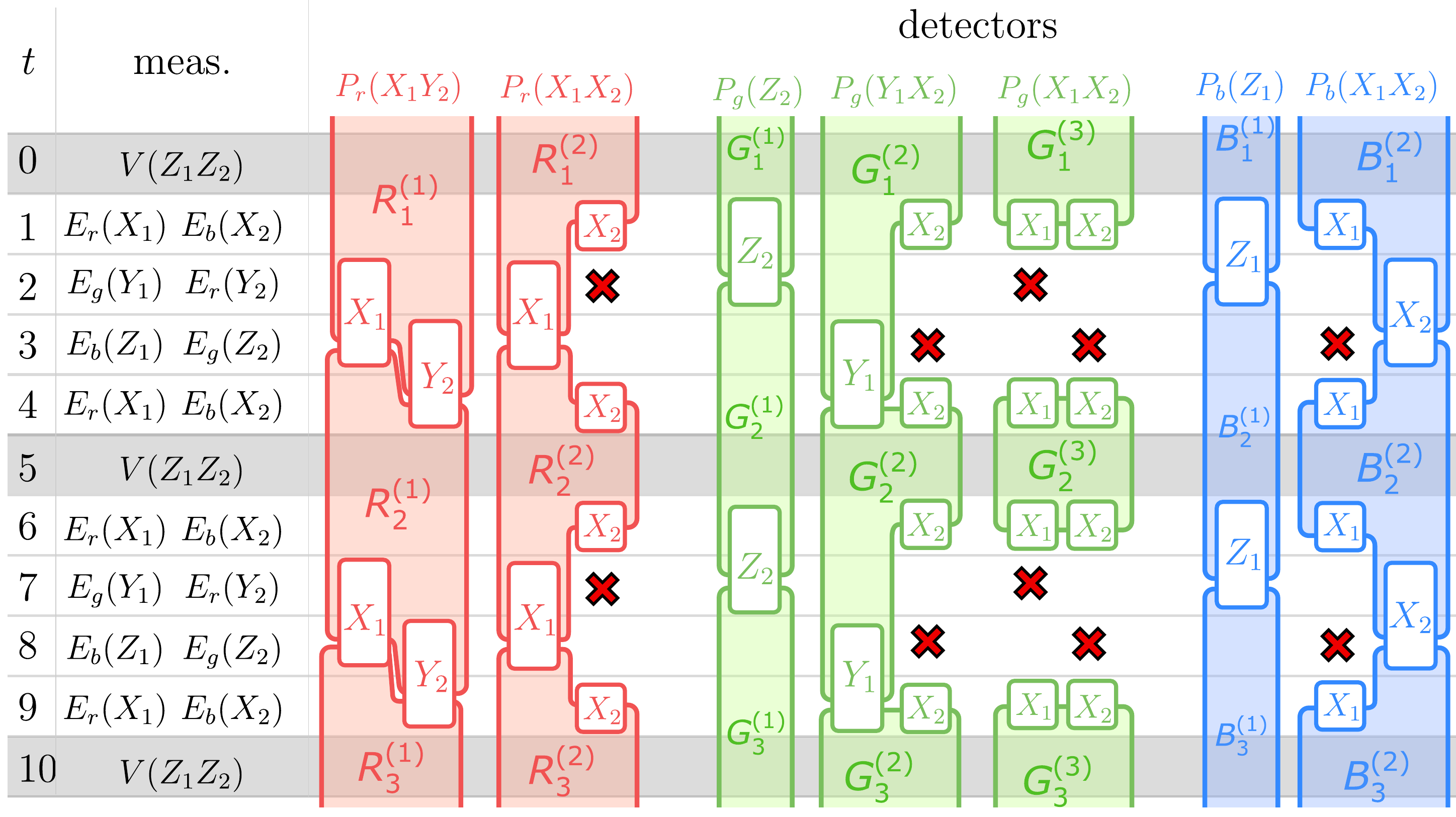}
\caption{Detectors in two layers for the periodic sequence where each period corresponds to the padding sequence implementing $\pi = \varphi_{(\texttt{xz})(\texttt{br})}$. Each colored shape marks the spacetime volume within a detector. Different types of detectors are labeled by plaquettes of respective colors that can belong to one of the layers or both layers. The white boxes show the times when the constituent plaquettes used to determine the value of the detector are inferred. The red crosses mark the rounds where the measurements in either of the layers randomize the values of the plaquettes entering a given type of detectors.}
\label{fig:3d3}
\end{figure}

Let us start by considering a periodic sequence of paddings (i.e. $\hat{M}(\pi) \circ \hat{M}(\pi) \circ \hat{M}(\pi) \circ \cdots$), which will allow us to find all detectors contained entirely within a padding sequence and detectors that might straddle between two padding sequences. A basis for detectors is shown in Fig.~\ref{fig:3d3}. A convenient feature of the $\hat{M}(\pi)$ sequence is that it looks like two decoupled layers of the honeycomb code sequence that are periodically merged together into a color code by $V(Z_1Z_2)$ measurements. If the $V(Z_1Z_2)$ rounds never occurred, the detectors would be those of two decoupled honeycomb codes.

When $V(Z_1Z_2)$ measurement rounds are added, the detectors are still, in some sense, inherited from the honeycomb code, but pairs of detectors have to be combined together (sometimes along with additional measurements) in order for the respective plaquettes to not be erased by measuring $V(Z_1Z_2)$ checks before they are re-measured again (i.e. before the detector is completed). As such, the detectors for the $P_g(Z_2)$ and $P_b(Z_1)$ plaquettes are nearly identical to those in the honeycomb code because adding $V(Z_1Z_2)$ measurements to the protocol does not erase these detectors. However, their temporal support is now larger due to the longer sequence length.

Since in the honeycomb code $X$ and $Y$-flavored detectors are never completed over the course of 4 rounds (the duration of decoupled honeycomb rounds), detectors for these plaquettes alone would be randomized by the $V(Z_1Z_2)$ measurements. However, plaquettes of flavors $X_1 X_2$, $Y_1 Y_2$, $X_1 Y_2$, and $Y_1 X_2$ (i.e. effective $X$ and $Y$ Pauli operators of the color code at $0 \mod 5$ rounds) commute with $V(Z_1Z_2)$ measurements. Therefore, once we infer these plaquettes, they survive the $V(Z_1Z_2)$ round and are inferred again during the next period of the protocol. It may appear strange that the measurements of such detectors are inferred in an ``out-of-time'' way, i.e. as an example, for $P_r(X_1Y_2)$, $P_r(X_1)$ is inferred at $t=3$ while $P_r(Y_2)$ is inferred at $t=4$, from which the values are combined to infer $P_r(X_1Y_2)$ at $t=4$. However, this is not problematic because outside of the $V(Z_1Z_2)$ rounds, the two layers are completely decoupled and the measurements are not ``causally related'' so long as they are performed between the same two consecutive $V(Z_1Z_2)$ measurement rounds. The shaded region of each detector in Fig.~\ref{fig:3d3} denotes the temporal support in each layer which is ``out-of-time'' for many detectors (including $P_r(X_1Y_2)$). 

\begin{figure}[!b] \centering
\centering
\includegraphics[width= 0.9\columnwidth]{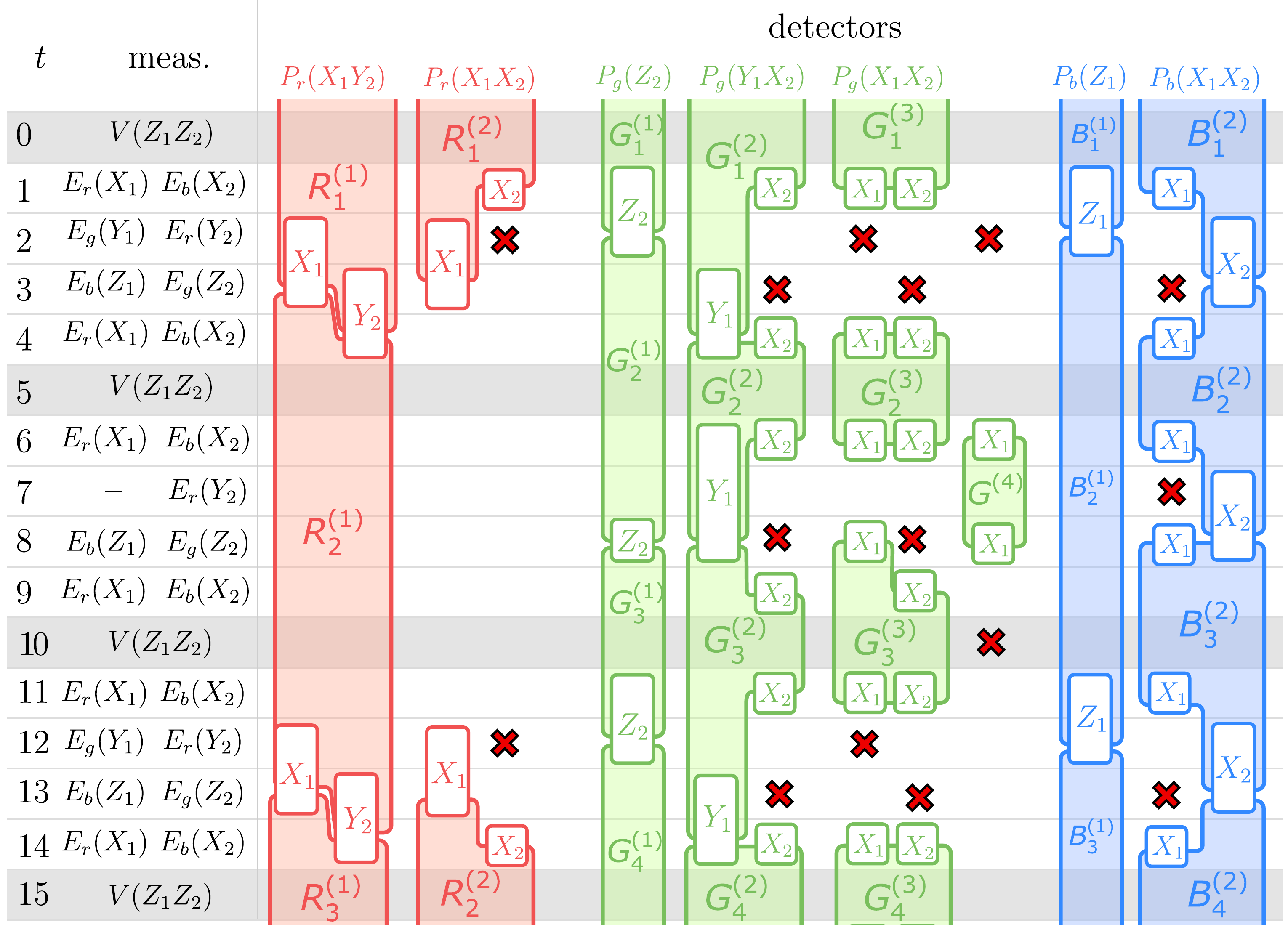}
\caption{Detectors in two layers corresponding to the padded $\varphi_{(\texttt{zr})(\texttt{yg})(\texttt{xb})}$ automorphism. The detectors that extend past the sequence are started or completed within another adjacent padding sequence. The notation is the same as in Fig.~\ref{fig:3d3}. The detectors for the padded $\varphi_{(\texttt{xr})(\texttt{yg})(\texttt{zb})}$ can be obtained by swapping the colors and layers accordingly to the measurement sequence. Additionally, instead of using padded $\varphi_{(\texttt{zr})(\texttt{yg})(\texttt{xb})}$ and $\varphi_{(\texttt{xr})(\texttt{yg})(\texttt{zb})}$ as two of the generators, one can use padded $\varphi_{(\texttt{zr})(\texttt{yg})(\texttt{xb})}$ and the padding itself repeated odd number of times. }
\label{fig:3d3b}
\end{figure}

\begin{figure}[!t] \centering
\centering
\includegraphics[width= 1\columnwidth]{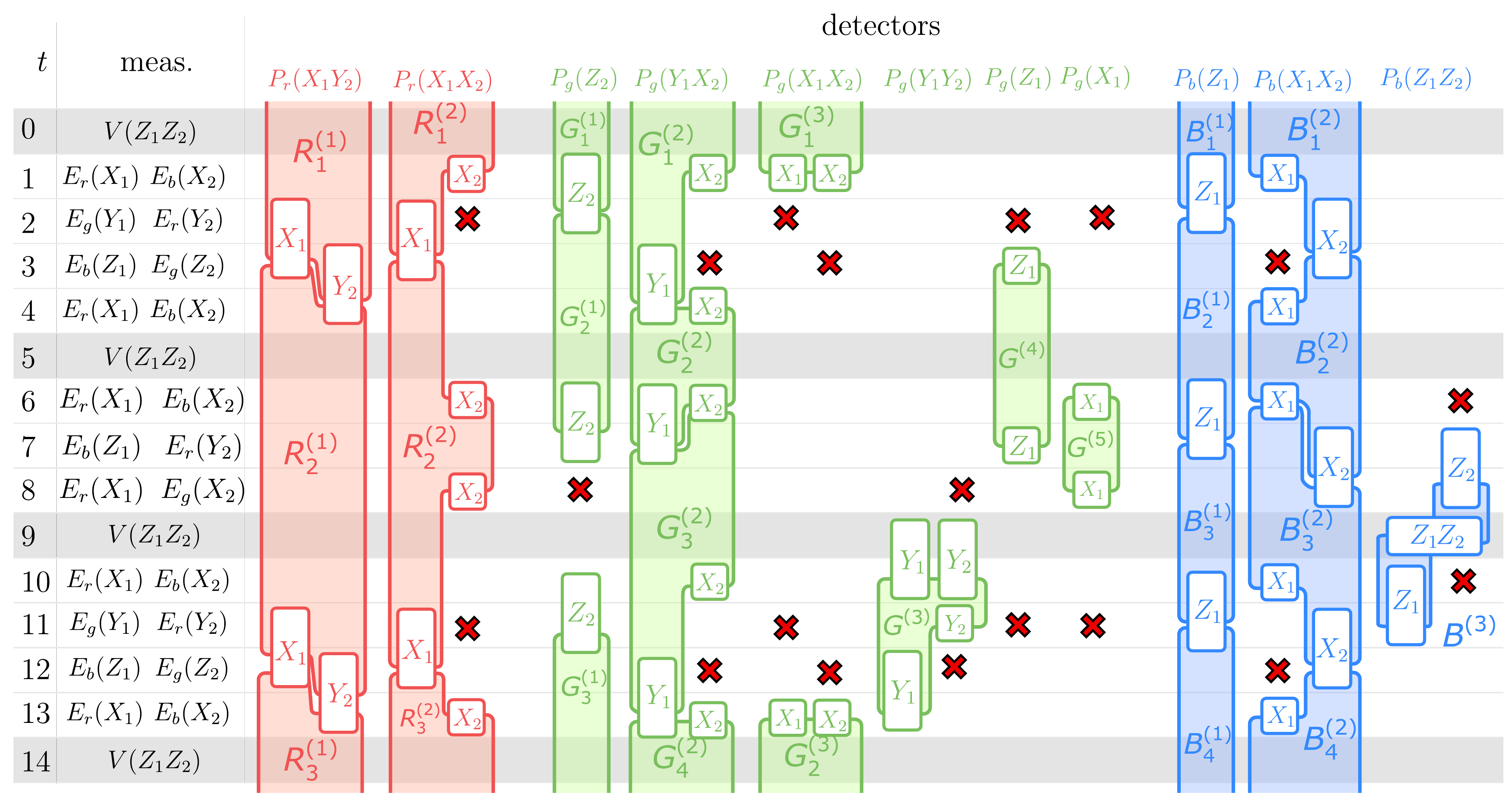}
\caption{Detectors in two layers  corresponding to the padded $\varphi_{(\texttt{gb})}$ automorphism. The detectors that extend past the sequence are started or completed within another adjacent padding sequence. The notation is the same as in Fig.~\ref{fig:3d3}.}
\label{fig:3d5}
\end{figure}

As we see, the structure of detectors in the DA color code is substantially more complex than in the honeycomb and CSS honeycomb codes. This is further exemplified by constructing detectors for sequences where we sandwich an automorphism with a padding sequence; in particular, we illustrate this for $\hat{M}(\pi) \circ \hat{M}(\varphi_{(\texttt{zr})(\texttt{yg})(\texttt{xb})}) \circ \hat{M}(\pi)$ (see Fig.~\ref{fig:3d3b}) and $\hat{M}(\pi) \circ \hat{M}(\varphi_{(\texttt{gb})}) \circ \hat{M}(\pi)$ (see Fig.~\ref{fig:3d5}). Notice that some detectors start before or end after the steps shown in the tables. One can check that these detectors start and end in the preceding or subsequent padding sequences, respectively.

There is a further subtlety concerning error correction in these codes. Specifically, we purposefully choose a basis of detectors such that no detectors involve the outcomes of $V(Z_1Z_2)$ measurements, whenever the measurement rounds preceding and following the $V(Z_1Z_2)$ round are the same. This is because, as we show below, the error correction will work up to correcting measurement errors or circuit-level errors equivalent to measurement errors at these $V(Z_1Z_2)$ rounds. However, for the exact reason that the preceding and following measurement rounds are the same, we can argue that measurement outcomes of $V(Z_1Z_2)$ measurements are not required to update the logical operators and stabilizers. Since these measurement outcomes are not used either for logical update, stabilizer update, or detectors, we can safely discard them. 

Let us discuss this subtlety in more detail now. Consider an error whose action flips the measurement outcome of the $V(Z_1Z_2)$ operator during round 5 of the protocol for the $\varphi_{(\texttt{gb})}$ automorphism. This error can be written as a pair of errors at rounds 4 and 6   $\varepsilon_4^\ell \varepsilon_6^\ell$, where the operators act on the same red(blue) edge in layer $\ell = 1$ ($\ell = 2$). In fact, the measurements in rounds 4 and 6 are identical, and it is therefore impossible to distinguish whether Pauli operators comprising this error occurred at round 4 or 6. 
One can understand the effect of an error $\varepsilon_4^\ell \varepsilon_6^\ell$ as a measurement error during the $V(Z_1Z_2)$ round because if we consider the action of such an error on a logical state, we find that it can be replaced by changing the sign of the intervening $V(Z_1Z_2)$ measurement outcome. Explicitly, consider a red edge on the lattice denoted by $\langle i,j\rangle$. Then labelling measurement outcomes by $m_t^{i}$ where $t$ denotes the time stamp and $i$ denotes the site in case of a $V(Z_1Z_2)$ check, we can write the action of the error $\varepsilon_{4,i}^1 \varepsilon_{6,i}^1$ on some logical state $\ket{L}$ of the code at time step 4 as 
\begin{align}
\ket{\varepsilon_{4,i}^1 \varepsilon_{6,i}^1} &= \varepsilon^1_{6,i} \lp 1 + m_6 X_{1,i} X_{1,j} \rp \lp 1 + m_5^i Z_{1,i} Z_{2,i} \rp \lp 1 + m_5^{j} Z_{1,j} Z_{2,j} \rp \varepsilon^1_{4,i}\lp 1 + m_4 X_{1,i} X_{1,j} \rp \ket{L} \nonumber\\
&= X_{1,i} \lp 1 + m_6 X_{1,i} X_{1,j} \rp \lp 1 + m_5^i Z_{1,i} Z_{2,i} \rp \lp 1 + m_5^{j} Z_{1,j} Z_{2,j} \rp X_{1,i} \lp 1 + m_4 X_{1,i} X_{1,j} \rp \ket{L} \nonumber \\
&= \lp 1 + m_6 X_{1,i} X_{1,j} \rp \lp 1 + m_5^i X_{1,i} Z_{1,i} X_{1,i} Z_{2,i} \rp \lp 1 + m_5^{j} Z_{1,j} Z_{2,j} \rp \lp 1 + m_4 X_{1,i} X_{1,j} \rp \ket{L} \nonumber \\
&= \lp 1 + m_6 X_{1,i} X_{1,j} \rp \lp 1 - m_5^i Z_{1,i} Z_{2,i} \rp \lp 1 + m_5^{j} Z_{1,j} Z_{2,j} \rp \lp 1 + m_4 X_{1,i} X_{1,j} \rp \ket{L}.
\end{align}
As we see, the effect of the error is to flip the sign of $m_5^i$. More generally, such a correlated error whose support is right before and after $V(Z_1Z_2)$ round will flip the sign of a $V(Z_1Z_2)$ measurement. In a spacetime picture, such an error creates a pair of anyons and annihilates them after two timesteps, and thus, forms a closed loop.

We circumvent this issue by discarding the outcomes of $V(Z_1Z_2)$ measurements whenever they are made. Although these rounds are necessary between the end of a previous and the next automorphism sequence, measurement outcomes during these rounds can be safely discarded as they are also not used to perform logical string updates, stabilizer updates, and error detection.

At the same time, when a $V(Z_1Z_2)$ measurement occurs via the same scenario as round 9 in the padded $\varphi_{(\texttt{gb})}$ sequence (see Fig.~\ref{fig:3d5}), in which the preceding and following measurements are different, there are no longer undetectable error chains and this subtlety no longer appears.

Next, let us discuss the detectors for the padded protocols for $\varphi_{(\texttt{zr})(\texttt{yg})(\texttt{xb})}$ and $\varphi_{(\texttt{gb})}$ shown in Figs.~\ref{fig:3d3b} and \ref{fig:3d5}, respectively. The sequence for the padded $\varphi_{(\texttt{zr})(\texttt{yg})(\texttt{xb})}$ looks identical to the honeycomb code in the second layer and performs three measurements (instead of four) in the first layer. Therefore, one might expect the detectors to look similar to those of the padding sequence itself (Fig.~\ref{fig:3d3}). The difference is that the detectors of type $R^{(2)}$ are temporarily discontinued (which does not turn out to negatively impact error detectability) and a short detector $G^{(4)}$ is added. For the padded $\varphi_{(\texttt{gb})}$ sequence, the protocol is different; some of the detectors change shape while others are added/removed. The most important distinction is that the outcomes of $V(Z_1Z_2)$ measurements at round 9 of this sequence can be collected and used for error correction, as previously discussed. This allows us to introduce new detectors $G^{(3)}$ and $B^{(3)}$ that involve the $V(Z_1Z_2)$ round of measurements.

Next, we consider every possible distinct single-qubit error, which corresponds to an examination of all distinct $\varepsilon^{\ell}_t$ and $\varepsilon^{\ell}_t \varepsilon^{\ell }_{t+1}$ errors. First consider a $\mathbb{F}_2$-valued vector space of errors $V_{\varepsilon}$. Single qubit errors such as $\varepsilon^{\ell}_t$ furnish a basis for $V_{\varepsilon}$, and will be denoted ${\hat{\varepsilon}}^{\ell}_t$. Further define an $\mathbb{F}_2$-valued vector space of detectors labeled by $V_{D}$, whose basis vectors are isomorphic to the basis detectors $\hat{D}^i$ (we added a hat to emphasize that we are dealing with the basis detectors). For each single-qubit error, we identify the set of detectors that have support on the respective timelike edge and light up if this error has occurred. We express this set as a vector $s({\hat{\varepsilon}}^{\ell}_t) \in V_{D}$. The inner product $\langle s({\hat{\varepsilon}}^{\ell}_t), \hat{D}^i\rangle = 1$ if error $\varepsilon^{\ell}_t$ lights up detector $D^i$ and $0$ otherwise. 

There exists a natural map $H:V_{\varepsilon} \to V_{D}$ defining a parity check matrix of a classical code. Its action is
\be
s({\hat{\varepsilon}}^{\ell}_t) = H {\hat{\varepsilon}}^{\ell}_t.
\ee
For each error of the type $\varepsilon^{\ell}_t$ occurring after a round corresponding to the decoupled toric codes, we find that at least two detectors of different colors `light up', and moreover, the intersection of the detectors give the time and the edge where the error has occurred (the exception are errors of the $\varepsilon^\ell_4 \varepsilon^\ell_6$ type, which, as we have already discussed, form an undetectable and inconsequential error. If an error occurred after a $V(Z_1Z_2)$ round, at least three detectors of three different colors light up, revealing the `vertical' edge where the error has occurred. For example, an error $\varepsilon_1^1$ in the sequence implementing $\pi$ lights up detectors $G_1^{(2)}$ and $B_1^{(1)}$. An error $\varepsilon_5^1$ lights up detectors $R_2^{(1)}$, $R_2^{(2)}$, $G_2^{(2)}$, $G_2^{(3)}$ and $B_2^{(2)}$.

In fact, we can show that the detectors shown in Figs.~\ref{fig:3d3}, \ref{fig:3d3b} and \ref{fig:3d5} constitute a basis for detecting all possible single-qubit Pauli errors. In particular, all of these detectors are independent because each of them includes at least one type of measurement that is not used for measuring plaquettes in any other detector. Moreover, from analyzing the rows of matrix $H$, we find that each single-qubit error of the type $\varepsilon^\ell_t$ can be detected and corrected (up to measurement outcome errors on the $V(Z_1Z_2)$ rounds) and moreover all possible pairwise products $\varepsilon^\ell_t \varepsilon^\ell_{t+1}$ can also be detected and corrected.

\subsection{Outlook on fault-tolerance of dynamic automorphism color code}
\label{sec:EC_decoder}

The decoding hypergraph of the DA color code is constructed by placing detectors at the (hyper)vertices and defining the (hyper)edges to run between detectors that can light up simultaneously in response to single-qubit Pauli errors (either $\varepsilon^\ell_n$ or $\varepsilon^\ell_n\varepsilon^\ell_{n+1}$).   The decoding hypergraph has a complicated and irregular structure, but the DA color code for any automorphism can be constructed from a few automorphism generators plus padding, which gives a finite number of kinds of error detection events. In fact, when showing that any single-qubit Pauli error can be corrected, we implicitly considered each type of detecting event that will occur on the decoding hypergraph (explicit calculation is bulky and is omitted for brevity, but is straightforward). 

The question of a proper definition of the distance in dynamic codes is still being discussed in the literature. Let us define the distance of a DA color code as the minimum between the weight of the smallest logical operator (where the minimum is taken over all time slices) and the weight of the smallest consequential but undetectable spacetime error chain. In fact, we treat the code distance and fault-tolerance similarly in spirit to how it is done in measurement- and fusion-based quantum computation~\cite{Bartolucci2023Feb,Bombin2023Apr}. In (2+1)D, we define the logical membrane for each logical operator $\mathcal L^{(i)}$ to be $\mathcal L_0^{(i)} \cup \mathcal L_1^{(i)} \cup ... \cup \mathcal L_T^{(i)}$, and the difference between $\mathcal L_{t}^{(i)}$ and $\mathcal L_{t+1}^{(i)}$ is an element of the ISG at round $t$ that is found according to the update rule for the respective logical operator. The distance $d$ of our code has to be $O(L)$, where $L$ is the linear size of the code, because the only undetectable objects that anticommute with logical membranes of the DA color code are the ones that form a homologically nontrivial loop around the torus in their spatial support.

Likely, one can obtain a lower bound on the threshold of these codes under the single-qubit i.i.d. error model by a mapping to a random-bond Ising model. This will lead to a generalized model with two- and three-body interaction terms that lacks periodicity in one of the three directions (the temporal one). We expect the phase transition of this generalized random-bond Ising model to give us the threshold of the maximum likelihood decoder~\cite{Kovalev2018}. For finding an efficient decoder, this would not be sufficient and one needs to understand the structure of the decoding hypergraph. Because of the equivalence between the ISGs of the 2D DA code and the stabilizer groups of the color code and copies of the toric code at different steps, and the correspondence between detectors and the stabilizer types, we find that the decoding hypergraph of the DA color code interpolates between those of a pair of decoupled toric codes and of the color code. This structure could be leveraged for construction of an appropriate decoder. Alternatively, because the code can be unitarily mapped to two copies of the toric code at each step, one can develop an efficient decoder using the decoder for copies of toric code, similarly to Refs.~\cite{Kubica2023efficient,Delfosse2014}. 
It would also be interesting to see if one can work out efficient decoders for DA codes based on other known decoders for the color code~\cite{Bombin2012,Sarvepalli2012,Chamberland_2020,Sahay2022}.  
In addition, when it comes to practical error correction for the DA color code, it should be possible to perform an exhaustive numerical search over the space of measurement sequences implementing a given automorphism and padding sequences in order to find the sequence maximizing the threshold. Such a search would also be useful for finding sequences that minimize the size of detectors, because larger detectors lead to a larger `entropic' factor (i.e. a larger combinatorial factor coming from the larger number of error configurations leading to the same syndrome) that causes reduction of the threshold.

In closing, we conjecture that using the ideas mentioned above, one can construct an efficient decoder with a threshold for the 2D DA color code.

\section{{TQFT perspective on dynamic automorphism codes}} \label{sec:tqft_sec}

In this section, we will study DA codes from the perspective of TQFTs, which allows us to study the universal topological properties of such codes. We will restrict the discussion to parent Pauli topological codes in two spatial dimensions on prime-dimensional $p$-qudits. In particular, this applies to qubits where $p=2$. Such Pauli topological codes have been shown to be equivalent up to a finite depth unitary circuit to a finite stack of $\mathbb Z_p$ toric codes \cite{Bombin2012, bombin2014structure,Haah2018a} under an extra assumption of translation invariance. For example, the color code defined on prime-dimensional qudits is equivalent to two copies of the $\mathbb Z_p$ toric code~\cite{Kubica_2015}. A full analysis of more general Abelian TQFTs in two dimensions will be given in an upcoming work~\cite{Aasen2023_TQFT}.

In Subsec.~\ref{sec:TQFTreview} we review the basic concepts that are needed for defining condensation sequences from a parent model that is equivalent to several copies of a prime-dimensional toric code. Next, in Subsec.~\ref{sec:allowed_bulk_cond}, we will introduce the concept of reversible condensations between two child theories obtained from the same parent theory. Reversible condensations are the building blocks for constructing sequences that implement automorphisms of a child theory.
In Subsec.~\ref{sec:tqft_transitions}, we explain how to find shared basis of deconfined anyons between two child theories that is conserved upon a reversible condensation between these theories. Given a sequence of reversible condensations, a shared basis is changed between each two consecutive pairs. Tracking this update allows us to determine the automorphism implemented by the sequence, which is explained in Subsec.~\ref{ref:tqft:computing_aut}.

In Subsec.~\ref{sec:condensation_complex}, we define the concept of the condensation graph, which charts the space of possible reversible condensations, and shows how to assign isomorphisms to its edges and automorphisms to its cycles. As an example, we explore condensation graphs for the cases when the parent theory is a single color code (i.e. the graph for the ``Floquet color code''~\cite{brown_2022,Davydova22}) and when the parent theory is two copies of a color code (i.e. the graph for the DA color code). Lastly, in Subsec.~\ref{sec:tqft_aut_construction} we explain a technique that, starting from a condensation sequence implementing an arbitrary automorphism, allows one to design sequences for other automorphisms. 

\subsection{TQFT review of Abelian anyons} \label{sec:TQFTreview}

We will assume that the parent theory is equivalent to the toric code of some Abelian group $\abelian = \prod_p \mathbb Z_p^{N_p}$, where the product runs over all primes $p$ and $N_p \in \mathbb Z_{\ge 0}$ denotes the number of copies of that prime-dimensional qudit toric code. The analysis in this section factors into each prime $p$ separately. Therefore, without loss of generality, we can further simplify and analyze the case $\abelian = \mathbb Z_p^{N_p}$ for some prime $p$.

An anyon in the parent theory can be specified by a vector $c \in M = \mathbb F_p^{2 N_p}$\footnote{In this section, we use non-calligraphic symbols ($M,A,\ldots)$ for the vector spaces to distinguish them from equivalent modular tensor category notation $(\mathcal M,\mathcal A, \ldots)$ used elsewhere in the paper. Moreover, fusion is replaced by the addition of vectors.} where the first $N_p$ unit vectors correspond to electric charges $\ttt e$ for each copy of the $\mathbb Z_p$ the toric code, and the last $N_p$ unit vectors are the respective magnetic charges $\ttt m$. Note that for clarity, we use the regular and not typewriter font for anyons in the vector notation. 

The self-statistics of an anyon is given by the $\mathbb{F}_p$-valued quadratic form 
\begin{align}
 q(c) &= c^T\mu c, &
 \mu &= \begin{pmatrix} 0_{N_p \times N_p} & \mathbbm 1_{N_p \times N_p}\\
0_{N_p \times N_p} &0_{N_p \times N_p} \end{pmatrix}
\end{align}
and the mutual statistics of two anyons $c_1$ and $c_2$ can be computed by the inner product 
\begin{align}
 \langle c_1, c_2\rangle &= q(c_1+c_2)-q(c_1)-q(c_2)= c_1^T \lambda c_2, &
 \lambda = \begin{pmatrix} 0_{N_p \times N_p} & \mathbbm 1_{N_p \times N_p}\\
\mathbbm 1_{N_p \times N_p} &0_{N_p \times N_p} \end{pmatrix}.
\end{align}
We remark that this is a special case of a more general statement that an Abelian modular tensor category can be fully characterized by an Abelian group along with a $U(1)$-valued quadratic form (see for example Ref.~\cite{Etingof2016}).

Recall that a choice of condensation is a set of bosons that all braid trivially with each other~\cite{Burnell18,Kong2014anyon}. In this formalism, we can specify such a condensation by a subspace $A \subset M$ which satisfies $q(a_1)=0$ and $\langle a_1,a_2 \rangle =0$ for all $a_1,a_2 \in A$. Let $|A| = p^{N_A}$ for some $N_A < N_p$.
Given such a condensation, consider the orthogonal subspace $A^\perp = \{ b\in M| \ \forall a \in A, \langle a,b\rangle=0\}$. Physically, this corresponds to anyons in $M$ that braid trivially with all anyons in $A$.
By definition, $A$ is a subspace of $A^\perp$ so we may quotient out $A$ to obtain $C=A^\perp/A$ which labels the deconfined anyons in the child theory\footnote{This will ultimately be associated to logical operators of the child code. }. These anyons correspond to equivalence classes in the parent theory $M$.
Note that $C \cong \mathbb F_p^{2N_C}$ where $N_C = N_p-N_A$. Thus an anyon in the child theory can labeled by a vector in $A^{\perp}$, up to an equivalence of addition by vectors in $A$.

It is useful to define a \textit{representative} of an anyon $c\in C$ within $M$, which we denote $\rho(c)$ (we will often abbreviate this as a representative of $c$). The representative is not unique and depends on the choice of the linear map $\rho: C\rightarrow M$.

To formalize the notion of a representative,
we first define the quotient map $\kappa: A^\perp \rightarrow A^\perp/A =C$. Next, we define a section to be any linear map $s: C\rightarrow A^\perp$ such that $\kappa \circ s = \text{id}_C$. The section, along with the inclusion map $\iota: A^\perp \rightarrow M$, can then be used to defined the representative map as $\rho = \iota \circ s$.

To conclude, we can define a representative $\rho(c)$ for any $c\in C$ and choice of map $\rho$. Note that since the elements of $\rho(C)$ represent every coset of $A$ in $A^\perp$, we have $\rho(C) + A = A^\perp$.

\subsection{Reversible condensations} \label{sec:allowed_bulk_cond}

\begin{figure} \centering
 \centering
 \includegraphics[width = \textwidth]{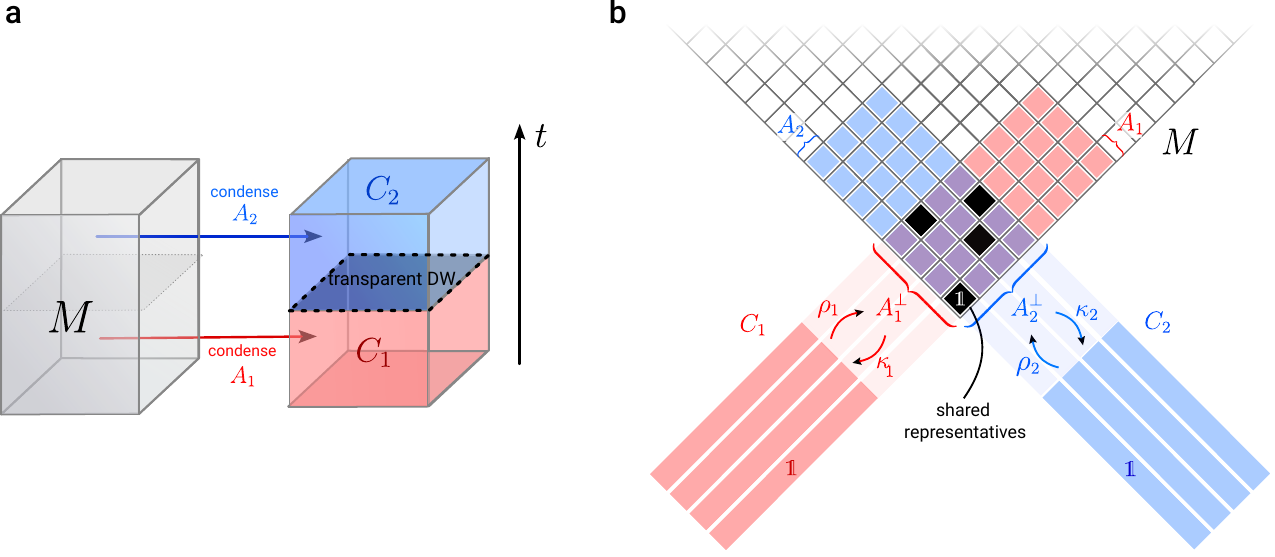}
 \caption{ Two definitions of a reversible pair of condensations $A_1$ and $A_2$ in $M$. (a) Informal definition: condensing $A_1$ in on region of spacetime and $A_2$ in the other region of a parent theory $M$ results in child theories $C_1$ and $C_2$ separated by a transparent interface. (b) Formal definition in the context of Abelian anyons: there exists a shared and complete set of representatives (black squares) for both $C_1$ and $C_2$ in $M$. }
 \label{fig:tqft_repr}
\end{figure}

Recently, Ref.~\cite{Aasen2023} formalized the notion of a reversible pair of stabilizer groups. Let us develop an analogous notion of a \textit{reversible pair of condensations}. First, we provide an informal definition. Take the parent theory $M$ and consider condensing $A_1$ in one region of spacetime and $A_2$ in the other region (see Fig.~\ref{fig:tqft_repr}(a)). This results in child theories $C_1$ and $C_2$ in the two regions separated by an interface. We say that $A_1$ and $A_2$ are a reversible pair of condensations (with respect to the parent $M$) if the resulting interface is transparent on both sides, by which we mean any anyon from $C_1$ can go through and become some anyon in $C_2$ and vice versa\footnote{Invertibility is a sufficient but not necessary condition. For example, the domain wall could host multiple channels for a single anyon, making it non-invertible.} 

The above definition sets some constraints on what pairs of condensations can be allowed for constructing meaningful sequences. For example, it requires that $A_1$ and $A_2$ are of equal size, since $C_1$ and $C_2$ have to be isomorphic. Henceforth, we will make the assumption that $|A_1| = |A_2|= p^{N_A}$ for some $N_A$.

Let us now make a formal definition in the context of Abelian anyons. $A_1$ and $A_2$ are a reversible pair of condensations if there exists a shared and complete set of representatives for both $C_1$ and $C_2$ in $M$. That is, there exist linear maps $\rho_1$ and $\rho_2$ such that $\rho_1(C_1) = \rho_2(C_2)$, $\rho_1(C_1) + A_1 = A_1^\perp$, and $\rho_2(C_2) + A_2 = A_2^\perp$. (see Fig.~\ref{fig:tqft_repr}(b)).

\subsection{Transitions between child theories}
\label{sec:tqft_transitions}

Consider a reversible pair of condensations, $A_1$ and $A_2$, and their corresponding child theories $C_1$ and $C_2$. For every anyon in $C_1$ there will exist a representative that will braid trivially with condensation $A_2$ and thus will become an anyon in child theory $C_2$. We call finding such a representative an ``update'' because starting with an arbitrary representative (in $M$) of a given anyon in $C_1$, we can ``update'' it by fusing with some anyon in $A_1$ such that it becomes an element of $A_2^\perp$. This allows us to consider it (project it to) an anyon of $C_2$. Reversibility guarantees that these anyons map onto complete $C_2$.

Let us now provide an explicit calculation for how to update representatives of anyons when going from one child theory to another by a reversible condensation. We note that this update procedure directly translates into updating logical operators in DA codes by considering logical strings instead of deconfined anyons and multiplying by measured operators instead of fusing with condensed anyons. 

Assume that we have a condensation $A_1$ and a map $\rho_1$ which realizes representatives $\rho_1(C_1)$. A priori, we are not guaranteed that $\rho_1(C_1) \subset A_{2}^\perp$. Thus, we would like to construct an alternative map $\tilde \rho_1(C_1)$ such that $\tilde \rho_1(C_1) \subset A_{2}^\perp$, and is moreover a complete set of representatives for $C_2$. This construction will allow us to determine an \textit{update map} between condensation rounds, which we will explain later.

We proceed as follows. Let us define $N_Q$ to be the number of generators of $Q=A_{1} \cap A_{2}$ and $N_{\overline Q} = N_A-N_Q$, where $N_A$ is the number of generators in $A_1$ (and $A_2$). First, we pick an arbitrary basis for $Q$:
\begin{align}
 Q = \text{span}\{ \hat{a}_Q^{(i)}\}
\end{align}
for $i = 1, \ldots, N_Q$. Next, we may pick a basis for the bosons in $A_{1}$ and $A_{2}$ as
\begin{align}
 A_{1}& = \text{span}\{ \hat{a}^{(i)}_{1} \} ,& A_{2}& = \text{span}\{ \hat{a}^{(i)}_{2} \}
\end{align}
where $i = 1, \ldots, N_A$, with the constraint
\begin{align}
 \hat{a}^{(i+N_{\overline Q})}_{1} = \hat{a}^{(i+N_{\overline Q})}_{2} = \hat{a}_Q^{(i)}
\end{align}
for $i = 1, \ldots, N_Q$, corresponding to a shared basis of bosons in $Q$.

Let us define the matrix $B$ which has components
\begin{align}
 B^{ij} = \langle\hat{a}^{(i)}_{2},\hat{a}^{(j)}_{1}\rangle
\end{align}
As a matrix, we find that it must take the form
\begin{align}
 B =\begin{pmatrix}
 \beta_{N_{\overline Q} \times N_{\overline Q}} & 0_{N_{\overline Q} \times N_Q}\\
 0_{N_Q \times N_{\overline Q}}& 0_{N_Q \times N_Q}
 \end{pmatrix}
\end{align}
for some matrix $\beta$. The zero matrices follow from the fact that $\langle A,Q \rangle = \langle B,Q \rangle = \langle Q,Q \rangle =0$.

Let $L\in \rho_1(C_1)$ be a given representative. For each representative, we would like to find a different representative $\widetilde{L}$ that also lives in $A_{2}^\perp$ and, moreover, $\text{span}\{\widetilde L\}$ forms a complete set of representatives for $C_2$. For each $L$, finding an appropriate $\widetilde L$ can be achieved by adding appropriate elements of $A_1$ to the original representative. That is, assume
\begin{align} \label{eq:logical_update}
\widetilde{L}= L + \sum_{i=1}^{N_A} \hat{a}^{(i)}_{1} W^{i}
\end{align}
for some $\mathbb F_p$-valued vector $W^{i}$. Since we want $\widetilde{L} \in A^\perp_2$, this implies that
\begin{align} \label{eq:connection_condition1}
 \langle \hat{a}^{(k)}_{2}, \widetilde{L} 
 \rangle = \langle \hat{a}^{(k)}_{2}, L \rangle + \sum_{i=1}^{N_A} B^{ki} W^{i} =0
\end{align}
for $k = 1,\ldots, N_A$. However, note that for $k=N_{\overline Q}+1,\ldots,N_{A}$, the equation is trivially satisfied since $\hat{a}^{(k)}_{2} \in Q$, so we actually only need to solve for $W^{ij}_t$ for $k=1, \ldots, N_{\overline Q}$. Thus, writing $W = \begin{pmatrix} \omega_{N_{\overline Q} }\\ 0_{N_Q }\\ \end{pmatrix}$ for some vector $\omega$, we instead can solve
\begin{align} 
 \langle \hat{a}^{(k)}_{2}, L \rangle + \sum_{i=1}^{N_{\overline Q}} \beta^{ki} \omega^{i} =0,
\end{align}
for $k=1, \ldots, N_{\overline Q}$. Assuming $\beta^{ki}$ is invertible over $\mathbb F_p$, the solution is
\begin{align} 
\omega^{i} = - \sum_{k=1}^{N_{\overline Q}}(\beta^{-1})^{ik} \langle \hat{a}^{(k)}_{2}, L \rangle. 
\end{align}
In conclusion, we have found a new set of representatives which generate $\tilde \rho_1(C_1)$ given by
\begin{align} \label{eq:update_formula}
 \widetilde{L} = L - \sum_{i=1}^{N_{\overline Q}} \hat{a}_{1}^{(i)} (\beta^{-1})^{ik} \langle \hat{a}^{(k)}_{2},L \rangle.
\end{align}
The desired representative map $\widetilde{\rho}_1$ is then defined via its action $\kappa_1(\widetilde{L}) \mapsto \widetilde{L}$.
Note that the above update Eq.~\eqref{eq:update_formula} is linear in $L$. Thus, we may use it to define the \textit{update map} $\nu_1: \rho_1(C_1) \rightarrow \widetilde{\rho}_1(C_1)$ which acts as $L \mapsto \widetilde{L}$. We will sometimes refer to Eq.~\ref{eq:update_formula} as an ``update rule'' or an ``update formula''.

We have found a complete set of representatives that is shared between $C_1$ and $C_2$. Namely, we have a map $\widetilde{\rho}_1$ such that $\widetilde{\rho}_1(C_1)+A_1 = A_1^\perp$ and $\widetilde{\rho}_1(C_1)+A_2 = A_2^\perp$. This can be used to define a canonical isomorphism between $C_1$ and $C_2$. Namely, we may define $\lambda_{21}:C_1 \rightarrow C_2$ where $\lambda_{21} = \kappa_2 \circ \widetilde{\rho}_{1}$.

We remark again that the update formula described here translates naturally to the update of logical operators in the corresponding DA code. Namely, the logical operators can be modified to commute with the next round of measurements by multiplying them with the measurements from the current round, or equivalently, by multiplying them by string operators of the condensed anyons.

\subsection{Computing automorphisms for a condensation sequence}
\label{ref:tqft:computing_aut}

Assume that we have a sequence of condensations $A_t$ for $t=0,\ldots, T$ with $A_0 = A_T$, where each $A_t$ and $A_{t+1}$ are a pair of reversible condensations for $t=0,\ldots, T-1$. We may now ask what automorphism is applied between the initial and final steps when we realize the same child theory the second time, i.e. we arrive at $C_T = C_0$.

\begin{figure} \centering
 \centering
\adjustbox{scale=0.9,center}{\begin{tikzcd}[scale = 0.9]
\cdots =&\rho_1(C_1) \arrow[rr, " \nu_1 "] \arrow[dddr,shift left,"\kappa_1 "] && 
\widetilde{\rho}_1(C_1) \arrow[dddl,shift left,"\kappa_1 "] &=&\rho_2(C_2) \arrow[rr, " \nu_2 "] \arrow[dddr,shift left,"\kappa_2 "] && 
\widetilde{\rho}_2(C_2) \arrow[dddl,shift left,"\kappa_2 "]&=\cdots
 \\
&&&&&&&&\\
&&&&&&&&\\
\cdots \arrow[rr,"\lambda_{10} "] && C_1 \arrow[uuul,"\rho_1 "] \arrow[uuur,"\tilde{\rho}_1 "] \arrow[rrrr,"\lambda_{21} "]&&&&C_2 \arrow[uuul,"\rho_2 "] \arrow[uuur,"\tilde{\rho}_2 "] \arrow[rr,"\lambda_{32} "] && \cdots
\end{tikzcd}}
 \caption{Summary of the representative updates, which can be used to compute automorphisms. Here, $\rho_t: C_t \rightarrow M$ maps the anyons of the child theory to their representatives in parent theory $M$. Anyons in $\rho_t(C_t)$ are updated via $\nu_t$ is defined via the update formula Eq.~\eqref{eq:update_formula}. This gives a complete set of shared representatives between $C_t$ and $C_{t+1}$, and moreover allows us to define of an isomorphism $\lambda_{t+1,t}$ between $C_t$ and $C_{t+1}$.}
 \label{fig:repupdatesummary}
\end{figure}
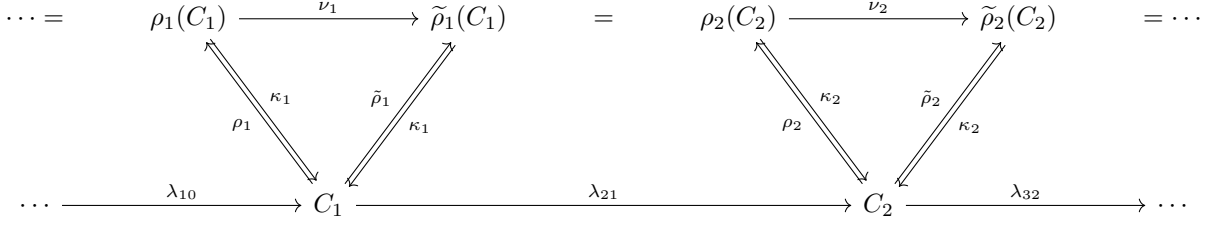

One possible way is to work directly with anyons in the parent code (see Fig.~\ref{fig:repupdatesummary}). Start with $C_0$ and pick any representative $\rho_0(C_0)$. We then may use the update formula in Eq.~\eqref{eq:update_formula} to update them to the next round. 

Each condensation $A_t$ will participate in two reversible pairs: one with $A_{t-1}$ and one with $A_{t+1}$. Thus we will need to represent it using two different bases during the calculation. Hence, define
$A_{t} = \text{span}\{ \hat{\texttt{a}}^{(i)}_{t,{\pm}} \} $ to be the basis for participating in the reversible condensation with $A_{t\pm1}$. Thus given $L_0 = \rho_0(c)$ for $c \in C$ we may solve for $L_T$ recursively via $L_{t+1} = \nu_t(L_t)$ where
\begin{align} \label{eq:logical_update_full}
 \nu_t(L_t) = L_t- \sum_{i,k=1}^{N_{\overline Q_{t+1,t}}} \hat{a}_{t,+}^{(i)} (\beta_{t+1,t}^{-1})^{ik} \langle \hat{a}^{(k)}_{t+1,-}, L_t \rangle 
\end{align}
where $N_{\overline Q_{t+1,t}}$ is the rank of the matrix $B_{t+1,t}^{ij} = \langle\hat{\texttt{a}}^{(i)}_{t+1,-},\hat{\texttt{a}}^{(j)}_{t,+}\rangle$ and $\beta_{t+1,t}$ is the restriction of the matrix to the invertible part.

The automorphism $\varphi$ in $C_0$ therefore acts as
\begin{align}
 \varphi = 
 \kappa_T \circ(\nu_{T-1}\circ \cdots \circ \nu_0)\circ\rho_0
\end{align}

Alternatively, suppose all the canonical isomorphisms $\lambda$ have been computed in advance, then one can readily compose them
to obtain the automorphism 
\begin{align}
 \varphi = \lambda_{T,T-1} \circ \cdots \circ \lambda_{21} \circ \lambda_{10} 
\end{align}

\subsection{Condensation graph} \label{sec:condensation_complex}

The emergence of automorphisms as a consequence of a sequence of reversible condensations motivates us to define the \textit{condensation graph}. We define a vertex for each condensation and connect a pair of vertices by an edge if the corresponding pair of condensations is reversible. We can furthermore assign an isomorphism to each edge given by $\lambda_{21}: C_1 \rightarrow C_2$ and its inverse $\lambda_{12} = \lambda_{21}^{-1}: C_2 \rightarrow C_1$. Fixing a base point, every cycle can be assigned an automorphism given by multiplying all the isomorphisms along the cycle.

\subsubsection{Example: condensing a parent color code}

Ref.~\cite{brown_2022} introduced the example where the parent tat is a single copy of the color code $\CC$ is turned into a single toric code theory $\TC(\ttt{c}\tsig)$ by condensing a single boson $\ttt{c}\tsig$. Let us study the corresponding condensation graph as our first example.

To do this, we first introduce a second, rearranged, magic square, complementary to the Mermin square in Eq.~\eqref{eq:magicsquare}. Recall that the color code contains nine non-trivial bosons and six fermions. The relation between the fermions and bosons can be summarized by the following ``fermion magic square'':
\begin{equation}
{\large
\renewcommand{\rc}[1]{{\color{red}{#1}}}
\renewcommand{\bc}[1]{{\color{blue}{#1}}}
\renewcommand{\gc}[1]{{\color{ForestGreen}{#1}}}
\renewcommand{\yc}[1]{{\color{amber}{#1}}}
\raisebox{0.2\height}{ \begin{tabular}{c|c|c|c|}
\multicolumn{1}{c}{}& \multicolumn{1}{c}{$\frbar$} & \multicolumn{1}{c}{$\fgbar$} & \multicolumn{1}{c}{$\fbbar$}\\ 
\cline{2-4}
$\fr$ &\ry & \bx & \gz \\ \cline{2-4}
$\fg$ &\bz & \gy & \rx \\ \cline{2-4}
$\fb$ &\gx & \rz & \by \\ \cline{2-4}
\end{tabular}} \triangleq 
\raisebox{0.2\height}{ 
 \begin{tabular}{c|c|c|c|}
\multicolumn{1}{c}{}& \multicolumn{1}{c}{\ttt{fe}} & \multicolumn{1}{c}{\ttt{1f}} & \multicolumn{1}{c}{\ttt{fm}}\\ 
\cline{2-4}
\ttt{mf} &\em & \mi & \EE \\ \cline{2-4}
\ttt{f1} &\ie & \ff & \im \\ \cline{2-4}
\ttt{ef} &\mm & \ei & \me \\ \cline{2-4}
\end{tabular}}
\label{eq:fermionmagicsquare}
}
\end{equation}
Here, the squares on the left and on the right correspond to the anyon labels for the color code and two copies of the toric code, respectively. The above magic squares satisfies the following properties:
\begin{enumerate}
 \item The product of three bosons on individual rows or columns is the corresponding fermion denoted on the side of the respective row or column.
 \item Each boson is a product of two of the fermions: one from the row label and one from the column label.
 \item The mutual statistics of two bosons that share the same row or column is $-1$. Otherwise, the mutual statistics is trivial. 
\end{enumerate}
The last point actually follows naturally from the viewpoint of the 3-fermion (3F) theory as follows. All fermions in a 3F theory have pairwise mutual statistics $-1$. 
Thus, if two bosons only differ by a column or a row, then using the property 2, they must differ by a single fermion, which means that they will braid non-trivially. On the other hand, if they differ by both a row and a column, then both the fermion in 3F and $\overline{\text{3F}}$ differ, which means they braid trivially.

Let us show that the condensation graph is exactly the ``dual'' of this fermion magic square. There are nine possible condensations corresponding to each boson in the magic square.
Thus, we can construct the condensation graph as follows. First, we assume the magic square is placed on periodic boundary conditions. Take the vertices to be the bosons that are to be condensed (plaquettes of the fermion magic square) and connect the vertices with an edge if the plaquettes in the magic square share an edge (which are exactly the bosons that have non-trivial mutual statistics).

\begin{figure}[t] \centering
 \centering
 \includegraphics[width=0.9 \columnwidth]{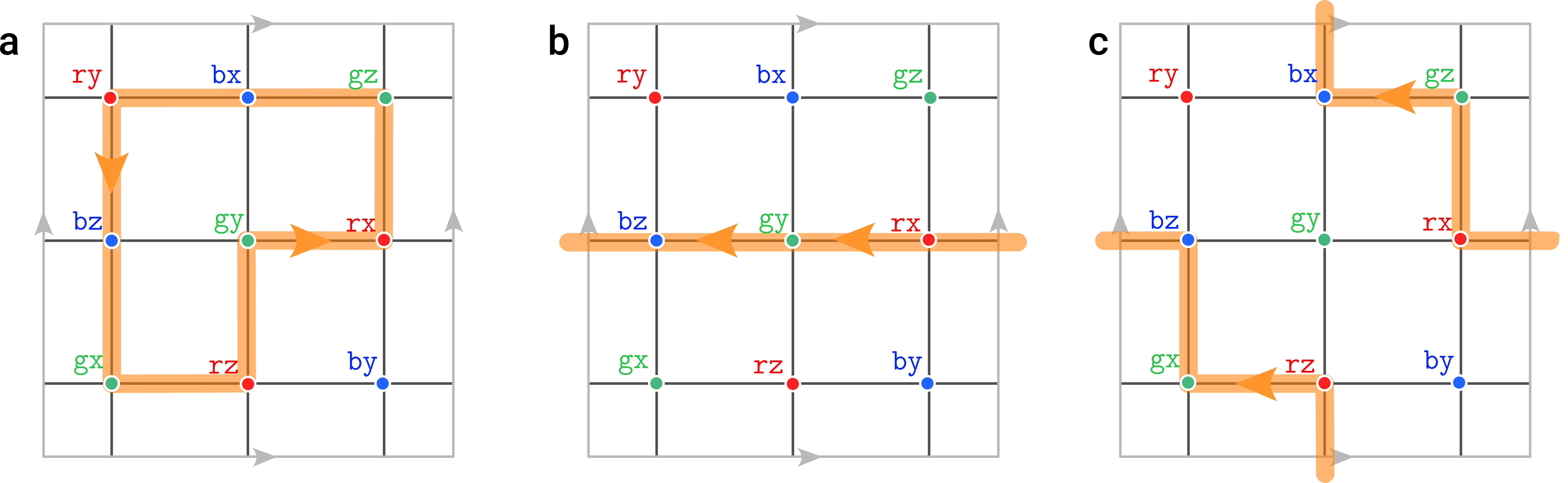}
 \caption{The condensation graph for the color code corresponds to the torus grid graph $T_{3,3}$. a) A contractible path of condensations gives rise to a trivial automorphism. b) A non-contractible path of condensations (such as the honeycomb sequence as shown) gives rise to an $e\rightarrow m$ automorphism. c) The CSS honeycomb sequence wraps around both cycles of the torus, resulting in a trivial automorphism.}
 \label{fig:f_iv}
\end{figure}
The resulting graph is called the torus grid graph $T_{3,3}$ because it can be depicted as a $3\times 3$ grid with periodic boundary conditions as shown in Fig.~\ref{fig:f_iv}. We can also alternatively denote the vertices with coordinates $(i,j) \in \mathbb Z_3^2$ where in the 3F notation, $i = -1,0,1$ correspond to the fermion $\fr,\fg,\fb$, and $j = -1,0,1$ correspond to $\overline{\ttt{f}}_{\ttt r},\overline{\ttt{f}}_{\ttt g}, \overline{\ttt{f}}_{\ttt b}$. In color code notation, the color label is given by mapping $i+j \in \{-1,0, 1\}$ to $\{ \ttt{b},\ttt{g},\ttt{r}\}$ and the Pauli flavor is given by mapping $i-j\in \{-1,0, 1\}$ to $\{ \ttt{x},\ttt{y},\ttt{z}\}$.

We can now study what automorphisms are assigned to the the different cycles of this graph.
In order to aid in the calculation of the automorphisms, it is helpful to make a choice for the representative anyons. Since the child anyons correspond a single toric code, we will make the following choice to identify anyons that braid trivially with the condensation $\ttt{c}\tsig$ and anyons as the the toric code
\begin{align} \label{eq:emchoicefortoriccode}
 \kappa(\ttt{c}\tsig') &=\ttt e \text{ for }\sigma' \ne \sigma & 
 \kappa(\ttt{c}'\tsig) &= \ttt{m} \text{ for } c' \ne c
\end{align}
With this choice, note that every reversible pair of condensations is of the form $\ttt{c}\tsig$ and $\ttt{c}'\tsig'$ for $c' \ne c$ and $\sigma' \ne \sigma$ (this is the ``tree code condition'' of Ref.~\cite{Davydova22}). This implies that $\ttt{c}\tsig'$ maps to $\ttt e$ in the child for $\ttt{c}\tsig$ but maps to $\ttt m$ in the child for $\ttt{c}'\tsig'$. Similarly, for $\ttt{c}'\tsig$. In conclusion, with this choice we find that every isomorphism on this graph swaps $\ttt e$ and $\ttt m$ of the child toric code!

The above isomorphisms can be used to conveniently compute all automorphisms on this condensation graph. First, consider an arbitrary contractible loop on this torus (Fig. \ref{fig:f_iv}a). Such a loop is always composed of an even number of edges, and therefore, the automorphism is trivial.

On the other hand, let us consider the honeycomb code measurement sequence, which corresponds to the cycle $\rx \rightarrow \gy \rightarrow \bz \rightarrow \rx$.
Notice that this path is a non-contractible cycle around the torus (Fig. \ref{fig:f_iv}a). Indeed, it is known that automorphism associated with this path is the $e - m$ automorphism \cite{Hastings2021}, which agrees with the fact that the sequence has an odd length.

The above two facts allow us to derive an intriguing property of this torus. Any contractible cycle on the torus will also result in a trivial automorphism.
While any non-contractible loop around the $(1,0)$ or $(0,1)$ cycles of the torus will result in the $\ttt e - \ttt m$ automorphism! More generally, any path $(n,m)$ implements the $\ttt e - \ttt m$ automorphism if $n+m$ is odd and the trivial automorphism if $n+m$ is even. A notable example is the sequence corresponding to the CSS honeycomb code \cite{Davydova22,brown_2022,Bombin23} wraps around the $(1,1)$ cycle of the torus (Fig. \ref{fig:f_iv}c). Thus, it implements a trivial automorphism. To conclude, we can colloquially say that each hole of the torus contains such an $\ttt e - \ttt m$ automorphism.

One might be concerned that choosing a particular choice of basis of $e$ and $m$ in the child will affect the final automorphism calculated. To illustrate that this is not the case, suppose we make a basis transformation by applying an $\ttt e - \ttt m$ automorphism to a vertex. This will modify the isomorphism for all edges entering that vertex to be the trivial isomorphism. However, note that this does not change the property of the product of isomorphisms around any closed loop, since any path entering that vertex must also exit it, thus the number of $\ttt e - \ttt m$ isomorphisms along the loop always changes by an even number of times.

Readers familiar with lattice gauge theory might find the above argument very familiar. Indeed, one can abstractly view this graph as having a $C$-valued bundle on every vertex. Isomorphisms connecting different vertices can be thought of as ``connections'', which are not ``gauge invariant''. However, the ``integration of a connection around a closed loop'' is gauge-invariant, corresponding to an automorphism. Choosing a basis for $C$ at each vertex can be thought of as ``gauge fixing'' while changing a basis can be thought of as a ``gauge transformation''.

\subsubsection{Products of condensation graphs}
To build up to describing the condensation graph for the DA color code, we first consider a subclass of condensations $\ttt{c}\tsig_1$ and $\ttt{c}'\tsig_2'$ in two parent color codes (compared to the DA color code, we are only missing the condensation to $\widetilde{\CC}$, which we will add back shortly). 
 The corresponding condensation graph can be simply described by taking a product of the two graphs.

More generally, suppose we take a product of two parent theories $M= M_1 \boxtimes M_2$ and only consider a subset of condensations of the form $A = \langle A_1 \boxtimes I_2, I_1 \boxtimes A_2 \rangle$. Reversible condensations to $A$ are of the form $A = \langle A_1' \boxtimes I_2, I_1 \boxtimes A_2 \rangle$ or $A = \langle A_1 \boxtimes I_2, I_1 \boxtimes A_2'\rangle$. Therefore, if the condensation graphs are $\mathcal G_1$ and $\mathcal G_2$, the condensation graph of the product is exactly the (Cartesian) product $\mathcal G_1 \Box\mathcal G_2$.

 Note that within the product graph, new cycles of the form
 \begin{equation}
\begin{tikzcd}
\langle A_1 \boxtimes I_2, I_1 \boxtimes A_2 \rangle \arrow[rr, " \text{id} \boxtimes \lambda_2 "] && 
\langle A_1 \boxtimes I_2, I_1 \boxtimes A_2' \rangle \arrow[ddd, "\lambda_1 \boxtimes\text{id}"]
 \\
&&\\
&&\\
\langle A_1' \boxtimes I_2, I_1 \boxtimes A_2 \rangle \arrow[uuu,"\lambda_1^{-1} \boxtimes\text{id} "] 
&&
\langle A_1' \boxtimes I_2, I_1 \boxtimes A_2 \rangle \arrow[ll, " \text{id} \boxtimes \lambda_2^{-1} "]
\end{tikzcd}
\end{equation} 
are introduced. However, we see that the corresponding automorphism is trivial. Thus, the only cycles of the graph that correspond to non-trivial automorphisms are those that are already present in $\mathcal G_1$ and $\mathcal G_2$.

\subsubsection{Condensation graph of the DA color code}

First, let us discuss the condensation graph where the parent code is two copies of the color code $\CC_1 \times \CC_2$ and condensations of the form $A = \text{span}\{\ttt{c}_1\tsig_1,\ttt{c}_2\tsig_2 \}$ which leads to the child $C=\TC(\ttt{c}_1\tsig_1) \times \TC(\ttt{c}_2\tsig_2)$. Because the condensations are products, we conclude that the corresponding condensation graph is $T_{3,3,3,3}=T_{3,3} \Box T_{3,3}$.

The above graph can be visualized on a 4-torus. One way to describe such a graph is to label the vertices by $(i,j,k,l) \in \mathbb Z_3^4$. Using the same basis for anyons as in the single color code, the edges and the corresponding isomorphisms are then
\begin{align}
(i,j,k,l) &\rightarrow (i+1,j,k,l) & (\texttt{xr})(\texttt{yg})(\texttt{zb})\\
(i,j,k,l) &\rightarrow (i,j+1,k,l) & (\texttt{xr})(\texttt{yg})(\texttt{zb})\\
(i,j,k,l) &\rightarrow (i,j,k+1,l) & (\texttt{zr})(\texttt{yg})(\texttt{xb})\\
(i,j,k,l) &\rightarrow (i,j,k,l+1) & (\texttt{zr})(\texttt{yg})(\texttt{xb})
\end{align}
for $i,j,k,l \in \mathbb Z_3$ and we remind that $(\texttt{xr})(\texttt{yg})(\texttt{zb})$ and $(\texttt{zr})(\texttt{yg})(\texttt{xb})$ in the color code convention corresponds to $\ttt{e}_\ttt{1} \leftrightarrow \ttt{m}_\ttt{1}$ and $\ttt{e}_\ttt{2} \leftrightarrow \ttt{m}_\ttt{2}$ in the toric code convention, respectively.

Now, we will add the vertex corresponding to condensing $A_{\widetilde{\CC}}=\langle \rzi\rzii,\bzi\bzii\rangle$ (which corresponds to measuring $V(Z_1Z_2)$ in the code). The other vertices in $T_{3,3,3,3}$ that share an edge with $A_{\widetilde{\CC}}$ are of the form $\langle \ttt{c}_1\tsig_1, \ttt{c}_2\tsig_2\rangle$ where $c_1 \ne c_2 \in \{ r,g,b \}$ and $\sigma_1, \sigma_2 \in \{ x,y \}$. Thus $A_{\widetilde{\CC}}$ connects to 24 other vertices.

\begin{table}[]
 \centering
 \begin{tabular}{|c|c|c||c|c|c|}
 \hline
 Isomorphism $(\lambda)$ & \multicolumn{2}{l||}{Condensation $(A)$} & Isomorphism $(\lambda)$ & \multicolumn{2}{c|}{Condensation $(A)$}\\
 \hline
 \multirow{2}{*}{$1$} & \hspace{8 pt} $\rxi$ \hspace{8 pt} & $\ryi$ & \multirow{2}{*}{$(\ttt{xy})$} & $\rxi$ & $\ryi$ \\
 & $\bxii$ & $\byii$ & & $\byii$ & $\bxii$\\
 \hline
 \multirow{2}{*}{ $(\ttt{rg})$} & $\gxi$ & $\gyi$ & \multirow{2}{*}{$(\ttt{xy})(\ttt{rg})$} & \hspace{8 pt} 
 $\gxi$ \hspace{8 pt} & $\gyi$ \\
 & $\bxii$ & $\byii$ & & $\byii$ & $\bxii$\\
 \hline
 \multirow{2}{*}{ $(\ttt{gb})$} & $\rxi$ & $\ryi$ & \multirow{2}{*}{$(\ttt{xy})(\ttt{gb})$} & $\rxi$ & $\ryi$ \\
 & $\gxii$ & $\gyii$ & & $\gyii$ & $\gxii$\\
 \hline
 \multirow{2}{*}{ $(\ttt{rb})$} & $\bxi$ & $\byi$ & \multirow{2}{*}{$(\ttt{xy})(\ttt{rb})$} & $\bxi$ & $\byi$ \\
 & $\rxii$ & $\ryii$ & & $\ryii$ & $\rxii$\\
 \hline
 \multirow{2}{*}{ $(\ttt{rgb})$} & $\gxi$ & $\gyi$ & \multirow{2}{*}{$(\ttt{xy})(\ttt{rgb})$} & $\gxi$ & $\gyi$ \\
 & $\rxii$ & $\ryii$ & & $\ryii$ & $\rxii$\\
 \hline
 \multirow{2}{*}{ $(\ttt{rbg})$} & $\bxi$ & $\byi$ & \multirow{2}{*}{$(\ttt{xy})(\ttt{rbg})$} & $\bxi$ & $\byi$ \\
 & $\gxii$ & $\gyii$ & & $\gyii$ & $\gxii$\\
 \hline
 \end{tabular}
 \caption{Isomorphisms $ \lambda: \widetilde{\CC} \rightarrow \TC(\ttt{c}_1\tsig_1) \boxtimes \TC(\ttt{c}_2\tsig_2). $ We have chosen a representative for the anyons according to Eq.~\eqref{eq:effCC_bosons} and for the toric codes as Eq.~\eqref{eq:emchoicefortoriccode}.}
 \label{tab:isomorphismfromVZZ}
\end{table}

The isomorphisms from $A_{\widetilde{\CC}}$ are calculated using the update formula Eq.~\eqref{eq:update_formula} and are summarized in Table~\ref{tab:isomorphismfromVZZ}. Because the table contains all the generators of all 72 automorphisms, this shows that we can get all of them. Recall that we have chosen a representative for the anyons of the effective color code according to Eq.~\eqref{eq:effCC_bosons} and for the toric codes as Eq.~\eqref{eq:emchoicefortoriccode}.

With this, let us derive the automorphisms shown in Table~\ref{tab:generators}.
Consider the first sequence
\begin{center}
\begin{tabular}{|c|c|c|c|c|c|}
\hline
 $t=$0 & 1 & 2 & 3 & 4 & 5 \\
\hline
 $\rzi \rzii$ & $\rxi$ & $\gyi$ & $\bzi$ & $\rxi$ & $\rzi \rzii$ \\
 $\bzi \bzii$ & $\bxii$ & $-$ & $\gzii$ & $\bxii$ & $\bzi \bzii$
 \\\hline
\end{tabular}
\end{center}
We have $\lambda_{10} = \lambda_{54}^{-1} = \text{id}$. From rounds $1-4$ we accrue $\varphi_{(\texttt{xr})(\texttt{yg})(\texttt{zb})}$ three times and $\varphi_{(\texttt{zr})(\texttt{yg})(\texttt{xb})}$ two times. Thus, combining all isomorphisms, we find that this sequence gives $\varphi_{(\texttt{xr})(\texttt{yg})(\texttt{zb})}$. The second sequence gives $\varphi_{(\texttt{zr})(\texttt{yg})(\texttt{xb})}$ through an identical calculation. Lastly, the third sequence 
\begin{center}
\begin{tabular}{|c|c|c|c|c|}
\hline
 $t=$0 & 1 & 2 & 3 & 4 \\
\hline
$\rzi \rzii$ & $\rxi$ & $\bzi$ & $\rxi$ & $\rzi \rzii$ \\
$\bzi \bzii$ & $\bxii$ & $\ryii$ & $\gxii$ & $\bzi \bzii$ \\\hline
\end{tabular}
\end{center}
has $\lambda_{10}= \text{id}$, $\lambda_{21} =\lambda_{32}= \varphi_{(\texttt{xz})(\texttt{rb})}$, $\lambda_{43}= \varphi_{(\texttt{gb})}$. Thus, this sequence gives the automorphism $\varphi_{(\texttt{gb})}$. We remark that this argument shows that we would also have gotten the same automorphism had we replaced the second condensation with $\bzi$ and $\rzii$, for example.

\subsection{A method to construct desired automorphisms}
\label{sec:tqft_aut_construction}

Let us finally provide a method to construct a sequence of condensations that produces the desired automorphism on a child anyon theory.
Let us first show that we can construct a desired isomorphism from a condensation sequence. Assume we have a reversible sequence of condensations $A_0, \ldots ,A_T$, where, unlike the previous section, we do not need to assume $A_0=A_T$. Since each consecutive pair of condensations is reversible, we can associate an isomorphism $\lambda$ from the child theory $C_0 = A_0^\perp/A_0$ to the child theory $C_T=A_T^\perp/A_T$.

Now, let us assume that there exists an automorphism $\varphi$ of the parent such that $\varphi(A_0)=A_0$ and $\varphi(A_T) = A_T$. Note that this induces well-defined automorphisms on the child models $C_0$ and $C_T$\footnote{To see that an automorphism that preserves a condensation $A$ defines an automorphism on the child theory $C$, first note that $\varphi$ restricts to the action on $A^\perp$ since if $b\in A^\perp$, the $\langle b,a\rangle=0$ for all $a\in A$, Applying $\varphi$ we have that $\langle \varphi(b),\varphi(a)\rangle=0$. But since $\varphi(a)\in A$ we conclude that $\varphi(b) \in A^\perp$ also. Next, we may quotient $A^\perp$ by $A$ to get $C = A^\perp/A$. Again, since $\varphi(A) =A$ this guarantees that $\varphi(b)$ is in the same coset as $\varphi(ba)$.}. Let us further assume that this induced automorphism is a trivial automorphism $\psi_0 = \text{id}_{C_0}$ on $C_0$ but a non-trivial automorphism $\psi_T$ on $C_T$.

We can apply this automorphism of the parent theory $\varphi$ to the entire sequence. This produces a new condensation sequence
$ \varphi(A_0), \varphi(A_1),\ldots,\varphi(A_{T-1}), \varphi(A_T) $. This sequence can be equivalently written as $ A_0, \varphi(A_1),\ldots,\varphi(A_{T-1}), A_T$. As a consequence of this modification, an isomorphism from $C_0$ to $C_T$ of the original sequence is now changed to $ \psi_T \circ \lambda \circ \psi_{0}^{-1} = \psi_T \circ \lambda$.

Thus, consider the sequence $A_T, A_{T-1}, \ldots, A_1,A_0, \varphi(A_1),\ldots,\varphi(A_{T-1}), A_T$ where we apply the original sequence in reverse followed by the new sequence. This induces an automorphism $\psi \circ \lambda \circ \lambda^{-1} =\psi$ on $C_T$.

To give an example, let us show how we can use this method to obtain an automorphism $\varphi_{(\texttt{xy})}$, which for the color code is equivalent to implementing a transversal $S$ gate. The parent is equivalent to two copies of the color code, and consider the condensation sequence
\begin{align}
 A_0 &= \{ 1, \gzi,\rzii, \bzi\rzii\}, & A_1 &= \{ 1,\rxi,\gxii,\rxi\gxii\}, & A_2 & =\{1, \rzi\rzii,\gzi\gzii ,\bzi\bzii \}.
\end{align}
We choose the automorphism of the parent theory $\varphi = {\varphi_{(\texttt{xy})}}_1 \boxtimes \text{id}_2$. That is, we only permute \ttt{x} and \ttt{y} type anyons in the first copy. This leaves $A_0$ and $A_2$ invariant. First consider the child at round zero $C_0 =\TC(\gzi) \times \TC(\rzii)$. The equivalence classes of the anyons in the first toric code copy are
\begin{align}
 [\ttt{e}_1] &= \{ \rzi,\bzi,\rzi\rzii,\bzi\rzii \} & [\ttt{m}_1]&= \{ \gxi,\gyi,\gxi\rzii,\gyi\rzii \}
\end{align}
which are invariant under the action of $\varphi$. Thus $\psi_0=\text{id}_{C_0}$ as desired. On the other hand, $C_2 =\widetilde{\CC}$, and $\varphi$ acts as $\psi_2 = \widetilde{\varphi}_{(\texttt{xy})}$.

This allows us to design a cycle of condensations that starts and ends with an effective color code and use the trick above to implement a desired automorphism. Specifically, based ont he sequence above, let us choose the cyclic sequence 
\begin{align}
 A_0' &= A_2, &A_1' &= A_1, & A_2' &= A_0, &A_3 &=\varphi(A_1) = \{ 1,\ryi,\gxii,\ryi\gxii\} & A_4' &= A_2
\end{align}
performs the automorphism $\widetilde{\varphi}_{(\texttt{xy})}$ on the child $\widetilde{\CC}$. In terms of measurements on stabilizer codes, this is exactly sequence 19 in Appendix~\ref{app:72}.

Later in Sec.\ref{sec:3DFCC}, we will also use this trick to design measurement sequences that perform the $(\ttt{ry})$ automorphism of the 3D color code (see table \ref{tab:3DDACC_table_short}) as well as the automorphism that is equivalent to implementing a transversal $T$ gate~\cite{Yoshida2015,Yoshida2017} (see Table~\ref{tab:3DDACC_table_T}).

\section{{Full Clifford group with dynamic automorphism color code on a triangle}} \label{sec:triangle}

In Sec.~\ref{sec:torus} we showed how to implement all automorphisms of the color code via a sequence of measurements.
On a torus, the action of the color code automorphisms on the logical operators furnishes a subgroup of the real Clifford group on four logical qubits. 
Here, we show that the full (complex) Clifford group can be achieved on two logical qubits by constructing a DA color code on a triangle with Pauli boundaries. By extension, the full Clifford group on $N$ logical qubits can be realized on $N$ layers of a triangular DA color code.

To achieve automorphisms that realize complex Clifford gates, one needs to consider a color code on a manifold with a boundary. 
For example, a color code on a triangle with Pauli boundaries (first introduced in \cite{Kesselring_2018}) encodes a single logical qubit and achieves automorphisms corresponding to all possible permutations of logical strings. 
In terms of gates, these automorphisms form a basis for all single-qubit Clifford gates. 
This color code can be considered as a child code of a parent bilayer color code with specific bulk and boundary condensations. 
Further taking two layers of the (child) color code on a Pauli triangle and augmenting the single qubit gates with an additional automorphism that realizes the iSWAP gate, we form a basis for the Clifford group on two logical qubits.

We start by showing in Subsec.~\ref{sec:tqft_bound} how a sequence of reversible condensations for a theory without a boundary plus a fixed boundary condition at one rounds gives rise to a boundary condensation sequence at all other rounds.
Such boundary condensation sequences can be translated into boundary measurement sequences upon which the corresponding logical strings can terminate without violating a boundary stabilizer.
The length of the logical operators for such a code at each round is proportional to the geometric distance between the boundaries of the same kind.
In Subsec.~\ref{sec:condensation_triangles} we work out (bulk and boundary) condensation sequences that produce the automorphisms corresponding to a full Clifford group on two logical qubits living on two (bilayer) Pauli triangles. These condensation sequences can be mapped to an explicit set of measurements on a triangular patch, which allows us to construct a DA color code with Pauli boundaries in Subsec.~\ref{sec:boundary_micro}. Lastly, we address the question of error correction of proposed protocols in Subsec.~\ref{sec:EC_triangle}; in particular, we discuss the basis of bulk and boundary detectors of the DA color code. 

Although this section addresses the DA color code that realizes color code ISGs on a Pauli triangle only, the case of the color triangle~\cite{Bombin2006} is discussed in Appendix~\ref{sec:SM_color_triangle}.

\subsection{Condensation sequences in the presence of boundaries}
\label{sec:tqft_bound}

Let us explain how we extend the concept of reversible condensations to child theories with boundaries. Roughly speaking, the boundaries are associated with condensing certain representatives of anyons in the parent theory, which imposes constraints on the sequence of allowed boundary measurements. A valid sequence of measurements near each boundary will allow logicals (and their updates) to terminate at their associated boundary during each round. 
Without loss of generality, we will explain the case of a single gapped boundary, but the construction readily generalizes to multiple gapped boundaries, as is the case of the color code on a triangle.

Consider a child theory $C$. A gapped boundary of $C$ can be constructed by first choosing a condensation $B \subset C$. $B$ has the further property that it is Lagrangian in $C$, namely $B=B^\perp$ where orthogonality is based on the inner product defined on $C$ (see Sec.~\ref{sec:TQFTreview}). Condensing such a Lagrangian subspace guarantees that there are no remaining deconfined anyons after the condensation (that is, the resulting theory is trivial). Thus, given $C$ defined on an infinite plane, a gapped boundary of $C$ can be constructed by condensing $B$ on, for example, the right half-plane. This realizes $C$ on the left half of the plane, with condensation to the vacuum on the right half.

Next, recall that $C$ itself is a child theory which resulted by starting with a parent $M$ and condensing $A$. We would now like to construct $C$ with a boundary starting from $M$. To do so, we find a representative for every anyon in $B$, which forms a subspace $\rho(B) \subset M$. Therefore, starting from $M$ on an infinite plane, we can condense $A$ on the left half of the plane and condense $A+ \rho(B)$ on the right half. Note that $A+ \rho(B)$ is Lagrangian in $M$\footnote{When $M$ is an Abelian modular category and $A$ is a condensible subset of anyons we have $|A^\perp||A| = |M|$, see Ref.~\cite{Mueger2002structure}.
We also have $|A^\perp/A| = |A^\perp|/|A|$. 
An algebra $B$ is Lagrangian in a modular category $C$ iff $|C| = |B|^2$. 
We wish to show $A + \rho(B)$ is Lagrangian in $M$.
Note that $A \cap \rho(B) = 0$ since $B$ is a non-trivial set of anyons in $C$ and $C = A^\perp/A$. 
We also have that $|B| = |\rho(B)|$.
Hence $|A + \rho(B)|^2 = |A |^2|\rho(B)|^2= |A|^2|B|^2 = |A|^2 |C| = |A|^2 \left| {A^\perp}/{A} \right| = |A| |A^\perp| = |M|$.
This implies that $A + \rho(B)$ is Lagrangian in $M$.}.
This matches the intuition that condensing both $A$ and $\rho(B)$ must result in the trivial theory. After the above condensation, the anyons in $B$ condense on this boundary to the vacuum.

We can now analyze the case when we have a pair of reversible condensations in the bulk, $A_1$ and $A_2$. Given a Lagrangian subspace $B_1 =B_1^\perp \subset C_1$, there exists $\rho_1$ such that $\rho_1(B_1) \subset A_2^\perp$. Moreover, $B_2 =\kappa_2 (\rho_1(B_1))$ is Lagrangian in $C_2$ since $|B_2|^2 =|B_1|^2 =|C_2|$. Let us define $A_{\partial,12} =\rho_1(B_1)$. 
We would like to show that the following condensations constitute a reversible pair:
\begin{enumerate}
 \item Condensing $A_1$ on the left and $A_1 + A_{\partial,12}$ on the right.
 \item Condensing $A_2$ on the left and $A_2 + A_{\partial,12}$ on the right.
\end{enumerate}
To see this, we can reinterpret the construction of the gapped boundary as follows. Since $A_{\partial,12}$ appears in both condensations and the condensations commute, let us first condense only $A_{\partial,12}$ on the right half-plane. This results in $M$ on the left plane and $C_{\partial,12} = A_{\partial,12}^\perp/A_{\partial,12}$ on the right plane. Let us now consider each half plane separately. On the left half-plane, the remaining condensation pair $A_1$ and $A_2$ is reversible by assumption. On the right plane, the condensation pair $\kappa_{\partial,12}(A_1)$ and $\kappa_{\partial,12}(A_2)$ are also reversible with respect to $C_{\partial,12}$, because there are no deconfined anyons left on either side. This shows our claim.

Let us discuss the implications of reversible condensations in the presence of a gapped boundary. Since $A_{\partial,12}$ are only shared among a pair of reversible condensations $A_1$ and $A_2$, this means that condensations must also be updated for each pair of condensation rounds that follow. In fact, since $A_{\partial,12} =\rho_1(B_1)$, where $B_1 \subset C_1$, this means that the boundary condensations must be updated in the exact same way as the representatives of child theory anyons are updated! Explicitly, given bulk condensations $A_0, A_1, \ldots A_T=A_0$ (i.e. on the left half-plane), and a Lagrangian boundary condensation $B_0 \subset C_0$, the condensations of the right half plane are fixed to be
\begin{align}
 A_0 &+\rho_0 (B_0); & &\\
 A_1 &+ A_{\partial,01}; & A_{\partial,01} &= \rho_0 (B_0)\\
 A_2 &+ A_{\partial,12}; & A_{\partial,12} &= \nu_1 (A_{\partial,01})\\
 &\vdots& &\vdots\\
 A_{T} &+ A_{\partial,T-1 \ T}; & A_{\partial,T-1 \ T} &= \nu_{T-1} (A_{\partial,T-2 \ T-1})
\end{align}
where $\nu$ updates the anyons according to the update formula, see Eq.\eqref{eq:update_formula}.

Going to dynamic automorphism codes, we recall that a measurement sequence can be translated from a condensation sequence by considering a Hamiltonian realization of the parent and each of the child theories and by translating condensations into measurements (see for example Eq.~\eqref{eq:anyontomeasurement} for the DA color code). Note that in doing so, we have the freedom of redefining the basis of anyons of $A_t + A_{\partial, t-1 t}$. In particular, we may choose a basis that corresponds to low-weight measurements or another that has advantageous error-correction properties.

The boundary update condition for condensations translates into the following condition on measurements: the boundary measurements have to be such that, if a logical string terminates (condenses) at a boundary at a given round, this logical string, updated according to the bulk update rule, has to condense at the same boundary in the subsequent round. Intuitively, one might think about this the following way: within the child code, each boundary can be put in a one-to-one correspondence with the set of logical strings that can terminate at this boundary. For each logical string, there exists a representative that is also a logical string of the next round. The condition demands that this representative terminates on the same boundary in the next round; in the presence of several boundaries, it means that the shape and position of the logical operators do not change (however, they might shift by an edge upon being updated). In turn, that means that 2+1D logical membranes remain ``continuous'' in spacetime and do not rotate, which could cause uncorrectable errors.

\subsection{Condensation protocols on single and double triangles with Pauli boundaries} \label{sec:condensation_triangles}

\subsubsection{Automorphisms on Pauli triangles}

\begin{figure}[b] \centering
\vspace{0pt}
\centering
\vspace{0pt}
\includegraphics[width= 1\columnwidth]{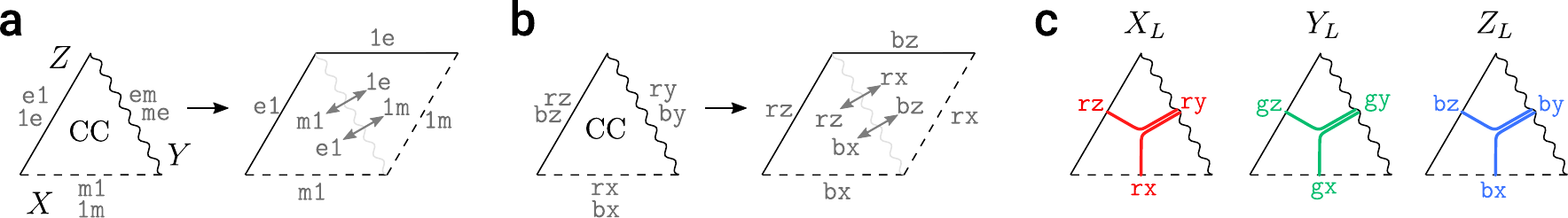}
\caption{Unfolding the Pauli triangle color code into a pair of toric codes in (a) double toric code and (b) color code notation. Here and in subsequent drawings, the solid boundary is of $Z$ type, the dashed one is of $X$ and the wavy one is the $Y$ boundary. here is an $\ttt e - \ttt m$ domain wall along the line where the toric codes join, and the transformations of anyons crossing this domain wall are shown on both sides of the arrows. 
Each Pauli boundary corresponds to condensing a Lagrangian algebra object generated by a corresponding Pauli column in the color code magic square. The bosons condensed at each boundary are shown in gray.
(c) The logical operator strings of the Pauli triangle color code and our convention for the logical Pauli operators. We display the trivalent junction as an ${\texttt{rx}}$ and a ${\texttt{rz}}$ logical strings that join to form ${\texttt{ry}}$ before meeting the $ Y$ boundary; similar logic applies to the other colors.
}
\label{fig:41}
\end{figure}

\begin{figure}[!b] \centering
\vspace{0pt}
\centering
\vspace{0pt}
\includegraphics[width= 1\columnwidth]{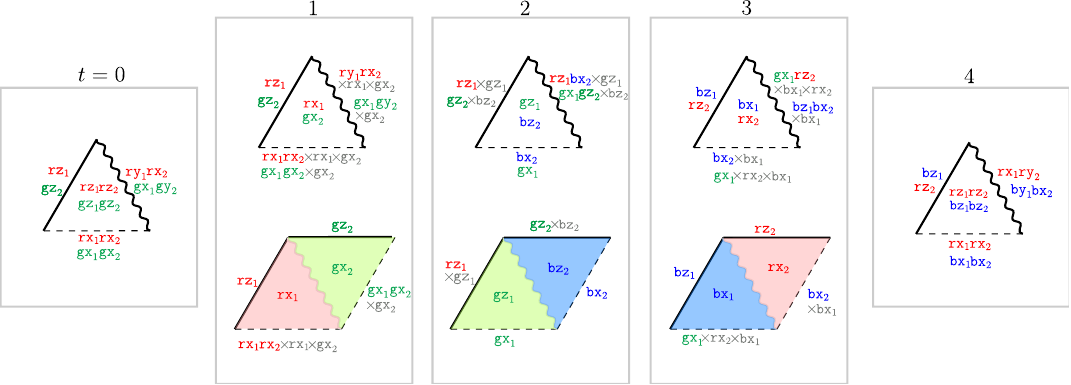}
\caption{A sequence of anyon condensations from the parent bilayer color code that implements $\varphi_{(\texttt{rbg})}$ automorphism in a DA color code on a triangle. Specifically, the condensations in the bulk correspond to the $\varphi_{(\texttt{rbg})}$ protocol from Table~\ref{tab:S3_color}. At $t=0$, we start in an effective color code with $\mathtt{rz_1 rz_2}$ and $\mathtt{gz_1 gz_2}$ condensations in the bulk and appropriate condensations at the boundary that give Pauli boundaries of the effective color code.
Between rounds 1 and 3, the top row shows the triangle picture with boundary condensations worked out as described in Sec.~\ref{sec:tqft_bound}, such that each transition preserves the logical information. Rounds 1 and 3 correspond to a toric code with an $\ttt e - \ttt m$ domain wall at each round.
The bottom row at these rounds shows the usual unfolding picture for these rounds. 
The other generating sequence for a single triangle, corresponding to the $\varphi_{(\texttt{rb})}$ automorphism, as well as the double triangle sequence can be worked out analogously and the result is shown in Table~\ref{tab:TQFTcondensationdata}.
 }
\label{fig:411}
\end{figure}

First, let us consider a color code on a triangle with Pauli boundaries as shown in Fig.~\ref{fig:41}. There are three such boundaries labeled by $\sigma= X,Y,Z$ where the $\sigma$ boundary is obtained by condensing the Lagrangian subgroup $\{1, \ttt{r}\tsig, \ttt{g}\tsig , \ttt{b}\tsig\}$ (i.e. all the bosons in the $\sigma$ column of the magic square) at the boundary. In the presence of these boundaries, the automorphisms must leave the Lagrangian subgroup for each boundary invariant. 
Therefore, the only boundary-preserving automorphisms are the color-permuting automorphisms, for which there are six of them in total.

Starting from two layers with the parent color code theory on a triangle as described above, let us show how to construct a single child color code on a triangle with Pauli boundaries. For now, assume that we work on an infinite plane, and consider bulk condensations acting everywhere on the plane and boundary condensations acting on disjoint thirds of the plane. The partitioning of the plane is chosen so that a triangular patch is formed from the set of sites that involves bulk measurements but no boundary measurements. Thus, an effective (child) color code can be obtained by condensing $\rzi \rzii$ and $\bzi \bzii$ (in which case $\gzi \gzii$ is also condensed) everywhere in space and then forming effective $X, Y$, and $Z$ boundaries by condensing appropriate bosons of the effective color code (shown in Eq.~\eqref{eq:effCC_bosons}) at the respective boundaries. Such a condensation pattern is shown at $t=0$ in Fig.~\ref{fig:411}.

The distinct measurement sequences giving rise to color-permuting automorphisms in the bulk are shown in the first two rows of Table~\ref{tab:S3_color}. The first two sequences in Table~\ref{tab:S3_color} correspond to $\varphi_{(\texttt{rb})}$, which swaps green and blue anyon labels, and the $\mathbb Z_3$ automorphism $\varphi_{(\texttt{rbg})}$ that cyclically permutes the colors. In fact, any of the 6 automorphisms can be obtained from these generating measurement sequences. Alternatively, we can obtain sequences realizing other automorphisms by simply using the sequences in Table~\ref{tab:S3_color} and appropriately permuting the colors in the measurement sequences.

\begin{table}[!b]
{
\renewcommand{\rc}[1]{{\color{red}{#1}}}
\renewcommand{\bc}[1]{{\color{blue}{#1}}}
\renewcommand{\gc}[1]{{\color{ForestGreen}{#1}}}
\renewcommand{\yc}[1]{{\color{amber}{#1}}}
 \centering
\resizebox{\textwidth}{!}{ \begin{tabular}{|c|c|c|c|c|c|c|c| }
 \hline
 \multirow{2}{*}{Gate/Aut} & \multicolumn{7}{c|}{$t$} \\ \cline{2-8} 
 & 0 & 1 & 2 & 3 & 4 & 5 & 6\\
 \hline
 \multirow{2}{*}{$H \sim \varphi_{(\texttt{rb})} $} & \multirow{2}{*}{ $V(Z_1Z_2)$} & $\Eb(X_1)$ & $\Eg(Z_1) $ & $\Er(X_1) $ & $\Eb(Z_1)$ & $\Er(X_1)$ & \multirow{2}{*}{ $V(Z_1Z_2)$} \\
 & & $\Er(X_2)$ & $\Eb(Z_2)$ & $\Eg(X_2)$ & $\Er(Z_2)$ & $\Eb (X_2)$ & \\
 \hline
 \multirow{2}{*}{$SH \sim \varphi_{(\texttt{rbg})}$} &\multirow{2}{*}{ $V(Z_1Z_2)$} & $\Er(X_1)$ & $\Eg(Z_1)$ & $\Eb(X_1)$ & \multirow{2}{*}{ $V(Z_1Z_2)$} & & \\
 & & $\Eg(X_2)$ & $\Eb(Z_2)$ & $\Er(X_2)$ & & & \\ 
 \hline
 \hline
 \multirow{4}{*}{iSWAP} &\multirow{2}{*}{ $V(Z_1Z_2)$} & $\Er(X_1)$ & \multirow{2}{*}{ $V(Z_1Z_3)$} & $\Eb(X_1)$ & \multirow{2}{*}{ $V(Z_1Z_2)$} & & \\
 & & $\Eb(X_2)$ & & $\Eg(X_2)$ & & & \\ 
 &\multirow{2}{*}{ $V(Z_3Z_4)$} & $\Eb(X_3)$ & \multirow{2}{*}{ $V(Z_2Z_4)$} & $\Eg(X_3)$ & \multirow{2}{*}{ $V(Z_1Z_2)$} & & \\
 & & $\Er(X_4)$ & & $\Eb(X_4)$ & & & \\
 \hline
 \end{tabular}}
 \caption{The first two rows show the bulk measurement sequences for the two generators of the $S_3$ subgroup of color permutations of the color code automorphisms. These sequences are used in a Pauli triangle dynamic color code, which generates the Clifford group on one qubit. To obtain the full Clifford group, we augment it with the two-qubit logical iSWAP gate, shown in the last row. We discuss the boundary measurements in subsection \ref{sec:boundary_micro} and how the two-qubit gate is identified is explained in subsection~\ref{sec:logicals_triangle}.}
 \label{tab:S3_color}
}
\end{table}

The sequences for automorphisms shown in Table~\ref{tab:S3_color} were chosen such that the corresponding logical gates on a triangle have a nice form. We also restrain ourselves to CSS protocols in the bulk (i.e. only $X$ and $Z$ measurements) because the codes obtained from them are easier to analyze. Moreover, as we demonstrate later on, the chosen protocols for a single Pauli triangle DA color code are error-correcting both in the bulk and on the boundaries. 

The automorphisms shown in the first two rows of Table~\ref{tab:S3_color} allow us to realize all single-qubit Clifford gates. To complete the generators for a multi-qubit Clifford group, we need to achieve a two-qubit entangling gate. The last row of the table shows a sequence that yields an automorphism corresponding to an iSWAP gate. Together with the single qubit generators, this furnishes a Clifford group on two qubits. For the iSWAP gate, we require two triangle DA color codes, each of which contributes one logical qubit. This sequence will be discussed in greater detail in Subsec.~\ref{sec:boundary_micro}.

\subsubsection{Condensation sequences for the dynamic automorphism color code on a triangle}\label{sec:cond_seq}

The sequences of condensations producing generators for color permutations can be worked out according to Sec.~ \ref{sec:tqft_bound}. We start with the effective color code on a Pauli triangle having performed $\rzi\rzii$ and $\bzi\bzii$ bulk condensations (corresponding to measuring $V(Z_1Z_2)$), as well as boundary condensations that produce effective $X,$ $Y$ and $Z$ boundaries, as shown in Fig.~\ref{fig:411} at $t=0$. Next, we choose the sequences of bulk condensations corresponding to the measurements shown in Table~\ref{tab:S3_color}, and work out the boundary condensations in all rounds from the requirement that a representative of each logical operator has to be passed to each next round and has to condense on the same boundary each time. As a remark, we keep the notations for the $X$, $Y$ and $Z$ boundaries static for simplicity. The true boundary labels for the measurement rounds when the bulk theory is that of the toric code are ``rough'' and ``smooth'' and are permuted each round. 

\begin{table}[!t]
{%
\renewcommand{\rc}[1]{{\color{red}{#1}}}
\renewcommand{\bc}[1]{{\color{blue}{#1}}}
\renewcommand{\gc}[1]{{\color{ForestGreen}{#1}}}
\renewcommand{\yc}[1]{{\color{amber}{#1}}}
 \centering
 \resizebox{1\textwidth}{!}{
 \begin{tabular}{|c|c|c|rl|rl|rl|rl|rl|rl|rl| }
 \hline
 \multicolumn{3}{|c|}{$t$} & \multicolumn{2}{c|}{0} & \multicolumn{2}{c|}{1} & \multicolumn{2}{c|}{2} & \multicolumn{2}{c|}{3} & \multicolumn{2}{c|}{4} & \multicolumn{2}{c|}{5} & \multicolumn{2}{c|}{6} \\
 \hline
 \multirow{9}{*}{$H \sim \varphi_{(\texttt{rb})}$} &\multicolumn{2}{c|}{\multirow{2}{*}{ Bulk }}& \multicolumn{2}{c|}{$\rzi\rzii$} & \multicolumn{2}{c|}{$\bxi$} & \multicolumn{2}{c|}{$\gzi$} & \multicolumn{2}{c|}{$\rxi$} & \multicolumn{2}{c|}{$\bzi$}& \multicolumn{2}{c|}{$\rxi$}& \multicolumn{2}{c|}{$\rzi\rzii$} \\
 &\multicolumn{2}{c|}{}&\multicolumn{2}{c|}{$\bzi\bzii$} & \multicolumn{2}{c|}{$\rxii$} & \multicolumn{2}{c|}{$\bzii$} & \multicolumn{2}{c|}{$\gxii$} & \multicolumn{2}{c|}{$\rzii$} & \multicolumn{2}{c|}{$\bxii$} & \multicolumn{2}{c|}{$\bzi\bzii$} \\
 \cline{2-17} 
 & \multirow{6}{*}{Bdry} &\multirow{2}{*}{ $X$} & $\widetilde{\rx}$ &$\sim \rxi\rxii$ & $\rxi\rxii$& $\gt \bxig\rxiig$& $\gxi$& & $\gxi$& $\gt \rxig$&$\bxi$ & &$\bxi$ & $\gt \bxiig$ & $\bxi\bxii$&$ \sim \widetilde{\bx}$ \\
&& & $\widetilde{\bx}$ &$\sim\bxi\bxii$ & $\bxi\bxii$&$\gt\bxig$ &$\bxii$& & $\bxii$&$\gt\gxiig$& $\rxii$ &&$\rxii$ & $\gt \rxig$ & $\rxi\rxii$ &$\sim \widetilde{\rx}$ \\
\cline{3-17}
&& \multirow{2}{*}{$Y$} & $\widetilde{\ry}$ & $\sim\rxi\ryii$ & $\rxi\ryii$ & $\gt\bxig\rxiig$ & $\gxi\rzii$& $\gt \bziig$&$\gxi\gzii$&$\gt\rxig$& $\bxi\gzii$ & $\gt \rziig$ & $\bxi\bzii$ & $\gt \bxiig$ & $\bxi\byii$ &$ \sim \widetilde{\by}$\\
&& & $\widetilde{\by}$&$\sim\byi\bxii$ & $\byi\bxii$&$\gt\bxig$&$\bzi\bxii$&$\gt \gzig$&$\rzi\bxii$ &$\gt\gxiig$ &$\rzi\rxii$ & & $\rzi\rxii$ & $\gt \rxig$ & $\ryi\rxii$ &$\sim \widetilde{\ry}$ \\
\cline{3-17}
&& \multirow{2}{*}{$Z$} &$\widetilde{\rz}$& $\sim\rzii$ & $\rzii$& & $\rzii$& $\gt \bziig$ & $\gzii$&&$\gzii$&$\gt \rziig$ & $\bzii$ & &$\bzii$&$\sim \widetilde{\bz}$\\
&& & $\widetilde{\bz}$&$\sim\bzii$ &$\bzi$&&$\bzi$&$\gt \gzig$& $\rzi$&&$\rzi$ & &$\rzi$ & &$\rzi$ & $\sim \widetilde{\rz}$ \\ 
\hline
 \multirow{9}{*}{$SH \sim \varphi_{(\texttt{rbg})}$} &\multicolumn{2}{c|}{\multirow{2}{*}{ Bulk }}& \multicolumn{2}{c|}{$\rzi\rzii$} & \multicolumn{2}{c|}{$\rxi$} & \multicolumn{2}{c|}{$\gzi$} & \multicolumn{2}{c|}{$\bxi$} & \multicolumn{2}{c|}{$\rzi\rzii$} & & &&\\
 &\multicolumn{2}{c|}{}&\multicolumn{2}{c|}{$\bzi\bzii$} & \multicolumn{2}{c|}{$\gxii$} & \multicolumn{2}{c|}{$\bzii$} & \multicolumn{2}{c|}{$\rxii$} & \multicolumn{2}{c|}{$\bzi\bzii$} & & &&\\
 \cline{2-17} 
 & \multirow{6}{*}{Bdry} &\multirow{2}{*}{ $X$} & $\widetilde{\rx}$ &$\sim \rxi\rxii$& $\rxi\rxii$& $\gt \rxig\gxiig$& $\bxii$& & $\bxii$& $\gt \bxig$&$\bxi\bxii$ &$\sim \widetilde{\bx}$ & & &&\\
&& & $\widetilde{\gx}$ &$\sim \gxi\gxii$ & $\gxi\gxii$&$\gt\gxiig$ &$\gxi$& & $\gxi$&$\gt\bxig\rxiig$& $\rxi\rxii$ &$\sim \widetilde{\rx}$ & & &&\\
\cline{3-17}
&& \multirow{2}{*}{$Y$} & $\widetilde{\ry}$ &$\sim \ryi\rxii$ & $\ryi\rxii$ & $\gt\rxig\gxiig$ & $\rzi\bxii$& $\gt \gzig$&$\bzi\bxii$&$\gt\bxig$& $\byi\bxii$ & $\sim \widetilde{\by}$ & & &&\\
&& & $\widetilde{\gy}$ &$\sim \gxi\gyii$ & $\gxi\gyii$&$\gt\gxiig$&$\gxi\gzii$&$\gt \bziig$&$\gxi\rzii$ &$\gt\bxig\rxiig$ &$\rxi\ryii$ & $\sim \widetilde{\ry}$ & & &&\\
\cline{3-17}
&& \multirow{2}{*}{$Z$} & $\widetilde{\rz}$ &$\sim \rzi$ & $\rzi$& & $\rzi$& $\gt \gzig$ & $\bzi$&&$\bzi$ &$\sim \widetilde{\bz}$ & & &&\\
&& & $\widetilde{\gz}$ &$\sim \gzii$ &$\gzii$&&$\gzii$&$\gt \bziig$& $\rzii$&&$\rzii$ & $\sim \widetilde{\rz}$ & & &&\\ 
\hline
\hline
\multirow{18}{*}{iSWAP} &\multicolumn{2}{c|}{\multirow{4}{*}{ Bulk }}& \multicolumn{2}{c|}{$\rzi\rzii$} & \multicolumn{2}{c|}{$\rxi$} & \multicolumn{2}{c|}{$\rzi\rziii$} & \multicolumn{2}{c|}{$\bxi$} &$\rzi\rzii$ & & & && \\
 &\multicolumn{2}{c|}{}&\multicolumn{2}{c|}{$\bzi\bzii$} & \multicolumn{2}{c|}{$\bxii$} & \multicolumn{2}{c|}{$\rzii\rziv$} & \multicolumn{2}{c|}{$\gxii$} &$\bzi\bzii$ & & & & &\\
 &\multicolumn{2}{c|}{} & \multicolumn{2}{c|}{$\rziii\rziv$} & \multicolumn{2}{c|}{$\bxiii$} & \multicolumn{2}{c|}{$\bzi\bziii$} & \multicolumn{2}{c|}{$\gxiii$} & $\rziii\rziv$ & & & &&\\
 & \multicolumn{2}{c|}{}&\multicolumn{2}{c|}{$\bziii\bziv$} & \multicolumn{2}{c|}{$\rxiv$} & \multicolumn{2}{c|}{$\bzii\bziv$} & \multicolumn{2}{c|}{$\bxiv$} & $\bziii\bziv$ & & & && \\
 \cline{2-17} 
& \multirow{13}{*}{Bdry} &\multirow{4}{*}{ $X$} &$\widetilde{\rxi}$ &$\sim\rxi\rxii$& $\rxi\rxii$ & $\gt \rxig\rxivg$& $\rxii\rxiv$&& $\rxii\rxiv$ & $\gt \bxig\gxiig \gxiiig\bxivg$ & $\bxi\bxii\gxiii\gxiv$ & $\sim\widetilde{\bxi}\widetilde{\gxii}$ & & &&\\
&& & $\widetilde{\bxi}$&$\sim\bxi\bxii$& $\bxi\bxii$ & $\gt \bxiig\bxiiig$&$\bxi\bxiii$&& $\bxi\bxiii$& $\gt \bxig \bxivg$&$\bxiii\bxiv$ &$\sim\widetilde{\bxii}$ & & & &\\
&& & $\widetilde{\rxii}$ &$\sim\rxiii\rxiv$& $\rxiii\rxiv$ & $\gt\rxig\rxivg$ &$\rxi\rxiii$&& $\rxi\rxiii$& $\gt \bxig\gxiig \gxiiig\bxivg$ & $\gxi\gxii\bxiii\bxiv$ & $\sim\widetilde{\gxi}\widetilde{\bxii}$ & & & &\\
&& &$\widetilde{\bxii}$ &$\sim\bxiii\bxiv$& $\bxiii\bxiv$ & $\gt \bxiig\bxiiig$&$\bxii\bxiv$ && $\bxii\bxiv$&$\gt \bxig\bxivg$&$\bxi\bxii$ &$\sim\widetilde{\bxi}$ & & & & \\
\cline{3-17}
&&\multirow{4}{*}{$Y$}& $\widetilde{\ryi}$ &$\sim\ryi\rxii$& $\ryi\rxii$ & $\gt \rxig\rxivg$& $\rzi\rxii\rxiv$&$\gt \gzig\gziiig$& $\bzi\rxii\gziii\rxiv$ & $\gt \bxig\gxiig \gxiiig\bxivg$ & $\byi\bxii\gyiii\gxiv$ &$\sim\widetilde{\byi}\widetilde{\gyii}$ & & & & \\
&& & $\widetilde{\bxi}$ &$\sim\bxi\byii$& $\bxi\byii$ & $\gt \bxiig\bxiiig$&$\bxi\bzii\bxiii$&$\gt \bziig\bzivg$& $\bxi\bxiii\bziv$& $\gt \bxig \bxivg$&$\bxiii\byiv$ &$\sim\widetilde{\byii}$ & & & &\\
&& & $\widetilde{\rxii}$ &$\sim\ryiii\rxiv$& $\ryiii\rxiv$ & $\gt\rxig\rxivg$ &$\rxi\rxiii\rziv$&$\gt \gziig\gzivg$& $\rxi\gzii\rxiii\bziv$& $\gt \bxig\gxiig \gxiiig\bxivg$ & $\gxi\gyii\bxiii\byiv$ &$\sim\widetilde{\gyi}\widetilde{\byii}$ & & & &\\
&& & $\widetilde{\bxii}$ &$\sim\bxiii\bxiv$& $\bxiii\byiv $ & $\gt \bxiig\bxiiig$&$\bxii\bziii\bxiv$ &$\gt\bzig\bziiig$& $\bzi\bxii\bxiv$&$\gt \bxig \bxivg$&$\byi\bxii$ &$\sim\widetilde{\byi}$ & & & &\\
\cline{3-17}
&& \multirow{4}{*}{$Z$} &$\widetilde{\rzi}$ &$\sim \rzi$& $\rzi$&&$\rzi$&$\gt \gzig\gziiig$&$\bzi\gziii$&&$\bzi\gziii$ & $\sim \widetilde{\bzi}\widetilde{\gzii}$ & & & &\\
&& & $\widetilde{\bzi}$ &$\sim \bzii$&$\bzii$&&$\bzii$&$\gt \bziig\bzivg$&$\bziv$&&$\bziv$ &$\sim \widetilde{\bzii}$ & & & &\\
&& &$\widetilde{\rzii}$ &$\sim \rziv$& $\rziv$&& $\rziv$& $\gt \gziig\gzivg$&$\gzii\bziv$&&$ \gzii\bziv $ & $\sim \widetilde{\gzi}\widetilde{\bzii}$ & & & &\\
&& & $\widetilde{\bzii}$ &$\sim \bziii$&$\bziii$&&$\bziii$& $\gt \bzig\bziiig$&$\bzi$&&$\bzi$ & $\sim \widetilde{\bzi}$ & & & &\\
\hline

 \end{tabular}}
 \caption{Condensation data for the generating automorphisms for the Pauli triangle DA color code and the double triangle DA color code. Anyons that are condensed at the $X$ and $Z$ boundaries fuse into anyons at the $Y$ boundary up to anyons condensed in the bulk. }
 \label{tab:TQFTcondensationdata}
 }
\end{table}

Before discussing the boundary condensations in detail, we first address the unfolding picture between the regular color code and decoupled toric codes in more detail, as it provides a pedagogical example. Upon applying the unfolding map, the layers become two triangles glued across the $Y$ boundary. This is shown in Fig.~\ref{fig:41}(a) and (b) using the two toric codes and the color code notations for the anyons, respectively. The unfolded rectangle is a patch of toric code with a single logical qubit encoded on it with an $\ttt e - \ttt m$ domain wall running along the gluing region ($Y$ boundary before unfolding). Going back to condensation sequences, at the rounds when the parent doubled color code is condensed into a single effective color code in the bulk, this corresponds to a Pauli triangle (the folded picture). When the parent code is condensed into two decoupled toric codes in the bulk, this corresponds to two triangle toric codes glued together at the $Y$ boundary (the unfolded picture).

The sequence of bulk and boundary condensations for the protocol resulting in the $\varphi_{(\texttt{rbg})}$ automorphism according to Table~\ref{tab:S3_color} is shown in Fig.~\ref{fig:411}. Note that the actual boundary condensations are the union of bulk condensations and logical string condensations, but for brevity, we only indicate the logical strings that condense at each boundary and it is implicitly assumed that the bulk condensation occurs everywhere in space. The generators of the boundary condensations of the first round $t=0$ that are shown in the first column commute with the bulk condensations of both rounds 0 and 1 and thus are passed to the next round. At round $1$, the boundary condensations are carried over from round $0$ (these condensations are shown in color) and correspond to a representative for an equivalence class of boundary condensations. Members within an equivalence class are related by multiplication with bulk condensations of the given round. Therefore, we modify the representative at round $1$ (which will be the representative at round $2$) by multiplying by certain bulk condensations indicated in gray. We iterate this process during each subsequent round, updating representative boundary condensations by multiplication with suitable bulk condensations.

The boundary condensations for the $\varphi_{(\texttt{rb})}$ automorphism and the double-triangle iSWAP gate can be worked in a similar way. The condensation sequences both in the bulk and at the boundary are summarized in Table~\ref{tab:TQFTcondensationdata} for all the generating automorphisms. These will serve as a starting point which is crucial to working out the full the measurement protocol on the lattice.

\subsubsection{Logical strings evolution and gates} \label{sec:logicals_triangle}

In this part, we review how the logical strings are transformed from round to round for each of the condensation sequences discussed in Subsec.~\ref{sec:cond_seq} and how the automorphisms map to logical gates.

\begin{figure}[!b] \centering
\vspace{0pt}
\centering
\vspace{0pt}
\includegraphics[width= 1\columnwidth]{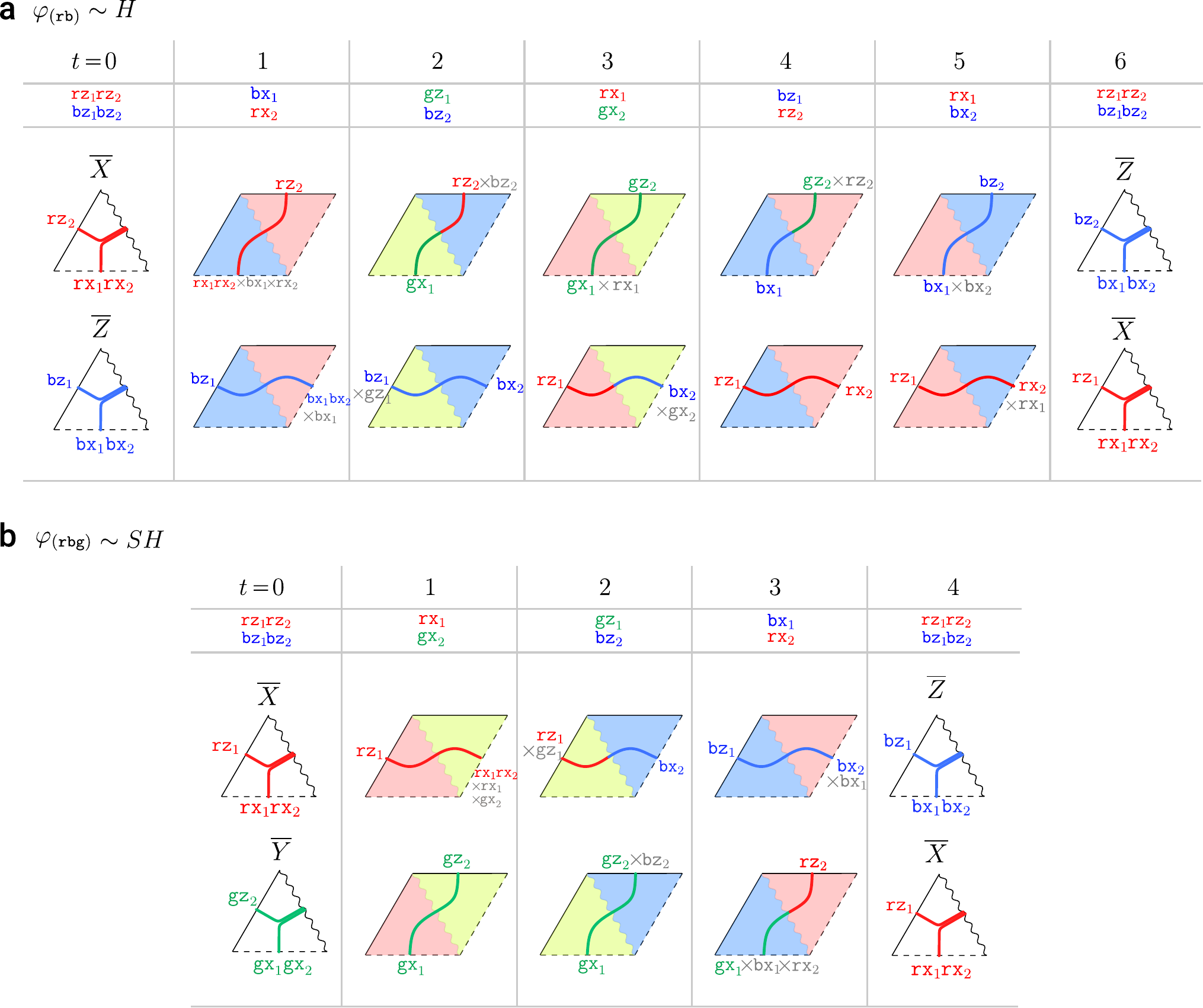}
\caption{Evolution of logical strings under condensation sequences that realize (a) $\varphi_{(\texttt{rb})}$ and (b) $\varphi_{(\texttt{rbg})}$ automorphisms according to Table~\ref{tab:S3_color} in the bulk and to the boundary condensations shown in Fig.~\ref{fig:41}. 
At each step, one representative logical string is shown and, where necessary, another representative (equivalent up to condensations of the current round) is indicated. 
At round $t=0$, we start with an effective color code with effective Pauli ${X}$, ${Y}$ and ${Z}$ boundaries and respective Y-shaped logical strings of fixed colors. 
The unfolded picture for the toric code rounds is shown for clarity at the intermediate steps.
The logical Pauli operators of the effective color code are shown above the logical strings at the beginning and the end of the sequence.}
\label{fig:42}
\end{figure}

The logical strings and the mapping to logical Paulis for the color code on a triangle are introduced in Fig.~\ref{fig:41}(c). 
Explicitly, the logical operators are labeled by a junction of three strings of the same color $({\texttt{cx}},{\texttt{cy}},{\texttt{cz}})$.
Each string originates on its respective Pauli boundaries, with all of them meeting at the middle of the triangle, forming a Y-junction. Here, we choose the convention that the logical $X$, $Y$, and $Z$ of the triangle correspond to the triples $( {\rx}, {\ry}, {\rz})$, $( {\gx}, {\gy}, {\gz})$, and $( {\bx}, {\by}, {\bz})$, respectively. 
One can check the Pauli algebra is satisfied by these logical operators. 
Let us also see what the logical strings look like upon unfolding the color code into a parallelogram. A logical string of this parallelogram is a continuous string which (as an example) forms an $\ttt e$-string in the toric code on one of the triangles and an $\ttt m$ string on the other triangle, due to the $\ttt e - \ttt m$ domain wall separating both triangles. Folding back into a triangle thus gives this logical string a Y-junction shape.

\begin{figure}[!b] \centering
\vspace{0pt}
\centering
\vspace{0pt}
\includegraphics[width= 0.85\columnwidth]{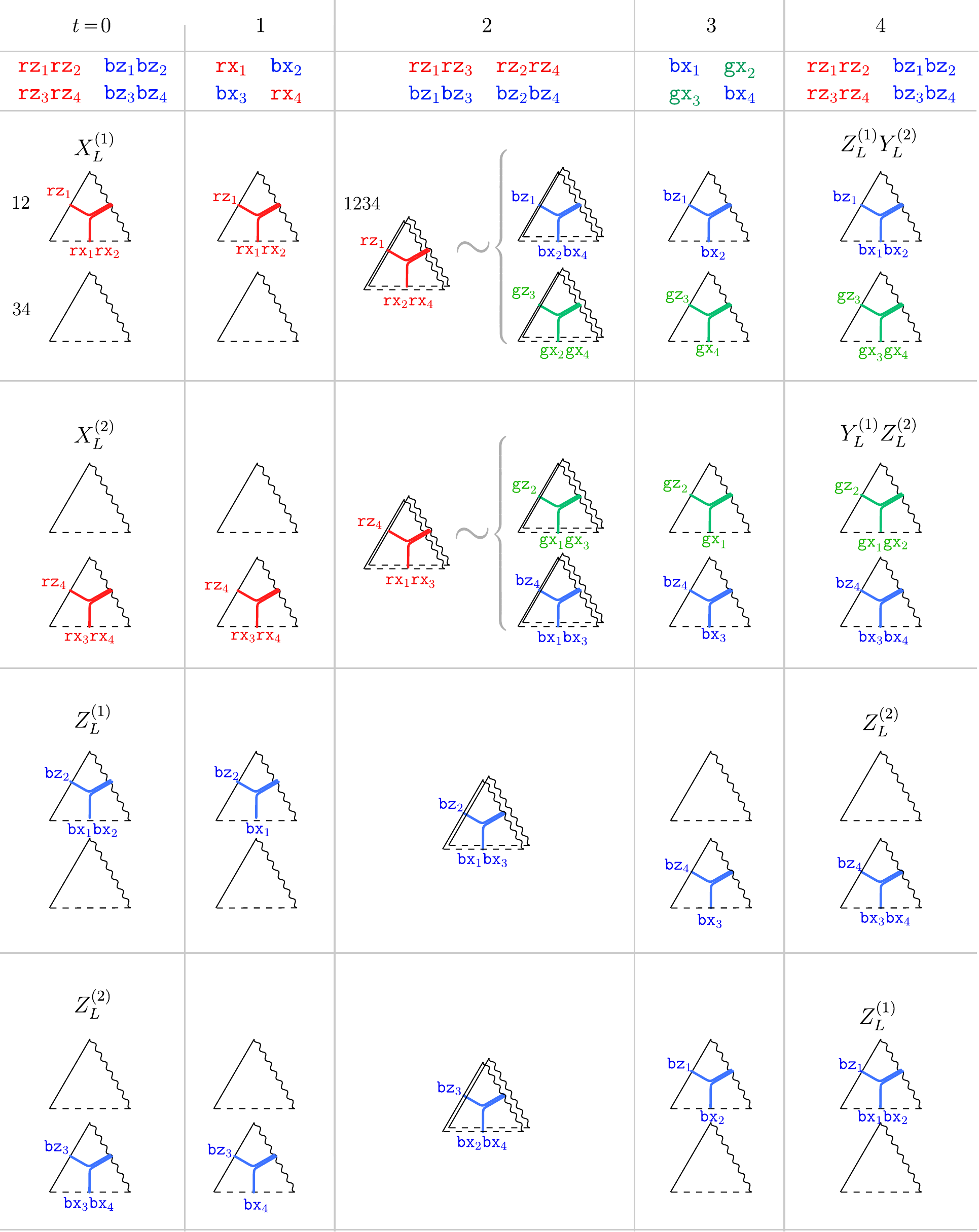}
\caption{Evolution of logical strings of double triangle color code under the iSWAP sequence according to Table~\ref{tab:TQFTcondensationdata}. At rounds 0, 1, 3, and 4 the top triangle contains layers 1 and 2 and the bottom triangle contains layers 3 and 4. At round 2, all four layers are coupled and sometimes we represent the logical string as a product of two other strings as shown by the equivalence. This explains why this condensation sequence produces an entangling gate on two qubits. The logical Paulis of the two logical qubits are indicated at the first and last rounds, which allows us to read off the iSWAP gate. }
\label{fig:421}
\end{figure}

As discussed in Sec.~\ref{sec:tqft_bound}, the evolution of each logical string under the sequence of bulk condensations is the same as the evolution of the respective boundary condensation, because of the one-to-one correspondence between the logical strings and the boundary condensations. At a given boundary, there is a set of logical strings that terminate at this boundary at round $t$. There exists a set of representatives of these logical strings that commutes with the bulk condensations of the next round, and we demand that these representatives condense at the same geometric boundary in the next round. This defines the boundary condensation of the next round. The evolution of logical operators on a single Pauli triangle is shown in Fig.~\ref{fig:42} (a,b) for the $\varphi_{(\texttt{rb})}$ and $\varphi_{(\texttt{rbg})}$ automorphisms, respectively. These two automorphisms generate all single-qubit Clifford gates. The correspondence between gates and automorphisms can be deduced immediately, without following the evolution of the logical strings throughout all the rounds, because the logical Pauli operators are in one-to-one correspondence with colors. Because the automorphisms permute colors, their action on the logical Pauli operators can be read off immediately. Under the $\varphi_{(\texttt{rb})}$ automorphism we observe $X_L \leftrightarrow Z_L$, which is the action of the logical Hadamard gate. Similarly, under $\varphi_{(\texttt{rbg})}$, we have a cyclic permutation $X_L \rightarrow Z_L \rightarrow Y_L$, which is the logical $SH$ gate. This action is also obtained by tracking the evolution of the logical strings explicitly, as shown in 
Fig.~\ref{fig:42}.

The evolution of the logical operators under the two-qubit iSWAP protocol is shown in Fig.~\ref{fig:421}. In this case, we have two triangle codes, (1,2) and (3,4) respectively, and start with two effective color codes on these layers. At round 2, all four layers become coupled, which entangles the logical strings on the two triangles. For example, the red logical string at round 2 in the first row is equivalent to the product of blue and green strings which are passed to the next round (where the triangles are again decoupled). This evolves a logical string on a single triangle into a product of two logical strings on different triangles and therefore produces an entangling gate needed to complete the Clifford group. The only condensation round that is qualitatively different from what we encountered before is round 2. It can be checked again that the boundary condensation for the double triangle gate at round 2 shown in Table~\ref{tab:TQFTcondensationdata} yields the logical strings that have support on three layers as shown in Fig.~\ref{fig:421}. 

\subsection{Measurement schedules for the triangular dynamic automorphism color code} \label{sec:boundary_micro}

From the condensation sequences described above, we can now construct respective measurement sequences on single and double triangles. All of the bulk condensations shown in Table~\ref{tab:TQFTcondensationdata} turn into two-qubit measurements. The measurements corresponding to the boundary condensations depend on the way the lattice terminates at the boundary, and we are able to reduce them to two-qubit measurements in all cases apart from round 2 for the iSWAP gate, where we use three-qubit measurements at one of the boundaries. We consider these protocols in detail in this subsection and show that the single-triangle protocols generate a full topological stabilizer group and, together with the protocol for the iSWAP gate, they implement the full Clifford group. Finally, we explore the error correction of these codes.

\subsubsection{Single Pauli triangle DA color code}

\begin{figure}[!t] \centering
\vspace{0pt}
\centering
\vspace{0pt}
\includegraphics[width= 0.6\columnwidth]{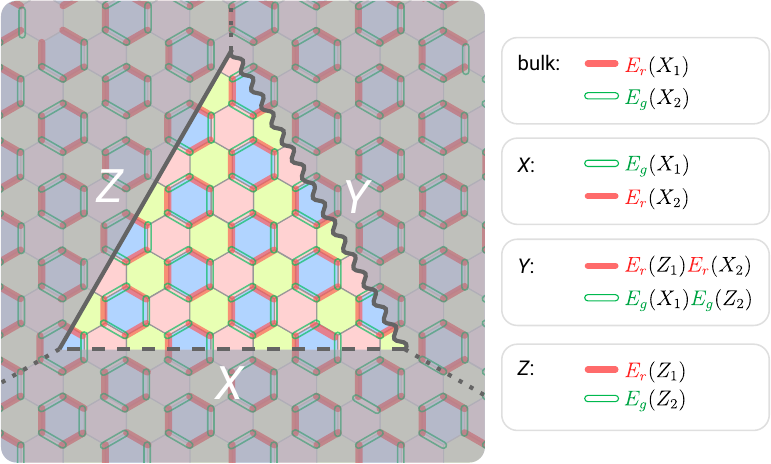}
\caption{The naive measurements of step $t=1$ of $\varphi_{(\texttt{rbg})}$ sequence derived from the condensation sequence shown in Table~\ref{tab:TQFTcondensationdata} and Fig.~\ref{fig:411} using the prescription in Eq.~\eqref{eq:anyontomeasurement}. The measurements are performed on an infinite plane where the bulk measurements occur everywhere on a plane, and the additional boundary measurements are indicated on each side of the triangular region and occur in thirds of the plane minus the triangular patch. The labeling of the boundaries inherits that from the color code step, i.e. $X$, $Y$ and $Z$. All the measurements are two-qubit ones apart from the $Y$ boundary, where they are four-qubit ones. Later on, where we show how to truncate the triangular patch in a way that the measurements on the $Y$ boundary reduce to weight-2 measurements.
} 
\label{fig:43}
\end{figure}

We start by considering a lattice with two layers of qubits on an infinite plane. We perform bulk condensations everywhere in space while performing boundary condensations in thirds of the infinite plane outside of the triangle; a visual depiction of this is Fig.~\ref{fig:43}, showing the $t=1$ round of the $\varphi_{(\texttt{rbg})}$ sequence. The protocol that we obtain using this method will, unfortunately, involve measuring weight-4 check operators in the $Y$-boundary region at colored rounds. Although this is more efficient than measuring plaquette stabilizers of the color code, it is still more practically desirable for all of our measurements to be two-qubit. We introduce a technique called {\it docking} which allows us to reduce the weight-4 measurements to native weight-2 measurements without adding ancillae.

Let us now translate the sequences of condensations shown in Table~\ref{tab:TQFTcondensationdata} into measurement sequences on a plane. In the bulk, the measurement sequences simply follow Table~\ref{tab:S3_color}. At the boundaries, the condensations are translated into measurements in the following way:
\begin{equation}
\begin{split}
\texttt{c}\tsig_\ell &\rightarrow E_c(\sigma_\ell),\\
\texttt{c}\tsig_1 \texttt{c}'\tsig_2 &\rightarrow E_c(\sigma_1)E_{c'}(\sigma_2),\\
\{\texttt{c}\tsig_\ell, \texttt{c}'\tsig_\ell \} &\rightarrow V(\sigma_\ell),\\
\{\texttt{c}\tsig_1\texttt{c}\tsig_2, \texttt{c}'\tsig_1\texttt{c}'\tsig_2 \} &\rightarrow V(\sigma_1\sigma_2),
\\
\{\texttt{c}\tsig_1\texttt{c}\tsig_2\texttt{c}\tsig_3, \texttt{c}'\tsig_1\texttt{c}'\tsig_2 \texttt{c}'\tsig_3 \} &\rightarrow V(\sigma_1\sigma_2\sigma_3),
\end{split}
\label{eq:anyontomeasurement}
\end{equation}
where the colors are labeled as ${c},{c}'$, Pauli flavors are $\tsig$, and subscripts $1,2,\ell$ denote layer indices. As a reminder, $E_c$ stands for a two-qubit measurement performed on an edge of color $c$, $V$ stands for a measurement at a vertex (or an interlayer two-qubit measurement), and whenever a product of measurements is written it corresponds to measuring a product of the respective edge/vertex two-qubit operators. All single-triangle sequences shown in Table~\ref{tab:TQFTcondensationdata} can be shown to be topological ISG-generating both in the bulk and at the boundary. An example of such a transation for round $t=1$ is shown in Fig.~\ref{fig:43}.

Such an approach suffers from two issues. First, in practice, one has to truncate the infinite plane to a finite-sized triangular patch. Second, the measurement sequences descend from simultaneous condensations of anyons on different layers, which correspond to measuring weight-4 operators on the $Y$ boundary---which is not practical.
Both of these problems can be resolved.

\begin{figure}[t] \centering
\vspace{0pt}
\centering
\vspace{0pt}
\includegraphics[width= 0.9\columnwidth]{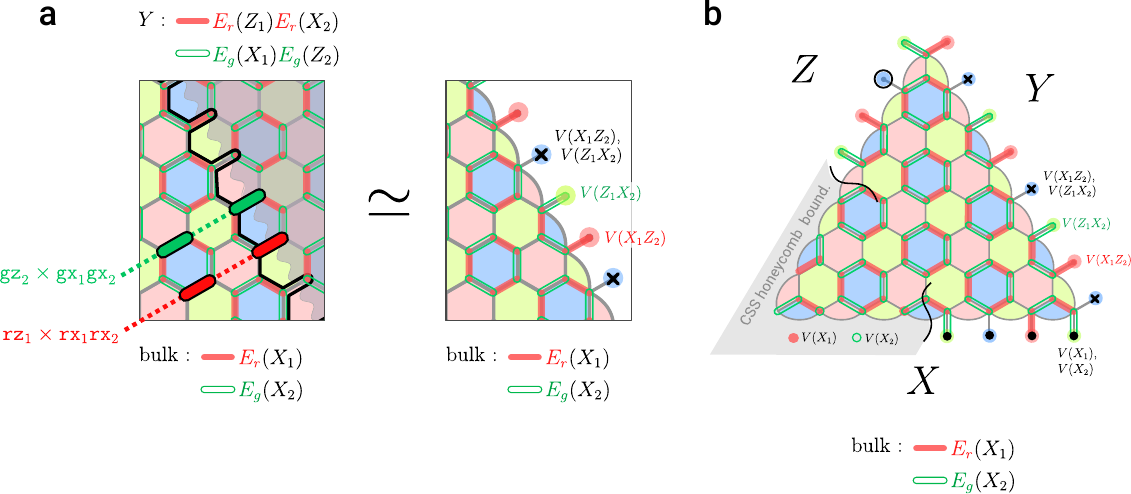}
\caption{ (a) Termination of the lattice at the $Y$ boundary is shown by the black outline on the left and its result is shown on the right. The measurements correspond to those at $t=0$ for the $\varphi_{(\texttt{rbg})}$ sequence shown in Fig.~\ref{fig:43} earlier. The additional layer of qubits to the other side of the boundary are the auxiliary docking qubits. The four-qubit measurements occurring on the left are ``cut in half'' and become two-qubit measurements on the right, and they are responsible for condensing the correct logical strings. The crossed qubits are ``dead'', i.e. decoupled from the rest of the system by measurements. These strings are shown in the left panel.
(b) A finite triangular patch at $t=0$ round of $\varphi_{(\texttt{rbg})}$ sequence. No additional measurements are performed at $Z$ boundary; the measurements at $X$ and $Y$ boundaries are indicated in the drawing. The crossed qubits and the ones with the black dots are decoupled from the rest of the system by measurements. Moreover, the $X$ and $Z$ boundaries are equivalent to the boundaries of the CSS honeycomb code, and the boundaries of Ref.~\cite{brown_2022} can be recovered if we remove the auxiliary layer as shown for the half of $X$ and $Z$ boundaries. This requires us to use single-qubit measurements at $X$ boundary as shown at the bottom left of the triangle. }
\label{fig:431}
\end{figure}

We will now discuss how to simultaneously resolve both of these problems. Let us consider the specific example of the $Y$ boundary at $t=1$ of the $\varphi_{(\texttt{rbg})}$ sequence, where the aforementioned issue arises. Consider terminating the lattice so that there is a single row of auxiliary qubits along each boundary as explained in Fig.~\ref{fig:431}. Each auxiliary vertex is marked by color depending on the color of the logical string that touches the $Y$ boundary at this vertex and terminates there. We treat these auxiliary qubits as {\it docks} for the logical strings. We also refer to them as ``docking qubits'' or ``docking vertices'' of a given color. 
The logical strings that condense at the $Y$ boundary at $t=1$ of the $\varphi_{(\texttt{rbg})}$ are shown in the respective cell in Table~\ref{tab:TQFTcondensationdata}. 
These are $\mathcal L (\ryi \rxii) \simeq \mathcal L (\rzi) \times \mathcal L (\rxi \rxii)$ and $\mathcal L (\gxi \gyii) \simeq \mathcal L (\gxi \gxii) \times \mathcal L (\gzii)$, and both are products of two strings that terminate on two other boundaries that join before touching the $Y$ boundary. On an infinite plane, the four-qubit boundary measurements $E_g(X_1Z_2)$ (they occur in the gray shaded region only) glue together $\mathcal L (\gxi \gxii)$ and $\mathcal L (\gzii)$ into a single logical string. Similarly, $E_r(Z_1X_2)$ checks glue together $\mathcal L (\rzi) $ and $\mathcal L (\rxi \rxii)$. When the lattice terminates with an auxiliary row of qubits, as shown in Fig.~\ref{fig:431}(a) on the right, the green string is only capable of entering the green ``dock'' and the red one, similarly, only enters the red one. The green docking qubits were once part of an edge $E_r(Z_1X_2)$ on an infinite plane; now, we only keep half of this edge, resulting in measuring $V(Z_1X_2)$ on the green docking vertex. Similarly, we measure $V(X_1Z_2)$ on the red docking vertex. Because the blue vertices would support both boundary measurements, one measures both $V(Z_1X_2)$ and $V(X_1Z_2)$ on the blue vertices, turning them into ``dead'' qubits (indicated by a black X in the Figure) which are decoupled from the rest of the system.

One can explicitly check that the logical strings will behave exactly in the same way as on an infinite plane and the correct strings will condense on the $Y$ boundary with this prescription. Moreover, the boundary ISG is appropriately generated as we discuss below and summarize in Table~\ref{fig:44}. Thus, we can realize a topological code on a planar patch.

\begin{figure}[t] \centering
\vspace{0pt}
\centering
\vspace{0pt}
\includegraphics[width= 1\columnwidth]{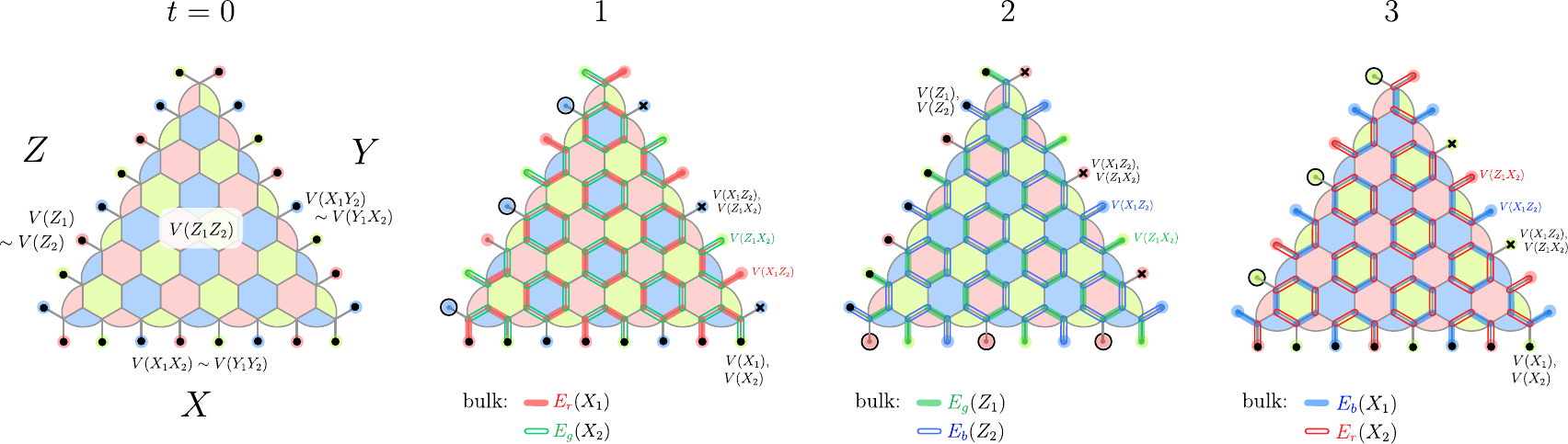}
\caption{The measurement sequence for $\varphi_{(\texttt{rbg})}$ DA color code on a Pauli triangle patch. The entire measurement sequence is derived from the condensation sequence shown in Table~\ref{tab:TQFTcondensationdata} and also in Fig.~\ref{fig:411}. Round $r=4$ is equivalent to $r=0$. 
} 
\label{fig:432}
\end{figure}

Having introduced the docking trick, we are ready to consider the measurement sequences in detail. Fig.~\ref{fig:432} shows the complete set of bulk and boundary measurements on a triangular patch derived from the condensation sequence shown in Table~\ref{tab:TQFTcondensationdata}, which is also illustrated in Fig.~\ref{fig:411}. At the color code rounds ($t=0$ in Fig.~\ref{fig:432}), the bulk measurements are $V(Z_1Z_2)$ on all qubits. The logical strings for each round of this sequence can be readily translated from the TQFT perspective shown in Fig.~\ref{fig:42}. 
 On the (folded) triangle, the logical operators look like Y-junctions, where the part that terminates on the $Z$ boundary (which has $V(Z_1)\sim V(Z_2)$ measurements) is a $Z_1\sim Z_2$ Pauli string (the equivalence is set by the bulk measurement). The $X$ boundary has extra $V(X_1X_2)\sim V(Y_1Y_2)$ measurements and allows the $X_1 X_2 \sim Y_1 Y_2$ logical strings to end. Finally, the $Y$ boundary has extra $V(X_1Y_2) \sim V(Y_1 X_2)$ measurements that allows the $X_1 Y_2 \sim Y_1 X_2$ string to terminate. All of these boundary measurements are shown by black dots in the figure. 

At round $t=1$ (see Fig.~\ref{fig:432}) we have an effective toric code on each layer and the layers coupled together at the $Y$ boundary. We have already considered the two-qubit measurements at the $Y$ boundary in detail. At the $X$ boundary, we need to allow $\mathcal L(\rxi \rxii) \simeq \mathcal L(\rxii)$ and $\mathcal L(\gxi \gxii) \simeq \mathcal L(\gxi)$ logical strings to terminate, which is achieved by measuring single-qubit Pauli $X$ operators in each layer shown by black dots. Finally, at the $Z$ boundary we need $\mathcal L(\rzi)$ and $\mathcal L(\gzii)$ logical strings to terminate, which is achieved if we do not measure anything at this boundary (apart from the bulk condensations that generically happen everywhere). Notice that the logical strings in the two layers are decoupled in the bulk and at $X$ and $Z$ boundaries, but not at $Y$ boundary. Below, we will see that the ISG is similarly decoupled everywhere apart from the $Y$ boundary.

The boundary condensations at any other colored round can be worked out completely analogously to how it was described for $t=1$ above up to appropriate color and $X \leftrightarrow Z$ flavor changes, and are shown explicitly in Fig.~\ref{fig:432}. The measurement details for the $\varphi_{(\texttt{rb})}$ and iSWAP sequences are worked out analogously and are summarized in Appendix \ref{sec:SM_triangle_rb} and in Table \ref{fig:45}, respectively. 

\begin{table}[!t]
\vspace{0pt}
\centering
\vspace{0pt}
\includegraphics[width= 1\columnwidth]{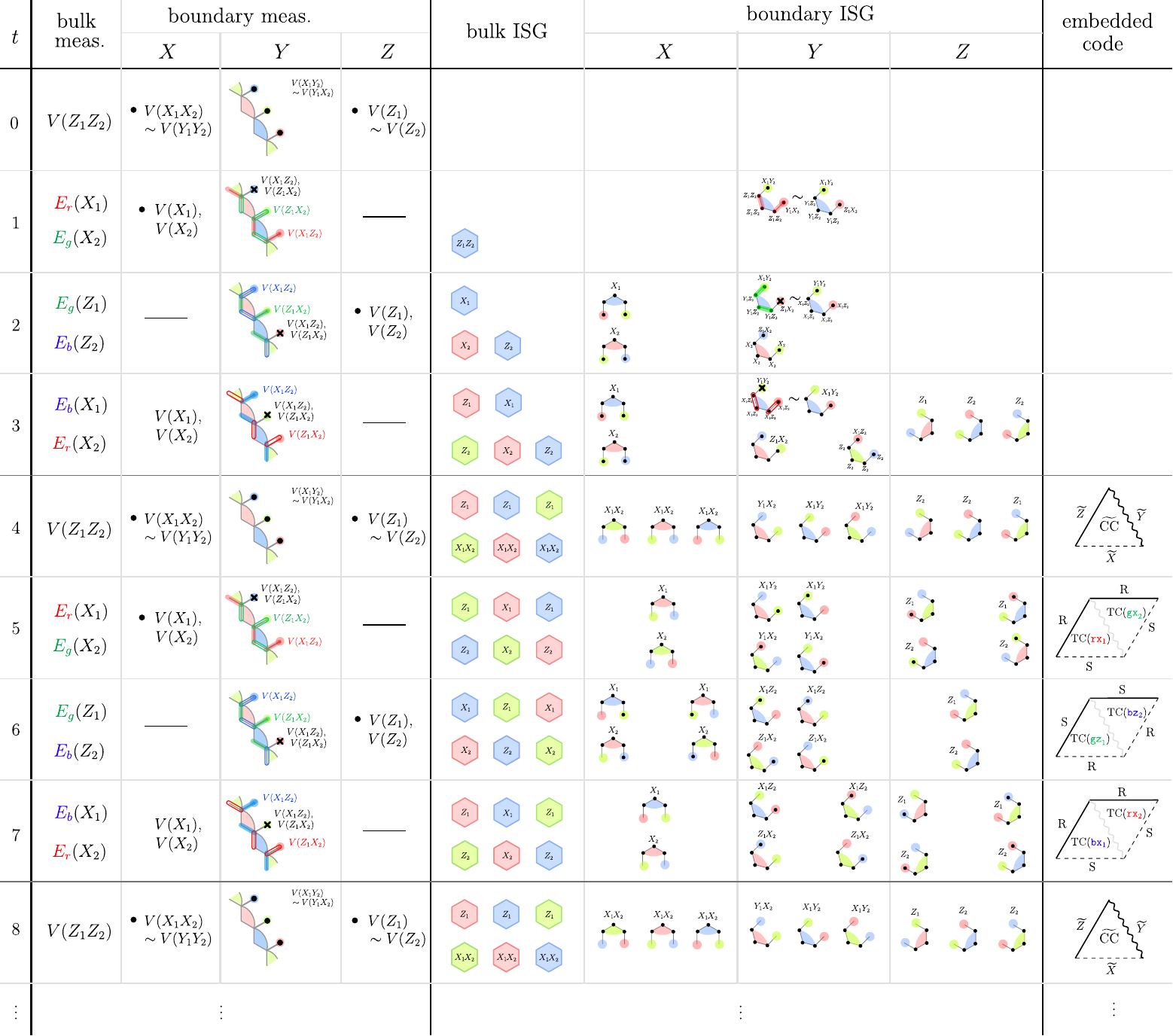}
\caption{ Complete summary of the measurement sequence for $\varphi_{(\texttt{rbg})}$ corresponding to the lattice termination and measurements shown in Fig.~\ref{fig:432}. The first four columns show bulk and boundary measurements. The full topological ISG is measured by step 4, at which the effective color code on the triangular patch is achieved for the first time (indicated in the last column). The bulk ISG plaquettes are shown as plaquettes of respective colors with the Pauli flavor of the plaquette denoted inside. The boundary plaquettes are shown schematically where the qubits in the support of the boundary plaquette stabilizer are marked by black dots. The Pauli flavors on these plaquettes are indicated where necessary (if it is indicated only once, it applies to all qubits with black dots). }
\label{fig:44}
\end{table}

Now, we discuss the generation of the instantaneous stabilizer group (ISG) and handling of the plaquettes in the ISG during our sequences. The measurements and ISG elements are summarized in Fig.~\ref{fig:44} for the $\varphi_{(\texttt{rbg})}$ sequence, in table ~\ref{fig:45} for iSWAP and in Appendix \ref{sec:SM_triangle_rb} for the $\varphi_{(\texttt{rb})}$ sequence. 
The boundary ISG elements are worked out analogously to the recipe used on a torus in section~\ref{sec:torus}. For a bulk (boundary) plaquette in an ISG at round $t$, it can be (if necessary) multiplied by a combination of bulk and boundary checks from the current round $t$ such that it will commute with the bulk and boundary checks of the next round $t+1$.
All plaquettes in Table~\ref{fig:44} have been formed in this fashion. Displayed plaquettes are sometimes already multiplied by the checks of the current round so that it is easier to see that they represent the stabilizers of the effective code at that step. The entire topological ISG is measured by round 4, and starting from this round we have an effective topological code.

\begin{table}[!t]
\vspace{0pt}
\centering
\vspace{0pt}
\includegraphics[width= 1\columnwidth]{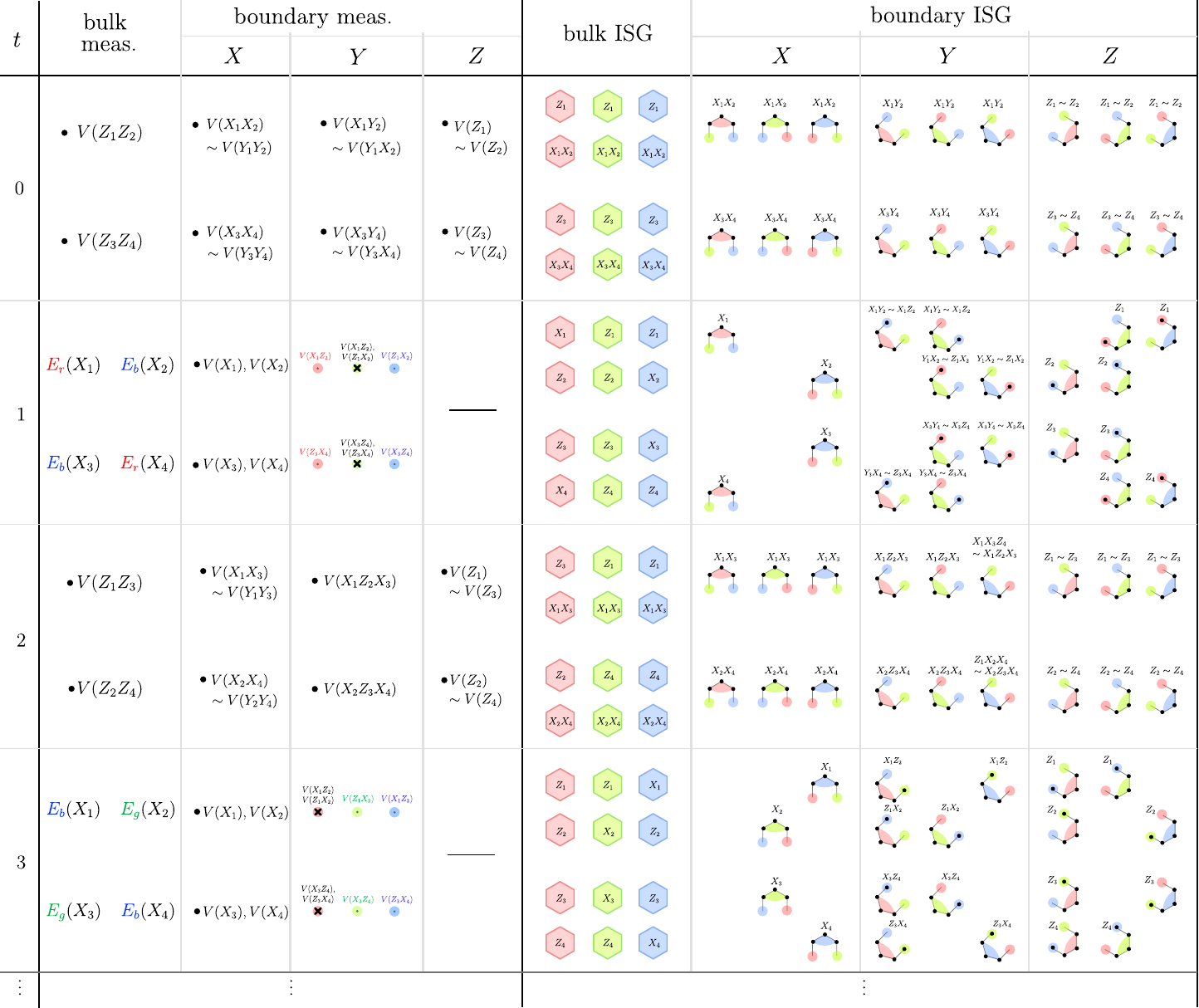}
\caption{Summary of the measurement sequence for iSWAP gate on two Pauli triangles (four layers) corresponding to the condensation protocol shown in Table~\ref{tab:TQFTcondensationdata} modified by the lattice termination. We assume that we start at $t=0$ with two decoupled Pauli triangles with effective color codes and with ISG prepared in both codes. The boundary plaquettes are shown schematically where the qubits in the support of the boundary plaquette stabilizer are marked by black dots. The Pauli flavors on these plaquettes are indicated where necessary (if it is indicated only once, it applies to all qubits with black dots). }
\label{fig:45}
\end{table}

Next, let us consider the measurement sequence for the two-qubit gate iSWAP realized on a double Pauli triangle shown in Table~\ref{fig:45}. First of all, on each separate triangle, we are using the same geometry as for the single triangle, i.e. with a single layer of docking qubits near each edge. We assume that at round 0 we start with two triangle patches realizing effective color codes with bulk measurements $V(Z_1Z_2)$ and $V(Z_3 Z_4)$ and corresponding boundary measurements. Only layers within the pairs (1,2) and (3,4) are coupled during rounds 0, 1, and 3, and therefore, these steps operate analogously to the single-triangle steps that we've seen before. Step 2 is substantially different because it couples the two triangles and, hence, 4 layers of qubits. The boundary measurements at this step are similarly derived from the condensations shown in Table~\ref{tab:TQFTcondensationdata}. On a finite patch, the measurements at that round at the $Z$ boundary turn into $V(Z_1)\sim V(Z_3)$ and $V(Z_2)\sim V(Z_4)$ checks; the measurements at the $X$ boundary are $V(X_1X_3)$ and $V(X_2X_4)$. The measurements on the $Y$ boundary can be chosen (up to bulk measurements) to be $\ttt{cx}_1\ttt{cz}_2\ttt{cx}_3$ and $\ttt{cx}_2\ttt{cz}_3\ttt{cx}_4$ for $\ttt{c}=\ttt{r},\ttt{g},\ttt{b}$. Since all three colors are present, according to the last line in the mapping shown in Eq.~\eqref{eq:anyontomeasurement}, the corresponding measurements can be chosen to be weight-3 vertex measurements $V( X_2 Z_3 X_4)$ and $V(X_1 Z_2 X_3)$. This is consistent with the handling of logical strings as shown in Fig.~\ref{fig:421}, and moreover, it preserves the rank of the ISG and obeys local reversibility. 

To conclude, we have translated our condensation sequence into DA color codes on Pauli triangle patches. We have also shown that logical qubits can be dynamically generated, the measurement sequence is locally reversible, and the measurement sequences implement generators of the Clifford group on two qubits. This can be achieved by 2-body measurements in the case of single triangle protocols and 2-body measurements, apart from one round of 3-body measurements at the $Y$ boundary. We note that in principle, the 3-body checks can be further decreased in weight using ancilla qubits.

\subsection{Error correction and fault tolerance}
\label{sec:EC_triangle}

\begin{figure}[!h] \centering
\vspace{0pt}
\centering
\vspace{0pt}
\includegraphics[width= 1\columnwidth]{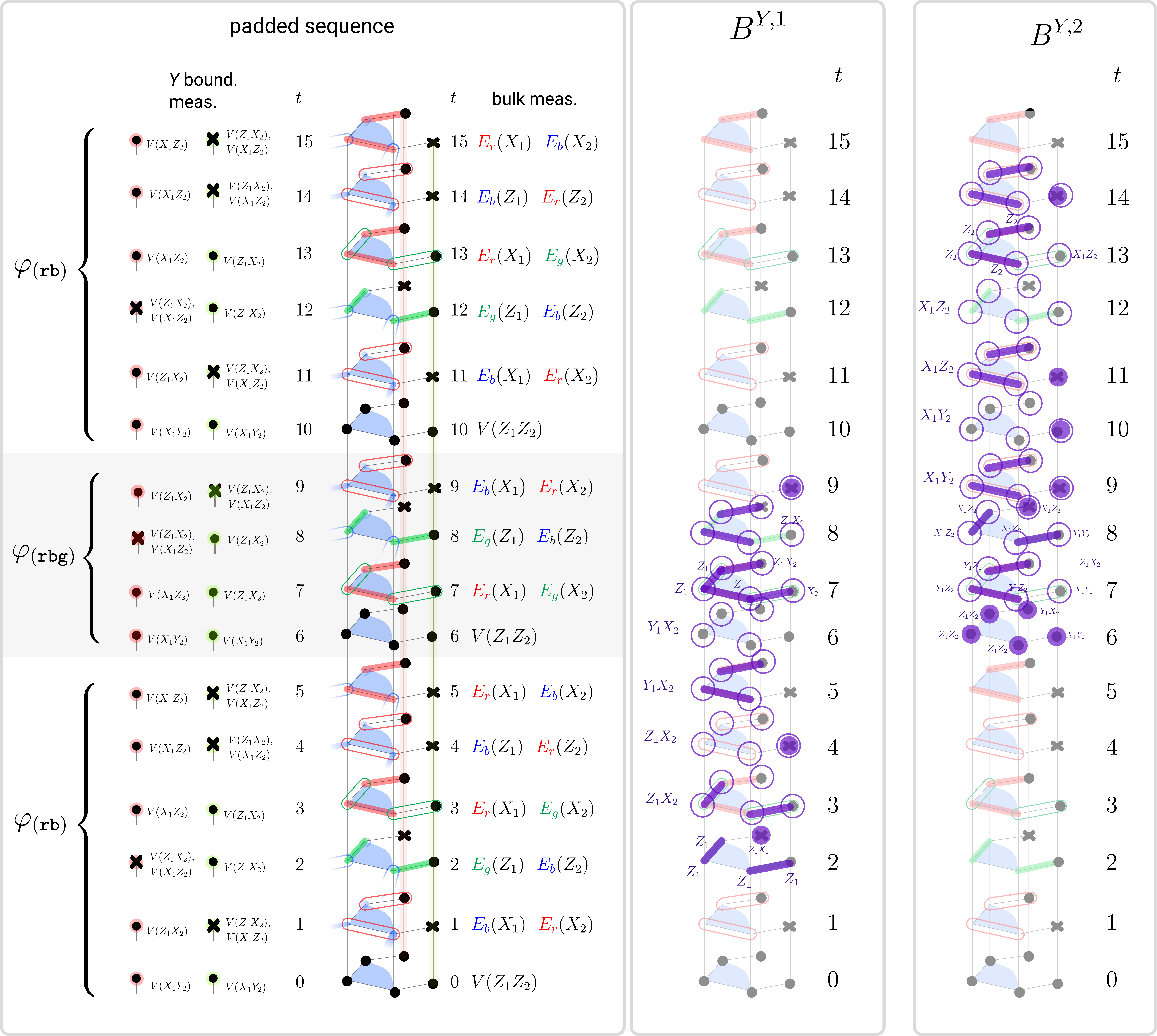}
\caption{
Detailed depiction of the two types of detectors at the $Y$ boundary for the padded measurement sequence corresponding to $\varphi_{(\texttt{rbg})}$ (the padding corresponds to the $\varphi_{(\texttt{rb})}$ sequence). The figure shows two detectors, $B^{Y,1}$ and $B^{Y,2}$, of the blue boundary plaquettes, and the detectors for other colors can be found in Appendix~\ref{sec:SM_triangle_EC_bondary}. The measurements at each round are summarized in the left panel. The measurements comprising the detectors are shown by the edges and vertices highlighted in purple in the second and third panels. The kind of measurement used to form the detector can be read off by comparing the purple highlighted edges/vertices with the left panel. The hollow circles in the detector are used to demarcate its spacetime support. Specifically, the timelike edge $E_{t,t+1}$ at a given qubit belongs to the support of the detector if its end at $t+1$ has a hollow circle. The instantaneous flavor of the detector (i.e. its flavor on given qubits at a given time) after the necessary measurements of the current round have been added to the detector is indicated either by text near each vertex separately or by a single flavor on the left if the flavor is the same on all vertices. }
\label{fig:46}
\end{figure}

The error correction in the bulk can be approached analogously to the torus case. For single-qubit gates, we treat the sequences for $\varphi_{(\texttt{rb})}$ and $\varphi_{(\texttt{rbg})}$ as generating sequences. We chose the measurement sequence for the $\varphi_{(\texttt{rb})}$ automorphism (see Appendix \ref{sec:SM_triangle_rb}) so that it can be used for padding on top of that.

The bulk detectors are constructed in accordance with our discussion in Section~\ref{sec:EC}. We use the same simplified error basis as introduced in Sec.~\ref{sec:EC_error_basis}. We find a full basis of bulk detectors and detectors at $X$ and $Z$ boundaries for padded $\varphi_{(\texttt{rb})}$ and $\varphi_{(\texttt{rbg})}$ gates, using $\varphi_{(\texttt{rb})}$ sequence as padding. The basis for the bulk detectors for these sequences is shown in Appendix~\ref{sec:SM_triangle_EC}. In fact, the detectors at $X$ and $Z$ boundaries are the same as $X$ and $Z$ flavored detectors in the bulk respectively; these detectors are supported on boundary plaquettes of respective colors. These are not shown separately in Appendix~\ref{sec:SM_triangle_EC} because they can be easily deduced from the bulk detectors that are already shown.

The detectors at the $Y$ boundary have a much more complicated structure because the flavor of the detector needs to be $V(X_1Y_2)$ or $V(Y_1X_2)$ at $V(Z_1Z_2)$ measurement steps. We find all the detectors at this boundary but we leave the laborious task of showing that they form a complete basis of detectors for error correction to future work. 
We use the error basis from Subsec.~\ref{sec:EC_error_basis}, and find detectors again by searching for all measurements whose product would give a constant in the absence of any errors. 
In Fig.~\ref{fig:46} we show an example of two blue detectors for the $\varphi_{(\texttt{rbg})}$ padded sequence. The measurements that are used to form the detector are highlighted purple and the support of the detector on timelike edges can be determined from the hollow circles (a timeline edge $E_{t,t+1}$ at a given qubit is in the support of the detector if its end at $t+1$ is marked with a hollow circle). One also has to carefully keep track of which error flavors the detector is sensitive to at different rounds and locations. The rest of the detectors at $Y$ boundary for both padded generating sequences are shown in Appendix~\ref{sec:SM_triangle_EC_bondary}.

Finally, we leave the analysis of error correction of the iSWAP sequence to future work. We remark that there exist infinitely many sequences realizing a two-qubit gate that would complete the two-qubit Clifford group. One might need to exhaustively search for alternative sequences for two-qubit gates before one successfully finds an error-correcting one that is also compatible with the single-qubit sequences chosen here. This may require a rather sophisticated software implementation but would be a worthwhile endeavor to pursue.

\section{{Towards universal quantum computation: the 3D dynamic automorphism color code}}
\label{sec:3DFCC}

In previous sections, we showed how the two-dimensional dynamic color code can be used to achieve the full Clifford logical gate group. For universal quantum computation, we need to supplement this set with at least one non-Clifford gate. In this section, we make the first step in this direction by constructing a three-dimensional dynamic automorphism color code, whose ISG admits a transversally implemented non-Clifford logical gate.

There are several natural methods to achieve non-Clifford gates in color codes. A non-Clifford transversal gate can be achieved in three dimensions in the 3D color code~\cite{Bombin2007}, the gauge color code~\cite{Bombin2015}, or in copies of 3D toric codes~\cite{Vasmer_2019,iverson2020aspects}. Alternatively, one can ``pull'' the quasi-2D code through a three-dimensional virtual code while applying a $T$-gate in order to fault-tolerantly implement a non-Clifford gate in 2D~\cite{Bombin2018} (similar results can be achieved in copies of 3D toric code \cite{Brown2020,Scruby_2022}). In this method, the fault tolerance is achieved by just-in-time decoding. Finally, it is of course possible to perform magic state injection~\cite{Landahl2014quantum}. It would be also interesting to see how much other methods for achieving the universal gateset, such as pieceable fault-tolerance~\cite{Yoder2016}, can be adapted for general DA codes.

In this section, we develop a 3D dynamic automorphism color code, which periodically realizes the 3D color code (3DCC) ISG, and realizes variants of the ISG of three decoupled three-dimensional toric codes (3DTC) at intermediate rounds. We similarly introduce condensation paths for this and explore a limited subclass of measurement sequences for this class of codes and show that we can implement some of the charge- and flux-permuting automorphisms of the 3DCC. Moreover, taking a specific sequence of 3D dynamic automorphism color code would, by definition, be a 3D Floquet color code, which has not appeared in the literature before. 

The three-dimensional dynamic color code is a reasonable starting point toward achieving universal quantum computation in dynamic automorphism codes. Having it might enable the development of a dimensional jump in dynamic automorphism codes, allowing to interface the two-dimensional DA color code with a three-dimensional one. It would also be interesting to develop just-in-time decoder-like techniques \cite{Bombin2018,Brown2020,Scruby_2022} in the future. Additionally, this new dynamic code might be interesting by itself as well because it would be an intriguing challenge to design the boundaries for it and explore and classify the complete space of condensation sequences.

The rest of this section is structured as follows. In Subsec.~\ref{sec:3DCC_review} we review the 3D color code. In Subsec.~\ref{sec:3DCC_measurement_transition} we explain different ways to condense from a 3D color code (3DCC) to a single copy of a 3D toric code (3DTC) using two-qubit measurements, which we will use later to construct our code.  In Subsec.~\ref{sec:3DCC_parent} we introduce the three-dimensional parent code, which is three copies of 3DCC, and show how to obtain a single effective 3DCC or three decoupled 3DTCs by condensation in a logical information-preserving way. We also show how the transition between an effective 3DCC and three decoupled 3DTC copies works and how it relates to the many ways of unfolding the 3DCC~\cite{Kubica_2015}. In Subsec.~\ref{sec:3DFCC_schedule} we introduce measurement protocols for several generators of automorphisms for the 3D dynamic automorphism color code, showing that it dynamically generates logical qubits (i.e. generates the complete topological ISG). Finally, we address the application of a non-Clifford gate by measurements in Subsec.~\ref{sec:3DFCC_nonClifford}. 
\subsection{Review of 3D color code}
\label{sec:3DCC_review}
Let us first review the three-dimensional color code~\cite{Bombin2007}. The 3DCC can be defined on any 4-valent volume 4-colorable cellulation of a 3-manifold $M$~\cite{Bombin2007}, and in this paper, we assume that $M$ is closed, and the qubits are located at the vertices of the lattice. 
A cellulation of a 3-manifold $M$ is volume 4-colorable if every 3-cell can be colored with one of four colors $c \in \{ {\rc r},{\gc g}, {\bc b},{\yc y}\}$ so that no two adjacent 3-cells (volumes) have the same color.
The coloring of the 3-cells induces a coloring of the 1-cells (edges) and 2-cells (plaquettes). 
We follow the colex (``cell complex'') convention of Ref.~\cite{Bombin2007} to name the plaquettes and logical membranes, where each plaquette is labeled by the two volumes that it interfaces. Therefore, there are six types of plaquettes corresponding to the six color pairs: ${\rc r}{\gc g},{\rc r}{\bc b},{\rc r}{\yc y},{\gc g}{\bc b},{\gc g}{\yc y},{\bc b}{\yc y}$. Notice that each plaquette of color ${\rc r}{\gc g}$ is bordered by edges of the complementary colors ${\bc b}{\yc y}$.
For this reason, it is convenient to introduce the shorthand notation $\overline{rg}$ to mean the complementary colors of ${\rc r}{\gc g}$, i.e.,
\be \nonumber
\overline{{\rc r}{\gc g}} \triangleq {\bc b}{\yc y}.
\ee
On the other hand, we use a different color convention for the 1-cells (edges): instead of labeling an edge by the colors of the three volumes it interfaces, it is named after the color of the volumes connected by this edge.\footnote{This mixed convention for 1-cells and 2-cells provides more intuitive fusion rules for both charges and fluxes as shown in Eqs.~\eqref{eq:chargefusion3D} and~\eqref{eq:3Dfluxfusionrules}.
This contrasts the convention for the 2D color code where both $x$ and $z$ logical operators live on 1-cells (edges) and we use the color convention that is complementary to the colex one for both, and consequently, the commutation relations between the logical operators follow different rules from those in 3D. Namely, $x$- and $z$- logical operators anticommute in 3D if they share a color, while in 2D, they commute if they share a color.
}

\begin{figure}[t] \centering
\vspace{0pt}
\centering
\vspace{0pt}
\includegraphics[width= 1\columnwidth]{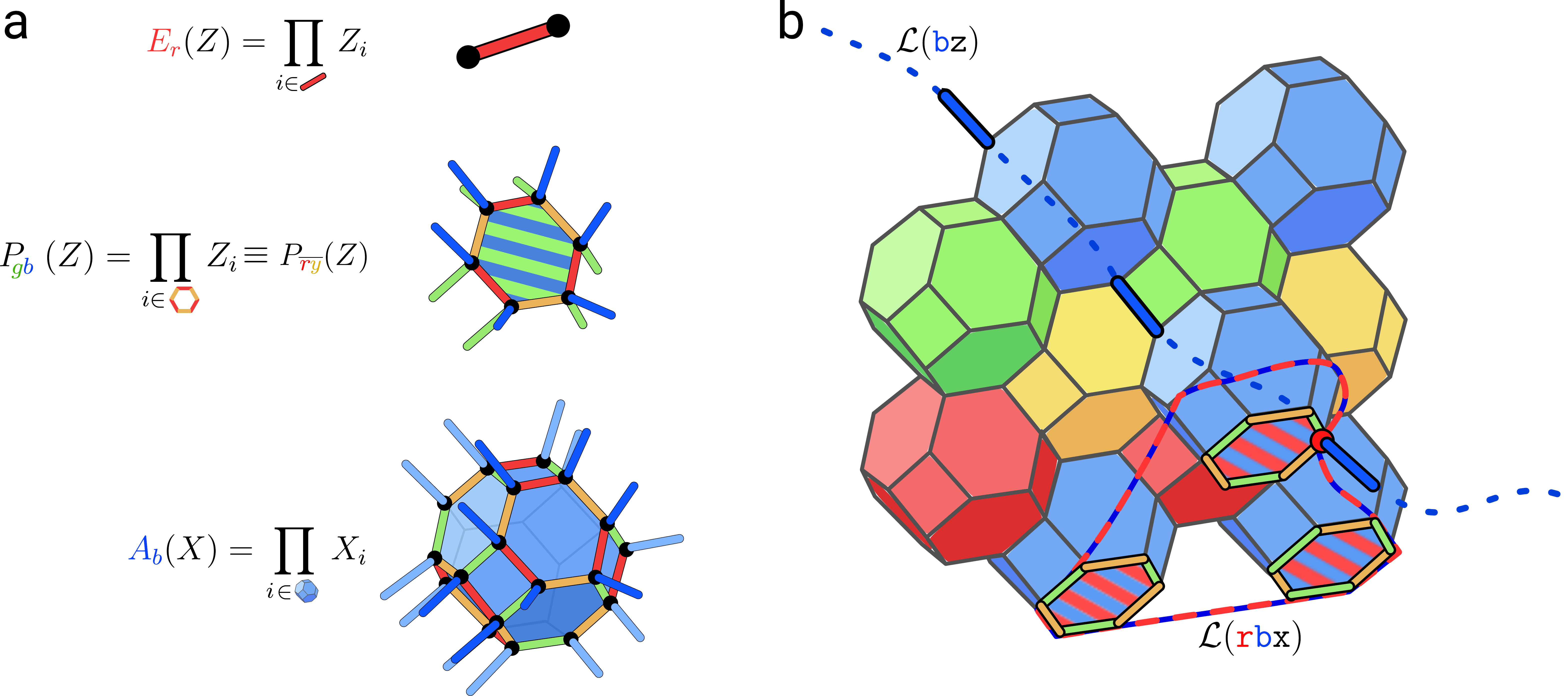}
\caption{(a) Representatives of an edge $E_r(Z)$, plaquette $P_{gb}(Z) = P_{\overline{ry}}(Z)$ and a volume $\vol_b(X)$ operators on for a four-colorable bitruncated cubic honeycomb lattice. The qubits are shown as black dots. For plaquette and volume, all four edges coming from the visible vertices are shown. (b) An example of logical string $\mathcal L(\texttt{bz})$ and logical membrane $\mathcal L(\texttt{rbx})$ operators. }
\label{fig:3d_1}
\end{figure}

The stabilizer group of the 3D color code has volume and plaquette terms.
Let us define $\vol_c(X)$ to be the group of volume stabilizers, where the generators are the 3-cells with color $c \in\{ {\rc r},{\gc g}, {\bc b},{\yc y}\}$. Each generator is a product of Pauli $X$ operators acting on the vertices belonging to the volume.
Similarly, let us define $P_{cc'}(Z)$ to be the group of plaquette stabilizers, where there is one generator per plaquette where each plaquette of color $cc'$ is straddled by volumes of color $c$ and $c'$. Each plaquette generator is a product of Pauli $Z$ operators acting on all vertices at the boundary of that plaquette. Representatives of plaquette and volume operators are shown in Fig.~\ref{fig:3d_1}(a) on a four-colorable lattice that is shown in Fig.~\ref{fig:3d_1})(b). 
To summarize, the full stabilizer group of the 3D color code is given by 
\begin{align} \label{eq:3DCC_ISG}
 \mathcal S(\CC)&=\langle \volr(X), \volg(X), \volb(X), \voly(X), \Prg(Z), \Prb(Z), \Pry(Z), \Pgb(Z), \Pgy(Z),\Pby(Z) \rangle.
\end{align}

The 3DCC is a topological code, known to be finite depth local unitarily equivalent to three copies of the 3DTC~\cite{Kubica_2015}.
The 3DCC has string-like and membrane-like logical operators.
A string operator is found by selecting a path along edges connecting volumes of the same color. 
A Pauli $Z$ operator is then applied to every qubit on that path.
We denote these strings by $\rz$, $\gz$, $\bz$, and $\yz$ in the algebraic theory, and the corresponding logical operators for the code corresponding to homologically nontrivial strings are denoted by $\mathcal{L}(\rz), \mathcal{L}(\gz), \mathcal{L}(\bz)$, and $\mathcal{L}(\yz)$. In the following text, we will use convention the $\mathcal{L}(\rz)$ interchangeably with $\rz$ where it doesn't cause confusion.

We can terminate each logical operator in two natural ways. 
For example, a terminated $\mathcal{L}(\rz)$ can either anti-commute with a single volume stabilizer of color ${\rc r}$, or anticommute with three volume stabilizers colored ${\gc g},{\bc b}$, and ${\yc y}$.
Similarly for the $\mathcal{L}(\gz)$, $\mathcal{L}(\bz)$, and $\mathcal{L}(\yz)$ logical operators.
This is a reflection of the fact that applying a single Pauli-$Z$ operator on any qubit will violate the four adjacent volume stabilizers which have support on that qubit. 
Therefore, the string operators satisfy the fusion rule 
\be 
\rz \otimes \gz \otimes \bz \otimes \yz \cong \id. 
\ee
We can also define the fusion of two string operators of different colors as
{
\renewcommand{\rc}[1]{{\color{red}{#1}}}
\renewcommand{\bc}[1]{{\color{blue}{#1}}}
\renewcommand{\gc}[1]{{\color{ForestGreen}{#1}}}
\renewcommand{\yc}[1]{{\color{amber}{#1}}}
\begin{equation}
\begin{split}
 \rgz &\triangleq \rz \otimes \gz \cong \bz \otimes \yz\\
 \rbz &\triangleq \rz \otimes \bz \cong \gz \otimes \yz\\ 
 \gbz &\triangleq \gz \otimes \bz \cong \rz \otimes \yz 
\end{split}
\label{eq:chargefusion3D}
\end{equation}}Altogether, there are seven distinct isomorphism classes of non-trivial string operators, i.e. $\{\rz,\gz,\bz,\yz,\rgz,\rbz, \gbz \}$. Fig.~\ref{fig:3d_1}(b) shows a representative of logical string $\mathcal L(\ttt{bz})$.

The membrane operators for the 3D color code are defined in Ref.~\cite{Bombin2007} and are more complicated. Each membrane generator is labeled by a pair of colors, and the membrane consists of a product of plaquette operators of the same color labels.
For example, the $\rc{\ttt{r}}\gc{\ttt{g}}$ membrane operator which we denote $\mathcal{L}(\rgx)$ in the code is given by a product of generators in $ \Prg(X)$.
If the $\rgx$ membrane crosses an $\rz$ or $\gz$ string operator an odd number of times, the operators will anti-commute. The $\rgx$ membrane always commutes with the strings of complementary colors (i.e. $\bz$ and $\yz$). 
The corresponding logical membrane and string operators of the code obey the same relations, for example a homologically non-trivial logical membrane operator $\mathcal{L}(\rgx)$ will anti-commute with a homologically non-trivial logical string operator $\mathcal{L}(\gz)$ if they intersect an odd number of times.

The membranes are denoted as $\ttt{cc}'\ttt{x}$ with $\ttt{c} \neq \ttt{c}'$ in the algebraic theory and $\mathcal{L}(\ttt{cc}'\ttt{x})$ in the code. On the lattice, the color of the membrane corresponds to the color of the plaquettes that form the membrane. The fusion rules for the membranes are given by
\be
\ttt{c}'\ttt{c}\ttt{x} \otimes \ttt{cc}''\ttt{x} \cong \ttt{c}'\ttt{c}''\ttt{x}.
\label{eq:3Dfluxfusionrules}
\ee
Notice that 
{
\renewcommand{\rc}[1]{{\color{red}{#1}}}
\renewcommand{\bc}[1]{{\color{blue}{#1}}}
\renewcommand{\gc}[1]{{\color{ForestGreen}{#1}}}
\renewcommand{\yc}[1]{{\color{amber}{#1}}}
\begin{align}
\rgbyx \triangleq \rgx \otimes \byx \cong \rbx \otimes \gyx \cong \ryx \otimes \gbx
\end{align}
}is distinct from any of the six membrane generators. Altogether we have seven distinct isomorphism classes of membrane operators $\{\rgx,\rbx,\ryx,\gbx,\gyx,\byx, \rgbyx \}$. A logical membrane and a string operator anti-commute if they share a color and the support of their intersection is odd.

Fig.~\ref{fig:3d_1}(b) shows a representative of the logical membrane $\mathcal L(\ttt{rbx})$; the qubit where this membrane operator intersects with a string $\mathcal L(\ttt{bz})$ is shown in red, indicating that these operators anticommute.

The topological content of the 3DCC can be understood from four copies of 3DTC along with a condensation.
Label the four copies of 3DTC as $\TC^r \boxtimes \TC^g \boxtimes \TC^b \boxtimes \TC^y$. 
Label the string operators $\ttt{e}_{\rc{\ttt{r}}},\ttt{e}_{\gc {\ttt{g}}}, \ttt{e}_{\bc {\ttt{b}}}$, and $ \ttt{e}_{\yc {\ttt{y}}}$ and similarly for the membranes $\ttt{m}_{\rc{\ttt{r}}},\ttt{m}_{\gc {\ttt{g}}}, \ttt{m}_{\bc {\ttt{b}}}$, and $ \ttt{m}_{\yc {\ttt{y}}}$, 
then condense $\ttt{e}_{\rc{\ttt{r}}} \ttt{e}_{\gc {\ttt{g}}} \ttt{e}_{\bc {\ttt{b}}} \ttt{e}_{\yc {\ttt{y}}} 
\triangleq \ttt{e}_{\rc{\ttt{r}}} \otimes \ttt{e}_{\gc {\ttt{g}}} \otimes \ttt{e}_{\bc {\ttt{b}}} \otimes \ttt{e}_{\yc {\ttt{y}}}$.
All string-like logical operators survive the condensation, but we now have a new equivalence relation $\ttt{e}_{\rc{\ttt{r}}} \ttt{e}_{\gc {\ttt{g}}} \ttt{e}_{\bc {\ttt{b}}} \ttt{e}_{\yc {\ttt{y}}} \cong \id $, indicating that we only have three independent string operators remaining, just as in the 3DCC. 
One can identify these string operators with those of the 3DCC via $\ttt{cz} \leftrightarrow \ttt{e}_\ttt{c}$.
Only pairs of membrane operators survive the condensation.
For example $\ttt{m}_{\ttt{\rc r}}\ttt{m}_{\ttt{\gc g}}$ braids trivially with $\ttt{e}_{\rc{\ttt{r}}} \ttt{e}_{\gc {\ttt{g}}} \ttt{e}_{\bc {\ttt{b}}} \ttt{e}_{\yc {\ttt{y}}}$ and therefore survives.
Identifying these membrane operators with those of the 3DCC is straightforward, yielding $\ttt{cc}'\ttt{x} \leftrightarrow \ttt{m}_{\ttt{c}}\ttt{m}_{\ttt{c}'}$.
 We summarize the correspondence in Table~\ref{tab:3DCCto3TC}.

\begin{table}[h]
{
\renewcommand{\rc}[1]{{\color{red}{#1}}}
\renewcommand{\bc}[1]{{\color{blue}{#1}}}
\renewcommand{\gc}[1]{{\color{ForestGreen}{#1}}}
\renewcommand{\yc}[1]{{\color{amber}{#1}}}
 \centering
 \resizebox{\textwidth}{!}{
 \begin{tabular}{|c|c|c|c|c|c|c|c||c|c|c|c|c|c|c|}
 \hline
 Notation & \multicolumn{7}{|c||}{Logical string classes, $\mathcal L(\cdot)$} & \multicolumn{7}{|c|}{Logical membranes classes, $\mathcal L(\cdot)$} \\
 \hline
 $\TC^r\boxtimes \TC^g \boxtimes \TC^b \boxtimes \TC^y/\langle 1 \cong \ttt{e}_{\ttt{r}} \ttt{e}_{\ttt{g}} \ttt{e}_{\ttt{b}} \ttt{e}_{\ttt{y}} \rangle $ &$ \ttt{e}_\ttt{r}$ &$\ttt{e}_\ttt{g}$ &$\ttt{e}_\ttt{b}$& $\ttt{e}_\ttt{r}\ttt{e}_\ttt{g}$& $\ttt{e}_\ttt{r}\ttt{e}_\ttt{b}$&$\ttt{e}_\ttt{g}\ttt{e}_\ttt{b}$&$\ttt{e}_\ttt{r}\ttt{e}_\ttt{g}\ttt{e}_\ttt{b}$
 & $\ttt{m}_\ttt{r}\ttt{m}_\ttt{y}$ &$\ttt{m}_\ttt{g}\ttt{m}_\ttt{y}$ &$\ttt{m}_\ttt{b}\ttt{m}_\ttt{y}$& $\ttt{m}_\ttt{r}\ttt{m}_\ttt{g}$& $\ttt{m}_\ttt{r}\ttt{m}_\ttt{b}$&$\ttt{m}_\ttt{g}\ttt{m}_\ttt{b}$&$\ttt{m}_\ttt{r}\ttt{m}_\ttt{g}\ttt{m}_\ttt{b}\ttt{m}_\ttt{y}$
 \\
 \hline
 $\TC^{(1)}\boxtimes \TC^{(2)} \boxtimes \TC^{(3)}$ &$\ttt{e}_\ttt{1}$ &$\ttt{e}_\ttt{2}$ &$\ttt{e}_\ttt{3}$& $\ttt{e}_\ttt{1}\ttt{e}_\ttt{2}$& $\ttt{e}_\ttt{1}\ttt{e}_\ttt{3}$&$\ttt{e}_\ttt{2}\ttt{e}_\ttt{3}$&$\ttt{e}_\ttt{1}\ttt{e}_\ttt{2}\ttt{e}_\ttt{3}$
 & $\ttt{m}_\ttt{1}$ &$\ttt{m}_\ttt{2}$ &$\ttt{m}_\ttt{3}$& $\ttt{m}_\ttt{1}\ttt{m}_\ttt{2}$& $\ttt{m}_\ttt{1}\ttt{m}_\ttt{3}$&$\ttt{m}_\ttt{2}\ttt{m}_\ttt{3}$&$\ttt{m}_\ttt{1}\ttt{m}_\ttt{2}\ttt{m}_\ttt{3}$
 \\
 \hline
 CC& $\rz$ & $\gz$ & $\bz$ & $\rgz$ & $\rbz$ & $\gbz$ & $\yz$ & $\ryx$ & $\gyx$ & $\byx$ & $\rgx$ & $\rbx$ & $\gbx$ & $\rgbyx$ \\ 
 
 \hline
 \end{tabular}
 }
 \caption{Correspondence between the logical strings and membranes of the 3D color code, three copies of the 3D toric code and four copies of the 3D toric code with $\ttt{e}_{\ttt{r}} \ttt{e}_{\ttt{g}} \ttt{e}_{\ttt{b}} \ttt{e}_{\ttt{y}}$ condensation. 
 We have used the shorthand $\ttt{e} \ttt{e}' \triangleq \ttt{e} \otimes \ttt{e}'$, and similarly for membranes.}
 \label{tab:3DCCto3TC}
 }
\end{table}

\subsection{Condensation transitions from 3D color code to 3D toric code}
\label{sec:3DCC_measurement_transition}
Similarly to the two-dimensional case, the construction of the 3D DA color code relies on taking a parent code, which is three copies of 3D color code, and following a path of reversible condensations which implement different automorphisms.
Condensation paths can be directly turned into measurement sequences thanks to the relation between the Hamiltonian lattice picture and the topological codes; we further choose sequences that generate the full topological ISG in order to design a proper DA color code.
The child codes in our constructions are either a single effective color code or decoupled copies of 3D toric codes. 
However, in three dimensions there are more ways to turn a single copy of 3D color code into the toric code, and different ways to do so will be instrumental for further discussion because these condensations will be later used for constructing 3D DA color code protocols. Therefore, we consider different ways to condense a single 3D color code into a single 3D toric code in this subsection. We first show how to perform these condensations in algebraic language, and then explain the translation to measurements in the code language. In fact, these considerations alone allow us to construct a generalization of the honeycomb codes to three dimensions presented in App.~\ref{sec:3D_HH}.

In the Hamiltonian picture, the excitations of the 3D color code are point-like (charges) and loop-like (fluxes), corresponding to the boundaries of string-like and membrane-like operators, respectively. They fall into the same seven isomorphism classes that we introduced earlier for string and membrane operators. However, there exist other types of (invertible) excitations that one might consider condensing, such as twist strings~\cite{Yoshida2017,Roumpedakis23,Barkeshli23,Barkeshli2022higher}. The string created from truncating the transversal $S$ gate of the color code on an open surface is such an excitation. It would be interesting to work out a complete picture of possibilities as well as more general classes of 3D DA color codes coming from such condensations. However, here we narrow our focus to simply condensing charges and fluxes only and leaving full exploration of possible three-dimensional DA color code sequences to future work.

\begin{figure}[t] \centering
\vspace{0pt}
\centering
\vspace{0pt}
\includegraphics[width= 1 \columnwidth]{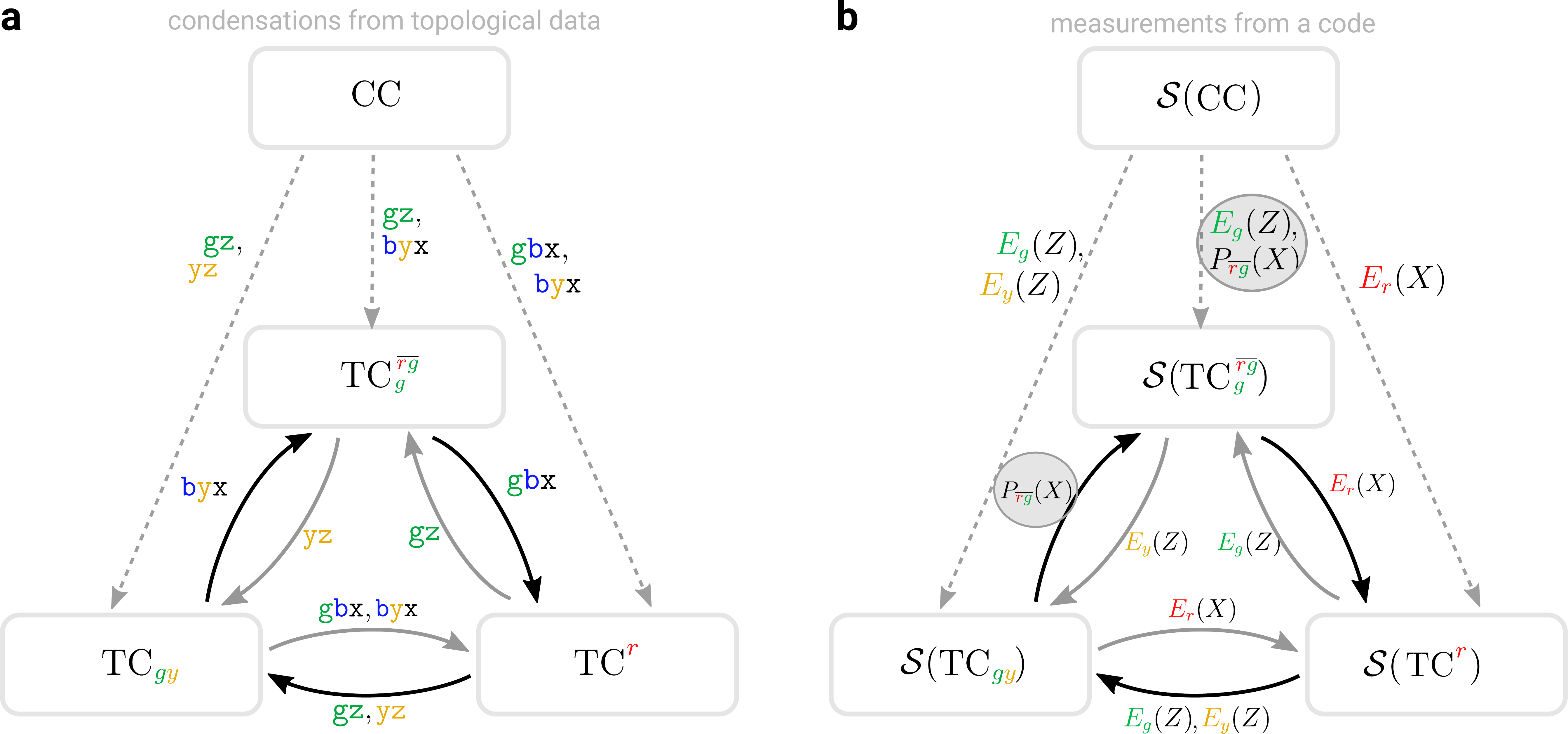}
\caption{ (a) The dashed lines show examples corresponding to different types of condensations from 3D color code to a single copy of 3D toric code, which is explained in detail in the text. The solid lines show child-child condensations which can take us from one child toric code to another. (b) Going between the instantaneous stabilizer groups of respective codes by measurements. 
It is always possible to perform the transitions between the child codes using two-body measurements only with the exception of two transitions which are circled in grey. 
}
\label{fig:3d_2}
\end{figure}

Thus, to obtain a 3D toric code by condensing a set of objects within a 3D color code, we condense either: (i) two charges, (ii) two fluxes, or (iii) one charge and one flux.
These three options are shown in Fig.~\ref{fig:3d_2}(a). There also exist options when, for example, pairs of charges or pairs of fluxes can be condensed (recall that there are 7 isomorphism classes for charges and fluxes, and any class can be condensed). However, in sequences considered here, we strive to use two qubit measurements only; when translating condensations of pairs of objects into measurements, we cannot straightforwardly avoid measuring higher-weight operators. Therefore, classifying all automorphisms and types of sequences in 3D DA color codes is left to future work. 

The dashed gray arrows in Fig.~\ref{fig:3d_2}(a) show the condensations of (i), (ii), and (iii) types from the parent model down to the child models, which are always equivalent to a toric code.
The solid lines are reversible child-child transitions, i.e. the ones implementing a transition from one child theory to another. We will address them later, and for now, let us focus on condensations from a parent 3DCC to a toric code. We denote the outcome of the condensations ``TC'' with a superscript of colors of the condensed fluxes and the subscript is the union of the colors of all condensed charges. When appropriate, we will also often instead denote the colors by their complementary ones appearing with an overline. For example, the complement of $gby$ is $\overline r$ and the complement of $by$ is $\overline{rg}$. Let us consider each of the options explicitly. 
\begin{enumerate}[(i)]
 \item Condensing two charges of colors $c$ and $c'$ produces TC$_{cc'}$. In this case,
 the remaining deconfined charges $\ttt{c}'' \ttt{z}$ with $c'' \neq c',c$ form a single isomorphism class. The only flux excitation that braids trivially with both condensed charges $\ttt{cz}$ and $\ttt{c}'\ttt{z}$ is of complementary colors to both of them, i.e. it is $\overline{\ttt{cc}'\ttt{x}}$. For example, Fig.~\ref{fig:3d_2}(a), we condense $\gz$ and $\yz$. In the child toric code, the charge that is deconfined is $\rz \cong \bz$, and the deconfined flux is $\rbx$ which clearly braids non-trivially the deconfined charge as expected. 
 \item Condensing two classes of fluxes such that a complementary color to them is $c$ produces TC$^{\overline c}$ (two different fluxes will always have a single complementary color). In this case, the only charge that braids trivially with both condensed fluxes has to be complementary in color to both fluxes: $\ttt{cz}$. This is the only deconfined charge. Similarly, all fluxes of colors $\ttt{cc}'\ttt{x}$ will remain deconfined, but will belong to the same equivalence class (they become isomorphic up to fusions by the condensed membranes). The fluxes in this class clearly braid nontrivially with the deconfined charge, which explains why we are left with a toric code. In the example in Fig.~\ref{fig:3d_2}(a), we condense $\gbx$, $\byx$ (and their product $\gyx$). The complementary color is therefore red and so in TC$^{\overline r}$ the deconfined charge is $\rz$ and the deconfined flux is $\rgx \sim \rbx \sim \ryx$. 
 \item Condensing a single charge of color $c$ and a flux with colors $\overline{cc'}$ produces TC$_c^{\overline{cc'}}$. Because $\ttt{cz}$ is condensed, only one charge that braids trivially with the condensed flux remains, which is $\ttt{c}'\ttt{z}$. The colors of deconfined fluxes have to be complementary to $c$ because they have to braid trivially with the condensed charge. We find that the remaining fluxes form a single isomorphism class $\overline{\ttt{c}\ttt{c}''\ttt{x}}$ where $c'' \neq c'$ (otherwise the flux coincides with the confined one). In Fig.~\ref{fig:3d_2}(a), we condense $\gz$ and $\byx$. The complementary color to the fluxes are red and green, therefore we label this toric code $\TC^{\overline{rg}}_g$. The deconfined charge will be $\rz$ and the deconfined flux corresponds to $\rbx \cong \ryx$.
\end{enumerate}
Explicit examples of passing logical operators corresponding to homologically nontrivial strings (membranes) generated by deconfined charges (fluxes) from the parent code down to each of these toric codes are shown in Table~\ref{table:3Dlogicals}. We will show explicit stabilizer groups for each of these types of toric code later on. Fig.~\ref{fig:3d_2}(a) shows specific choices of colors for the illustration. Any combination of colors is allowed, as long as all the objects that are being condensed braid trivially with each other and the total number of independent classes that are being condensed is two, in which case we obtain a single 3D toric code from a 3D color code.

{
\renewcommand{\rc}[1]{{\color{red}{#1}}}
\renewcommand{\bc}[1]{{\color{blue}{#1}}}
\renewcommand{\gc}[1]{{\color{ForestGreen}{#1}}}
\renewcommand{\yc}[1]{{\color{amber}{#1}}}
\begin{table}[!b]
\centering
$\underbrace{\text{
\begin{tabular}{cccc} 
$\rz$ & $\gz$ & $\bz$ & $\yz$ \\
\hline 
& $\rgx$ & $\rbx$ & $\ryx$\\
&&$\gbx$ & $\gyx$\\
&&& $\byx$ \\
\end{tabular} }
}_{\text{$\CC$ }}$
$\xrightarrow{\gz, \yz}$
$\underbrace{\text{
\begin{tabular}{cccc} 
$\rz$ & \color{gray}{$\ttt{gz}$} & $\bz$ & \color{gray}{$\ttt{yz}$} \\
\hline 
& \sout{\color{gray}{$\ttt{rgx}$}} & $\rbx$ & \sout{\color{gray}{$\ttt{ryx}$}}\\
&&\sout{\color{gray}{$\ttt{gbx}$ }} & \sout{\color{gray}{$\ttt{gyx}$}}\\
&&& \sout{\color{gray}{$\ttt{byx}$}} \\
\end{tabular}}
}_{\text{$\TC_{\gc{g}\yc{y}}$ }}$ \\ \hspace{95 pt}  CC $\xrightarrow{\gbx,\gyx, \byx}$ $\underbrace{\text{
\begin{tabular}{cccc} 
$\rz$ & \sout{\color{gray}{$\ttt{gz}$}} & \sout{\color{gray}{$\ttt{bz}$}} & \sout{\color{gray}{$\ttt{yz}$}} \\
\hline 
& $\rgx$ & $\rbx$ & $\ryx$\\
&&\color{gray}{$\ttt{gbx}$} & \color{gray}{$\ttt{gyx}$}\\
&&& \color{gray}{$\ttt{byx}$} \\
\end{tabular} 
}}_{\text{$\TC^{\overline{\rc r}}$ } }$
\\
\hspace{105 pt} CC 
$\xrightarrow{\gz, \byx}$
$\underbrace{\text{
\begin{tabular}{cccc} 
$\rz$ & \color{gray}{$\ttt{gz}$} & \sout{\color{gray}{$\ttt{bz}$}} & \sout{\color{gray}{$\ttt{yz}$}} \\
\hline 
& \sout{\color{gray}{$\ttt{rgx}$}} & $\rbx$ & $\ryx$\\
&&\sout{\color{gray}{$\ttt{gbx}$ }} & \sout{\color{gray}{$\ttt{gyx}$}}\\
&&& {\color{gray}{$\ttt{byx}$}} \\
\end{tabular}}
}_{\text{$\TC^{{\rc r} {\gc g} }_{\gc g}$ }}$
\\
\caption{
The deconfined charges and fluxes of three types of child toric codes that are obtained by condensing different sets of objects in the 3D color code. 
The grayed-out elements correspond to condensed objects, and striked-out elements are confined. 
In each case, the remaining fluxes belong to a single isomorphism class, related by fusion with trivialized (gray) fluxes.  Similarly, the remaining charges belong to a single isomorphism class. 
}
\label{table:3Dlogicals}
\end{table}
}

Next, the solid black and gray lines in Fig.~\ref{fig:3d_2}(a) show transitions between child theories. We are especially concerned with rank-preserving transitions that preserve the topological order of the child theory and the logical information. A transition between child theories (i)-(iii) is allowed if the following conditions are satisfied: if an object that we want to condense in the next step commutes with all the objects condensed in the previous step, it has to already be among the condensations in the previous step. 
Analogously to the condensations from the parent 3D color code, one can consider how charges and fluxes, and consequently logical operator representatives, behave under such reversible condensations. For allowed pairs of reversible condensations, there is always a single shared logical string representative (or a single isomorphism class of them) and a single shared logical membrane representative (or a single isomorphism class of them). Thus, we can observe that each pair of toric codes in Table~\ref{table:3Dlogicals} shares an $\rz$ string representative. TC$_{gy}$ and TC$^{\overline r}$, and similarly, TC$_{gy}$ and TC$_g^{\overline{rg}}$ share $\rbx$ logical membrane. TC$^{\overline r}$ and TC$_g^{\overline{rg}}$ share a class of membranes $\rbx \sim \ryx$ that are equivalent up to a shared condensation $\byx$. Notice that the objects condensed in a child-child transition always coincide with the ones condensed when going from the parent code to the respective child codes. This is always the case for allowed transitions apart from the situation when the two child codes share a condensed object, which is already condensed and then doesn't appear in the list of objects that need to be condensed in order to perform the transition.

Finally, let us show how the same transitions can be performed between respective topological codes by few-qubit measurements. 
Let us define the stabilizer group $E_c(X)$ as the group generated by the two-qubit edge operators acting on edges of color $c\in\{{\rc r},{\gc g},{\bc b},{\yc y}\}$. Such edges are shown in Fig.~\ref{fig:3d_1}.
Each edge operator is given by $XX$ acting on each endpoint of the edge.

Then, there exists the following correspondence between the operators condensing the charges and fluxes in the Hamiltonian model and measurements in topological codes:
\be
\begin{split}
\ttt{cz} &\rightarrow E_c(Z)
\\
\ttt{cz}, \overline{\ttt{cc}'}\ttt{x} &
 \rightarrow \{E_c(Z), P_{\overline{\ttt{cc}'}}(X) \}\\
\overline{\ttt{cc}'}\ttt{x}, 
\overline{\ttt{cc}''}\ttt{x}, \overline{\ttt{c}\ttt{c}'''}\ttt{x} &\rightarrow E_c(X)
\end{split}
\ee
Fig.~\ref{fig:3d_2}(b) translates condensations shown in panel (a) into measurements according to this prescription. The only two instances where it is not possible to perform a transition by two-body measurements are circled in gray.

The stabilizer groups for each of the toric code examples can be obtained by starting with the parent stabilizer group for the 3D color code in Eq.~\eqref{eq:3DCC_ISG} and measuring the respective operators identified in Fig.~\ref{fig:3d_2}(b). The stabilizer group obtained this way consists of the measured checks and the elements of the parent stabilizer group that commute with the measurements. This way, we obtain the stabilizer groups for each of the toric code types:
\begin{align}
\mathcal{S}(\TC_{cc'}) = \langle E_c(Z), E_{c'}(Z), \vol_{c''}(X), \vol_{c'''}(X), P_{\overline{c''c'''}}(Z)\rangle
\end{align}
where $c''$ and $c'''$ are the two complementary colors to the pair$c,c'$. For example:
\begin{align}
\mathcal{S}(\TC_{gy}) = \langle E_g(Z), E_{y}(Z), \vol_{r}(X), \vol_{b}(X), P_{\overline{rb}}(Z)\rangle
\end{align}
The other example of the toric code was obtained from condensing two fluxes:
\begin{align}
\mathcal{S}(\TC^{\overline{c}}) = \langle E_c(X), \vol_c(X), P_{\overline{cc'}}(Z) \rangle_{c'\neq c \in \{r,g,b,y\}}
\end{align}
The specific toric code from this family that appeared in the figures in this section has the following ISG:
\begin{align}
\mathcal{S}(\TC^{\overline{\rc r}}) = \langle \Er(X), \volr(X), \Pbarrg(Z), \Pbarrb(Z), \Pbarry(Z) \rangle
\end{align}
Finally, one more example is:
\begin{align}
 \mathcal{S}(\TC^{\overline{cc'}}_c)
 = \langle E_c(Z), P_{\overline{cc'}}(X), \vol_r(X),P_{\overline{c'c''}}(Z), P_{\overline{c' c'''}}(Z)\rangle 
\end{align}
Where $c \neq c'$, $c' \neq c'',c'''$.
If we pick specific colors as in figures in this section, we have:
\begin{align}
 \mathcal{S}(\TC^{\overline{rg}}_g)
 = \langle E_g(Z), P_{\overline{rg}}(X), \vol_r(X),P_{\overline{rb}}(Z), P_{\overline{ry}}(Z) \rangle.
\end{align}

Apart from 3D DA color code, these transitions by measurements and examples of the toric code stabilizer groups can be used to generalize of the honeycomb code to three dimensions presented in App.~\ref{sec:3D_HH}.

\subsection{Parent code for 3D dynamic automorphism color code}
\label{sec:3DCC_parent}

Finally, we are ready to present the parent code for the 3D Floquet color code and 3D DA color code, and show how it overarches the principle at work for these codes.
We present measurement schedules that realize a subclass of automorphisms present in the 3D CC which yield an $S_4$ permutation symmetry of the individual charges and fluxes.
We also provide a measurement sequence in Sec.~\ref{sec:3DFCC_nonClifford} which implements the automorphism related to the transversal $T$ gate in the usual color code.
We anticipate that more general automorphisms may be realized with other measurement sequences but leave their discovery to future work.

First, let us briefly talk about the 3D DA color code in the condensation language. 
The parent theory is three copies of 3D color code, and similarly to the 2D case, we can condense products of charges across pairs of layers i.e. $\{ \ttt{c}\ttt{z}_1\ttt{c}\ttt{z}_2,\ttt{c}\ttt{z}_2\ttt{c}\ttt{z}_3 \}_{c = r,g,b,y}$, where the subscript indicates that these objects are condensed for each color. There are four distinct isomorphism classes of logical strings left after this condensation, one for each color: $\{ \rzi, \gzi, \bzi, \yz_1\}$ whose equivalent representatives can live in any layer. 
A membrane of a given color in one layer $\ttt{cc}'\ttt{x}_\ell$ will braid non-trivially with the condensate. However, a product of the same kind of membrane in all layers $\ttt{cc}'\ttt{x}_1\ttt{cc}'\ttt{x}_2\ttt{cc}'\ttt{x}_3$ braids trivially with the condensed objects, which is true for every pair of colors $c,c'$. One can explicitly check that these are all the deconfined objects left after condensation. It is easy to see that the theory that we obtain this way is isomorphic to the 3D color code. 

The other kind of condensations that we will use are those that independently map each copy of the 3D color code (in the parent code) to different 3D toric codes. These condensations were summarized in detail in the previous subsection.

\begin{figure}[t] \centering
\vspace{0pt}
\centering
\vspace{0pt}
\includegraphics[width= 1\columnwidth]{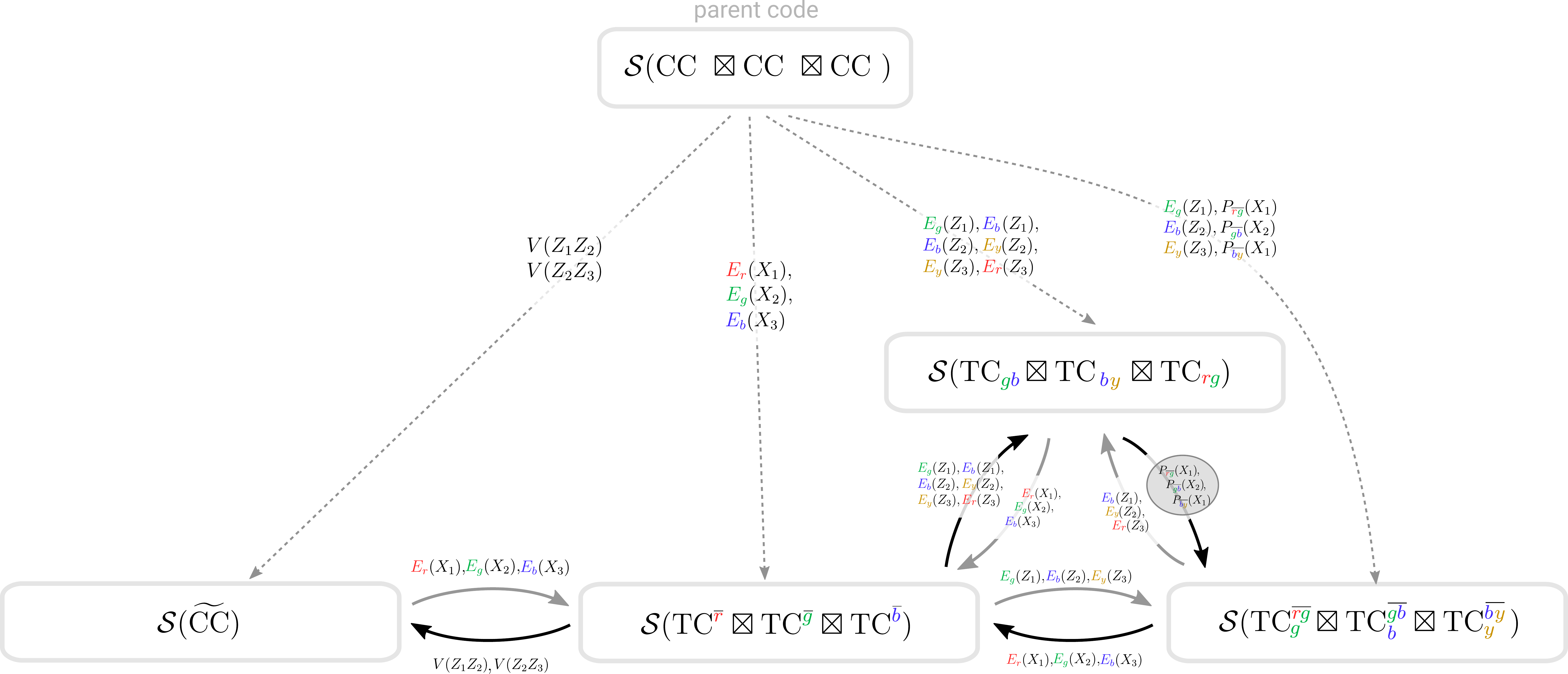}
\caption{
Example of measurement transitions from the parent code (three copies of the 3D color code) and some of the child codes, including a single effective copy of the 3D color code, as well as transitions by measurements between the child codes. }
\label{fig:3d_3}
\end{figure}

Finally, we discuss transitions between the child models.  We find that transitions are possible between triples of toric codes whenever the three toric codes correspond to the same class of condensations from the previous subsection, i.e. charge-charge, charge-flux or flux-flux, and the colors of all three of them are different. The transitions between the child models are associated with condensing the same objects as the ones needed in order to obtain the given child code from the parent code. Some of these scenarios correspond to new kinds of unfolding maps that have not been explicitly constructed in literature before. 

In the topological codes language, the condensation from the parent code down to a single effective 3DCC is achieved by measuring $V(Z_1Z_2)$ and $V(Z_2Z_3)$ operators analogously to the 2D case, namely:
\begin{align}
\mathcal{S}(\CC \boxtimes \CC \boxtimes \CC )\xrightarrow{V(Z_1Z_2), V(Z_2Z_3)} \mathcal{S}(\widetilde{\CC}) = 
\langle V(Z_1Z_2), V(Z_2Z_3), \vol_c(X_1 X_2 X_3), P_{cc'}(Z_1) \rangle 
\end{align}
with $c \in \{ {\rc r},{\gc g},{\bc b}, {\yc y} \}$ and $ cc' \in \{ {\rc r}{\gc g},{\rc r}{\bc b},{\rc r} {\yc y},{\gc g}{\bc b},{\gc g}{\yc y},{\bc b}{\yc y}\}$.
In terms of the stabilizer group, the measurements $V(Z_1Z_2), V(Z_2Z_3)$ simply bind together $X$ type operators of $\CC^3$, and identify $Z$ type operators from separate layers with one another.

A transition between $\widetilde{\CC}$ and $\TC^{\overline{\rc r}} \boxtimes \TC^{\overline{\gc g}} \boxtimes \TC^{\overline{\bc b}}$ is given by:
\begin{align}
\label{eq:CCeff_toTC3}
\mathcal{S}(\widetilde{\CC}) \xrightarrow{\Er(X_1), \Eg(X_2),\Eb(X_3)} \mathcal{S}(\TC^{\overline{\rc r}} \boxtimes \TC^{\overline{\gc g}} \boxtimes \TC^{\overline{\bc b}}).
\end{align}
The reverse transition is:
\begin{align} 
\label{eq:TC3_toCCeff}
\mathcal{S}(\TC^{\overline{\rc r}} \boxtimes \TC^{\overline{\gc g}} \boxtimes \TC^{\overline{\bc b}}) \xrightarrow{V(Z_1Z_2), V(Z_2Z_3)} \mathcal{S}(\widetilde{\CC})
\end{align}
Of course, the choice of ${\rc r}{\gc g}{\bc b}$ was arbitrary, we could have used any one triple from $$\{{\rc r}{\gc g}{\bc b},{\rc r}{\bc b}{\yc y},{\rc r}{\gc g}{\yc y},{\bc b}{\gc g}{\yc y}\}$$ instead.

For illustration, we exemplify some of the transitions that will occur later in our codes in Fig.~\ref{fig:3d_3}.
Lastly, we emphasize that for each logical operator class there exists a representative that survives upon a transition, whenever this transition is one of the rank-preserving kinds that satisfies the conditions that we introduced in this and previous subsection. Thus, all the ``allowed'' transitions will be logical information-preserving.

\subsection{3D dynamic automorphism color code}
\label{sec:3DFCC_schedule}

\begin{table}[!b]
{
\renewcommand{\rc}[1]{{\color{red}{#1}}}
\renewcommand{\bc}[1]{{\color{blue}{#1}}}
\renewcommand{\gc}[1]{{\color{ForestGreen}{#1}}}
\renewcommand{\yc}[1]{{\color{amber}{#1}}}
 \centering
 \resizebox{\textwidth}{!}{\begin{tabular}{|c|c|c|c|c|c|c|c| }
 \hline
 \multirow{2}{*}{Gate/Aut} & \multicolumn{4}{c|}{$t$} \\ \cline{2-5} 
 & 1 & 2 & 3 & 4 \\
 \hline
 \multirow{2}{*}{$\rc r \rightarrow \yc y \rightarrow \bc b \rightarrow \gc g \rightarrow \rc r$} & $\Er(X_1)$ & $\Eg(Z_1), \Eb(Z_1)$ & $\Ey(X_1)$ & \multirow{2}{*}{ $V(Z_1Z_2)$}\\
 & $\Eg(X_2) $ & $\Ey(Z_2), \Eb(Z_2)$ & $\Er(X_2)$ & \\ 
 (ISG generating)& $\Ey(X_3)$ & $\Er(Z_3), \Eg(Z_3)$ & $\Eb(X_3) $ & $V(Z_2Z_3)$ \\
 \hline
 Eff. code & TC$^{\overline{\rc r} } \boxtimes $TC$^{\overline{\gc g} } \boxtimes $TC$^{\overline{\yc y} } $ & TC$_{\gc g \bc b} \boxtimes$ TC$_{\yc y \bc b} \boxtimes$TC$_{\rc r \gc g} $ & TC$^{\overline{\yc y} } \boxtimes $ TC$^{\overline{\rc r} } \boxtimes $ TC$^{\overline{\bc b} }$ & $\widetilde{\text{CC}}$ 
 \\
 \hline
 \hline
 \multirow{2}{*}{$\rc r \leftrightarrow \yc y $} & $\Er(X_1)$ & $\Eg(Z_1), \Eb(Z_1)$ & $\Ey(X_1)$ & \multirow{2}{*}{ $V(Z_1Z_2)$}\\
 & $\Eg(X_2) $ & $\Er(Z_2), \Ey(Z_2)$ & $\Eg(X_2)$ & \\ 
 & $\Ey(X_3)$ & $\Eg(Z_3), \Eb(Z_3)$ & $\Er(X_3) $ & $V(Z_2Z_3)$ \\
 \hline
 Eff. code & TC$^{\overline{\rc r} } \boxtimes $TC$^{\overline{\gc g} } \boxtimes $TC$^{\overline{\yc y} } $ & TC$_{\gc g \bc b} \boxtimes$ TC$_{\rc r \bc b} \boxtimes$TC$_{\gc g \bc b} $ & TC$^{\overline{\yc y} } \boxtimes $ TC$^{\overline{\gc g} } \boxtimes $ TC$^{\overline{\rc r} }$ & $\widetilde{\text{CC}}$ 
 \\
 \hline
 \end{tabular}}
 \caption{ Two-qubit measurement sequences generating color permutations in 3D dynamic automorphism color code on a three-torus (and, consequently, SWAP gates between logical qubits). The first sequence permutes colors cyclically, and, as we show below, generates the full topological ISG (shorthand `ISG-generating') after 5 steps, and the second one swaps two colors and does not generate the full topological stabilizer group. Swapping any other two colors can be obtained by an analogous measurement sequence with an appropriate color exchange. The child (effective) code realized at each given step is shown below the respective measurements. }
 \label{tab:3DDACC_table_short}
}
\end{table}

In this paper, we construct a subclass of codes belonging to the class of the 3D dynamic automorphism color code, using three copies of 3D color code as a parent code. Specifically, in this subsection, we show the protocols for the DA color codes that perform the group of color permutations (which is an $S_4$ subgroup of the Clifford group of the nine logical qubits on a three-torus) using two-body Pauli measurements only. 
Pauli measurements should in principle allow us to construct a larger subgroup of the logical Clifford group, but here we narrow down our scope to transitions that can certainly be implemented by two-qubit measurements only, which are shown in Fig.~\ref{fig:3d_3}. As a result, we design short period-4 sequences that generate color-permuting and syndrome-measuring automorphisms of the 3D color code and can be achieved by two-qubit measurements only.
Finally, in the next subsection, we discuss the possibility of using Clifford two-qubit measurements in order such that a non-Clifford gate will be applied by the sequence.

Table~\ref{tab:3DDACC_table_short} shows two measurement sequences that we can use as a basis to generate the ISG of the topological code and also any color-permuting automorphism. Moreover, they are in one-to-one correspondence with respective condensations which was explained in the previous subsections. Both sequences shown in the table contain only transitions that preserve logical information and the ISG rank. 
The first sequence in the table shows a cyclic color permutation $\rc r \rightarrow \yc y \rightarrow \bc b \rightarrow \gc g \rightarrow \rc r$ that generates the full stabilizer group of the 3D color code in 4 rounds. The second sequence does not generate the full stabilizer group but nevertheless preserves it (similarly to the generating sequences for 2D DAC).
It swaps two colors, and a similar sequence exchanging any pair of colors can be easily obtained from it.

The automorphisms realized from charge and flux condensations can be easily understood analogously to the 2D case. Table~\ref{tab:3DCC_cyclic_logicals} shows the transformation of logical operators under the condensation/measurement sequence for the $\rc r \rightarrow \yc y \rightarrow \bc b \rightarrow \gc g \rightarrow \rc r$ automorphism. For each pair of condensation rounds, we find a shared logical string and a logical membrane representative that is inherited to the next round, which is shown by an entry in Table~\ref{tab:3DCC_cyclic_logicals}. 
If the logical membrane representative is to be taken to the subsequent round, we need to find another representative by multiplying the current one by the condensations of the current round such that it commutes with all measurements of the subsequent round, this is shown with the respective cells in gray in Table~\ref{tab:3DCC_cyclic_logicals}. One can explicitly confirm that exchanges of colors in logical stings and membranes coincide with the $\rc r \rightarrow \yc y \rightarrow \bc b \rightarrow \gc g \rightarrow \rc r$ action of the automorphism. Passing the logical operators within measurement sequences for topological 3D DA color codes is analogous, where the logical string and membranes are represented by respective Pauli strings and membranes, and condensations are replaced with Pauli measurements.

\begin{table}[t]
\resizebox{\textwidth}{!}{
\renewcommand{\rc}[1]{{\color{red}{#1}}}
\renewcommand{\bc}[1]{{\color{blue}{#1}}}
\renewcommand{\gc}[1]{{\color{ForestGreen}{#1}}}
\renewcommand{\yc}[1]{{\color{amber}{#1}}}
\begin{tabular}{| c | c | c | c | c|}
\hline
Time & Condense & Resp. meas. & string rep. & log. membrane rep.\\
\hline
0 & $\rzi \rzii, \gzi \gzii, \bzi \bzii$ &$V(Z_1Z_2)$& \multirow{2}{*}{${\rzi},{\gzii},{\yziii}$} & \multirow{2}{*}{${\widetilde{\rbx}},{\widetilde{\gbx}},{\widetilde{\byx}}$}\\
& $\rzii \rziii, \gzii \gziii, \bzii \bziii$& $V(Z_2Z_3)$& & \\
\hline
1& $\gbx_1, \gyx_1, \byx_1$ & $\Er(X_1)$ & ${\rzi}$& ${\rbx_1} {\color{gray} \times \ttt{byx}_1}$ \\
& $\rbx_2, \ryx_2, \byx_2$& $\Eg(X_2)$ & ${\gzii}$ & ${\gbx_2}{\color{gray}\times \ttt{rbx}_2}$ \\
& $\rgx_3, \rbx_3, \gbx_3$& $\Ey(X_3)$ & ${\yziii}$& ${\byx_3}$\\ 
\hline 
2& $\gzi, \bzi$ & $\Eg(Z_1),\Eb(Z_1)$& ${\rzi} {\color{gray} {\times \ttt{gz}_1 \times \ttt{bz}_1 }}$ & ${\ryx_1}$\\
& $\yzii, \bzii$& $\Ey(Z_2),\Eb(Z_2)$& ${\gzii} {\color{gray} \times \ttt{yz}_2 \times \ttt{bz}_2 }$ & ${\rgx_2}$\\
& $\rziii, \gziii$& $\Er(Z_3),\Eg(Z_3)$& ${\yziii} {\color{gray} \times \ttt{rz}_3 \times \ttt{gz}_3}$ & ${\byx_3}$\\ 
\hline 
3& $\rgx_1, \rbx_1, \gbx_1$& $\Ey(X_1)$& ${\yzi}$ & ${\ryx_1 {\color{gray} \times \ttt{rgx}_1}}$\\
& $\gbx_2, \gyx_2 ,\byx_2$& $\Er(X_2)$& ${\rzii}$& ${\rgx_2}$\\
& $\rgx_3, \ryx_3, \gyx_3$& $\Eb(X_3)$& ${\bziii}$& ${\byx_3 {\color{gray} \times \ttt{gyx}_3}}$\\ 
\hline 
4& $\rzi \rzii, \gzi \gzii, \bzi \bzii$& $V(Z_1Z_2)$& \multirow{2}{*}{${\yzi},{\rzii},{\bziii}$} & \multirow{2}{*}{${\widetilde{\gyx}},{\widetilde{\rgx}},{\widetilde{\gbx}}$}\\
& $\rzii \rziii, \gzii \gziii, \bzii \bziii$ & $V(Z_2Z_3)$& & \\
\hline
\end{tabular}
}
\caption{ Evolution of logical strings and membranes for the condensation sequence producing $\rc r \rightarrow \yc y \rightarrow \bc b \rightarrow \gc g \rightarrow \rc r$ automorphism. In the steps when two fluxes are condensed, we write out two independent condensations and the third condensation which is a product of the first two and thus also is condensed. This is convenient for tracking the evolution of logical membranes. 
} \label{tab:3DCC_cyclic_logicals}
\end{table}

\begin{table}[t]
\scalebox{1}{
\renewcommand{\rc}[1]{{\color{red}{#1}}}
\renewcommand{\bc}[1]{{\color{blue}{#1}}}
\renewcommand{\gc}[1]{{\color{ForestGreen}{#1}}}
\renewcommand{\yc}[1]{{\color{amber}{#1}}}
\begin{tabular}{| c | c | l | l | }
\hline
Time &\text{Measure} & ISG plaquettes & ISG volumes \\
\hline
0&$V(Z_1Z_2)$& & \\
&$V(Z_2Z_3)$& & \\
\hline
1&$\Er(X_1)$& $\Pbarrg(Z_1Z_2)$ & \\
&$\Eg(X_2)$ & $\Pbargy(Z_2Z_3)$ & \\
&$\Ey(X_3)$ &$\Pbarry(Z_1Z_3)$ & \\ 
\hline 
2&$\Eg(Z_1),\Eb(Z_1)$&$\Pbarry(Z_1)$& $\vol_{\yc y}(X_1)$ \\
&$\Ey(Z_2),\Eb(Z_2)$&$\Pbarrg(Z_2)$& $\vol_{\rc r}(X_2)$ \\
&$\Er(Z_3),\Eg(Z_3)$& & $\vol_{\bc b}(X_3)$\\ 
\hline 
3&$\Ey(X_1)$& $\Pbarry(Z_1),\Pbargy(Z_1),\Pbarby(Z_1)$& $\vol_{\yc y}(X_1)$\\
&$\Er(X_2)$& $\Pbarrg(Z_2),\Pbarrb(Z_2),\Pbarry(Z_2)$& $\vol_{\rc r}(X_2)$\\
&$\Eb(X_3)$& $\Pbarrb(Z_3),\Pbargb(Z_3)$& $\vol_{\bc b}(X_3)$\\ 
\hline 
4&$V(Z_1Z_2)$&$\Pbarrg(Z_2), \Pbarrb(Z_2), \Pbarry(Z_2)$ & $\vol_{\rc r}(X_1X_2X_3)$,$\vol_{\gc g}(X_1X_2X_3)$ \\
&$V(Z_2Z_3)$& $\Pbargb(Z_3) ,\Pbargy(Z_1), \Pbarby(Z_1)$& $\vol_{\bc b}(X_1X_2X_3)$,$\vol_{\yc y}(X_1X_2X_3)$ \\
\hline
\end{tabular}
}
\caption{ISG generation by the measurement sequence producing a 3D DA color code with $\rc r \rightarrow \yc y \rightarrow \bc b \rightarrow \gc g \rightarrow \rc r$ automorphism. The third and fourth columns show the generation of ISG plaquettes and volumes for a period of the measurement sequence. 
} \label{tab:3DCC_cyclic_ISG}
\end{table}

In terms of gates, the automorphisms achieved by the generators that our sequences can produce realize the color-permuting $S_4$ subgroup of all automorphisms of the 3D color code. If we assign the logical qubits to the string and membrane operators in the effective 3D color code as follows. For each principal direction on the torus, we will have 3 logical operators and we define the $Z$-logical Pauli operators as strings along a given principal direction, i.e. 
\be
\overline Z_{1,2,3}^{x} = \mathcal{L}(\widetilde{\rz}), \mathcal{L}(\widetilde{\gz}), \mathcal{L}(\widetilde{\bz}) \text{ in $x$-direction},
\ee
and the conjugate logical $X$ operators as membranes:
\be
\overline X_{1,2,3}^{x} = \mathcal{L}(\widetilde{\ryx}), \mathcal{L}(\widetilde{\gyx}), \mathcal{L}(\widetilde{\byx}) \ \ \perp \text{ to $x$-direction}.
\ee
In that case, the automorphisms will have the same action on each triple of logical qubits, and the generating gates for the $S_4$ subgroup will be SWAP$_{12}$, SWAP$_{13}$ and CNOT$_{21}\times$CNOT$_{31}$.

Note that the sequences shown in Table~\ref{tab:3DDACC_table_short} use two types of 3D toric codes introduced in previous subsections; we provide an example of a protocol that implements a trivial automorphism that involves the third type of toric code in Appendix~\ref{sec:app_3DCC_full}. The 3D Floquet color code is a byproduct of our 3D DA color code construction, if the latter is run periodically in time.

Table~\ref{tab:3DCC_cyclic_ISG} breaks down the operation of the $\rc r \rightarrow \yc y \rightarrow \bc b \rightarrow \gc g \rightarrow \rc r$ measurement sequence for the 3D DA color code. The plaquettes and volumes that appear in the ISG are products of checks in the previous round that commute with the measurements of the current round. This way, the complete topological ISG is measured by round 4, and because the measurements are locally reversible, the ISG is further preserved. The logical information is preserved at each round. The last two columns of Table~\ref{tab:3DCC_cyclic_ISG} track the logical string and membrane operators throughout all rounds assuming that we started with an effective color code at round $0$. For each two consequent rounds, there is a class of logical strings and a class of logical membranes that is shared between both rounds. This can be worked out for each pair of subsequent rounds analogously to Table~\ref{table:3Dlogicals}. When the logical operator needs to be multiplied by measurements of a given round in order to survive to the next round, it is shown by gray multiplication in the table.

Thus, we have constructed a subclass of 3D DA color codes that can be natively performed on closed manifolds using two-qubit measurements only. It would be, however, interesting to understand a complete picture of measurement paths for 3D DA color codes, as well as study error correction in them.

\subsection{Non-Clifford gate through measurements}
\label{sec:3DFCC_nonClifford}

An important property of the 3D color code is that it admits a transversal $T = \text{diag}\left ( 1, e^{i \pi/4}\right )$ unitary gate~\cite{Bombin2007Tgate}. 
For the color code on a bipartite and 4-colorable lattice and appropriate boundary conditions, it is possible to implement a logical non-Clifford gate by transversally applying $T$ to each of the A-sublattice sites and $T^\dagger$ to each of the B-sublattice sites. We call such an application of a gate a ``transversal bipartite'' gate. 
For the 3D color code on the 3-torus, which encodes 9 logical qubits, a transversal bipartite $T$-gate realizes a logical CCZ gate between triples of logicals for each principal direction on the torus.
Similarly, on $\mathbb{R} \text{P}^3$ where 3 logical qubits are encoded, the transversal bipartite $T$-gate also realizes a logical CCZ gate. On a tetrahedron with colored boundaries, the transversal $T$-gate realizes a logical $T$-gate~\cite{Bombin2007Tgate}.
The 3D color code has these properties in particular because it is triorthogonal~\cite{Bravyi2012}.

The 3D DA color code admits a unitarily applied transversal bipartite $T$ gate whenever it embeds the ISG of the color code. At these steps, the measurements $V(Z_1Z_2)$ and $V(Z_2Z_3)$ define the effective qubit at each vertex, and the effective local $\widetilde{T}$ gate acts as $\id \otimes \id \otimes T$ on the A sublattice and $\id \otimes \id \otimes T^\dagger$ on the B sublattice. We will denote the global gate that is the product of such local gates as $\widetilde{\boldsymbol{T}} = \prod_i \widetilde T_i$.
Additionally, because transversality of the $T$ gate in the 3D color code and the associated CCZ gate it applies are related due to unfolding~\cite{Kubica_2015}, applying a transversal CCZ should also be possible in the 3D DA color code during the toric code steps. However, we only focus on the $T$ gate in our discussion.

Although the $T$ gate can be unitarily applied during the color code steps, here we pursue the application of a non-Clifford gate in a way that is ``native'' to DA codes, i.e. purely from a measurement sequence.
As a first step, we show that if all measurement outcomes are postselected to $+1$, the measurement sequence for the 3D DA color code can be modified in a simple way in order to implement the $T$ gate. Technically, the postselection is needed for $Z$-type measurements and stabilizers only. However, for conciseness, the argument below will assume that all stabilizers and measurements are postselected, and generalization to random $X$-measurement outcomes and stabilizer values is straightforward. Next, we will show how to relax this constraint, removing the need for postselection entirely by making the measurements adaptive\footnote{As a remark, one could also develop this method further to implement addressable Clifford gates.}.

\begin{table}[b]
{
\renewcommand{\rc}[1]{{\color{red}{#1}}}
\renewcommand{\bc}[1]{{\color{blue}{#1}}}
\renewcommand{\gc}[1]{{\color{ForestGreen}{#1}}}
\renewcommand{\yc}[1]{{\color{amber}{#1}}}
 \centering
 \begin{tabular}{|c|c|c|c|c|c|c|c| }
 \hline
 \multirow{2}{*}{Gate/Aut} & \multicolumn{4}{c|}{$t$} \\ \cline{2-5} 
 & 1 & 2 & 3 & 4 \\
 \hline
 \multirow{2}{*}{$\text{Identity}$} & $\Er(X_1)$ & $\Eg(Z_1), \Eb(Z_1)$ & $\Er(X_1)$ & \multirow{2}{*}{ $V(Z_1Z_2)$}\\
 & $\Eg(X_2) $ & $\Ey(Z_2), \Eb(Z_2)$ & $\Eg(X_2) $ & \\ 
 & $\Ey(X_3)$ & $\Er(Z_3), \Eg(Z_3)$ & $\Ey(X_3)$ & $V(Z_2Z_3)$ \\
 \hline
 Eff. code & TC$^{\overline{\rc r} } \boxtimes $TC$^{\overline{\gc g} } \boxtimes $TC$^{\overline{\yc y} } $ & TC$_{\gc g \bc b} \boxtimes$ TC$_{\yc y \bc b} \boxtimes$TC$_{\rc r \gc g} $ & TC$^{\overline{\rc r} } \boxtimes $ TC$^{\overline{\gc g} } \boxtimes $ TC$^{\overline{\yc y} }$ & $\widetilde{\text{CC}}$ 
 \\
 \hline
 \end{tabular}
 \caption{ Sequence generating a trivial automorphism which we later modify for implementing} the $T$ gate by measurements.
 \label{tab:3DDACC_table_identity}
}
\end{table}

First, let us choose a schedule that implements the identity automorphism as a starting point (before modifying the measurement sequence to implement the $T$-gate). This closely follows the construction in Sec.~\ref{sec:tqft_aut_construction} generalized to 3D.
This ``trivial'' sequence is shown in Table~\ref{tab:3DDACC_table_identity}; it does not generate the full topological stabilizer group, and thus it would have to be padded by a full ISG-generating sequence if applied in practice. Since we assume that all measurement outcomes are $+1$, the values of all of the stabilizers are also $+1$. 
Let us call $\hat \Pi_t^+$ the projector associated with the measurements of round $t$, and we use a `$+$' superscript to denote the postselection assumption. Then, if at round $t=0$ we start in a state $\ket{\widetilde{\text{CC}}}$ which is in the codespace of the effective 3D color code, after 4 rounds of measurements it will be:
\be
\ket{\widetilde{\text{CC}}^+} = \hat \Pi^+_4 \cdot \hat \Pi^+_3 \cdot \hat \Pi^+_2 \cdot \hat \Pi^+_1 \ket{\widetilde{\text{CC}}^+},
\ee
where we have used that the measurement sequence implements the identity gate, and thus, we arrive at the same state. For simplicity, we have postselected on every round, although we only need to postselect during $Z$-type measurement rounds. The $\ket{\widetilde{\text{CC}}^+}$ notation indicates a state in the codespace of the effective color code where all $Z$-type stabilizers are $+1$. 

Now, let us find the modification of the measurement sequence such that the resulting action is the same as the application of transversal $\widetilde{\boldsymbol T} = \left ( \prod_i \widetilde{T}_i \right ) $. Let us consider
\be \label{eq:T_mod_1}
\widetilde{\boldsymbol T} \ket{\widetilde{\text{CC}}^+} =\widetilde{\boldsymbol T} \, \hat \Pi_4^+ \cdot \hat \Pi_3^+ \cdot \widetilde{\boldsymbol T}^\dagger 
\widetilde{\boldsymbol T} \, \hat \Pi_2^+ \cdot \hat \Pi_1^+ \ket{\widetilde{\text{CC}}^+},
\ee
where we have inserted an additional identity unitary gate between $\hat \Pi_2$ and $\hat \Pi_3$. Now, because the measurements in $\hat \Pi_4^+$ all have $Z$-flavor, they commute with the application of $\widetilde{\boldsymbol T}$, and thus, we commute the first application of transversal $\widetilde{\boldsymbol T}$ to the right. We get
\be
\begin{split} \label{eq:3DDAC_modified}
\widetilde{\boldsymbol T} \ket{\widetilde{\text{CC}}^+} &= \hat \Pi_4^+ \cdot \left ( \widetilde{\boldsymbol T} \, \hat \Pi_3^+ \widetilde{\boldsymbol T}^\dagger \right ) \cdot 
 \widetilde{\boldsymbol T} \, \hat \Pi_2^+ \cdot \hat \Pi_1^+ \ket{\widetilde{\text{CC}}^+} = \\
 &=\hat \Pi_4^+ \cdot \hat {\widetilde{\Pi} }_3^+ \cdot 
 \hat \Pi_2^+ \cdot \hat \Pi_1^+ \ket{\widetilde{\text{CC}}^+} .
 \end{split}
\ee
Where we have defined $\hat {\widetilde{\Pi} }_3^+$ to be the measurements at step $t=3$ conjugated by transversal $\widetilde{\boldsymbol T}$ and used the fact that transversal $\widetilde{\boldsymbol T}$ acting on the state after $\hat \Pi_2^+$ leaves it invariant (will be shown below). Thus, the modified sequence of measurements $\hat \Pi_4^+ \cdot \hat {\widetilde{\Pi} }_3^+ \cdot 
\hat \Pi_2^+ \cdot \hat \Pi_1^+$, where the $X$-flavored measurements of the third rounds have been conjugated by $\widetilde{\boldsymbol T}$, will implement the same non-Clifford gate as a transversal unitary application of $\widetilde{\boldsymbol T}$. In the protocol, this is equivalent to measuring the operator $\left(\frac{X_3+Y_3}{\sqrt{2}}\right)\left(\frac{X_3-Y_3}{\sqrt{2}}\right)$ on yellow links at round 3, with the outcome postselected to be $+1$; the protocol is shown in Table~\ref{tab:3DDACC_table_T}.

 \begin{table}[b]
{
\renewcommand{\rc}[1]{{\color{red}{#1}}}
\renewcommand{\bc}[1]{{\color{blue}{#1}}}
\renewcommand{\gc}[1]{{\color{ForestGreen}{#1}}}
\renewcommand{\yc}[1]{{\color{amber}{#1}}}
 \centering
 \begin{tabular}{|c|c|c|c|c|c|c|c| }
 \hline
 \multirow{2}{*}{Gate/Aut} & \multicolumn{4}{c|}{$t$} \\ \cline{2-5} 
 & 1 & 2 & 3 & 4 \\
 \hline
 \multirow{2}{*}{$\varphi_{CCZ}$} & $\Er(X_1)$ & $\Eg(Z_1), \Eb(Z_1)$ & $\Er(X_1)$ & \multirow{2}{*}{ $V(Z_1Z_2)$}\\
 & $\Eg(X_2) $ & $\Ey(Z_2), \Eb(Z_2)$ & $\Eg(X_2) $ & \\ 
 & $\Ey(X_3)$ & $\Er(Z_3), \Eg(Z_3)$ & $\Ey\left(\widetilde{T}X_3 \widetilde{T}^\dagger\right)$ & $V(Z_2Z_3)$ \\
 \hline
 \end{tabular}
 \caption{ Sequence implementing a transversal $T$ gate by measurements in the postselected case. The shorthand $E_y(\widetilde{T}X_3 \widetilde{T}^\dagger)$ means measuring two-qubit Clifford operators on yellow edges where $ T X_3 T ^\dagger= (X_3+Y_3)/\sqrt{2}$ are measured on the $A$ sublattice and $ T^\dagger X_3 T = (X_3-Y_3)/\sqrt{2} $ are measured on the $B$ sublattice. }
 \label{tab:3DDACC_table_T}
}
\end{table}

What is left is to show why an application of transversal $\widetilde{\boldsymbol T}$ after step 2 of the measurement sequence leaves all logical states invariant. First notice that $\widetilde{\boldsymbol T}$ acts on the third layer only. The ISG for the third toric code at round 2 is given by $\mathcal{S}(\TC_{rg}) = \langle E_r(Z_3), E_{g}(Z_3), \vol_{b}(X_3), \vol_{y}(X_3), P_{\overline{by}}(Z_3)\rangle$. On a three-torus, we can write the logical $\ket{\overline{000}}_3$ state of this toric code as
\begin{equation}
\ket{\overline{000}}_3 = \prod_{i} \frac{1 + \vol_{b,i}(X_3)}{2} \prod_{j} \frac{1 + \vol_{y,j}(X_3)}{2} \ket{0^n}_3.
\end{equation}
where $n$ is the number of physical qubits in one copy of the lattice.

We can obtain other logical basis states of this toric code by applying a combination of logical $X$ membranes along different principal directions on the three-torus. We want to show that the action of $\widetilde{\boldsymbol T}$ is trivial on this subspace. We first compute 
\begin{align} \label{eq:Ts3}
\widetilde{\boldsymbol T}\ket{\overline{000}}_3 &= \prod_{i} \frac{1 + \vol_{b,i}(\widetilde{T} X_3 \widetilde{T}^{\dagger})}{2} \prod_{j} \frac{1 + \vol_{y,j}(\widetilde{T} X_3 \widetilde{T}^{\dagger})}{2} \widetilde{\boldsymbol T} \ket{0^n} \\
&= \prod_{i} \frac{1 + \vol_{b,i}(e^{\pm i \pi/4} X \widetilde{S}^{\dagger})}{2} \prod_{j} \frac{1 + \vol_{y,j}(e^{\pm i \pi/4} X \widetilde{S}^{\dagger})}{2} \ket{0^n}
\end{align}
where $\widetilde{S}$ is a bipartite application of the phase gate.
The phases appearing in the above equation are due to the relation $T X T^\dagger = e^{i \pi/4} X S^{\dagger}$, and the $e^{\pm i \pi/4}$ phases cancel because $K$ contains an equal number of sites in the $A$ and $B$ sublattices. We can expand out the product and commute all occurrences of $\widetilde{S}^{\dagger}$ to the right; since $\vol_b$ and $\vol_y$ can intersect at an even number of sites, commuting $\widetilde{S}^{\dagger}$ may result in additional $\widetilde{Z}^{\dagger}$ operators. However, the action of both of these operators on the vacuum $\ket{0^n}$ is trivial, so we find that
\begin{align}
\prod_{i} \frac{1 + \vol_{b,i}(e^{\pm i \pi/4} X \widetilde{S}^{\dagger})}{2} \prod_{j} \frac{1 + \vol_{y,j}(e^{\pm i \pi/4} X \widetilde{S}^{\dagger})}{2} \ket{0^n} &= \prod_{i} \frac{1 + \vol_{b,i}(X)}{2} \prod_{j} \frac{1 + \vol_{y,j}(X)}{2} \ket{0^n} \nonumber \\
& = \ket{\overline{000}}_3 .
\end{align}
Next, we must consider the action of $\widetilde{\boldsymbol T}$ on other logical states, formed by acting on $\ket{\overline{000}}_3$ with the logical membrane operator of the third copy of the toric code, i.e., with $\mathcal{L}(\byx_3)$. For clarity, we will label the three different orientations of this logical membrane by $\mathcal{X}_{1}$, $\mathcal{X}_2$, and $\mathcal{X}_3$. These have the property that $\text{supp}(\mathcal{X}_i) \cap \text{supp}(\mathcal{X}_j)$ and $\text{supp}(\mathcal{X}_1) \cap \text{supp}(\mathcal{X}_2) \cap \text{supp}(\mathcal{X}_3)$ have an even number of sites (i.e. $\mathcal{X}_i$ and $\mathcal{X}_j$ share an even number of sites, and $\mathcal{X}_1$, $\mathcal{X}_2$, $\mathcal{X}_3$ share an even number of sites). Furthermore, we can always choose a representative $\mathcal{X}_1$, $\mathcal{X}_2$, $\mathcal{X}_3$ such that their support is disjoint. It is simpler to use this choice to compute the following: 
\begin{align} \label{eq:T_act_post}
\widetilde{\boldsymbol T} \mathcal{X}_i \ket{\overline{000}}_3 &= (\widetilde{\boldsymbol T} \mathcal{X}_i \widetilde{\boldsymbol T}^{\dagger}) \widetilde{\boldsymbol T} \ket{\overline{000}}_3 = \mathcal{X}_i \widetilde{S}_{\mathcal{X}_i} \ket{\overline{000}}_3 \nonumber\\
\widetilde{\boldsymbol T} \mathcal{X}_i\mathcal{X}_j \ket{\overline{000}}_3 &= (\widetilde{\boldsymbol T} \mathcal{X}_i \widetilde{\boldsymbol T}^{\dagger})(\widetilde{\boldsymbol T} \mathcal{X}_j \widetilde{\boldsymbol T}^{\dagger}) \widetilde{\boldsymbol T} \ket{\overline{000}}_3 = \mathcal{X}_i \mathcal{X}_j \widetilde{S}_{\mathcal{X}_i\mathcal{X}_j} 
\ket{\overline{000}}_3 \nonumber\\
\widetilde{\boldsymbol T} \mathcal{X}_1 \mathcal{X}_2 \mathcal{X}_3 \ket{\overline{000}}_3 &= \mathcal{X}_1 \mathcal{X}_2 \mathcal{X}_3 \widetilde{S}_{\mathcal{X}_1\mathcal{X}_2\mathcal{X}_3} 
\ket{\overline{000}}_3.
\end{align}
The notation $\widetilde{S}_{\mathcal{O}}$ ($Z_{\mathcal{O}}$) indicates a transversal operator $\widetilde S$ ($Z$) whose support is the same as that of the object $\mathcal{O}$, and if $\mathcal{O} = \mathcal{X}_i\mathcal{X}_j$ this means the support is in $\mathcal{X}_i \cup \mathcal{X}_j$. Let us show that $\widetilde{S}_{\mathcal{O}}$ for specific operators $\mathcal O$ appearing above has trivial action on the $\ket{\overline{000}}_3$ logical state:
\begin{equation}
 \widetilde{S}_{\mathcal{O}} \ket{\overline{000}}_3 = \prod_{i} \frac{1 + \vol_{b,i}(e^{\pm i \vartheta(s) \pi/2} X Z^{\vartheta(s)})}{2} \prod_{j} \frac{1 + \vol_{y,j}(e^{\pm i \vartheta(s) \pi/2} X Z^{\vartheta(s)})}{2} \ket{0^n}
\end{equation}
where $s$ denotes a site on the lattice, and $\vartheta(s) $ is 1 if $s$ is in $\text{supp}(\mathcal{O})$ and 0 otherwise, and appears because of the relation $S X S^\dagger = i X Z$. 
The $\mp$ is applied to the A/B sublattices. As before, the additional phase $e^{\pm i \vartheta(s)}$ is inconsequential because the support of any $\mathcal O$ appearing in Eq.~\eqref{eq:T_act_post} overlaps with any volume on an even number of sites. The $Z^{\vartheta(s)}$ acts trivially on $\ket{0^n}$ so we find $\widetilde{S}_{\mathcal{X}}\ket{\overline{000}}_3 = \ket{\overline{000}}_3$. 
Put together, this means that a bipartite $\widetilde{\boldsymbol T}$ acts trivially on all logical states. Therefore, the measurement schedule we introduced indeed is equivalent to applying a transversal $T$ at the color code step.

Let us now show that a $\widetilde{\boldsymbol T}$-gate can be implemented by adaptive measurements if we do not postselect on $+1$ measurement outcomes. In particular, we show that this amounts to keeping track of an additional Clifford frame and virtually ``pushing'' it through the rest of the measurement rounds, which will conjugate these rounds by appropriate Clifford operators, thus requiring an adaptive measurement procedure. The correction can be determined by running a decoder right before the round with the measurement conjugated by the $\widetilde{\boldsymbol T}$ gate~\footnote{A similar approach as  presented in the section on error correction for the 2D DA color code applies to the 3D DA color code. For the same reasons, we believe it should be possible to develop an appropriate generalization of one of the color code decoders for this code; we leave this to future work. }. We then can keep pushing the Clifford unitary frame through all future measurements, which will then get conjugated by this unitary frame, and also keep track of how the Clifford frame is further modified.

As we mentioned above, for the purpose of applying the $\widetilde{ T}$-gate we only care about the randomness of measurement outcomes of $Z$-flavored operators. Therefore, let us express the state after round 2, where $ZZ$ measurements were performed, as:
\be
\ket{\psi_2} = \Pi_2 \cdot \hat \Pi_1 \ket {\widetilde{\text{CC}}} \triangleq \hat M_X^{(2)}\hat \Pi_2^+ \cdot \hat \Pi_1 \ket {\widetilde{\text{CC}}^+},
\ee
where we instead assume that the ``clean'' state $\hat \Pi_2^+ \cdot \hat \Pi_1 \ket {\widetilde{\text{CC}}^+}$ is stabilized by $Z$-plaquettes and $ZZ$ checks (i.e. they have $+1$ values) and the randomness of their signs can be accounted by some configuration of $X$ operators denoted by $M_X^{(2)}$, which act on the ``clean state''. Similarly, we can define the operator $M_X^{(4)}$ that absorbs the randomness of $\hat \Pi_4$ measurement round: $\hat \Pi_4 = M_X^{(4)} \hat \Pi_4^+$. Thus we can write: 
\be
\ket{\psi_4} = M_X^{(4)}\hat \Pi_4^+ \cdot \hat \Pi_3 \cdot M_X^{(2)}\hat \Pi_2^+ \cdot \hat \Pi_1 \ket {\widetilde{\text{CC}}^+} \triangleq \hat \Pi_4 \cdot \hat \Pi_3 \cdot \hat \Pi_2 \cdot \hat \Pi_1 \ket {\widetilde{\text{CC}}}.
\ee

Next, we can run a decoder (so long as it can be implemented in polynomial time) on the state $\ket{\psi_2}$ ($\ket{\psi_4}$) from known measurement and stabilizer outcomes, assuming that the ``correct'' values for them were $+1$. These are not technically error syndromes simply because they do not arise from physical errors in the code but rather from the fact that measurement outcomes in a dynamic code can be random; however, we can still use a decoder to fix their values to $+1$. The decoder gives us correction operators $\overline{M}_X^{(2)}$ and $\overline{M}_X^{(4)}$ which fixes the random $Z$-type measurement outcomes and stabilizers to be $+1$. The correction will be up to $X$-type elements of the stabilizer group of the model at each respective round, which form closed surfaces. As a result, the operators $\overline{M}_X^{(2)} {M}_X^{(2)}$ and $\overline{M}_X^{(4)} {M}_X^{(4)}$ are both sets of disjoint closed surfaces.

Knowing the corrections $\overline{M}_X^{(2,4)}$, we can turn our measurement protocol into an adaptive one, such that the state after round 4 will be a ``clean'' one (i.e. the one where $Z$ measurement outcomes and $Z$ stabilizers are all $+1$), which we call $\ket{\widetilde \psi_4}$. This state can be written as:
\be
\begin{split}
\ket{\widetilde \psi_4} &= \overline M_X^{(4)}\hat \Pi_4 \cdot \hat \Pi_3 \cdot \overline M_X^{(2)}\hat \Pi_2 \cdot \hat \Pi_1 \ket {\widetilde{\text{CC}}} =
\\
&=\overline M_X^{(4)} M_X^{(4)} \hat \Pi_4^+ \cdot \hat \Pi_3 \cdot \overline M_X^{(2)} M_X^{(2)} \hat \Pi_2^+ \cdot \hat \Pi_1 \ket {\widetilde{\text{CC}}^+} \\
&= \hat \Pi_4^+ \cdot \hat \Pi_3 \cdot \hat \Pi_2^+ \cdot \hat \Pi_1 \ket {\widetilde{\text{CC}}^+}.
\end{split}
\ee
where we used that $\overline{M}_X^{(2)} {M}_X^{(2)} \triangleq \boldsymbol{M}^{(2)}_X$ and $\overline{M}_X^{(4)} {M}_X^{(4)} \triangleq \boldsymbol{M}^{(4)}_X$ are stabilizers for the state at respective times. The last expression shows that it is manifestly equivalent to the postselected state.

Because the state $\ket{\widetilde \psi_4}$ is postselected, we know that the $\widetilde{\boldsymbol T}$ gate can be applied to it from our previous argument. $\widetilde{\boldsymbol T}$ applies a logical CCZ because the correction does not change the logical state. 

Lastly, we would like to absorb the action of the corrections and the $\widetilde{\boldsymbol T}$ unitary into the measurements and a Clifford frame $M_C$ (which later on can also be absorbed into measurements). For this, let us again ``push'' the $\widetilde{\boldsymbol T}$ to the right 
\be
\begin{split}
\widetilde{\boldsymbol T} 
 \ket{\widetilde \psi_4} &= \widetilde{\boldsymbol T} \overline M_X^{(4)}\hat \Pi_4 \cdot \hat \Pi_3 \cdot \boldsymbol{M}^{(2)}_X \hat \Pi_2^+ \cdot \hat \Pi_1 \ket {\widetilde{\text{CC}}^+} 
 \\
 &= \overline M_X^{(4)} \overline M_S^{(4)} \hat \Pi_4 \cdot \left ( \widetilde{\boldsymbol T} \hat \Pi_3 \widetilde{\boldsymbol T} ^\dagger \right) \cdot \boldsymbol{M}^{(2)}_X \boldsymbol{M}^{(2)}_S \widetilde{\boldsymbol T} \hat \Pi_2^+ \cdot \hat \Pi_1 \ket {\widetilde{\text{CC}}^+} \\
 &= \overline M_X^{(4)} \overline M_S^{(4)} \hat \Pi_4 \cdot \left ( \widetilde{\boldsymbol T} \hat \Pi_3 \widetilde{\boldsymbol T} ^\dagger \right) \cdot \boldsymbol{M}^{(2)}_X \hat \Pi_2^+ \cdot \hat \Pi_1 \ket {\widetilde{\text{CC}}^+}
 \end{split}
\ee
where the additional operators $\overline M_S^{(4)}=\overline M_X^{(4)} \widetilde{\boldsymbol T} \overline M_X^{(4)}\widetilde{\boldsymbol T}^\dagger $ and $\boldsymbol{M}^{(2)}_S=\boldsymbol{M}^{(2)}_X \widetilde{\boldsymbol T} \boldsymbol{M}^{(2)}_X \widetilde{\boldsymbol T}^\dagger $ arise because of ``pushing'' $\widetilde{\boldsymbol T}$ past respective operators. We have also used that the ``clean'' state after round 2 is stabilized by $\boldsymbol{M}^{(2)}_S$ (up to a phase that does not depend on the logical state), because this operator consists of a bipartite application of $S^{(\dagger)}$ around a closed surface.

Next, we absorb $M_X^{(2)}$ back into the state after round 2 and also commute all the unitaries to the left in order to turn them into a Clifford frame: 
\be
\begin{split}
\widetilde{\boldsymbol T} 
 \ket{\widetilde \psi_4} &= \overline M_X^{(4)} \overline M_S^{(4)} \hat \Pi_4 \cdot \left ( \widetilde{\boldsymbol T} \hat \Pi_3 \widetilde{\boldsymbol T} ^\dagger \right) \cdot \overline{M}^{(2)}_X {M}^{(2)}_X \hat \Pi_2^+ \cdot \hat \Pi_1 \ket {\widetilde{\text{CC}}^+}
 \\
 &=\overline M_X^{(4)} \overline M_S^{(4)} \hat \Pi_4 \cdot \left ( \widetilde{\boldsymbol T} \hat \Pi_3 \widetilde{\boldsymbol T} ^\dagger \right) \cdot \overline{M}^{(2)}_X \hat \Pi_2 \cdot \hat \Pi_1 \ket {\widetilde{\text{CC}}^+}
 \\
 &= M _C \, \hat \Pi_4 ^{\text{ad}} \cdot \hat \Pi_3 ^{\text{ad}} \cdot \hat \Pi_2 \cdot \hat \Pi_1 \ket {\widetilde{\text{CC}}}.
 \end{split}
\ee
We defined the Clifford frame $M_C = \overline M_X^{(4)} \overline M_S^{(4)} \overline M_X^{(2)} $ and 
\begin{equation} \label{eq:adaptive}
\hat \Pi_4 ^{\text{ad}} =\overline M_X^{(2)} \hat \Pi_4 \overline M_X^{(2)} , \quad \hat \Pi_3 ^{\text{ad}} =\overline M_X^{(2)} \widetilde{\boldsymbol T} \hat \Pi_3 \widetilde{\boldsymbol T} ^\dagger \overline M_X^{(2)}.
\end{equation}

Thus, the application of the $\widetilde{\boldsymbol T}$ gate in a case when there is no postselection can be achieved by running an adaptive measurement protocol and accounting for a Clifford frame, i.e. 
\be
\begin{split}
\widetilde{\boldsymbol T} 
 \ket{\widetilde \psi_4} = M _C \, \hat \Pi_4 ^{\text{ad}} \cdot \hat \Pi_3 ^{\text{ad}} \cdot \hat \Pi_2 \cdot \hat \Pi_1 \ket {\widetilde{\text{CC}}} ,
 \end{split}
\ee
where the adaptive measurements are defined in Eq.~\ref{eq:adaptive}. To find the set of adaptive measurements and the Clifford frame, one has to run a decoder after rounds 2 and 4. The adaptive measurements only rely on the output of the decoder, which can be determined rather efficiently. The argument does not change for different logical states because the $M_X$ membrane commutes with the $\mathcal X_i$ operators that toggle between the logical states. Additionally, if more measurement rounds are added afterward, one can deal with the Clifford frame $M_C$ by absorbing it into the new measurements.

Finally, we leave the question of error correction and fault tolerance in the 3D DA color code to future work.

\section{{Small dynamic automorphism color codes}}\label{sec:smallcodes}

Thus far, we have studied the DA color code on a large lattice, where we believe the distance is extensive and the number of encoded qubits is constant. It is more practically friendly to implement small versions of the DA color code, analogous to how the [[7,1,3]] Steane code is the smallest version of the color code. The number of physical qubits required for small DA codes will generally be much larger than for small stabilizer codes, but the small DA color code has an advantage of only requiring two-qubit measurements. 

\begin{figure}[h] \centering
\vspace{0pt}
\centering
\vspace{0pt}
\includegraphics[width= 0.75\columnwidth]{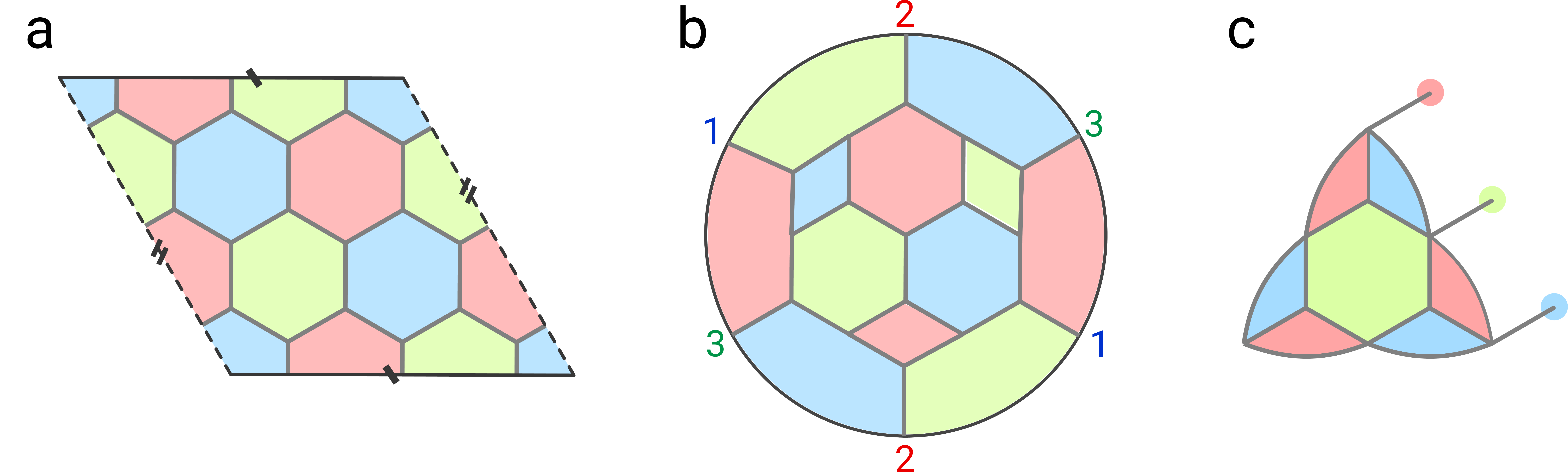}
\caption{Example of the smallest DA color code on (a) torus, (b) $\mathbb{R}\text{P}^2$ and (c) finite patch corresponding to [[36,4,4]], [[32,2,4]] and [[24,1,3]] codes respectively. The identified boundaries are marked with dashes for the torus, the identified edges are marked with numbers for the real projective space. In (c), the docks at $Z$ and $X$ boundaries were removed thanks to the correspondence with the CSS honeycomb boundaries, see Fig.~\ref{fig:431}.}
\label{fig:81}
\end{figure}

First, we find small DA color codes for lattices with no boundaries. For example, we can construct small DA color codes on closed $T^2$ and $\mathbb{R}\text{P}^2$ manifolds shown in Fig.~\ref{fig:81}(a,b). The DA color codes on the torus and $\mathbb{R}\text{P}^2$ are [[36,4,4]] and [[32,2,4]] codes, which are capable of correcting for single qubit errors as well as some two-qubit errors. With Pauli boundaries, the smallest code on a patch that we found so far is a [[24,1,3]] DA color code on a triangle (we have checked the distance numerically for the $\varphi_{\ttt {rb}}$ and $\varphi_{\ttt {rbg}}$ protocols). We have not been able to find an example of a small DA color code with color boundaries by considering a double layer Steane code layout. The measurement protocol for these small codes is entirely analogous to those for the large codes, and the small codes exhibit the same automorphisms as their large counterparts. It would be interesting to study the properties of concatenated small DA color codes, which along with magic state injection could present an alternative practical way of doing fault tolerant quantum computation.

\section{{Discussion}}

There are several significant open questions raised by this paper in regard to dynamic and Floquet codes as well as spacetime fault tolerance. 

An immediate question would be to present a more comprehensive TQFT picture for dynamic automorphism codes. 
We note that the results of the current paper generalize straightforwardly to ``prime anyon theories'' with $\mathcal D(\prod_{p_i}\mathbb Z_{p_i})$ topological order, where $\{p_i\}$ is a set of distinct primes. In an upcoming paper~\cite{Aasen2023_TQFT} we use the language of modular tensor categories to understand certain properties of general two-dimensional dynamic automorphism codes such as updating logical information, constraints on measurement sequences, and sequences for gapped boundaries. These results apply to general Abelian anyon theories. It would be interesting to consider reversible condensations for non-Abelian anyon theories and associated dynamic automorphism codes in two dimensions. It would also be interesting to develop a general theory for dynamic automorphism codes in higher dimensions. 

The general construction of gapped boundaries is another interesting question, which we make significant progress towards. On the more practical side, while this paper has primarily focused on the measurement realization of the DA color code with Pauli boundaries, we have also considered the condensation picture for the 2D DA color code on a color triangle, the details of which are presented in Appendix~\ref{sec:SM_color_triangle}. It would be important to work out the microscopic measurement sequences consistent with this picture. Understanding colored boundaries is also important for constructing the 3D color code on a tetrahedron. This in turn is necessary for interfacing the 2D triangular color code with the 3D tetrahedron color code to perform an analog of the dimensional jump~\cite{Bombin2016}. 

Perhaps the most important practical open question is further development of error correction in the DA color codes, especially finding more efficient sequences that are error-correcting and do not require detectors with very complicated shapes. We have not addressed error correction for the iSWAP gate and for the three-dimensional DA color code, and these are immediate questions that need resolution. Finally, developing an efficient generalized matching decoder and benchmarking the code is also important. We currently do not have a good understanding of constraints on condensation sequences that ensure a good error-correcting code.

One of the primary motivations for this work was to develop a model of universal fault-tolerant quantum computation using native tools for dynamic automorphism codes. In this paper, we achieved the full Clifford group and took the first step toward a native (in Floquet codes sense) realization of a non-Clifford gate. However, we can only apply the non-Clifford gate through native measurements by making them adaptive. Furthermore, fault-tolerantly combining the 3D DA color code with the 2D DA color code using a method similar to code switching (dimensional jump) is an important future direction. We emphasize that even if the specific measurement sequences considered in this paper do not eventually lead to fault-tolerant quantum computation, this paper can still serve as a blueprint for finding alternative measurement sequences for DA color code that do.

Finally, we believe there is a rich family of dynamic automorphism codes that have yet to be studied. For example, there should exist a generalization to twisted quantum doubles that admit a Pauli realization \cite{Ellison22stab}; examples of Floquet protocols for these have been recently proposed \cite{Bauer2023,Ellison23}. Dynamic automorphism codes could also be developed for other topological subsystem codes \cite{Ellison22subsystem,Kubica2022single,Bridgeman2023lifting}, and, hopefully, for LDPC codes with good parameters~\cite{Breuckmann_2021}. The latter would be interesting in its own right, and it might also lead to a new way of implementing gates in LDPC codes.

Interestingly, in the honeycomb code, the automorphism that is implemented by measurements cannot be otherwise implemented by an on-site transversal unitary gate~\footnote{The $\ttt{e-m}$   automorphism can nevertheless be implemented by a finite-depth local unitary~\cite{Barkeshli23}.}. Therefore, is it possible to find interesting examples of dynamic automorphism codes that fault-tolerantly apply automorphisms that are not implementable by on-site transversal gates? 

\section{Acknowledgments} 

We are grateful to Ben Brown, Jeongwan Haah, Matthew Hastings, Alexey Khudorozhkov, Roger Mong, Tom Scruby, and Zhenghan Wang for useful discussions. This research was supported in part by the National Science Foundation under grants No.PHY-1818914 and PHY-1748958, the Heising-Simons Foundation, and the Simons
Foundation (216179, LB). MD was also supported by the Air Force Office of Scientific Research under award number FA2386-21-1-4058. NT is supported by the Walter Burke Institute for Theoretical Physics at Caltech. SB was supported by the National Science Foundation Graduate Research Fellowship under Grant No.~1745302.

\normalem
\bibliographystyle{unsrtnat}

\clearpage
\pagebreak

\setcounter{equation}{0}
\setcounter{figure}{0}
\setcounter{table}{0}
\setcounter{section}{0}
\makeatletter
\renewcommand{\thesection}{Appendix \Alph{section}}
\renewcommand{\theequation}{A\arabic{equation}}
\renewcommand{\thefigure}{A\arabic{figure}}
\renewcommand{\thetable}{A\arabic{table}}



\appendix

\section{All 72 automorphisms of the color code in under length-5 sequences}\label{app:72}

In this Appendix, we list measurement sequences corresponding to all 72 automorphisms of the color code. Ignoring the trivial automorphism, among the remaining 71 automorphisms we find that 10 can be implemented in length 3 (Table~\ref{tab:length3sequence}), 41 in length 4 (Tables~\ref{tab:length4sequence1}-\ref{tab:length4sequence2}), and 20 in length 5 (Table~\ref{tab:length5sequence}). The ISG of the effective code at the reference timestamps is $\text{ISG}_0 = \mathcal S (\widetilde{\CC})= \langle V(Z_1 Z_2), P_c(X_1X_2), P_c(Z_1) \rangle$, where $c = r,g,b$. These sequences were obtained by exhaustive search over all sequences of a given length and explicitly applying eqs.~\eqref{eq:update_formula} and \eqref{eq:logical_update_full}.

\begin{table}[h]
 \centering
\begin{tabular}{|c|c|c|c|c|c|c|c|c|}
\hline
 \multirow{2}{*}{No.}& \multicolumn{4}{|c|}{ Automorphism ($\varphi$)} & \multicolumn{4}{c|}{$t$} \\
 \cline{2-9} 
 &$\varphi(\bx)$ & $\varphi(\rz)$ & $\varphi(\rx)$ & $\varphi(\bz)$ & 0 & 1 & 2 & 3 \\
\hline 
\multirow{2}{*}{1}&\multirow{2}{*}{$\bz$}&\multirow{2}{*}{$\rx$}&\multirow{2}{*}{$\rz$}&\multirow{2}{*}{$\bx$}& \multirow{2}{*}{ $V(Z_1Z_2)$} &$\Er(X_1)$ & $\Eg(Y_1)$& \multirow{2}{*}{ $V(Z_1Z_2)$} \\ &&&&&& $\Eg(X_2)$ & $\Er(Y_2)$ & \\ \hline
 \multirow{2}{*}{2}&\multirow{2}{*}{$\rz$}&\multirow{2}{*}{$\bx$}&\multirow{2}{*}{$\bz$}&\multirow{2}{*}{$\rx$}& \multirow{2}{*}{ $V(Z_1Z_2)$} &$\Er(X_1)$ & $\Eg(Y_1)$& \multirow{2}{*}{ $V(Z_1Z_2)$} \\ &&&&&& $\Eg(X_2)$ & $\Eb(Y_2)$ & \\ \hline
 \multirow{2}{*}{3}&\multirow{2}{*}{$\gz$}&\multirow{2}{*}{$\rx$}&\multirow{2}{*}{$\rz$}&\multirow{2}{*}{$\gx$}& \multirow{2}{*}{ $V(Z_1Z_2)$} &$\Er(X_1)$ & $\Eb(Y_1)$& \multirow{2}{*}{ $V(Z_1Z_2)$} \\ &&&&&& $\Eg(X_2)$ & $\Er(Y_2)$ & \\ \hline
 \multirow{2}{*}{4}&\multirow{2}{*}{$\bx$}&\multirow{2}{*}{$\gy$}&\multirow{2}{*}{$\by$}&\multirow{2}{*}{$\gx$}& \multirow{2}{*}{ $V(Z_1Z_2)$} &$\Er(X_1)$ & $\Eb(Y_1)$& \multirow{2}{*}{ $V(Z_1Z_2)$} \\ &&&&&& $\Eg(X_2)$ & $\Eg(X_2)$ & \\ \hline
 \multirow{2}{*}{5}&\multirow{2}{*}{$\bx$}&\multirow{2}{*}{$\rz$}&\multirow{2}{*}{$\bz$}&\multirow{2}{*}{$\rx$}& \multirow{2}{*}{ $V(Z_1Z_2)$} &$\Er(Y_1)$ & $\Er(Y_1)$& \multirow{2}{*}{ $V(Z_1Z_2)$} \\ &&&&&& $\Eg(Y_2)$ & $\Eb(X_2)$ & \\ \hline
 \multirow{2}{*}{6}&\multirow{2}{*}{$\gy$}&\multirow{2}{*}{$\rz$}&\multirow{2}{*}{$\gz$}&\multirow{2}{*}{$\ry$}& \multirow{2}{*}{ $V(Z_1Z_2)$} &$\Er(Y_1)$ & $\Er(Y_1)$& \multirow{2}{*}{ $V(Z_1Z_2)$} \\ &&&&&& $\Eg(X_2)$ & $\Eb(Y_2)$ & \\ \hline
 \multirow{2}{*}{7}&\multirow{2}{*}{$\ry$}&\multirow{2}{*}{$\gx$}&\multirow{2}{*}{$\rx$}&\multirow{2}{*}{$\gy$}& \multirow{2}{*}{ $V(Z_1Z_2)$} &$\Er(Y_1)$ & $\Eb(X_1)$& \multirow{2}{*}{ $V(Z_1Z_2)$} \\ &&&&&& $\Eg(X_2)$ & $\Eg(X_2)$ & \\ \hline
 \multirow{2}{*}{8}&\multirow{2}{*}{$\gz$}&\multirow{2}{*}{$\by$}&\multirow{2}{*}{$\gy$}&\multirow{2}{*}{$\bz$}& \multirow{2}{*}{ $V(Z_1Z_2)$} &$\Er(X_1)$ & $\Eg(Y_1)$& \multirow{2}{*}{ $V(Z_1Z_2)$} \\ &&&&&& $\Eb(X_2)$ & $\Eb(X_2)$ & \\ \hline
 \multirow{2}{*}{9}&\multirow{2}{*}{$\bz$}&\multirow{2}{*}{$\gx$}&\multirow{2}{*}{$\gz$}&\multirow{2}{*}{$\bx$}& \multirow{2}{*}{ $V(Z_1Z_2)$} &$\Er(X_1)$ & $\Eb(Y_1)$& \multirow{2}{*}{ $V(Z_1Z_2)$} \\ &&&&&& $\Eb(X_2)$ & $\Eg(Y_2)$ & \\ \hline
 \multirow{2}{*}{10}&\multirow{2}{*}{$\rz$}&\multirow{2}{*}{$\bx$}&\multirow{2}{*}{$\rx$}&\multirow{2}{*}{$\bz$}& \multirow{2}{*}{ $V(Z_1Z_2)$} &$\Er(Y_1)$ & $\Eg(X_1)$& \multirow{2}{*}{ $V(Z_1Z_2)$} \\ &&&&&& $\Eb(X_2)$ & $\Eb(X_2)$ & \\ \hline
\end{tabular}
\caption{All length 3 automorphisms. The anyons listed are those of the effective color code $\widetilde{\CC}$.}
\label{tab:length3sequence}
\end{table}

\begin{table}
 \centering
\begin{tabular}{|c|c|c|c|c|c|c|c|c|c|}
\hline
 \multirow{2}{*}{No.}& \multicolumn{4}{|c|}{ Automorphism ($\varphi$)} & \multicolumn{5}{c|}{$t$} \\
 \cline{2-10} 
 &$\varphi(\bx)$ & $\varphi(\rz)$ & $\varphi(\rx)$ & $\varphi(\bz)$ & 0 & 1 & 2 & 3 &4 \\
\hline
 \multirow{2}{*}{11}&\multirow{2}{*}{$\gy$}&\multirow{2}{*}{$\rx$}&\multirow{2}{*}{$\gx$}&\multirow{2}{*}{$\ry$}& \multirow{2}{*}{ $V(Z_1Z_2)$} &$\Er(X_1)$ & $\Er(X_1)$ & $\Eg(Y_1)$& \multirow{2}{*}{ $V(Z_1Z_2)$} \\ &&&&&& $\Eg(X_2)$ & $\Eb(Z_2)$ & $\Er(Y_2)$ & \\ \hline
 \multirow{2}{*}{12}&\multirow{2}{*}{$\gx$}&\multirow{2}{*}{$\ry$}&\multirow{2}{*}{$\gy$}&\multirow{2}{*}{$\rx$}& \multirow{2}{*}{ $V(Z_1Z_2)$} &$\Er(X_1)$ & $\Er(X_1)$ & $\Eg(Y_1)$& \multirow{2}{*}{ $V(Z_1Z_2)$} \\ &&&&&& $\Eg(X_2)$ & $\Eb(Z_2)$ & $\Er(X_2)$ & \\ \hline
 \multirow{2}{*}{13}&\multirow{2}{*}{$\by$}&\multirow{2}{*}{$\rx$}&\multirow{2}{*}{$\bx$}&\multirow{2}{*}{$\ry$}& \multirow{2}{*}{ $V(Z_1Z_2)$} &$\Er(X_1)$ & $\Er(X_1)$ & $\Eb(Y_1)$& \multirow{2}{*}{ $V(Z_1Z_2)$} \\ &&&&&& $\Eg(X_2)$ & $\Eb(Z_2)$ & $\Er(Y_2)$ & \\ \hline
 \multirow{2}{*}{14}&\multirow{2}{*}{$\bx$}&\multirow{2}{*}{$\ry$}&\multirow{2}{*}{$\by$}&\multirow{2}{*}{$\rx$}& \multirow{2}{*}{ $V(Z_1Z_2)$} &$\Er(X_1)$ & $\Er(X_1)$ & $\Eb(Y_1)$& \multirow{2}{*}{ $V(Z_1Z_2)$} \\ &&&&&& $\Eg(X_2)$ & $\Eb(Z_2)$ & $\Er(X_2)$ & \\ \hline
 \multirow{2}{*}{15}&\multirow{2}{*}{$\by$}&\multirow{2}{*}{$\gx$}&\multirow{2}{*}{$\bx$}&\multirow{2}{*}{$\gy$}& \multirow{2}{*}{ $V(Z_1Z_2)$} &$\Er(X_1)$ & $\Er(X_1)$ & $\Eb(Y_1)$& \multirow{2}{*}{ $V(Z_1Z_2)$} \\ &&&&&& $\Eg(X_2)$ & $\Eb(Z_2)$ & $\Eg(Y_2)$ & \\ \hline
 \multirow{2}{*}{16}&\multirow{2}{*}{$\gx$}&\multirow{2}{*}{$\rz$}&\multirow{2}{*}{$\rx$}&\multirow{2}{*}{$\gz$}& \multirow{2}{*}{ $V(Z_1Z_2)$} &$\Er(X_1)$ & $\Eg(Y_1)$ & $\Er(X_1)$& \multirow{2}{*}{ $V(Z_1Z_2)$} \\ &&&&&& $\Eg(X_2)$ & $\Er(Y_2)$ & $\Eb(X_2)$ & \\ \hline
 \multirow{2}{*}{17}&\multirow{2}{*}{$\gx$}&\multirow{2}{*}{$\by$}&\multirow{2}{*}{$\gy$}&\multirow{2}{*}{$\bx$}& \multirow{2}{*}{ $V(Z_1Z_2)$} &$\Er(X_1)$ & $\Eg(Y_1)$ & $\Eg(Y_1)$& \multirow{2}{*}{ $V(Z_1Z_2)$} \\ &&&&&& $\Eg(X_2)$ & $\Er(Y_2)$ & $\Eb(X_2)$ & \\ \hline
 \multirow{2}{*}{18}&\multirow{2}{*}{$\rx$}&\multirow{2}{*}{$\bz$}&\multirow{2}{*}{$\bx$}&\multirow{2}{*}{$\rz$}& \multirow{2}{*}{ $V(Z_1Z_2)$} &$\Er(X_1)$ & $\Eg(Y_1)$ & $\Eb(X_1)$& \multirow{2}{*}{ $V(Z_1Z_2)$} \\ &&&&&& $\Eg(X_2)$ & $\Er(Y_2)$ & $\Eg(X_2)$ & \\ \hline
 \multirow{2}{*}{19}&\multirow{2}{*}{$\by$}&\multirow{2}{*}{$\rz$}&\multirow{2}{*}{$\ry$}&\multirow{2}{*}{$\bz$}& \multirow{2}{*}{ $V(Z_1Z_2)$} &$\Er(X_1)$ & $\Eg(Z_1)$ & $\Er(Y_1)$& \multirow{2}{*}{ $V(Z_1Z_2)$} \\ &&&&&& $\Eg(X_2)$ & $\Er(Z_2)$ & $\Eg(X_2)$ & \\ \hline
 \multirow{2}{*}{20}&\multirow{2}{*}{$\gy$}&\multirow{2}{*}{$\rz$}&\multirow{2}{*}{$\ry$}&\multirow{2}{*}{$\gz$}& \multirow{2}{*}{ $V(Z_1Z_2)$} &$\Er(X_1)$ & $\Eg(Z_1)$ & $\Er(Y_1)$& \multirow{2}{*}{ $V(Z_1Z_2)$} \\ &&&&&& $\Eg(X_2)$ & $\Er(Z_2)$ & $\Eb(X_2)$ & \\ \hline
 \multirow{2}{*}{21}&\multirow{2}{*}{$\ry$}&\multirow{2}{*}{$\bz$}&\multirow{2}{*}{$\by$}&\multirow{2}{*}{$\rz$}& \multirow{2}{*}{ $V(Z_1Z_2)$} &$\Er(X_1)$ & $\Eg(Z_1)$ & $\Eb(Y_1)$& \multirow{2}{*}{ $V(Z_1Z_2)$} \\ &&&&&& $\Eg(X_2)$ & $\Er(Z_2)$ & $\Eg(X_2)$ & \\ \hline
 \multirow{2}{*}{22}&\multirow{2}{*}{$\ry$}&\multirow{2}{*}{$\bz$}&\multirow{2}{*}{$\rz$}&\multirow{2}{*}{$\by$}& \multirow{2}{*}{ $V(Z_1Z_2)$} &$\Er(X_1)$ & $\Eg(Z_1)$ & $\Eb(Y_1)$& \multirow{2}{*}{ $V(Z_1Z_2)$} \\ &&&&&& $\Eg(X_2)$ & $\Er(Y_2)$ & $\Er(Y_2)$ & \\ \hline
 \multirow{2}{*}{23}&\multirow{2}{*}{$\gy$}&\multirow{2}{*}{$\bx$}&\multirow{2}{*}{$\gx$}&\multirow{2}{*}{$\by$}& \multirow{2}{*}{ $V(Z_1Z_2)$} &$\Er(X_1)$ & $\Eg(Y_1)$ & $\Eg(Y_1)$& \multirow{2}{*}{ $V(Z_1Z_2)$} \\ &&&&&& $\Eg(X_2)$ & $\Er(Z_2)$ & $\Eb(Y_2)$ & \\ \hline
 \multirow{2}{*}{24}&\multirow{2}{*}{$\gx$}&\multirow{2}{*}{$\bz$}&\multirow{2}{*}{$\bx$}&\multirow{2}{*}{$\gz$}& \multirow{2}{*}{ $V(Z_1Z_2)$} &$\Er(X_1)$ & $\Eg(Y_1)$ & $\Eb(X_1)$& \multirow{2}{*}{ $V(Z_1Z_2)$} \\ &&&&&& $\Eg(X_2)$ & $\Eb(Y_2)$ & $\Er(X_2)$ & \\ \hline
 \multirow{2}{*}{25}&\multirow{2}{*}{$\gy$}&\multirow{2}{*}{$\bz$}&\multirow{2}{*}{$\by$}&\multirow{2}{*}{$\gz$}& \multirow{2}{*}{ $V(Z_1Z_2)$} &$\Er(X_1)$ & $\Eg(Z_1)$ & $\Eb(Y_1)$& \multirow{2}{*}{ $V(Z_1Z_2)$} \\ &&&&&& $\Eg(X_2)$ & $\Eb(Z_2)$ & $\Er(X_2)$ & \\ \hline
 \multirow{2}{*}{26}&\multirow{2}{*}{$\by$}&\multirow{2}{*}{$\rz$}&\multirow{2}{*}{$\bz$}&\multirow{2}{*}{$\ry$}& \multirow{2}{*}{ $V(Z_1Z_2)$} &$\Er(X_1)$ & $\Eg(Z_1)$ & $\Er(Y_1)$& \multirow{2}{*}{ $V(Z_1Z_2)$} \\ &&&&&& $\Eg(X_2)$ & $\Eb(Y_2)$ & $\Eb(Y_2)$ & \\ \hline
 \multirow{2}{*}{27}&\multirow{2}{*}{$\rx$}&\multirow{2}{*}{$\gz$}&\multirow{2}{*}{$\gx$}&\multirow{2}{*}{$\rz$}& \multirow{2}{*}{ $V(Z_1Z_2)$} &$\Er(X_1)$ & $\Eb(Y_1)$ & $\Eg(X_1)$& \multirow{2}{*}{ $V(Z_1Z_2)$} \\ &&&&&& $\Eg(X_2)$ & $\Er(Y_2)$ & $\Eb(X_2)$ & \\ \hline
 \multirow{2}{*}{28}&\multirow{2}{*}{$\ry$}&\multirow{2}{*}{$\gz$}&\multirow{2}{*}{$\gy$}&\multirow{2}{*}{$\rz$}& \multirow{2}{*}{ $V(Z_1Z_2)$} &$\Er(X_1)$ & $\Eb(Z_1)$ & $\Eg(Y_1)$& \multirow{2}{*}{ $V(Z_1Z_2)$} \\ &&&&&& $\Eg(X_2)$ & $\Er(Z_2)$ & $\Eb(X_2)$ & \\ \hline
 \multirow{2}{*}{29}&\multirow{2}{*}{$\ry$}&\multirow{2}{*}{$\gz$}&\multirow{2}{*}{$\rz$}&\multirow{2}{*}{$\gy$}& \multirow{2}{*}{ $V(Z_1Z_2)$} &$\Er(X_1)$ & $\Eb(Z_1)$ & $\Eg(Y_1)$& \multirow{2}{*}{ $V(Z_1Z_2)$} \\ &&&&&& $\Eg(X_2)$ & $\Er(Y_2)$ & $\Er(Y_2)$ & \\ \hline
 \multirow{2}{*}{30}&\multirow{2}{*}{$\by$}&\multirow{2}{*}{$\gz$}&\multirow{2}{*}{$\bz$}&\multirow{2}{*}{$\gy$}& \multirow{2}{*}{ $V(Z_1Z_2)$} &$\Er(X_1)$ & $\Eb(Z_1)$ & $\Eg(Y_1)$& \multirow{2}{*}{ $V(Z_1Z_2)$} \\ &&&&&& $\Eg(X_2)$ & $\Eg(X_2)$ & $\Eb(Y_2)$ & \\ \hline
\end{tabular}
\caption{All length 4 automorphisms. The anyons listed are those of the effective color code $\widetilde{\CC}$. }
\label{tab:length4sequence1}
\end{table}

\begin{table}
 \centering
\begin{tabular}{|c|c|c|c|c|c|c|c|c|c|}
\hline
 \multirow{2}{*}{No.}& \multicolumn{4}{|c|}{ Automorphism ($\varphi$)} & \multicolumn{5}{c|}{$t$} \\
 \cline{2-10} 
 &$\varphi(\bx)$ & $\varphi(\rz)$ & $\varphi(\rx)$ & $\varphi(\bz)$ & 0 & 1 & 2 & 3 &4 \\
\hline
\multirow{2}{*}{31}&\multirow{2}{*}{$\rx$}&\multirow{2}{*}{$\bz$}&\multirow{2}{*}{$\rz$}&\multirow{2}{*}{$\bx$}& \multirow{2}{*}{ $V(Z_1Z_2)$} &$\Er(Y_1)$ & $\Eg(X_1)$ & $\Eb(Y_1)$& \multirow{2}{*}{ $V(Z_1Z_2)$} \\ &&&&&& $\Eg(Y_2)$ & $\Er(X_2)$ & $\Er(X_2)$ & \\ \hline
 \multirow{2}{*}{32}&\multirow{2}{*}{$\rx$}&\multirow{2}{*}{$\gz$}&\multirow{2}{*}{$\rz$}&\multirow{2}{*}{$\gx$}& \multirow{2}{*}{ $V(Z_1Z_2)$} &$\Er(Y_1)$ & $\Eb(X_1)$ & $\Eg(Y_1)$& \multirow{2}{*}{ $V(Z_1Z_2)$} \\ &&&&&& $\Eg(Y_2)$ & $\Er(X_2)$ & $\Er(X_2)$ & \\ \hline
 \multirow{2}{*}{33}&\multirow{2}{*}{$\bx$}&\multirow{2}{*}{$\gz$}&\multirow{2}{*}{$\bz$}&\multirow{2}{*}{$\gx$}& \multirow{2}{*}{ $V(Z_1Z_2)$} &$\Er(Y_1)$ & $\Eb(Z_1)$ & $\Eg(Y_1)$& \multirow{2}{*}{ $V(Z_1Z_2)$} \\ &&&&&& $\Eg(Y_2)$ & $\Eg(Y_2)$ & $\Eb(X_2)$ & \\ \hline
 \multirow{2}{*}{34}&\multirow{2}{*}{$\ry$}&\multirow{2}{*}{$\bx$}&\multirow{2}{*}{$\rx$}&\multirow{2}{*}{$\by$}& \multirow{2}{*}{ $V(Z_1Z_2)$} &$\Er(Y_1)$ & $\Eg(X_1)$ & $\Eg(X_1)$& \multirow{2}{*}{ $V(Z_1Z_2)$} \\ &&&&&& $\Eg(X_2)$ & $\Er(Z_2)$ & $\Eb(X_2)$ & \\ \hline
 \multirow{2}{*}{35}&\multirow{2}{*}{$\gy$}&\multirow{2}{*}{$\bz$}&\multirow{2}{*}{$\gz$}&\multirow{2}{*}{$\by$}& \multirow{2}{*}{ $V(Z_1Z_2)$} &$\Er(Y_1)$ & $\Eg(Z_1)$ & $\Eb(Y_1)$& \multirow{2}{*}{ $V(Z_1Z_2)$} \\ &&&&&& $\Eg(X_2)$ & $\Er(Y_2)$ & $\Er(Y_2)$ & \\ \hline
 \multirow{2}{*}{36}&\multirow{2}{*}{$\gz$}&\multirow{2}{*}{$\rx$}&\multirow{2}{*}{$\gx$}&\multirow{2}{*}{$\rz$}& \multirow{2}{*}{ $V(Z_1Z_2)$} &$\Er(X_1)$ & $\Er(X_1)$ & $\Eg(Y_1)$& \multirow{2}{*}{ $V(Z_1Z_2)$} \\ &&&&&& $\Eb(X_2)$ & $\Eg(Z_2)$ & $\Er(Y_2)$ & \\ \hline
 \multirow{2}{*}{37}&\multirow{2}{*}{$\gz$}&\multirow{2}{*}{$\ry$}&\multirow{2}{*}{$\gy$}&\multirow{2}{*}{$\rz$}& \multirow{2}{*}{ $V(Z_1Z_2)$} &$\Er(X_1)$ & $\Er(X_1)$ & $\Eg(Y_1)$& \multirow{2}{*}{ $V(Z_1Z_2)$} \\ &&&&&& $\Eb(X_2)$ & $\Eg(Z_2)$ & $\Er(X_2)$ & \\ \hline
 \multirow{2}{*}{38}&\multirow{2}{*}{$\gz$}&\multirow{2}{*}{$\bx$}&\multirow{2}{*}{$\gx$}&\multirow{2}{*}{$\bz$}& \multirow{2}{*}{ $V(Z_1Z_2)$} &$\Er(X_1)$ & $\Er(X_1)$ & $\Eg(Y_1)$& \multirow{2}{*}{ $V(Z_1Z_2)$} \\ &&&&&& $\Eb(X_2)$ & $\Eg(Z_2)$ & $\Eb(Y_2)$ & \\ \hline
 \multirow{2}{*}{39}&\multirow{2}{*}{$\bz$}&\multirow{2}{*}{$\rx$}&\multirow{2}{*}{$\bx$}&\multirow{2}{*}{$\rz$}& \multirow{2}{*}{ $V(Z_1Z_2)$} &$\Er(X_1)$ & $\Er(X_1)$ & $\Eb(Y_1)$& \multirow{2}{*}{ $V(Z_1Z_2)$} \\ &&&&&& $\Eb(X_2)$ & $\Eg(Z_2)$ & $\Er(Y_2)$ & \\ \hline
 \multirow{2}{*}{40}&\multirow{2}{*}{$\bz$}&\multirow{2}{*}{$\ry$}&\multirow{2}{*}{$\by$}&\multirow{2}{*}{$\rz$}& \multirow{2}{*}{ $V(Z_1Z_2)$} &$\Er(X_1)$ & $\Er(X_1)$ & $\Eb(Y_1)$& \multirow{2}{*}{ $V(Z_1Z_2)$} \\ &&&&&& $\Eb(X_2)$ & $\Eg(Z_2)$ & $\Er(X_2)$ & \\ \hline
 \multirow{2}{*}{41}&\multirow{2}{*}{$\gx$}&\multirow{2}{*}{$\rz$}&\multirow{2}{*}{$\gz$}&\multirow{2}{*}{$\rx$}& \multirow{2}{*}{ $V(Z_1Z_2)$} &$\Er(X_1)$ & $\Eg(Z_1)$ & $\Er(Y_1)$& \multirow{2}{*}{ $V(Z_1Z_2)$} \\ &&&&&& $\Eb(X_2)$ & $\Eb(X_2)$ & $\Eg(Y_2)$ & \\ \hline
 \multirow{2}{*}{42}&\multirow{2}{*}{$\gx$}&\multirow{2}{*}{$\bz$}&\multirow{2}{*}{$\gz$}&\multirow{2}{*}{$\bx$}& \multirow{2}{*}{ $V(Z_1Z_2)$} &$\Er(X_1)$ & $\Eg(Z_1)$ & $\Eb(Y_1)$& \multirow{2}{*}{ $V(Z_1Z_2)$} \\ &&&&&& $\Eb(X_2)$ & $\Eb(X_2)$ & $\Eg(Y_2)$ & \\ \hline
 \multirow{2}{*}{43}&\multirow{2}{*}{$\bx$}&\multirow{2}{*}{$\gz$}&\multirow{2}{*}{$\gx$}&\multirow{2}{*}{$\bz$}& \multirow{2}{*}{ $V(Z_1Z_2)$} &$\Er(X_1)$ & $\Eb(Y_1)$ & $\Eg(X_1)$& \multirow{2}{*}{ $V(Z_1Z_2)$} \\ &&&&&& $\Eb(X_2)$ & $\Er(Y_2)$ & $\Eb(X_2)$ & \\ \hline
 \multirow{2}{*}{44}&\multirow{2}{*}{$\bz$}&\multirow{2}{*}{$\gy$}&\multirow{2}{*}{$\by$}&\multirow{2}{*}{$\gz$}& \multirow{2}{*}{ $V(Z_1Z_2)$} &$\Er(X_1)$ & $\Eb(Y_1)$ & $\Eb(Y_1)$& \multirow{2}{*}{ $V(Z_1Z_2)$} \\ &&&&&& $\Eb(X_2)$ & $\Er(Y_2)$ & $\Eg(X_2)$ & \\ \hline
 \multirow{2}{*}{45}&\multirow{2}{*}{$\by$}&\multirow{2}{*}{$\gz$}&\multirow{2}{*}{$\gy$}&\multirow{2}{*}{$\bz$}& \multirow{2}{*}{ $V(Z_1Z_2)$} &$\Er(X_1)$ & $\Eb(Z_1)$ & $\Eg(Y_1)$& \multirow{2}{*}{ $V(Z_1Z_2)$} \\ &&&&&& $\Eb(X_2)$ & $\Er(Z_2)$ & $\Eb(X_2)$ & \\ \hline
 \multirow{2}{*}{46}&\multirow{2}{*}{$\bz$}&\multirow{2}{*}{$\gx$}&\multirow{2}{*}{$\bx$}&\multirow{2}{*}{$\gz$}& \multirow{2}{*}{ $V(Z_1Z_2)$} &$\Er(X_1)$ & $\Eb(Y_1)$ & $\Eb(Y_1)$& \multirow{2}{*}{ $V(Z_1Z_2)$} \\ &&&&&& $\Eb(X_2)$ & $\Er(Z_2)$ & $\Eg(Y_2)$ & \\ \hline
 \multirow{2}{*}{47}&\multirow{2}{*}{$\rz$}&\multirow{2}{*}{$\gx$}&\multirow{2}{*}{$\rx$}&\multirow{2}{*}{$\gz$}& \multirow{2}{*}{ $V(Z_1Z_2)$} &$\Er(Y_1)$ & $\Eb(X_1)$ & $\Eb(X_1)$& \multirow{2}{*}{ $V(Z_1Z_2)$} \\ &&&&&& $\Eb(X_2)$ & $\Er(Z_2)$ & $\Eg(X_2)$ & \\ \hline
 \multirow{2}{*}{48}&\multirow{2}{*}{$\rz$}&\multirow{2}{*}{$\by$}&\multirow{2}{*}{$\ry$}&\multirow{2}{*}{$\bz$}& \multirow{2}{*}{ $V(Z_1Z_2)$} &$\Eg(X_1)$ & $\Er(Y_1)$ & $\Er(Y_1)$& \multirow{2}{*}{ $V(Z_1Z_2)$} \\ &&&&&& $\Eb(X_2)$ & $\Eg(Z_2)$ & $\Eb(Y_2)$ & \\ \hline
 \multirow{2}{*}{49}&\multirow{2}{*}{$\rx$}&\multirow{2}{*}{$\gy$}&\multirow{2}{*}{$\ry$}&\multirow{2}{*}{$\gx$}& \multirow{2}{*}{ $V(Z_1Z_2)$} &$\Eg(X_1)$ & $\Er(Z_1)$ & $\Eg(Y_1)$& \multirow{2}{*}{ $V(Z_1Z_2)$} \\ &&&&&& $\Eb(X_2)$ & $\Eb(X_2)$ & $\Er(Y_2)$ & \\ \hline
 \multirow{2}{*}{50}&\multirow{2}{*}{$\rx$}&\multirow{2}{*}{$\by$}&\multirow{2}{*}{$\ry$}&\multirow{2}{*}{$\bx$}& \multirow{2}{*}{ $V(Z_1Z_2)$} &$\Eg(X_1)$ & $\Er(Z_1)$ & $\Eb(Y_1)$& \multirow{2}{*}{ $V(Z_1Z_2)$} \\ &&&&&& $\Eb(X_2)$ & $\Eb(X_2)$ & $\Er(Y_2)$ & \\ \hline
 \multirow{2}{*}{51}&\multirow{2}{*}{$\rz$}&\multirow{2}{*}{$\gy$}&\multirow{2}{*}{$\ry$}&\multirow{2}{*}{$\gz$}& \multirow{2}{*}{ $V(Z_1Z_2)$} &$\Eg(X_1)$ & $\Eg(X_1)$ & $\Er(Y_1)$& \multirow{2}{*}{ $V(Z_1Z_2)$} \\ &&&&&& $\Eb(X_2)$ & $\Er(Z_2)$ & $\Eg(Y_2)$ & \\ \hline
 \end{tabular}
 \caption{All length 4 automorphisms (continued). The anyons listed are those of the effective color code $\widetilde{\CC}$. }
 \label{tab:length4sequence2}
\end{table}

\begin{table}
 \centering
\resizebox{\textwidth}{!}{\begin{tabular}{|c|c|c|c|c|c|c|c|c|c|c|}
\hline
 \multirow{2}{*}{No.}& \multicolumn{4}{|c|}{ Automorphism ($\varphi$)} & \multicolumn{6}{c|}{$t$} \\
 \cline{2-11} 
 &$\varphi(\bx)$ & $\varphi(\rz)$ & $\varphi(\rx)$ & $\varphi(\bz)$ & 0 & 1 & 2 & 3 &4 &5 \\
\hline
\multirow{2}{*}{52}&\multirow{2}{*}{$\rz$}&\multirow{2}{*}{$\by$}&\multirow{2}{*}{$\bz$}&\multirow{2}{*}{$\ry$}& \multirow{2}{*}{ $V(Z_1Z_2)$} &$\Er(X_1)$ & $\Er(X_1)$ & $\Eg(Y_1)$ & $\Eg(Y_1)$& \multirow{2}{*}{ $V(Z_1Z_2)$} \\ &&&&&& $\Eg(X_2)$ & $\Eb(Z_2)$ & $\Er(Y_2)$ & $\Eb(X_2)$ & \\ \hline
 \multirow{2}{*}{53}&\multirow{2}{*}{$\rz$}&\multirow{2}{*}{$\gy$}&\multirow{2}{*}{$\gz$}&\multirow{2}{*}{$\ry$}& \multirow{2}{*}{ $V(Z_1Z_2)$} &$\Er(X_1)$ & $\Er(X_1)$ & $\Eb(Y_1)$ & $\Eb(Y_1)$& \multirow{2}{*}{ $V(Z_1Z_2)$} \\ &&&&&& $\Eg(X_2)$ & $\Eb(Z_2)$ & $\Er(Y_2)$ & $\Eg(X_2)$ & \\ \hline
 \multirow{2}{*}{54}&\multirow{2}{*}{$\rz$}&\multirow{2}{*}{$\gx$}&\multirow{2}{*}{$\gz$}&\multirow{2}{*}{$\rx$}& \multirow{2}{*}{ $V(Z_1Z_2)$} &$\Er(X_1)$ & $\Er(X_1)$ & $\Eb(Y_1)$ & $\Eb(Y_1)$& \multirow{2}{*}{ $V(Z_1Z_2)$} \\ &&&&&& $\Eg(X_2)$ & $\Eb(Z_2)$ & $\Er(X_2)$ & $\Eg(Y_2)$ & \\ \hline
 \multirow{2}{*}{55}&\multirow{2}{*}{$\gz$}&\multirow{2}{*}{$\ry$}&\multirow{2}{*}{$\rz$}&\multirow{2}{*}{$\gy$}& \multirow{2}{*}{ $V(Z_1Z_2)$} &$\Er(X_1)$ & $\Er(X_1)$ & $\Eb(Y_1)$ & $\Eb(Y_1)$& \multirow{2}{*}{ $V(Z_1Z_2)$} \\ &&&&&& $\Eg(X_2)$ & $\Eb(Z_2)$ & $\Eg(Y_2)$ & $\Er(X_2)$ & \\ \hline
 \multirow{2}{*}{56}&\multirow{2}{*}{$\bz$}&\multirow{2}{*}{$\ry$}&\multirow{2}{*}{$\rz$}&\multirow{2}{*}{$\by$}& \multirow{2}{*}{ $V(Z_1Z_2)$} &$\Er(X_1)$ & $\Eg(Y_1)$ & $\Er(Z_1)$ & $\Eg(Y_1)$& \multirow{2}{*}{ $V(Z_1Z_2)$} \\ &&&&&& $\Eg(X_2)$ & $\Er(Y_2)$ & $\Eg(Z_2)$ & $\Er(X_2)$ & \\ \hline
 \multirow{2}{*}{57}&\multirow{2}{*}{$\gz$}&\multirow{2}{*}{$\bx$}&\multirow{2}{*}{$\bz$}&\multirow{2}{*}{$\gx$}& \multirow{2}{*}{ $V(Z_1Z_2)$} &$\Er(X_1)$ & $\Eg(Y_1)$ & $\Eb(X_1)$ & $\Er(Y_1)$& \multirow{2}{*}{ $V(Z_1Z_2)$} \\ &&&&&& $\Eg(X_2)$ & $\Er(Y_2)$ & $\Eg(X_2)$ & $\Eb(Y_2)$ & \\ \hline
 \multirow{2}{*}{58}&\multirow{2}{*}{$\gz$}&\multirow{2}{*}{$\by$}&\multirow{2}{*}{$\bz$}&\multirow{2}{*}{$\gy$}& \multirow{2}{*}{ $V(Z_1Z_2)$} &$\Er(X_1)$ & $\Eg(Y_1)$ & $\Eb(Z_1)$ & $\Er(Y_1)$& \multirow{2}{*}{ $V(Z_1Z_2)$} \\ &&&&&& $\Eg(X_2)$ & $\Er(Y_2)$ & $\Eg(Z_2)$ & $\Eb(X_2)$ & \\ \hline
 \multirow{2}{*}{59}&\multirow{2}{*}{$\bz$}&\multirow{2}{*}{$\gy$}&\multirow{2}{*}{$\gz$}&\multirow{2}{*}{$\by$}& \multirow{2}{*}{ $V(Z_1Z_2)$} &$\Er(X_1)$ & $\Eg(Y_1)$ & $\Eb(Z_1)$ & $\Er(Y_1)$& \multirow{2}{*}{ $V(Z_1Z_2)$} \\ &&&&&& $\Eg(X_2)$ & $\Eb(Y_2)$ & $\Er(Z_2)$ & $\Eg(X_2)$ & \\ \hline
 \multirow{2}{*}{60}&\multirow{2}{*}{$\by$}&\multirow{2}{*}{$\rx$}&\multirow{2}{*}{$\ry$}&\multirow{2}{*}{$\bx$}& \multirow{2}{*}{ $V(Z_1Z_2)$} &$\Er(Y_1)$ & $\Er(Y_1)$ & $\Er(Y_1)$ & $\Eg(X_1)$& \multirow{2}{*}{ $V(Z_1Z_2)$} \\ &&&&&& $\Eg(X_2)$ & $\Eb(Y_2)$ & $\Eg(Z_2)$ & $\Er(X_2)$ & \\ \hline
 \multirow{2}{*}{61}&\multirow{2}{*}{$\ry$}&\multirow{2}{*}{$\bx$}&\multirow{2}{*}{$\by$}&\multirow{2}{*}{$\rx$}& \multirow{2}{*}{ $V(Z_1Z_2)$} &$\Er(Y_1)$ & $\Er(Y_1)$ & $\Er(Y_1)$ & $\Eg(X_1)$& \multirow{2}{*}{ $V(Z_1Z_2)$} \\ &&&&&& $\Eg(X_2)$ & $\Eb(Y_2)$ & $\Eg(Z_2)$ & $\Eb(X_2)$ & \\ \hline
 \multirow{2}{*}{62}&\multirow{2}{*}{$\gy$}&\multirow{2}{*}{$\rx$}&\multirow{2}{*}{$\ry$}&\multirow{2}{*}{$\gx$}& \multirow{2}{*}{ $V(Z_1Z_2)$} &$\Er(Y_1)$ & $\Er(Y_1)$ & $\Er(Y_1)$ & $\Eb(X_1)$& \multirow{2}{*}{ $V(Z_1Z_2)$} \\ &&&&&& $\Eg(X_2)$ & $\Eb(Y_2)$ & $\Eg(Z_2)$ & $\Er(X_2)$ & \\ \hline
 \multirow{2}{*}{63}&\multirow{2}{*}{$\ry$}&\multirow{2}{*}{$\gx$}&\multirow{2}{*}{$\gy$}&\multirow{2}{*}{$\rx$}& \multirow{2}{*}{ $V(Z_1Z_2)$} &$\Er(Y_1)$ & $\Er(Y_1)$ & $\Eb(X_1)$ & $\Eb(X_1)$& \multirow{2}{*}{ $V(Z_1Z_2)$} \\ &&&&&& $\Eg(X_2)$ & $\Eb(Y_2)$ & $\Er(Z_2)$ & $\Eg(X_2)$ & \\ \hline
 \multirow{2}{*}{64}&\multirow{2}{*}{$\gy$}&\multirow{2}{*}{$\bx$}&\multirow{2}{*}{$\by$}&\multirow{2}{*}{$\gx$}& \multirow{2}{*}{ $V(Z_1Z_2)$} &$\Er(Y_1)$ & $\Eg(Z_1)$ & $\Eb(X_1)$ & $\Er(Y_1)$& \multirow{2}{*}{ $V(Z_1Z_2)$} \\ &&&&&& $\Eg(X_2)$ & $\Er(Z_2)$ & $\Eg(X_2)$ & $\Eb(Y_2)$ & \\ \hline
 \multirow{2}{*}{65}&\multirow{2}{*}{$\bx$}&\multirow{2}{*}{$\ry$}&\multirow{2}{*}{$\rx$}&\multirow{2}{*}{$\by$}& \multirow{2}{*}{ $V(Z_1Z_2)$} &$\Er(Y_1)$ & $\Eg(X_1)$ & $\Er(Z_1)$ & $\Eg(Y_1)$& \multirow{2}{*}{ $V(Z_1Z_2)$} \\ &&&&&& $\Eg(X_2)$ & $\Er(Z_2)$ & $\Eg(Y_2)$ & $\Er(X_2)$ & \\ \hline
 \multirow{2}{*}{66}&\multirow{2}{*}{$\rx$}&\multirow{2}{*}{$\by$}&\multirow{2}{*}{$\bx$}&\multirow{2}{*}{$\ry$}& \multirow{2}{*}{ $V(Z_1Z_2)$} &$\Er(Y_1)$ & $\Eg(X_1)$ & $\Er(Z_1)$ & $\Eg(Y_1)$& \multirow{2}{*}{ $V(Z_1Z_2)$} \\ &&&&&& $\Eg(X_2)$ & $\Er(Z_2)$ & $\Eg(Y_2)$ & $\Eb(X_2)$ & \\ \hline
 \multirow{2}{*}{67}&\multirow{2}{*}{$\gx$}&\multirow{2}{*}{$\ry$}&\multirow{2}{*}{$\rx$}&\multirow{2}{*}{$\gy$}& \multirow{2}{*}{ $V(Z_1Z_2)$} &$\Er(Y_1)$ & $\Eg(X_1)$ & $\Er(Z_1)$ & $\Eb(Y_1)$& \multirow{2}{*}{ $V(Z_1Z_2)$} \\ &&&&&& $\Eg(X_2)$ & $\Er(Z_2)$ & $\Eg(Y_2)$ & $\Er(X_2)$ & \\ \hline
 \multirow{2}{*}{68}&\multirow{2}{*}{$\rx$}&\multirow{2}{*}{$\gy$}&\multirow{2}{*}{$\gx$}&\multirow{2}{*}{$\ry$}& \multirow{2}{*}{ $V(Z_1Z_2)$} &$\Er(Y_1)$ & $\Eg(X_1)$ & $\Er(Z_1)$ & $\Eb(Y_1)$& \multirow{2}{*}{ $V(Z_1Z_2)$} \\ &&&&&& $\Eg(X_2)$ & $\Er(Z_2)$ & $\Eb(Y_2)$ & $\Eg(X_2)$ & \\ \hline
 \multirow{2}{*}{69}&\multirow{2}{*}{$\gx$}&\multirow{2}{*}{$\by$}&\multirow{2}{*}{$\bx$}&\multirow{2}{*}{$\gy$}& \multirow{2}{*}{ $V(Z_1Z_2)$} &$\Er(Y_1)$ & $\Eg(X_1)$ & $\Eb(Z_1)$ & $\Er(Y_1)$& \multirow{2}{*}{ $V(Z_1Z_2)$} \\ &&&&&& $\Eg(X_2)$ & $\Er(Z_2)$ & $\Eg(Y_2)$ & $\Eb(X_2)$ & \\ \hline
 \multirow{2}{*}{70}&\multirow{2}{*}{$\by$}&\multirow{2}{*}{$\gx$}&\multirow{2}{*}{$\gy$}&\multirow{2}{*}{$\bx$}& \multirow{2}{*}{ $V(Z_1Z_2)$} &$\Er(Y_1)$ & $\Eg(Z_1)$ & $\Eb(X_1)$ & $\Er(Y_1)$& \multirow{2}{*}{ $V(Z_1Z_2)$} \\ &&&&&& $\Eg(X_2)$ & $\Eb(Z_2)$ & $\Er(X_2)$ & $\Eg(Y_2)$ & \\ \hline
 \multirow{2}{*}{71}&\multirow{2}{*}{$\bx$}&\multirow{2}{*}{$\gy$}&\multirow{2}{*}{$\gx$}&\multirow{2}{*}{$\by$}& \multirow{2}{*}{ $V(Z_1Z_2)$} &$\Er(Y_1)$ & $\Eg(X_1)$ & $\Eb(Z_1)$ & $\Er(Y_1)$& \multirow{2}{*}{ $V(Z_1Z_2)$} \\ &&&&&& $\Eg(X_2)$ & $\Eb(Z_2)$ & $\Er(Y_2)$ & $\Eg(X_2)$ & \\ \hline
\end{tabular}}
\caption{All length 5 automorphisms.The anyons listed are those of the effective color code $\widetilde{\CC}$. }
\label{tab:length5sequence}
\end{table}

\clearpage
\newpage

\section{Lattice realization of $\varphi_{(\texttt{rb})}$ for dynamic automorphism color code with Pauli boundaries} \label{sec:SM_triangle_rb}

Below, we show an explicit measurement sequence in both the bulk and boundary which realizes the $\varphi_{(\texttt{rb})}$ automorphism (in the main text we presented the sequence for the $\varphi_{(\texttt{rbg})}$ automorphism; together, these form the generators of the full automorphism group).

\begin{table}[h]
\vspace{0pt}
\centering
\vspace{0pt}
\includegraphics[width= 1\columnwidth]{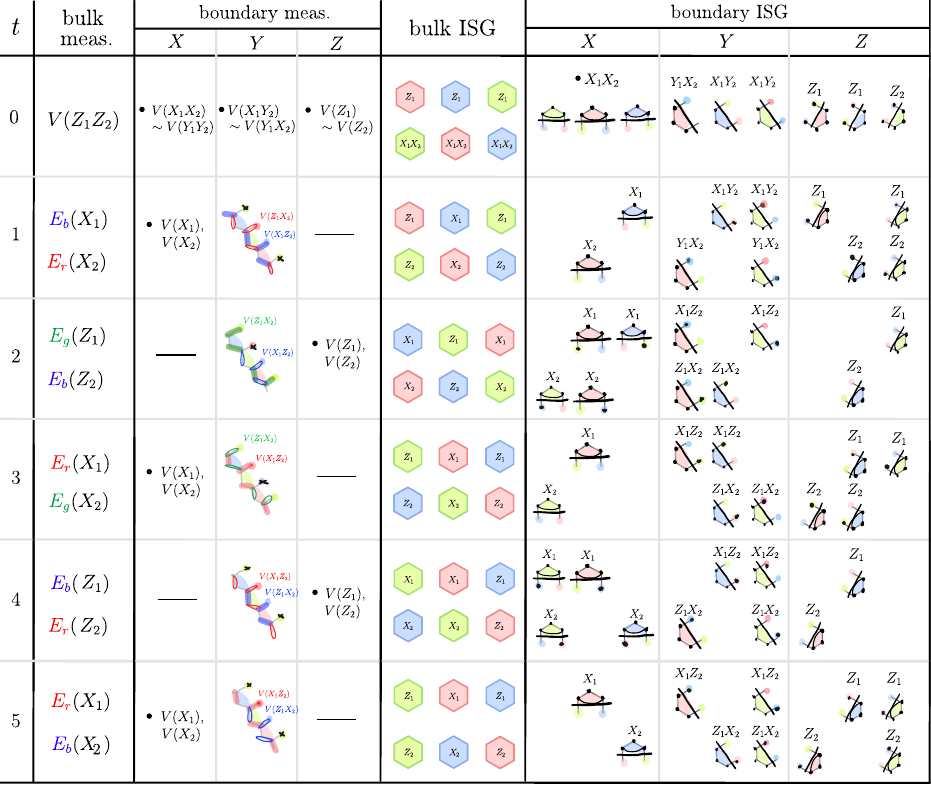}
\caption{ Summary of the measurement sequence for $\varphi_{(\texttt{rb})}$ corresponding to the lattice termination same as shown in the main text and measurements corresponding to the condensations in Table~\ref{tab:TQFTcondensationdata}. The first four columns show bulk and boundary measurements. The bulk ISG plaquettes are shown as plaquettes of respective colors with the Pauli flavor of the plaquette denoted inside. The boundary plaquettes are shown schematically where the qubits in the support of the boundary plaquette stabilizer are marked by black dots. The Pauli flavors on these plaquettes are indicated where necessary (if it is indicated only once, it applies to all qubits with black dots). }
\label{tab:SM_rb_sequence}
\end{table}

\clearpage
\section{Dynamic automorphism color code with color boundaries}
\label{sec:SM_color_triangle} 
\begin{figure}[b] \centering
\vspace{0pt}
\centering
\vspace{0pt}
\includegraphics[width= 0.9\columnwidth]{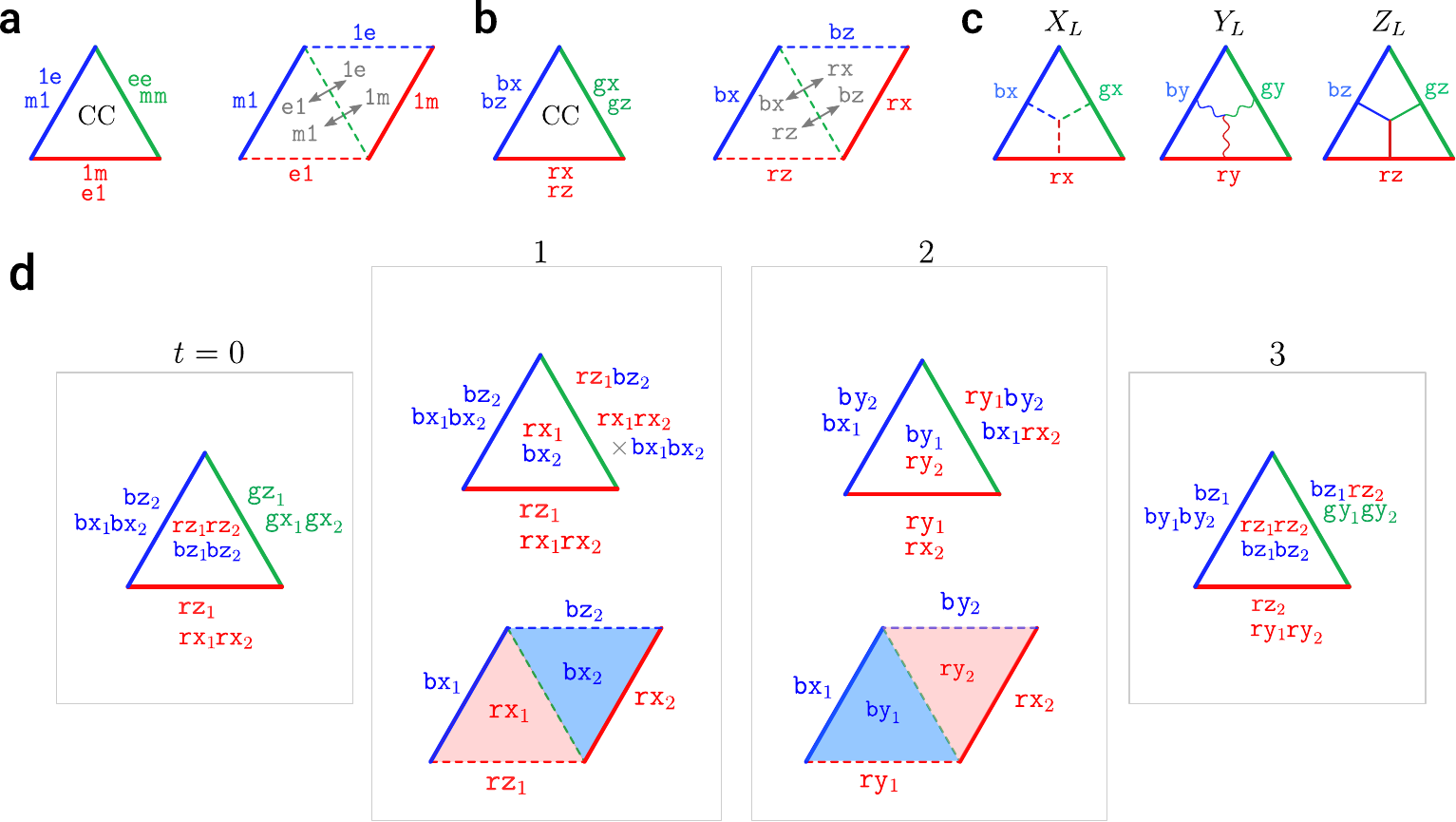}
\caption{Unfolding the color code triangle with color boundaries into a pair of toric codes with a color domain wall shown with boundary condensations in (a) double toric code and (b) color code notation. 
Each colored boundary corresponds to condensing all bosons of the corresponding color (a row in the magic square). 
(c) The logical operator strings of the color boundary triangle and our convention for the logical Pauli operators. We display the trivalent junction as, for example an $\widetilde{\texttt{rx}}$ and a $\widetilde{\texttt{bx}}$ logical strings that overlap and become a $\widetilde{\texttt{gx}}$ string before meeting the green boundary; similar logic applies to the other strings.
(d) A sequence of anyon condensations from the parent bilayer color code that implements $\varphi_{(\texttt{xz})}$ automorphism in a DA color code on a triangle. Specifically, the condensations in the bulk correspond to the $\varphi_{(\texttt{xz})}$ protocol, and the boundary condensations are chosen such that conservation of logical information is ensured. At $t=0$, we start in an effective color code with $\mathtt{rz_1 rz_2}$ and $\mathtt{gz_1 gz_2}$ condensations in the bulk and appropriate condensations at the boundary that give color boundaries of the effective color code.
Between rounds 1 and 2, the top row shows the triangle picture with boundary condensations worked out as described in Sec.~\ref{sec:tqft_bound}, such that each transition preserves logical information and the shape of the logical strings is not affected.
The bottom row at these rounds shows the respective unfolded picture. At each step, this corresponds to a toric code with an color domain wall at each round.
The other generating sequence for a single triangle, corresponding to the $\varphi_{(\texttt{xz})}$ automorphism, as well as the double triangle sequence can be worked out analogously.
 }
\label{fig:SM_color_bound}
\end{figure}

The DA color code on a color boundary triangle can be worked out similarly to the Pauli triangle. For example, Fig.~\ref{fig:SM_color_bound} (a,b) shows how such a triangle is unfolded into a toric code with a trivial domain wall. That is, in round one, the domain wall connecting $\TC(\rxi)$ to $\TC(\rxii)$ along the glued boundary is the trivial isomorphism. This should be contrasted with the Pauli triangle case where the toric codes are glued by an $\ttt e - \ttt m$ permuting isomorphism. Panel (c) shows the logical strings on a color triangle, which correspond to trivalent junctions of a given Pauli flavor, and each branch of a given color terminates on a boundary of the same color.

The bulk sequences generating the permutation group of Paulis $(\ttt{xz})$ and $(\ttt{xyz})$ are shown in Table~\ref{tab:SM_color_sequences}. These sequences respect the color boundaries and, if properly implemented in the presence of boundaries, they will generate the single-qubit Clifford group for the logical qubit on such a triangle.

The condensation sequence in the presence of a boundary can be worked out as prescribed in Sec.~\ref{sec:tqft_bound}.

\begin{table}[h]
{
\renewcommand{\rc}[1]{{\color{red}{#1}}}
\renewcommand{\bc}[1]{{\color{blue}{#1}}}
\renewcommand{\gc}[1]{{\color{ForestGreen}{#1}}}
\renewcommand{\yc}[1]{{\color{amber}{#1}}}
 \centering
 \begin{tabular}{|c|c|c|c|c|c|c| }
 \hline
 \multirow{2}{*}{Gate/Aut} & \multicolumn{6}{c|}{$t$} \\ \cline{2-7} 
 & 0 & 1 & 2 & 3 & 4 & 5 \\
 \hline
 \multirow{2}{*}{$H \sim \varphi_{(\texttt{xz})}$} & \multirow{2}{*}{ $V(Z_1Z_2)$} & $\Er(X_1)$ & $\Eb(Y_1)$ & \multirow{2}{*}{ $V(Z_1Z_2)$} & $-$ & $-$ \\
 & & $\Eb(X_2)$ & $\Eb(Y_2)$ & &$-$ & $-$\\
 \hline
 \multirow{2}{*}{$HS \sim \varphi_{(\texttt{xyz})}$} &\multirow{2}{*}{ $V(Z_1Z_2)$} & $\Er(Y_1)$ & $\Eg(Z_1)$ & $\Eb(X_1)$ & $\Er(Y_1)$&\multirow{2}{*}{ $V(Z_1Z_2)$} \\
 & & $\Eg(X_2)$ & $\Eb(Z_2)$ & $\Er(X_2)$ & $\Eg(Y_2)$ & \\ 
 \hline
 \end{tabular}
 \caption{The bulk measurement sequences for the two generators of the $S_3$ subgroup of Pauli permutations of the color code bosons for the color triangle DA color code. These sequences generate the Clifford group on one qubit. }
\label{tab:SM_color_sequences}
}
\end{table}

\clearpage
\section{Bulk detectors for padded sequences on Pauli triangles} \label{sec:SM_triangle_EC}

Below, we provide detectors for measurement sequences in the bulk (preceded and concluded by the padding sequences), which become further constrained when Pauli boundaries are considered. The bulk detectors alone do not furnish a full basis of detectors as we also have to consider boundary detectors, which we discuss in the next Appendix.

\begin{figure}[h] \centering
\centering
\includegraphics[width= 1\columnwidth]{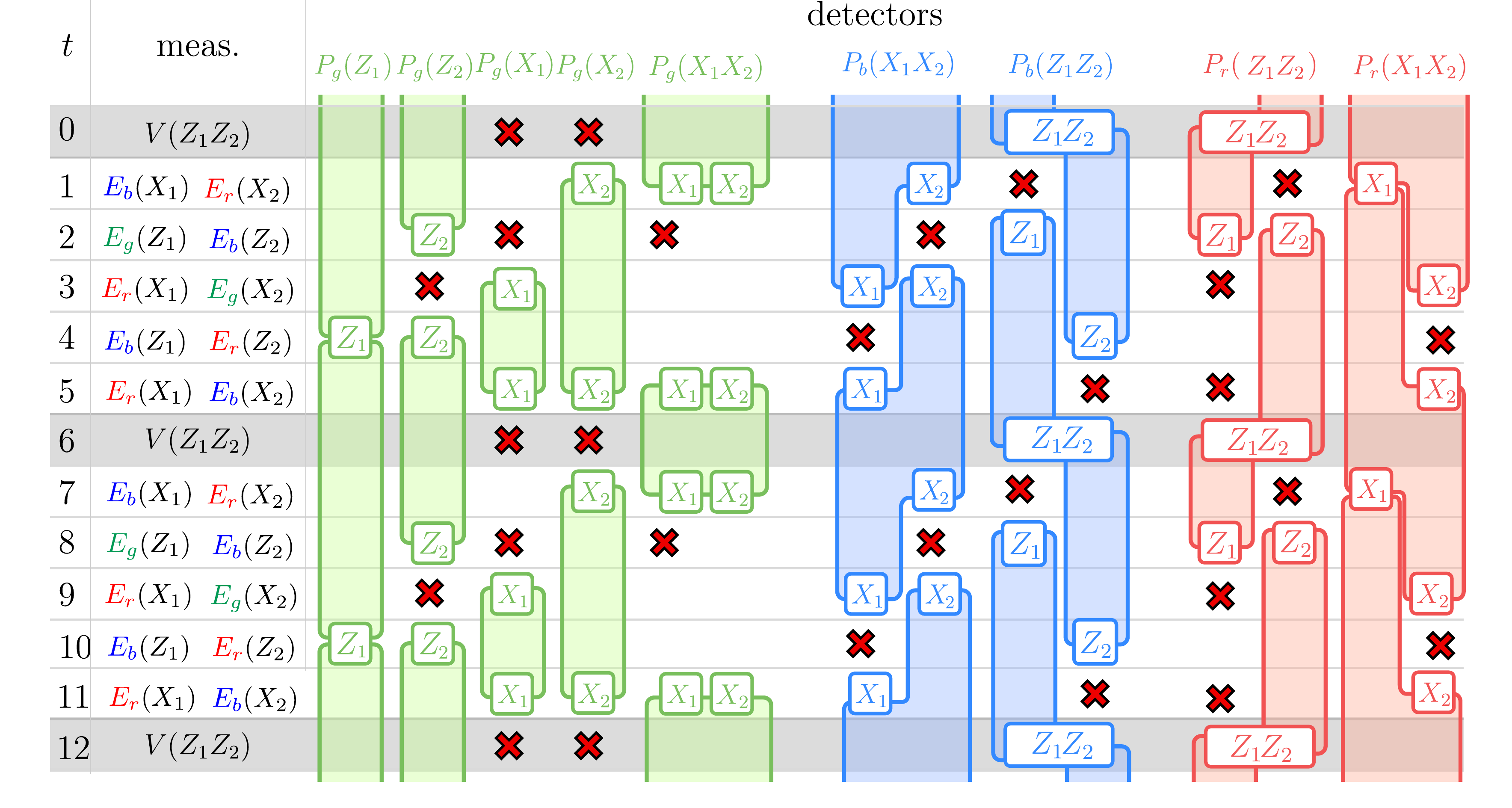}
\caption{Bulk error correction for the padding sequence for the Pauli triangle DA color code. Detectors for the periodic sequence where each period corresponds to the $\varphi_{(\texttt{rb})}$ automorphism. Repeated twice, this sequence is inserted between each pair of consequent gates in a computation, serving as a padding sequence. The same sequence repeated an odd number of times gives one of the two generating gates on a Pauli triangle, corresponding to $\varphi_{(\texttt{rb})}$. Each colored region represents a class of detectors, each having a plaquette of respective color in its support that can belong to one of the layers or both layers. The white boxes show the times when the plaquettes in the support of the detector in respective layers are inferred. The red crosses mark the rounds where the measurements in either of the layers randomize the values of the plaquettes forming the detector and are shown only where it helps explain the shape of the support of the detector. }
\label{fig:fSM1}
\end{figure}

\begin{figure}[t] \centering
\centering
\includegraphics[width= 1\columnwidth]{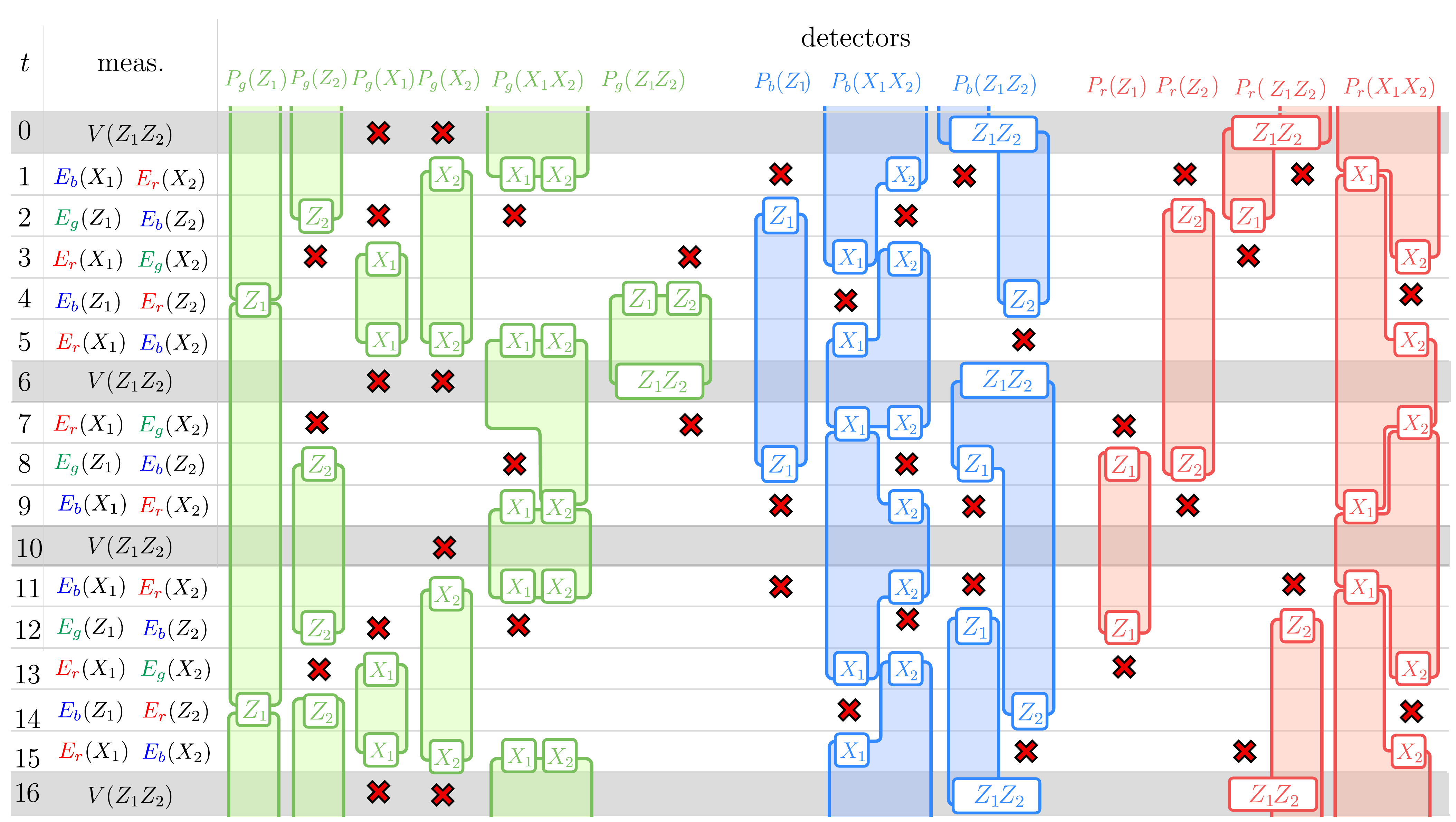}
\caption{Bulk error correction for the padding sequence for the Pauli triangle DA color code. Detectors for the corresponding to the padded $\mathcal{M}(\varphi_{(\texttt{rbg})})$ automorphism. The padding is chosen to be $\hat{M}(\varphi_{(\texttt{gb})})$. The detectors that extend past the sequence are started or completed within another adjacent padding sequence. }
\label{fig:fSM3}
\end{figure}

\clearpage
\section{Boundary detectors for padded sequences on Pauli triangles} \label{sec:SM_triangle_EC_bondary}

Here, we show the boundary detectors for the two measurement sequences of the DA color code on a triangle with Pauli boundaries (preceded and concluded by the padding sequences). Together with the bulk detectors, these constitute the candidate set for the full basis of detectors. 

\begin{figure}[h] \centering
\centering
\includegraphics[width= 1\columnwidth]{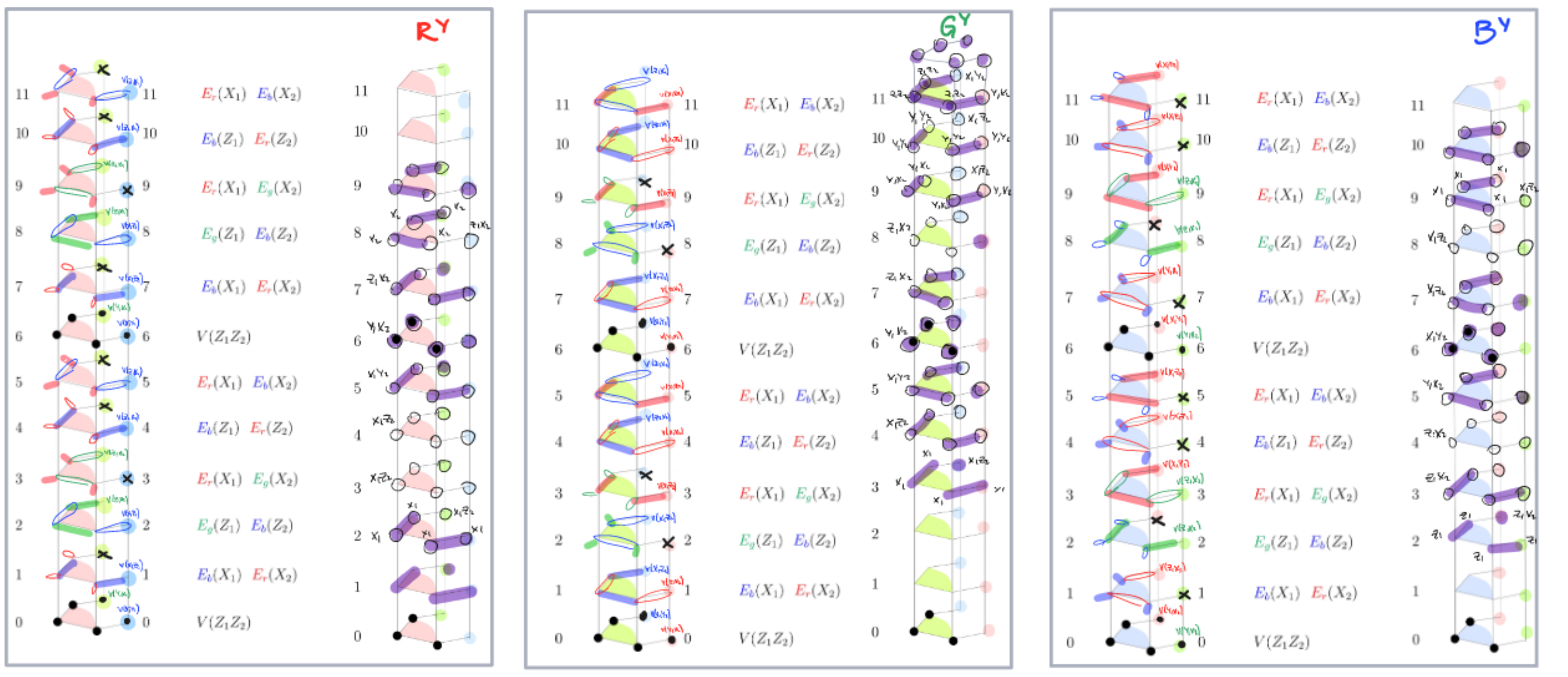}
\caption{$Y$-boundary detectors for the padding sequence $\varphi_{(\texttt{rb})}$ for the Pauli triangle DA color code. 
The figure shows three detectors, one for each color of boundary plaquettes. In each box, the measurements at each round are summarized in the left picture and the measurements comprising the detectors are shown by the edges and vertices highlighted in purple in the right picture. The hollow circles are used to demarcate its spacetime support. Specifically, the timelike edge $E_{t,t+1}$ at a given qubit belongs to the support of the detector if its end at $t+1$ has a hollow circle. The instantaneous flavor of the detector (i.e. its flavor on given qubits at a given time) is shown near respective qubits.}
\label{fig:SM_Y_rb}
\end{figure}

\begin{figure}[t] \centering
\centering
\includegraphics[width= 0.8\columnwidth]{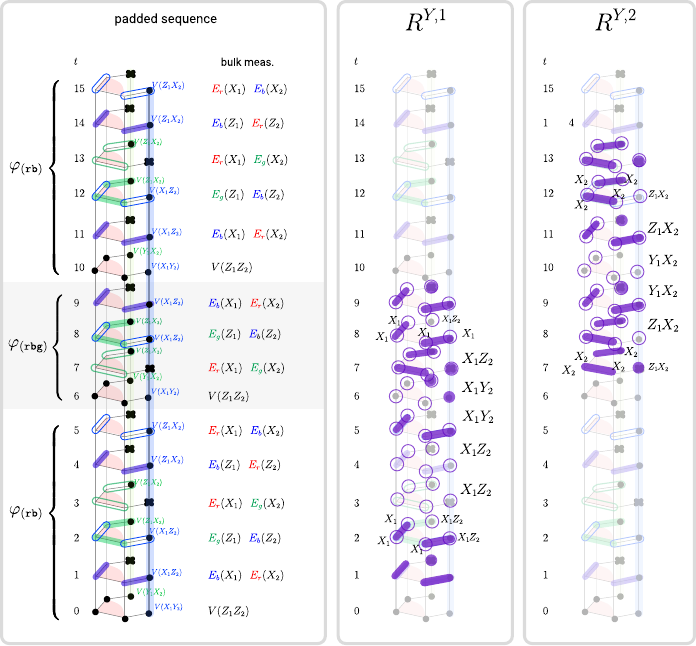}
\caption{$Y$-boundary detectors for the padded sequence $\varphi_{(\texttt{rb})}$ for the Pauli triangle DA color code. The figure shows two detectors for the red-colored boundary plaquettes. The detectors for the blue plaquettes can be found in the main text, Fig.~\ref{fig:46}, and for the green ones are shown in the next figure. In each box, the measurements at each round are summarized in the left picture and the measurements comprising the detectors are shown by the edges and vertices highlighted in purple in the right picture. The hollow circles are used to demarcate its spacetime support. Specifically, the timelike edge $E_{t,t+1}$ at a given qubit belongs to the support of the detector if its end at $t+1$ has a hollow circle. The instantaneous flavor of the detector (i.e. its flavor on given qubits at a given time) is shown near respective qubits. }
\label{fig:SM_Y_rbg}
\end{figure}

\begin{figure}[t] \centering
\centering
\includegraphics[width= 0.8\columnwidth]{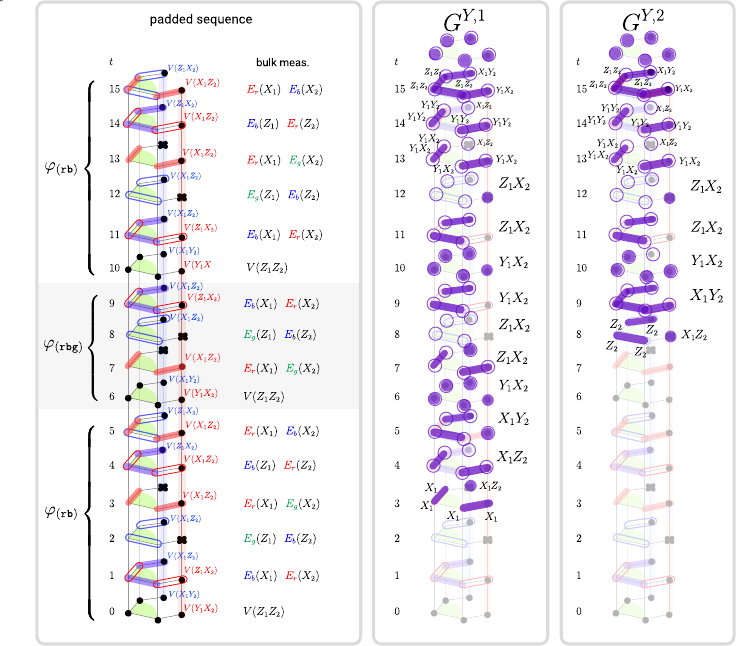}
\caption{$Y$-boundary detectors supported on the green plaquettes for the padded sequence $\varphi_{(\texttt{rb})}$ for the Pauli triangle DA color code. See Fig.~\ref{fig:SM_Y_rbg} for the rest of the caption.  }
\label{fig:SM_Y_rbg2}
\end{figure}

\clearpage
\section{3D generalization of the honeycomb code}
\label{sec:3D_HH}

In this Appendix, we use the set of reversible condensation transitions from Sec.~\ref{sec:3DCC_measurement_transition} to generalize the honeycomb code to three dimensions. In this generalization, we require that instantaneous topological codes are versions of a 3D toric code. All the toric codes are child codes of the same parent color code appearing in Table.~\ref{fig:3d2}. 

\begin{figure}[h] \centering
 \centering
 \includegraphics[width=1 \columnwidth]{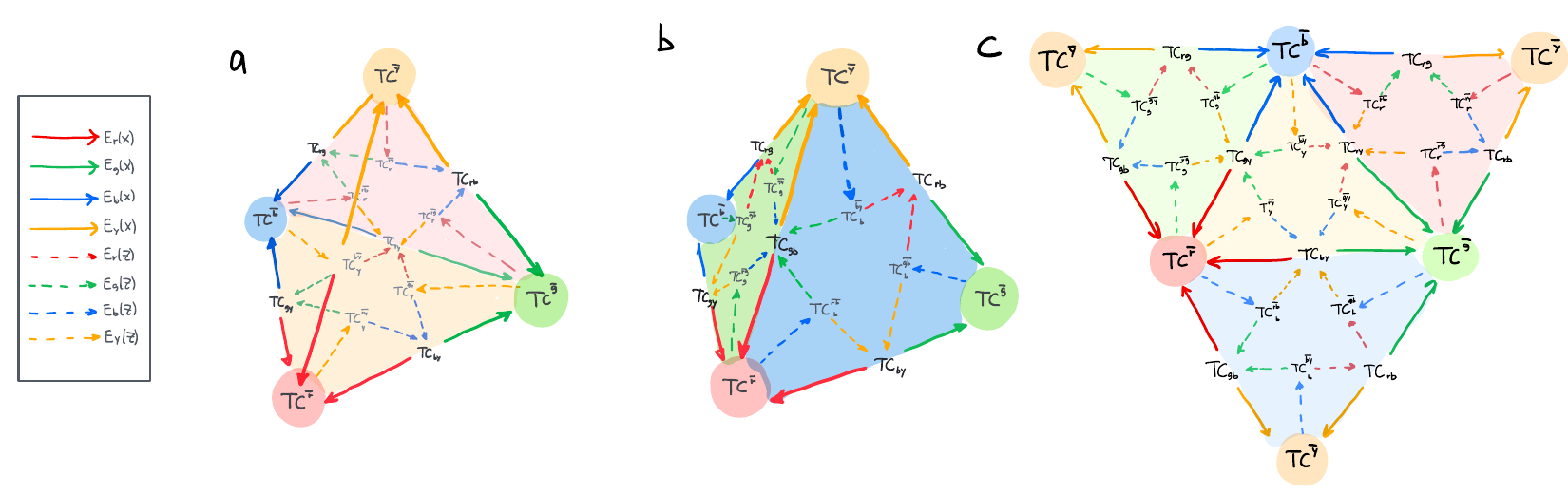}
 \caption{The space of condensation paths for a three-dimensional generalization of the honeycomb code, whose instantaneous stabilizer groups can be considered as vertices of a tetrahedron. In particular, these instantaneous stabilizer groups were defined in Sec.~\ref{sec:3DCC_measurement_transition}, and are all different realizations of the 3D toric code that can be obtained by condensation from a parent 3D color code. Panel (a) shows the red and yellow (bottom and back) faces of the tetrahedron and panel (b) shows the green and blue (left and right) faces. The solid lines correspond to $E_c(X)$-type measurements, and the dashed lines to $E_c(Z)$ ones. (c) Planar representation of the tetrahedron. Technically, each arrow corresponds to a reversible measurement path, in which case the direction of the arrow shows the directions which have to be followed to construct weight-2 measurement sequences only. If one wishes to use reverse directions, in some cases those would require performing plaquette stabilizer measurements. } 
 \label{fig:3DHH1}
\end{figure}

In our generalization, there are four toric codes, $\TC^{\bar c}$ with $c = \{ r,g,b,y\}$, that are obtained by condensing fluxes only. The condensation graph, shown in Fig.~\ref{fig:3DHH1}, has a shape of a tetrahedron with $\TC^{\bar c}$ at its corners. Each closed cycle corresponds to a sequence of reversible condensations that implements the trivial automorphism. 

\begin{figure}[h] \centering
 \centering
 \includegraphics[width=1 \columnwidth]{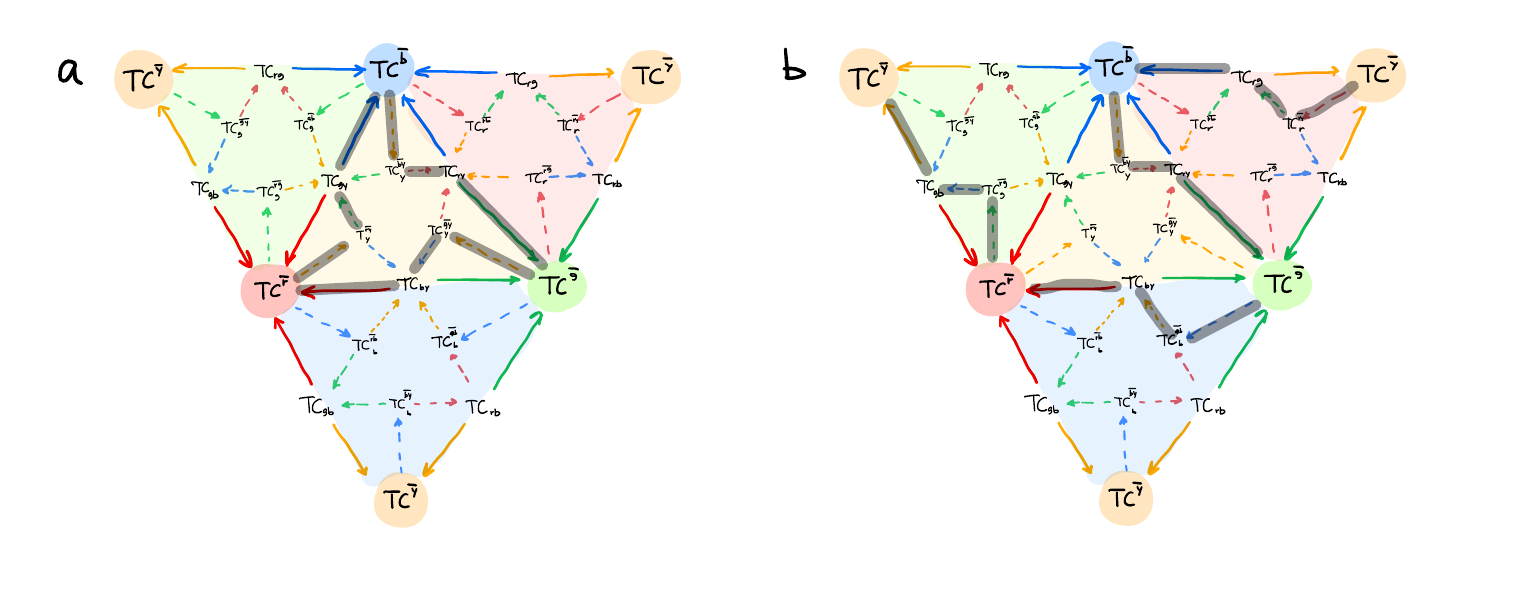}
 \caption{(a) Length-9 and (b) length-12 measurement sequences defining different realizations of the 3D honeycomb code protocol. In both sequences, it takes 6 rounds to generate the full topological stabilizer group. }
 \label{fig:3DHH2}
\end{figure}

\clearpage
\section{Alternative sequence for the 3D dynamic automorphism color code with trivial automorphism}
\label{sec:app_3DCC_full}

In this Appendix we present a measurement sequence which infers all stabilizers of the effective 3D color code.
This measurement schedule consists of four sets of six measurements with a total measurement period of 24.
It uses only two kinds of the toric code unfoldings and realizes a trivial automorphism using two-body measurements only while assembling a full effective color code ISG.
The relevant part of the condensation graph takes the following form,
\begin{equation}
\begin{tikzcd}
\widetilde{\CC } \arrow[rr, "\scalebox{1}{$\begin{array}{c}
E_{c_1} (X_1) \\ E_{c_2}(X_2)\\ E_{c_3}(X_3)\end{array}$} "] & & 
TC^{\overline{c_1}} \boxtimes TC^{\overline{c_2}} \boxtimes TC^{\overline{c_3}} 
\arrow[ll, "V(Z_1Z_2) {,} V(Z_2Z_3)", bend left] \arrow[rr, "\scalebox{1}{$\begin{array}{c} 
E_{c_1'} (Z_1) \\ E_{c_2'}(Z_2) \\ E_{c_3'}(Z_3) \end{array}$}"] 
&& \arrow[ll, "\scalebox{1}{$\begin{array}{c}E_{c_1} (X_1) \\ E_{c_2}(X_2)\\ E_{c_3}(X_3)\end{array}$}", bend left=32] 
TC^{\overline{c_1 c_1'}}_{c_1'} \boxtimes TC^{\overline{c_2 c_2'}}_{c_2'} \boxtimes TC^{\overline{c_3 c_3'}}_{c_3'}\\
&&&&\\
&&&&\\
\CC^3\arrow[uuu, dashrightarrow,"\scalebox{1}{$\begin{array}{c}
V(Z_1Z_2) \\ V(Z_2Z_3) \end{array}$} "] 
&&
\CC^3
\arrow[uuu, dashrightarrow, "\scalebox{1}{$\begin{array}{c}
E_{c_1} (X_1) \\ E_{c_2}(X_2)\\ E_{c_3}(X_3)\end{array}$} "']
&&\CC^3
\arrow[uuu, dashrightarrow,"\scalebox{1}{$\begin{array}{c}
E_{c_1'} (Z_1), P_{\overline{c_1c_1'}}(X_1) \\ E_{c_2'}(Z_2), P_{\overline{c_2c_2'}}(X_2)\\ E_{c_3'}(Z_3), P_{\overline{c_3c_3'}}(X_3)\end{array}$} "'].
\end{tikzcd}
\end{equation} 
We see that the horizontal arrows are conjugate measurements, and the path of measurements which we take simply traverses back and forth. 
Each of the six-step measurement sequences will infer one of the four volume stabilizers of the effective color code. 
The remaining plaquette stabilizers are measured more frequently.

Each of the four six-step measurement sequences look very similar and measures many of the same plaquette operators; each set of six measurements are the same as the previous six measurements along with a cyclic permutation of colors $\cdots \rightarrow \rc{r} \rightarrow \gc{g} \rightarrow \yc{y} \rightarrow \bc{b} \rightarrow \rc{r}\rightarrow \cdots$.
The key difference between each set of six measurements is that each one measures a new type of volume stabilizer, and since there are four types of volume stabilizers in the effective color code, our schedule has four sets of six measurements.

The full six step sequence from $\widetilde{\CC}$ to $\widetilde{\CC}$ takes the form
{
\renewcommand{\rc}[1]{{\color{red}{#1}}}
\renewcommand{\bc}[1]{{\color{blue}{#1}}}
\renewcommand{\gc}[1]{{\color{ForestGreen}{#1}}}
\renewcommand{\yc}[1]{{\color{amber}{#1}}}
\begin{align}
\label{eq:sixstep}
\widetilde{\CC}
\xrightarrow{
\scalebox{.65}{$
\begin{array}{c} \Er(X_1)\\ \Eg(X_2) \\ \Ey(X_3) \end{array}
$}
}
\begin{array}{c} \TC^{\overline{\rc r}}\\ \TC^{\overline{\gc g}} \\ \TC^{\overline{\yc y}} \end{array}
\xrightarrow{\scalebox{.65}{$
\begin{array}{c} \Ey(Z_1)\\ \Eb(Z_2) \\ \Er(Z_3) \end{array}
$}
}
\begin{array}{c} TC^{\overline{\rc r \yc y}}_{\yc y} \\ TC^{\overline{\gc g \bc b}}_{\bc b} \\ \TC^{\overline{\yc y \rc r}}_{\rc r} \end{array}
\xrightarrow{
\scalebox{.65}{$
\begin{array}{c} \Er(X_1)\\ \Eg(X_2) \\ \Ey(X_3) \end{array}
$}
}
\begin{array}{c} \TC^{\overline{\rc r}}\\ \TC^{\overline{\gc g}} \\ \TC^{\overline{\yc y}} \end{array}
\xrightarrow{
\scalebox{.65}{$
\begin{array}{c} \Eb(Z_1)\\\Er(Z_2) \\ \Eg(Z_3) \end{array}
$}
}
\begin{array}{c} TC^{\overline{\rc r \bc b}}_{\bc b} \\ \TC^{\overline{{\gc g} {\rc r}}}_{\rc r} \\TC^{\overline{\yc y \gc g}}_{\gc g} \end{array}
\xrightarrow{
\scalebox{.65}{$
\begin{array}{c} \Er(X_1)\\ \Eg(X_2) \\ \Ey(X_3) \end{array}
$}
}
\begin{array}{c} \TC^{\overline{\rc r}}\\ \TC^{\overline{\gc g}} \\ \TC^{\overline{\yc y}} \end{array}
\xrightarrow{
\scalebox{.65}{$
\begin{array}{c} V(Z_1Z_2)\\V(Z_2Z_3) \end{array}
$}
}
\widetilde{\CC}.
\end{align}
}
Note we have abused notation and left out the $\mathcal{S}(*)$ for each instantaneous stabilizer group.
Denote the sequence of measurements within layer $(1)$ by
\begin{align}\xrightarrow{C_{{\rc r} {\yc y} {\rc r} {\bc b}{\rc r}}^{(1)} } = \xrightarrow{\Er(X_1)} 
 \xrightarrow{\Ey(Z_1)}\xrightarrow{\Er(X_1)} \xrightarrow{\Eb(Z_1)}
\xrightarrow{\Er(X_1)}.
\end{align}
So that the entire sequence of measurements appearing in Eq.~\eqref{eq:sixstep} can be succinctly summarized as,

\begin{align}
\label{eq:one_step_of3DFCC}
\mathcal{S}(\widetilde{\CC}) \xrightarrow{C_{{\rc r}{\yc y} {\rc r} {\bc b}{\rc r} }^{(1)} \boxtimes C_{{\gc g} {\bc b} {\gc g}{\rc r}{\gc g} }^{(2)} \boxtimes C_{{\yc y}{\rc r} {\yc y}{\gc g} {\yc y}}^{(3)}} \mathcal{S}(\TC^{\overline{\rc r}} \boxtimes \TC^{\overline{\gc g}} \boxtimes \TC^{\overline{\yc y}} ) \xrightarrow{V{Z_1Z_2}, V{Z_2 Z_3}} \mathcal{S}(\widetilde{\CC}).
\end{align}
We use the $\boxtimes$ to denote the tensor product of measurement circuits.

With this notation at hand we can describe the full 24 step measurement schedule,
{
\renewcommand{\rc}[1]{{\color{red}{#1}}}
\renewcommand{\bc}[1]{{\color{blue}{#1}}}
\renewcommand{\gc}[1]{{\color{ForestGreen}{#1}}}
\renewcommand{\yc}[1]{{\color{amber}{#1}}}
\begin{equation}
\begin{tikzcd}
\cdots \arrow[rrr,"V(Z_1Z_2){,} V(Z_2Z_3)"]&&&\mathcal{S}(\widetilde{\CC}) \arrow[rrrr,"C_{{\rc r}{\yc y}{\rc r}{\bc b}{\rc r}}^{(1)} 
\boxtimes C_{{\gc g}{\bc b}{\gc g}{\rc r}{\gc g}}^{(2)} 
\boxtimes C_{{\yc y}{\rc r}{\yc y}{\gc g}{\yc y}}^{(3)}"] && \arrow[dd, phantom, ""{coordinate, name=Z},yshift = 0.15cm] && \mathcal{S}(\TC^{\overline{\rc r}} \boxtimes \TC^{\overline{\gc g}} \boxtimes \TC^{\overline{\yc y}} ) \arrow[ddllll,"V(Z_1Z_2){,}V(Z_2Z_3)", rounded corners,
to path={-- ([xshift=2ex]\tikztostart.east)|- (Z) [pos =.25]\tikztonodes-| ([xshift=-2ex]\tikztotarget.west)-- (\tikztotarget)}] &&&\\
&&&&&&&&&&\\
&&&\mathcal{S}(\widetilde{\CC}) \arrow[rrrr,"C_{{\gc g}{\bc b}{\gc g}{\rc r}{\gc g}}^{(1)} 
\boxtimes C_{{\yc y}{\rc r}{\yc y}{\gc g}{\yc y}}^{(2)} 
\boxtimes C_{{\bc b}{\gc g}{\bc b}{\yc y}{\bc b}}^{(3)}"] && \arrow[dd, phantom, ""{coordinate, name=ZZ},yshift = 0.15cm] && 
\mathcal{S}(\TC^{\overline{\gc g}} \boxtimes \TC^{\overline{\yc y}} \boxtimes \TC^{\overline{\bc b}})\arrow[ddllll,"V(Z_1Z_2){,}V(Z_2Z_3)",rounded corners,to path={ -- ([xshift=2ex]\tikztostart.east)|- (ZZ) [pos =.25]\tikztonodes-| ([xshift=-2ex]\tikztotarget.west)-- (\tikztotarget)}]\\
&&&&&&&&&&\\
&&&\mathcal{S}(\widetilde{\CC}) \arrow[rrrr,"C_{{\yc y}{\rc r}{\yc y}{\gc g}{\yc y}}^{(1)} 
\boxtimes C_{{\bc b}{\gc g}{\bc b}{\yc y}{\bc b}}^{(2)} 
\boxtimes C_{{\rc r}{\yc y}{\rc r}{\bc b}{\rc r}}^{(3)}"] && \arrow[dd, phantom, ""{coordinate, name=ZZZ},yshift = 0.15cm] && 
\mathcal{S}(\TC^{\overline{\yc y}} \boxtimes \TC^{\overline{\bc b}} \boxtimes \TC^{\overline{\rc r}} )\arrow[ddllll,"V(Z_1Z_2){,}V(Z_2Z_3)",rounded corners,to path={ -- ([xshift=2ex]\tikztostart.east)|- (ZZZ) [pos =.25]\tikztonodes-| ([xshift=-2ex]\tikztotarget.west)-- (\tikztotarget)}]\\
&&&&&&&&&&\\
&&&\mathcal{S}(\widetilde{\CC}) \arrow[rrrr,"C_{{\bc b}{\gc g}{\bc b}{\yc y}{\bc b}}^{(1)} 
\boxtimes C_{{\rc r}{\yc y}{\rc r}{\bc b}{\rc r}}^{(2)} 
\boxtimes C_{{\gc g}{\bc b}{\gc g}{\rc r}{\gc g}}^{(3)}"] && {} && 
\mathcal{S}(\TC^{\overline{\bc b}} \boxtimes \TC^{\overline{\rc r}} \boxtimes \TC^{\overline{\gc g}} )\arrow[rrr,"V(Z_1Z_2){,} V(Z_2Z_3)"]&&&{\cdots}\\
\end{tikzcd}
\end{equation}
}
The same pattern of single-layer measurements appears multiple times, only with layer indices changing.
In Tables~\ref{table:fullISG_App},\ref{table:fullISG_App1} we compute the first 24 ISGs in this measurement protocol and show all stabilizers are measured.

\begin{table}[t]
\renewcommand{\rc}[1]{{\color{red}{#1}}}
\renewcommand{\bc}[1]{{\color{blue}{#1}}}
\renewcommand{\gc}[1]{{\color{ForestGreen}{#1}}}
\renewcommand{\yc}[1]{{\color{amber}{#1}}}
\centering
\begin{tabular}{ c | c | l | l }
Time &\text{Measurement} & ISG plaquettes & ISG volumes\\
\hline
1&$\Er(X_1)$& & \\
&$\Eg(X_2)$& & \\
&$\Ey(X_3)$& & \\ 
\hline 
2&$\Ey(Z_1)$&$\Pbarry(X_1)$& \\
&$\Eb(Z_2)$&$\Pbargb(X_2)$& \\
&$\Er(Z_3)$&$\Pbarry(X_3)$& \\ 
\hline 
3&$\Er(X_1)$& $\Pbarry(Z_1)$&\\
&$\Eg(X_2)$& $\Pbargb(Z_2)$&\\
&$\Ey(X_3)$& $\Pbarry(Z_3)$&\\ 
\hline 
4&$\Eb(Z_1)$& $\Pbarrb(X_1), \Pbarry(Z_1)$ & \\
&$\Er(Z_2)$& $\Pbarrg(X_2), \Pbargb(Z_2)$ & \\
&$\Eg(Z_3)$& $\Pbargy(X_3), \Pbarry(Z_3)$ & \\ 
\hline 
5&$\Er(X_1)$& $\Pbarrb(Z_1), \Pbarry(Z_1)$ & \\
&$\Eg(X_2)$& $\Pbarrg(Z_2), \Pbargb(Z_2)$ & \\
&$\Ey(X_3)$& $\Pbargy(Z_3), \Pbarry(Z_3)$ & \\
\hline
6&$V(Z_1Z_2)$&
$\Pbarrg(Z_1),\Pbarrb(Z_1), \Pbarry(Z_1)$
& $\volb(X_1 X_2 X_3)$\\
&$V(Z_2Z_3)$&$\Pbargb(Z_1),\Pbargy(Z_1)$&\\
&&&\\
\hline
7 &$\Eg(X_1)$& $\Pbarrg(Z_1), \Pbargb(Z_1),\Pbargy(Z_1)$& $\volb(X_1 X_2 X_3)$\\
&$\Ey(X_2)$& $\Pbarry(Z_2),\Pbargy(Z_2),\Pbarby(Z_2Z_3)$& \\
&$\Eb(X_3)$& $\Pbarrb(Z_3),\Pbargb(Z_3) $& \\ 
\hline 
8 &$\Eb(Z_1)$&$\Pbargb(X_1), \Pbarrg(Z_1),\Pbargy(Z_1)$& $\volb(X_2 X_3)$\\
&$\Er(Z_2)$&$\Pbarry(X_2),\Pbargy(Z_2),\Pbarby(Z_2Z_3)$& \\
&$\Eg(Z_3)$&$\Pbargb(X_3), \Pbarrb(Z_3) $& \\ 
\hline 
9&$\Eg(X_1)$&$\Pbargb(Z_1), \Pbarrg(Z_1),\Pbargy(Z_1)$& $\volb(X_2 X_3)$\\
&$\Ey(X_2)$&$\Pbarry(Z_2), \Pbargy(Z_2),\Pbarby(Z_2Z_3)$& \\
&$\Eb(X_3)$&$\Pbargb(Z_3),\Pbarrb(Z_3)$& \\ 
\hline 
10 &$\Er(Z_1)$&$\Pbarrg(X_1),\Pbargb(Z_1),\Pbargy(Z_1)$ &$\volb(X_2 X_3)$\\
&$\Eg(Z_2)$&$\Pbargy(X_2),\Pbarry(Z_2),\Pbarby(Z_2)$ &\\
&$\Ey(Z_3)$&$\Pbarby(X_3),\Pbargb(Z_3),\Pbarrb(Z_3)$ &\\ 
\hline 
11 &$\Eg(X_1)$&$\Pbarrg(Z_1),\Pbargb(Z_1),\Pbargy(Z_1)$ &$\volb(X_2 X_3)$\\
&$\Ey(X_2)$&$\Pbargy(Z_2),\Pbarry(Z_2),\Pbarby(Z_2)$ &\\
&$\Eb(X_3)$& $\Pbarby(Z_3),\Pbargb(Z_3),\Pbarrb(Z_3)$&\\
\hline
12 &$V(Z_1Z_2)$&$\Pbarrg(Z_1),\Pbarrb(Z_1),\Pbarry(Z_1)$ &$\volr(X_1 X_2 X_3)$\\
&$V(Z_2Z_3)$&$\Pbargb(Z_1),\Pbargy(Z_1)$&$ \volb(X_1 X_2 X_3)$ \\
&&$\Pbarby(Z_1)$&\\
&&&
\end{tabular}
\caption{\label{table:fullISG_App}
The ISG for the first cycle of measurements for the first 12 rounds of the 24-round 3D DA color code protocol (see next table for the rest of the protocol).  
We only include stabilizers that are not generated by the most recent measurements.
We guide the eye by choosing bases that are localized to their respective layer. 
This isn't always possible, such as in steps 7, 8, and 9 where $\Pbarby(Z_2Z_3)$ appears. 
Similarly, we separate the plaquette stabilizers and the volume stabilizers in the middle column and rightmost column respectively.
To guide the eye, we separate the measurement steps with a horizontal line and enumerate the measurement step on the left.
There are four blocks of measurements in the circuit, (1-6), (7-12), (13-18), and (19-24). 
Each block determines one new volume stabilizer. 
The full set of plaquette stabilizers is inferred by step 12.
One can check that each stabilizer is inferred at least once per measurement period.
}
\end{table}

\begin{table}[t]
\renewcommand{\rc}[1]{{\color{red}{#1}}}
\renewcommand{\bc}[1]{{\color{blue}{#1}}}
\renewcommand{\gc}[1]{{\color{ForestGreen}{#1}}}
\renewcommand{\yc}[1]{{\color{amber}{#1}}}
\centering
\begin{tabular}{ c | c | l | l }
Time &\text{Measurement} & ISG plaquettes & ISG volumes\\
\hline
13&$\Ey(X_1)$& $\Pbarry(Z_1), \Pbargy(Z_1), \Pbarby(Z_1)$&$\volr(X_1 X_2 X_3)$ \\
&$\Eb(X_2)$& $\Pbarrb(Z_2), \Pbargb(Z_2), \Pbarby(Z_2)$&$\volb(X_1 X_2 X_3)$ \\
&$\Er(X_3)$& $\Pbarrg(Z_3), \Pbarrb(Z_3), \Pbarry(Z_3)$& \\ 
\hline 
14&$\Er(Z_1)$&$\Pbarry(X_1), \Pbargy(Z_1), \Pbarby(Z_1)$&$\volr(X_2 X_3)$ \\
&$\Eg(Z_2)$&$\Pbarrb(Z_2), \Pbargb(X_2), \Pbarby(Z_2)$&$\volb(X_1 X_2 X_3)$ \\
&$\Ey(Z_3)$&$\Pbarrg(Z_3), \Pbarrb(Z_3), \Pbarry(X_3)$& \\ 
\hline 
15&$\Ey(X_1)$& $\Pbarry(Z_1), \Pbargy(Z_1), \Pbarby(Z_1)$&$\volr(X_2 X_3)$\\
&$\Eb(X_2)$& $\Pbarrb(Z_2), \Pbargb(Z_2), \Pbarby(Z_2)$&$\volb(X_1 X_2 X_3)$\\
&$\Er(X_3)$& $\Pbarrg(Z_3), \Pbarrb(Z_3), \Pbarry(Z_3)$&\\ 
\hline 
16&$\Eg(Z_1)$& $\Pbarry(Z_1), \Pbargy(X_1), \Pbarby(Z_1)$ &$\volr(X_2 X_3)$ \\
&$\Ey(Z_2)$& $\Pbarrb(Z_2), \Pbargb(Z_2), \Pbarby(X_2)$ & $\volb(X_1 X_2)$\\
&$\Eb(Z_3)$& $\Pbarrg(Z_3), \Pbarrb(X_3), \Pbarry(Z_3)$ & \\ 
\hline 
17&$\Ey(X_1)$& $\Pbarry(Z_1), \Pbargy(Z_1), \Pbarby(Z_1)$ &$\volr(X_2 X_3)$ \\
&$\Eb(X_2)$& $\Pbarrb(Z_2), \Pbargb(Z_2), \Pbarby(Z_2)$ &$\volb(X_1 X_2)$ \\
&$\Er(X_3)$& $\Pbarrg(Z_3), \Pbarrb(Z_3), \Pbarry(Z_3)$ & \\
\hline
18&$V(Z_1Z_2)$&
$\Pbarrg(Z_1),\Pbarrb(Z_1),\Pbarry(Z_1)$ 
& $\volr(X_1 X_2 X_3)$\\
&$V(Z_2Z_3)$&$\Pbargb(Z_1),\Pbargy(Z_1)$& $\volg(X_1 X_2X_3)$\\
&&$\Pbarby(Z_1)$& $\volb(X_1 X_2X_3)$\\
\hline
19 &$\Eb(X_1)$& $\Pbarrb(Z_1), \Pbargb(Z_1), \Pbarby(Z_1)$& $\volr(X_1 X_2 X_3)$\\
&$\Er(X_2)$& $\Pbarrg(Z_2), \Pbarrb(Z_2), \Pbarry(Z_2)$& $\volg(X_1 X_2X_3)$ \\
&$\Eg(X_3)$& $\Pbarrg(Z_3), \Pbargb(Z_3),\Pbargy(Z_3)$& $\volb(X_1 X_2X_3)$ \\ 
\hline 
20 &$\Eg(Z_1)$&$\Pbarrb(Z_1), \Pbargb(X_1), \Pbarby(Z_1)$& $\volr(X_1 X_2 X_3)$\\
&$\Ey(Z_2)$&$\Pbarrg(Z_2), \Pbarrb(Z_2), \Pbarry(X_2)$&$\volg(X_2X_3)$ \\
&$\Eb(Z_3)$&$\Pbarrg(Z_3), \Pbargb(X_3),\Pbargy(Z_3)$& $\volb(X_1 X_2)$ \\ 
\hline 
21&$\Eb(X_1)$&$\Pbarrb(Z_1), \Pbargb(Z_1), \Pbarby(Z_1)$ & $\volr(X_1 X_2 X_3)$\\
&$\Er(X_2)$&$\Pbarrg(Z_2), \Pbarrb(Z_2), \Pbarry(Z_2)$ & $\volg(X_2X_3)$\\
&$\Eg(X_3)$&$\Pbarrg(Z_3), \Pbargb(Z_3),\Pbargy(Z_3)$& $\volb(X_1 X_2)$ \\ 
\hline 
22 &$\Ey(Z_1)$&$\Pbarrb(Z_1), \Pbargb(Z_1), \Pbarby(X_1)$ &$\volr(X_1 X_2 )$\\
&$\Eb(Z_2)$&$\Pbarrg(Z_2), \Pbarrb(X_2), \Pbarry(Z_2)$ &$\volg(X_2X_3)$\\
&$\Er(Z_3)$&$\Pbarrg(X_3), \Pbargb(Z_3),\Pbargy(Z_3)$ &$\volb(X_1 )$\\ 
\hline 
23 &$\Eb(X_1)$&$\Pbarrb(Z_1), \Pbargb(Z_1), \Pbarby(Z_1)$ &$\volr(X_1 X_2 )$\\
&$\Er(X_2)$&$\Pbarrg(Z_2), \Pbarrb(Z_2), \Pbarry(Z_2)$ &$\volg(X_2X_3)$\\
&$\Eg(X_3)$& $\Pbarrg(Z_3), \Pbargb(Z_3),\Pbargy(Z_3)$&$\volb(X_1 )$\\
\hline
24 &$V(Z_1Z_2)$&$\Pbarrg(Z_1),\Pbarrb(Z_1),\Pbarry(Z_1)$ &$\volr(X_1 X_2 X_3)$\\
&$V(Z_2Z_3)$&$\Pbargb(Z_1),\Pbargy(Z_1)$& $\volg(X_1 X_2X_3)$\\
&&$\Pbarby(Z_1)$& $\volb(X_1 X_2X_3)$\\
&&& $\voly(X_1 X_2X_3)$\\
\end{tabular}
\caption{\label{table:fullISG_App1}
The ISG for the first cycle of measurements for the rounds 13 through 24 of the 24-round 3D DA color code protocol.
}
\end{table}

\end{document}